\documentclass[final]{ukthesis}
\usepackage[utf8x]{inputenc}	
\usepackage[T1]{fontenc}
\usepackage[colorlinks,bookmarks]{hyperref}
\hypersetup{linkcolor=blue,
            citecolor=blue,
            urlcolor=blue,
            hypertexnames=false,
            breaklinks=true,
            pdfauthor={Firstname Lastname},
            pdftitle={PhD thesis - Title Of The Thesis},
            pdfsubject={Dissertation for PhD},
            pdfkeywords={condensed matter physics},
            pdfproducer={pdfLatex with hyperref},
            pdfcreator={pdflatex},
            pdfpagemode=UseOutlines,
            bookmarksopen=true,
            bookmarksnumbered=true}
\pdfcatalog{/PageMode (/UseOutlines)}
\usepackage{memhfixc}
\usepackage{multirow}

\usepackage[pdftex]{graphicx} 	
\graphicspath{{graphics/}}		

\usepackage{subfig}
\usepackage{lineno}

\usepackage{longtable} 		
\usepackage{threeparttable} 	

\usepackage{amsmath}
\usepackage{amssymb}
\usepackage{amsthm}
\usepackage{mathrsfs}

\usepackage{textcomp}
\usepackage{sidecap}			
\usepackage{url}
\usepackage{tipa}
\usepackage{rotating}

\usepackage{upquote}
						
\usepackage{capt-of}		

\maxsecnumdepth{subsection}
\changetocdepth{2}

\usepackage{lipsum}	

\usepackage{braket}

\usepackage[section]{placeins}

\usepackage{afterpage}

\usepackage{setspace}

\usepackage{bigstrut}
\usepackage{gensymb}
\usepackage{epstopdf}

\usepackage{titlesec}
\titleformat{\part}[display]{\centering\normalfont\bfseries}{\MakeUppercase{\partname} \thepart}{20pt}{\normalfont\bfseries}

\makeatletter
\renewcommand{\counterwithin}{\@ifstar{\@csinstar}{\@csin}}
\makeatother

\makeindex

     {\begin{list}{}%
             {\setlength{\leftmargin}{#1}}%
             \item[]%
     }
     {\end{list}}
\begin{document}

\author{James Kevin Adkins}
\title{STUDYING TRANSVERSE MOMENTUM DEPENDENT DISTRIBUTIONS IN POLARIZED PROTON COLLISIONS VIA AZIMUTHAL SINGLE SPIN ASYMMETRIES OF CHARGED PIONS IN JETS}
\abstract{A complete, fundamental understanding of the proton must include knowledge of the underlying spin structure. The transversity distribution, $h_1\left(x\right)$, which describes the transverse spin structure of quarks inside of a transversely polarized proton, is only accessible through channels that couple $h_1 \left(x\right)$ to another chiral odd distribution, such as the Collins fragmentation function ($\Delta^N D_{\pi/q^\uparrow}\left(z,j_T\right)$). Significant Collins asymmetries of charged pions have been observed in semi-inclusive deep inelastic scattering (SIDIS) data. These SIDIS asymmetries combined with $e^+e^-$ process asymmetries have allowed for the extraction of $h_1\left(x\right)$ and $\Delta^N D_{\pi/q^\uparrow}\left(z,j_T\right)$. However, the current uncertainties on $h_1\left(x\right)$ are large compared to the corresponding quark momentum and helicity distributions and reflect the limited statistics and kinematic reach of the available data. In transversely polarized hadronic collisions, Collins asymmetries may be isolated and extracted by measuring the spin dependent azimuthal distributions of charged pions in jets. This thesis will report on the first statistically significant Collins asymmetries extracted from $\sqrt{s}=200$ GeV hadronic collisions using $14$ $pb^{-1}$ of transversely polarized proton collisions at 57\% average polarization.}

\advisor{Renee Fatemi}
\dgs{Christopher Crawford}

\frontmatter
\maketitle

\begin{dedication}
For my grandparents:\\
Junior and Nell Adkins\\
Oakley and Bell Huff\\
\end{dedication}

\begin{acknowledgments}
I am most thankful for my small family. To my wife, Stacy, thank you for your love, support, and encouragement throughout my graduate school career, I would have never finished without you. To my daughter, Avery June, thank you for keeping me on my toes and teaching me about the most important things in life. Finally, to my favorite fuzzy dog, Maxwell, thank you for your companionship during many long nights of coding.

I feel very lucky to have the unconditional support of my mom and brother always behind me. To my mom, Gwenda Huff, thank you for pushing me to dream big and convincing me I can reach those dreams. To my brother, Dr. Josh Adkins, thank you for always encouraging me and understanding the problems of a graduate student.

During my time as a graduate student, I have had excellent guidance from my advisor, Dr. Renee Fatemi. Thank you, Renee, for your patience and commitment to my education. Developing this analysis has been very exciting and rewarding, thank you for giving me the opportunity to work on it.

I have received a lot of research and service task support from many people within the STAR collaboration. To Dr. Carl Gagliardi and Dr. James Drachenberg, thank you for being very generous with your time and sharing your analysis wisdom. To Dr. Zilong Chang, thank you for working with me on multiple projects, and generously sharing your coding expertise. To Suvarna Ramachandran, Dr. Will Jacobs, Dr. Scott Wissink, Dr. Jerome Lauret, and Dr. Gene Van Buren, thank you for your input and assistance during my time as the BEMC Software Coordinator.

Throughout my life, I have had many instructors that have been instrumental to my success. To Michael Flannery, thank you for developing my interest and giving me a solid foundation in mathematics at Elliott County High School. To Dr. Capp Yess, Dr. Jennifer Birriel, Dr. Ignacio Birriel, and Dr. Kent Price at Morehead State University, thank you for giving me an excellent foundation in physics and preparing me to seamlessly transition into graduate school, and for your continued friendship.

I am thankful to have an excellent group of friends who offer support outside of research. To Wade Conn and Brandon Adkins, thank you for the time we have spent on the water and in the woods, hunting and fishing was always a welcome distraction. To Mike Brown, thank you for our lunchtime debates, time on the golf course, and burning the midnight oil with me way back when we spent all of our time on homework problems.
\end{acknowledgments}
 
\tableofcontents\clearpage
\listoffigures\clearpage
\listoftables\clearpage


\newcommand{\figuremacro}[3]{
	\begin{figure}[htbp]
		\centering
		\includegraphics[width=1\textwidth]{#1}
		\caption[#2]{\textbf{#2} - #3}
		\label{#1}
	\end{figure}
}

\newcommand{\figuremacroW}[4]{
	\begin{figure}[htbp]
		\centering
		\includegraphics[width=#4\textwidth]{#1}
		\caption[#2]{\textbf{#2} - #3}
		\label{#1}
	\end{figure}
}

\newcommand{\figuremacroN}[3]{
	\begin{wrapfigure}{o}{0.5\textwidth}
		\centering
		\includegraphics[width=0.48\textwidth]{#1}
		\caption[#2]{{\small\textbf{#2} - #3}}
		\label{#1}
	\end{wrapfigure}
}

\newcommand{\PdfPsText}[2]{
  \ifpdf
     #1
  \else
     #2
  \fi
}

\newcommand{\IncludeGraphicsH}[3]{
  \PdfPsText{\includegraphics[height=#2]{#1}}{\includegraphics[bb = #3, height=#2]{#1}}
}

\newcommand{\IncludeGraphicsW}[3]{
  \PdfPsText{\includegraphics[width=#2]{#1}}{\includegraphics[bb = #3, width=#2]{#1}}
}

\newcommand{\InsertFig}[3]{
  \begin{figure}[!htbp]
    \begin{center}
      \leavevmode
      #1
      \caption{#2}
      \label{#3}
    \end{center}
  \end{figure}
}


\mainmatter

\begin{DoubleSpace}



\chapter{Motivation and Theoretical Background}

\ifpdf
    \graphicspath{{ch1_Introduction/figures/PNG/}{ch1_Introduction/figures/PDF/}{ch1_Introduction/figures/}}
\else
    \graphicspath{{ch1_Introduction/figures/EPS/}{ch1_Introduction/figures/}}
\fi


It is simply stunning to realize that all visible matter shares a common denominator. Multiple identical copies of protons, neutrons and electrons make up not only the human body, but the planet we inhabit and the food and water we must have to sustain our existence.

The study of subatomic particles has come a long way since the discovery that protons and neutrons, or nucleons, are composite particles \cite{ref:protonMag1,ref:protonMag2}. Initial experiments focused on discovering and understanding the nature of the nucleon's constituents, now commonly known as quarks and gluons, collectively called partons. The theory which describes the interactions of quarks and gluons is called quantum chromodynamics (QCD). In an effort to gain a better understanding of QCD, decades of experiments have been performed to study fundamental quantities such as the partonic charge, momentum and spin distributions inside the proton.

The spin structure of the proton's constituents is described using polarized parton distribution functions, which give the number density of partons that are aligned versus anti-aligned in a longitudinally or transversely polarized proton. Studying the proton's spin structure is one way to directly test QCD, and has been under investigation by multiple experiments for the better part of three decades. The main focus of this thesis will be to investigate the transverse spin structure of quarks inside of transversely polarized protons, which is currently much less understood than the helicity structure of partons.

This chapter is designed to motivate the experimental analysis through discussion of the theoretical concepts and experimental history. Section \ref{sec:spin} gives a brief introduction to spin before discussing the standard model in Section \ref{sec:standardmodel}, and how to experimentally study the proton's properties in Section \ref{sec:properties}. Section \ref{sec:ffandpdf} outlines the parton distribution and fragmentation functions used to describe the proton's structure and quark fragmentation, and we finally wrap up with a review of current theoretical and experimental results on transverse spin structure in Section \ref{sec:probingspin}.

\section{Wait, What is Spin?}
\label{sec:spin}
Before delving into a discussion about the spin structure of the proton, spin must be properly introduced. What is it? 

The first thing to understand is that the name is a misnomer. One should not associate the quantum mechanical spin of particles with the spinning of an object about an axis like a top or bullet shot from a rifle. Rather, one should think of quantum mechanical spin as the form of angular momentum a particle possesses when prepared in a state of zero linear momentum, meaning its wave function has no dependence upon spatial coordinates. In this state, acting on the wave function with any component of the angular momentum operator $\vec{L}$ will return zero, but experimentally measuring the angular momentum returns a nonzero result \cite{ref:Shankar}. Thus, spin is an intrinsic angular momentum.

The value of a particle's spin is a fixed fundamental quantity just like rest mass or charge. All particles carry either integer or half integer values of intrinsic spin, allowing them to be categorized as either bosons or fermions respectively. 

\section{The Standard Model}
\label{sec:standardmodel}
The theory that describes the elementary building blocks of matter, as well as the forces that govern them, is called the standard model. The particles may be arranged neatly into a small table shown in Figure \ref{StandardModel}. This depiction shows the spin-$1/2$ fermions which includes both the quarks and the leptons, the spin-1 force carrying gauge bosons, and the spin-0 Higgs boson.

\figuremacroW{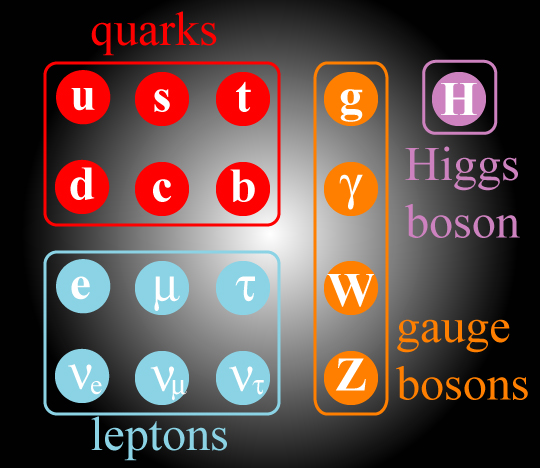}{Fundamental Particles in the Standard Model}{The fundamental elementary particles arranged nicely into a table. The fermion generations (columns) are in order of discovery and increasing mass from left to right.}{0.5}

\subsection{Quarks and Leptons}
Quarks form the foundation for a class of particles called hadrons, which are made up of three valence quarks (baryons) or a quark plus an antiquark (mesons). There are six so-called flavors, or types, of quarks that are grouped into three generations, and each flavor carries a fractional electric charge of $+\frac{2}{3}$ or $-\frac{1}{3}$. The first generation is up and down quarks, the second is charm and strange quarks, and the third generation is top and bottom quarks.

Since all quarks are fermions, the possibility of a three quark hadron immediately presents an issue: how do particles made of the same three ground state quarks not violate the Pauli exclusion principle? The answer is a new quantum number called color, and this is what makes quarks unique since they are the only fermions to carry a color ``charge''. Red, green, and blue are the arbitrarily used names for quark color, and each quark may assume any color. While the introduction of color explains the existence of hadrons that would otherwise violate the exclusion principle, it adds the restriction that quarks cannot be isolated in nature, since every particle we see or detect carries no net color charge. It is very clear then why known hadrons are subdivided into baryons (red, green and blue in equal parts is colorless) and mesons (a color and its anticolor combine to make colorless). Unfortunately, this consequence means a quark can never be plucked from a hadron and studied as a free particle, since they are confined to the hadron. 

Unlike quarks, leptons are colorless particles. Rather than forming composite particles, leptons combine with nucleons to form atoms, such as the electron combining with a proton to form hydrogen. There are also three generations of leptons each containing a charged lepton and a neutral neutrino, paralleling the three quark generations. The first generation contains the electron and electron neutrino, the second contains the muon and muon neutrino, and the third generation has the tau and tau neutrino. As in the quark sector, the electron, muon, and tau mass increase with each generation. The discussion of neutrino mass is complicated by the fact that their lepton flavor states are not their mass states. The experimental determination of neutrino masses is the focus of active research and currently we can only place upper limits of $\sim1$ $eV/c^2$. 

It is interesting to note that the first generation particles include the up and down quarks, and the electron. The valence quark structure of the proton and neutron, which in turn form the nucleus of all of the elements, is formed exclusively from up and down quarks. When the nucleus is combined with electrons, all elements of the periodic table may be formed. Thus, the first generation particles of the standard model form the skeleton of all visible matter in the universe, and particles from higher generations will eventually decay back to the first generation.

\subsection{Fundamental Forces and Gauge Bosons}
There are four fundamental forces of nature: gravitational, electromagnetic, weak, and strong. At this time the gravitational force is not included in the standard model. However, this fact has no bearing on the research in this thesis because the masses of the particles are so small that gravitational effects are negligible and it is completely reasonable to ignore them as we move forward. Standard model particles do feel the effects of the remaining three forces as outlined in Figure \ref{StandardModelForces}. The effect of these forces is carried, or mediated, by the force carrying gauge bosons shown in Figure \ref{StandardModel}.

\figuremacroW{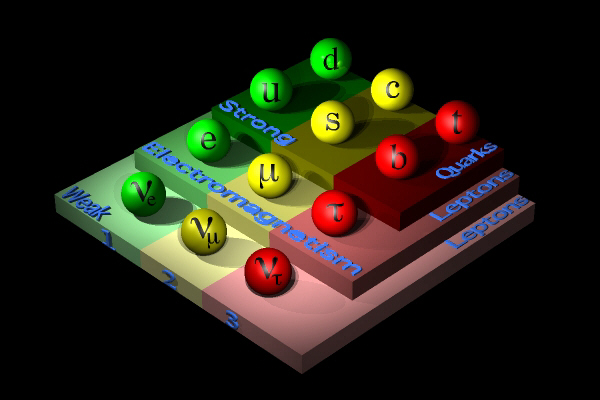}{Standard Model Forces}{The three standard model forces, and which particles experience the force. Particles on each level feel the force on that level, and the force of each level under it. For instance, quarks are subject to the effects of the weak, electromagnetic and strong forces.}{0.75}

The weak force is, as one may suspect, the weakest force in the standard model with a strength given by the weak coupling constant\footnote{The weak, electromagnetic, and strong coupling constants all scale with the energy of the mediating gauge boson. Therefore, the quoted size of the coupling constants should be used qualitatively to understand the relative strength of the forces, and one should keep in mind these numbers are not constant!} $\alpha_W \approx 10^{-6} - 10^{-7}$. The weak force is the only force which is mediated by multiple massive particles ($\sim80-90$ $GeV/c^2$), through charged and neutral current channels, and the only force that is experienced by neutrinos. Charged current interactions occur when there is an exchange of a $W^+$ or $W^-$, and the flavor of the involved particles changes. For instance, the interaction $\mu^- + \nu_e \rightarrow e^- + \nu_\mu$ is a charged current interaction where the flavor of the muon and the electron neutrino are changed in the final state into a muon neutrino and electron. Flavor changing interactions also occur in the quark sector.  A classic example is beta decay ($n \rightarrow p + e^- + \bar{\nu}_e$) which occurs when a down quark in a neutron transforms into an up quark by emitting a $W^-$ boson which then decays into an electron and an anti-electron neutrino. Neutral current interactions are mediated by the $Z^0$ boson where no charge is exchanged, and particle flavor stays the same. A simple example of this kind of interaction would be lepton-neutrino scattering such as $\mu^- + \nu_e \rightarrow \mu^- + \nu_e$.

The photon is the mediator of the electromagnetic force, which describe interactions between particles that carry electric charge. The electromagnetic force is at the heart of quantum electrodynamics which describes the electromagnetic interaction between particles, such as the repulsion between two electrons or the scattering of electrons from protons. The strength of the electromagnetic force is the second strongest force in the standard model, and is given by the fine structure constant $\alpha_{EM}\approx1/137$. While it is not responsible for forming quark bound states, the electromagnetic attraction between electrons and protons is the reason that electrons stay within the cloud surrounding a nucleus.

The strong force is, in fact, the strongest force in the standard model, bearing the responsibility of binding together quarks to form hadrons. This force is experienced by both quarks and its mediator, the gluon. This is in contrast to photons, which do not experience the electromagnetic force because they do not carry electric charge. The gluon mediator is a massless particle that, unlike the other gauge bosons, also carries a color factor. In fact, rather than three colors, the gluons carry a color and an anticolor and are arranged into a color octet leading to eight total gluon states in the standard model \cite{ref:GriffithsParticle}. Being that they carry color, singular gluons do not exist in nature, although combinations of gluons, or glueballs, can be colorless and therefore may exist in nature.

The strong coupling constant is a function of the energy of the exchanged gluon in a strong interaction \cite{ref:SkandsQCD}, but within the nucleon the coupling constant is of the order $\alpha_S \approx 1$. The methods of perturbative quantum chromodynamics, which use an expansion in orders of $\alpha_S$, do not converge when the coupling constant is this large. Therefore, it is not possible to gain a fundamental quantitative understanding of what is happening inside the proton, for instance, via perturbative QCD calculations alone.

Gluons that are constantly being exchanged between quarks may pair produce inside of hadrons, such as in the proton, contributing to the overall quark content. In addition to the valence quarks which are always present---two up quarks and one down quark for a proton---there is the so-called ``quark sea'' that consists of quarks and antiquarks originating from this pair production. The exchange of gluons also gives rise to motion of quarks inside of a hadron, as they recoil from emitting and accepting gluons. This gives rise to intrinsic quark transverse momentum, $k_T$, in hadrons which are traveling at relativistic speeds, such as in a collider. At these relativistic speeds the quarks are traveling longitudinally in a collinear way along the direction of motion and each carrying a fraction, $x$, of the momentum of the hadron. 

\section{Studying Properties of Partons}
\label{sec:properties}
If quarks and gluons---collectively referred to as partons---may not exist in nature outside of hadrons, how is it possible to study their properties? While color confinement means a single parton cannot be isolated and investigated, annihilation and scattering experiments at particle accelerators make it possible to study the properties of partons. 

Annihilation experiments take advantage of the fact that both leptons and quarks feel the effects of the electromagnetic force. These collide oppositely charged leptons---such as a positron and an electron since they are very stable and long lived---at a center of mass energy high enough to produce a quark-antiquark pair rapidly flying away from each other. These are initially ``free'' particles, but feel the overwhelming effect of the strong force once they reach a separation distance of $\sim1\times 10^{-15}$ m---roughly the radius of the proton. The further they fly, the more energy is stored in their interaction field and it becomes energetically favorable for a gluon to generate a new quark-antiquark pair that will minimize the energy stored in the field \cite{ref:GriffithsParticle}. Now there are two sets of two quarks flying away from each other and the process repeats itself until there are multiple quarks flying in opposite directions from each other. This generation of quark-antiquark pairs is known as fragmentation. Quarks cannot exist as free particles so they begin to combine into various colorless hadrons, a process called hadronization.

Scattering experiments work similarly, but rather than generating a quark-antiquark pair, a quark inside of a hadron is struck by either an incoming lepton or another hadron. In this case, the two particles collide with enough energy to break apart the target hadron. So a struck parton is given enough energy to overcome the hadronic boundary and fly off as a free particle. The fragmentation and hadronization processes follow the same progression as with the annihilation experiment. 

In both cases ``jets'' of particles are detected in the final state along the direction of the primordial parton. A jet is a collection of final state particles which spray out in a cone-like shape from the initial collision vertex and serve as a proxy for the initial parton that created it \cite{ref:AliJets}. Therefore, analyzing jets reconstructed from the detected particles allows for extraction of initial state parton properties. Jets will play a central role in the asymmetry analysis presented in this thesis.

\section{PDFs and FFs: How Do the Protons Constituents Behave?}
\label{sec:ffandpdf}
The analysis in this thesis is geared towards gaining a more fundamental understanding of partonic behavior in the proton that is not theoretically calculable using standard perturbative QCD methods. These topics include the internal structure and dynamics of the proton's constituents, as well as the quark fragmentation process and final state hadron distributions. While any hadronic target could be used to study these concepts in QCD, protons are a natural choice because they are abundant and very stable. Therefore, from this point on this thesis will focus on proton distribution functions and proton collisions.

As noted above, processes involving proton-proton collisions cannot be fully described using the tools of perturbative QCD. The cross section for the process may be factorized into two parts: perturbative and non-perturbative. These concepts are explored in Figure \ref{ProtonXsec}, where a final state pion is produced from an initial collision of two protons. The perturbative and non-perturbative effects can be further investigated by looking at the cross section given by:

\begin{equation}
\label{eq:piprod}
d\sigma^{pp\rightarrow \pi X} = \sum_{f_1,f_2,f}\int dx_1 dx_2 dz f_1^p\left(x_1\right) f_2^p\left(x_2\right) \times d\hat{\sigma}^{f_1 f_2 \rightarrow fX^\prime}\left(\hat{s},\hat{t},\hat{u}\right) D_f^\pi\left(z\right)
\end{equation}

\noindent where the indices 1 and 2 represent the different incoming protons in the initial collision. In this expression, $f_1^p\left(x_1\right)$ and $f_2^p\left(x_2\right)$ are the parton distribution functions (PDFs) which give the probability density of selecting a parton with flavors $f_1$ and $f_2$ from the initial state protons. Each parton carries a fraction $x_1$ and $x_2$, respectively, of the proton's momentum. $D_f^\pi\left(z\right)$ is the fragmentation function (FF), the probability of finding a pion in the final state carrying momentum fraction $z$ from the parton $f$. Both of these concepts encompass the non-perturbative effects in the reaction. The only perturbatively calculable part of the reaction is the ``hard'' partonic cross section, $\hat{\sigma}^{f_1 f_2 \rightarrow fX^\prime}$, between initial state partons $f_1$ and $f_2$ that produces final state parton $f$ that fragments into the pion, and some partons $X^\prime$ which are ignored for this process. For this example the PDFs and FF are unpolarized, however polarized functions do exist and will be discussed later in this section.

\figuremacroW{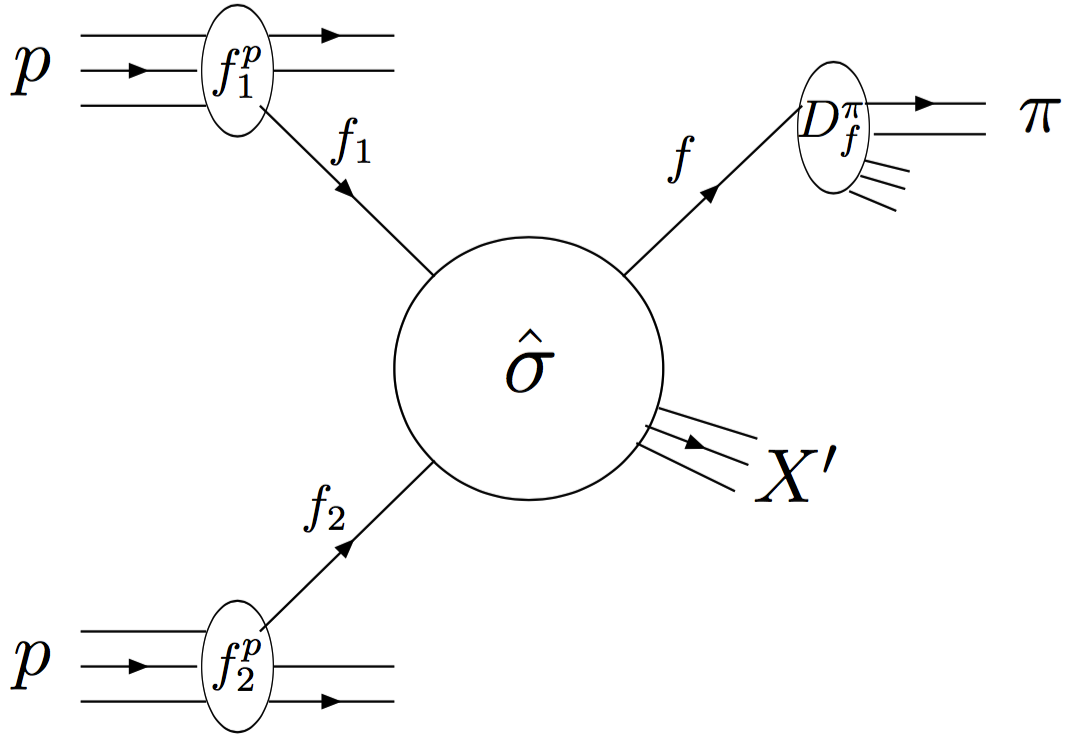}{$\pi$ Production in $pp$ Collisions}{This cartoon depicts the separation of the non-perturbative functions used to describe the partonic initial and final state, and the perturbative function used to describe the interaction when a pion is produced in a proton-proton collision \cite{ref:BunceProtonXsec}.}{0.65}

Looking at Equation \ref{eq:piprod}, it is straightforward to understand the broad definition of factorization, meaning the perturbative and non-perturbative effects of a complicated process are separated from each other mathematically. The hard scattering cross section can be calculated using perturbative QCD, however the PDFs and FFs must be extracted from data results. It is important to note that in this example the partons and pion in Figure \ref{ProtonXsec} carry no transverse momentum, therefore Equation 
\ref{eq:piprod} has no transverse momentum dependence. This is called a collinear factorization scheme, where all of the initial state partons and final state hadrons only carry longitudinal momentum, and transverse momenta are integrated over. Collinear factorization schemes work very well for describing data from collisions where the protons are either unpolarized or polarized longitudinally. However, when the protons are transversely polarized to their direction of motion, collinear factorization schemes fail to theoretically reproduce the very large single spin asymmetries (i.e. only one polarized proton in the collision) like those found in Figure \ref{TransAsym_Historical}. The significant single spin asymmetries show almost no dependence on the center-of-mass energy, $\sqrt{s}$, across a broad range of energies, but were predicted to be nearly zero using collinear factorization. These results, where theory fails to reproduce the data, demonstrate the importance of spin observables in testing the robustness of any QCD framework.

\figuremacroW{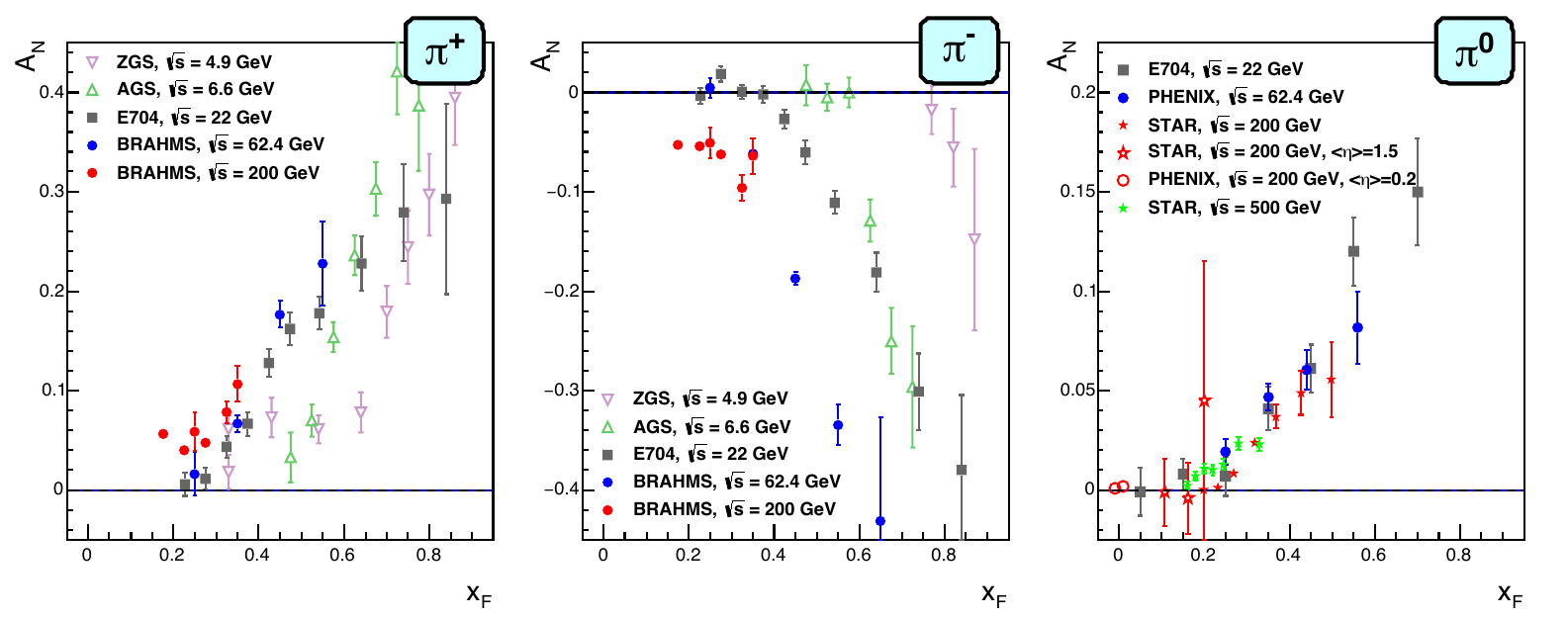}{Transverse Single Spin Asymmetries}{Inclusive pion single spin asymmetries collected from various experiments and plotted as a function of Feynman-$x$ \cite{ref:RHIC_QCDplan}.}{1}

There are currently two different approaches to describing transverse single spin asymmetries measured in collisions where one proton is polarized transverse to the direction of motion. These approaches depend upon the twist of the process, or how many vertices are considered in the Feynman diagram. Leading twist, or twist-2, results from a single interaction between two partons whereas twist-3 arises when one of these partons interacts with the proton remnant which adds another interaction vertex to the diagram \cite{ref:PitonyakThesis}. The higher twist and transverse momentum dependent factorization schemes began to be explored once the leading twist collinear scheme failed to explain the large transverse single spin asymmetries and predict a high transverse momentum ($q_T$) of final state $e^+e^-$ pairs in Drell-Yan experiments. In an effort to describe these non-perturbative effects, three types of functions have been explored:

\begin{itemize}
\item Twist-2 collinear PDFs
\item Twist-2 transverse momentum dependent (TMD) PDFs and FFs which depend upon the initial state parton $k_T$. This is more commonly called the TMD approach.
\item Twist-3 collinear correlators
\end{itemize}

The leading twist collinear factorization framework is sufficient to describe many observables. For instance, the twist-2 factorization is useful as a leading order approximation to fully inclusive deep inelastic scattering or $p+p$ collisions where only jets are detected in the final state and fragmentation is not characterized. These PDFs may be recovered by $k_T$ integration of the TMD PDFs.

The TMD factorization approach is sought in processes where there is a produced hadron in the final state, such as in semi-inclusive deep inelastic scattering (SIDIS) or in $p+p$ collisions where final state hadrons are recorded. The hadrons, for the most part, carry a transverse momentum ($P_\perp^h$) which is connected to the intrinsic $k_T$ of the primordial quark. Using this momentum, the TMD approach is said to be valid when $\Lambda_{QCD} \leq P_\perp^h \ll Q$ (with $\Lambda_{QCD} \approx 0.2$ GeV), where Q is the momentum transfer and thus sets the hard scattering scale.

The twist-3 approach does not require two momentum scales, only the hard scale, thus this factorization approach is used when $\Lambda_{QCD} \ll Q$. This means there is an overlap region where both the TMD and twist-3 factorization methods are equally valid, which occurs when $\Lambda_{QCD} \ll P_\perp^h \ll Q$ \cite{ref:JiMomentumScales}.

\subsection{Twist-2 Collinear}
This approach ignores the possibility of internal transverse momentum of the fragmenting parton, sometimes it is referred to as $k_T$-integrated. To characterize the structure of the proton in this approach, there are three PDFs. These are the unpolarized distribution $f$$\left(x\right)$, the helicity distribution $\Delta f$$\left(x\right)$, and the transversity distribution $h_1$$\left(x\right)$. Though not explicit in the definitions, each of these distributions are functions of the hard scale $Q^2$ in addition to their $x$ dependence. Each of these functions have a very straightforward interpretation:

\begin{itemize}
\item $f$$\left(x\right)$ describes the number density of partons in an unpolarized proton with flavor $f$ that carry a fraction $x$ of the proton's momentum.
\item $\Delta f$$\left(x\right)$ describes the number density of partons in a longitudinally polarized proton with flavor $f$ that carry a fraction $x$ of the proton's momentum with spin aligned with the proton's spin minus the number with spin anti-aligned.
\item $h_1$$\left(x\right)$ describes the transverse spin structure of quarks in a transversely polarized proton.
\end{itemize}
 
Note here that $h_1$$\left(x\right)$ only applies to quarks in a proton. This is because transversity is ``chiral-odd'', meaning it is the imaginary part of a forward Compton scattering amplitude which is an off diagonal matrix element in a helicity basis \cite{ref:Barone}, which is the most common choice of basis since the helicity operator commutes with the free Dirac hamiltonian. Taken in the limit of massless quarks, this amplitude is associated with a helicity flip of both the parton and proton. Being spin-1, a gluon would not satisfy conservation of helicity coming from the spin-1/2 proton, and thus gluons in the proton (or any spin-1/2 hadron) carry no transversity. 

\subsection{Twist-2 TMDs}
Not integrating over the intrinsic $k_T$ gives a more realistic picture of the proton, since all quarks are constantly receiving transverse kicks to their momentum from exchanging gluons. But the additional degree of freedom must now be characterized. Thus in the TMD picture of the proton, there are eight PDFs to describe the internal structure, outlined in Table \ref{table:tmdPDFs}.

\begin{table}[htp]
\begin{center}
\begin{tabular}{|c | c|}
\hline
PDF & Name \\
\hline
$f_1$$\left(x,k_T\right)$ & Unpolarized density\\
\hline
$g_{1L}$$\left(x,k_T\right)$ & Helicity \\
\hline
$h_{1T}$$\left(x,k_T\right)$ & Transversity \\
\hline
$f_{1T}^\perp$$\left(x,k_T\right)$ & Sivers \\
\hline
$h_1^\perp$$\left(x,k_T\right)$ & Boer-Mulders \\
\hline
$h_{1T}^\perp$$\left(x,k_T\right)$ & Pretzelocity \\ 
\hline
$h_{1L}^\perp$$\left(x,k_T\right)$ & Worm gear \\
\hline
$g_{1T}$$\left(x,k_T\right)$ & Worm gear \\
\hline
\end{tabular}
\end{center}
\caption{The eight leading twist TMD PDFs}
\label{table:tmdPDFs}
\end{table}%

Still working at leading twist, the distributions in Table \ref{table:tmdPDFs} may again be interpreted as number densities. The first three distributions have the same definition as the twist-2 collinear unpolarized, helicity, and transversity distributions, only with an additional $k_T$-dependence. The Sivers distribution describes the correlation between the spin of the proton and the $k_T$ of the quark or gluon \cite{ref:SiversFunction}. The Boer-Mulders distribution describes the relationship between the spin of a quark and its $k_T$ \cite{ref:BoerMuldersFunction}. Like the transversity distribution, the Boer-Mulders distribution does not apply to gluons. The pretzelocity distribution is a measure of the difference between helicity and transversity, and would be nonzero because relativistic boosts and rotations do not commute \cite{ref:AvakianPretzel,ref:MillerPretzel}. Finally, the so-called worm gear PDF $h_{1L}^\perp$$\left(x,k_T\right)$ describes the probability of finding a transversely polarized quark inside of a longitudinally polarized proton (and vice-versa for $g_{1T}$$\left(x,k_T\right)$) \cite{ref:HermesWormGear}. 

There are also eight leading twist TMD FFs, which are not dependent upon the intrinsic quark $k_T$, but do depend upon the final state hadron momentum that is transverse to the fragmenting quark momentum (henceforth referred to as $j_T$). Depending upon the experimental framework, $j_T$ may be defined differently. In the frame used by SIDIS, the hadron transverse momentum is defined with respect to the direction of the virtual photon. In proton collisions where a final state jet is desired, $j_T$ is defined as the transverse momentum of the hadron with respect to the jet axis.

If only fragmentation to final state $\pi^\pm$ from quarks is considered in transversely polarized proton collisions, these eight may be whittled down to two FFs \cite{ref:DAlesio}. These remaining two fragmentation functions are quite simple to understand. The first one, $D_{\pi/q}$$\left(z,j_T\right)$ describes the probability of an unpolarized quark to fragment into a pion. The second one, $\Delta^N D_{\pi/q^\uparrow}$$\left(z,j_T\right)$, called the Collins fragmentation function, describes the probability for a quark, polarized transversely with respect to its momentum, to fragment into a pion with a given value of $j_T$ \cite{ref:CollinsFF}. In both cases, the final state pion carries a fraction, $z$, of the primordial quark momentum and has a particular value of $j_T$. Thus the Collins function may be interpreted as the connection between the initial state quark spin and final state pion $j_T$, meaning the polarization direction of the primordial quark dictates the direction of the fragmenting pion.

\subsection{Twist-3 Collinear Correlators}
In the simplest terms, twist-3 interactions mean there is an additional vertex in the Feynman diagram where the struck parton simultaneously participates in more than one interaction. As an example, if the incoming parton $f_1$ in the leading twist interaction shown in Figure \ref{ProtonXsec} were to interact with its proton remnant in addition to the hard scattering, a twist-3 interaction would result. This additional interaction inserts another factor of $1/Q$ into the amplitude matrix elements, reducing the probability of these interactions when $Q$ is large. Using the twist-3 formalism as a theoretical extraction tool gives corrections to the leading twist PDFs and FFs. Unfortunately, because subleading twist, by definition, includes interactions with multiple partons, the PDFs and FFs can no longer be interpreted as single parton densities \cite{ref:PitonyakThesis}. If the subleading twist contributions are substantial then the leading twist functions represent only a partial picture of the proton's true spin structure.

\subsection{Evolution of the PDFs and FFs}
A current hot topic in QCD is how PDFs and FFs evolve with the center-of-mass energy, $\sqrt{s}$, or hard scale $Q$ \cite{ref:CollinsEvolution,ref:KangEvolution}. In a collinear framework, the parton distribution functions can be evolved to different hard scales using the Dokshitzer-Gribov-Lipatov-Altarelli-Parisi (DGLAP) evolution equations \cite{ref:dglap1,ref:dglap2,ref:dglap3}. However, this breaks down when including the transverse momentum dependence and other methods must be explored. The Solenoidal Tracker At RHIC (STAR) Collaboration is in an excellent position to offer results in different kinematic regimes by reporting Collins asymmetry results at both $\sqrt{s} = 200$ GeV and $500$ GeV. In both cases, a broad range of jet $p_T$, the $Q$ surrogate, will be investigated. The broad kinematic reach provided by the RHIC collider will offer excellent tests of the TMD evolution theoretical framework. Results from both center-of-mass energies will be reported and discussed in Chapter \ref{chapter:Results} of this thesis.

\section{Probing the Proton's Transverse Spin Structure}
\label{sec:probingspin}
Each of the TMD PDFs in Table \ref{table:tmdPDFs} contain information about partonic spin and momentum correlations in the proton. This section will introduce and discuss current results from the two effects that will be explored in this thesis: transversity and the Collins effect, and the Sivers effect. These effects can both be explored through single spin asymmetry extractions in collisions involving transversely polarized protons \cite{ref:DAlesio}.

\subsection{Transversity and the Collins Effect}
The transversity distribution has been defined previously as the function which describes the transverse spin structure of quarks in a transversely polarized proton. This distribution, as well as $f$$\left(x\right)$ and $\Delta f$$\left(x\right)$, is included in the TMD PDFs, but $k_T$ is often integrated over for the final result and the reported distribution is collinear.

Figures \ref{HERA_PDF_2015}-\ref{fig:transversity} show examples of recent results for the next-to-leading order collinear PDFs \cite{ref:HERAPDF,ref:NNPDF,ref:KangTransversity,ref:Anselmino2015}. The unpolarized distributions show very clearly that the proton is dominated by gluon and sea quark contributions when $x$ is small, and by the valence quarks when $x$ is large. The gluon distribution is clearly the largest (scaled here by $0.05$), comparable even with the valence quark distributions at large values of $x$. Although the size of the errors on the valence quark helicity PDFs are approaching those on the unpolarized distributions, the error band on the gluon helicity distribution is larger and increasing with decreasing $x$. The quark helicity distributions show that up quarks tend to align their spin with a longitudinally polarized proton, whereas down quarks like to anti-align. It is clear that compared to the unpolarized and helicity distributions the transversity distribution is quite unconstrained, even for the valence quark contributions. This is because transversity is a chiral-odd distribution that involves a spin flip of both the quark and proton, meaning it completely decouples from fully inclusive DIS where the majority of data for the other two leading twist distributions comes from.

\figuremacroW{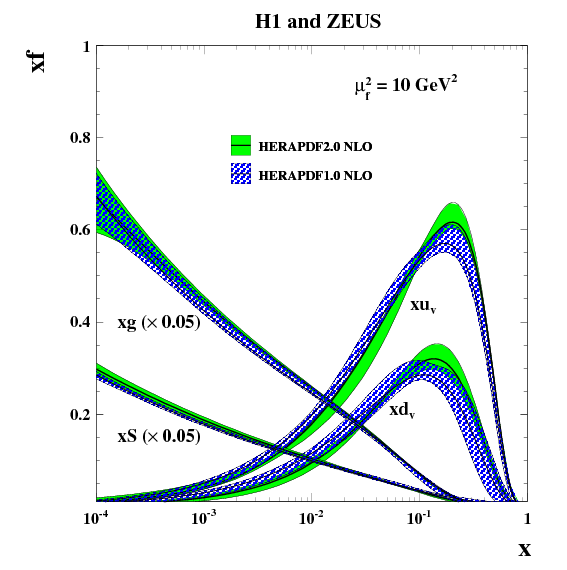}{Unpolarized PDFs}{Current results for the unpolarized parton distribution functions in the proton for up, down, gluon and quark sea contributions \cite{ref:HERAPDF}. Note that the quark sea and gluon contributions here are scaled by 0.05. This displays the important role of the quark and gluon sea at low $x$. However, above $x\approx0.03$ the valence quarks begin to dominate the proton.}{0.5}

\begin{figure}
    \centering
    \subfloat[]{{\includegraphics[width=6.5cm]{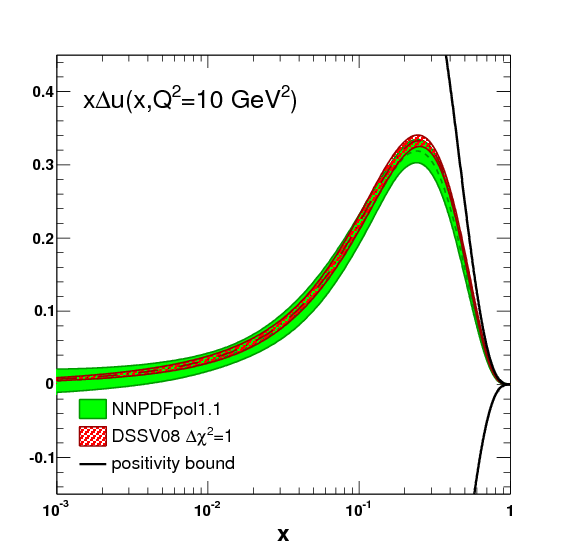} }}%
    \subfloat[]{{\includegraphics[width=6.5cm]{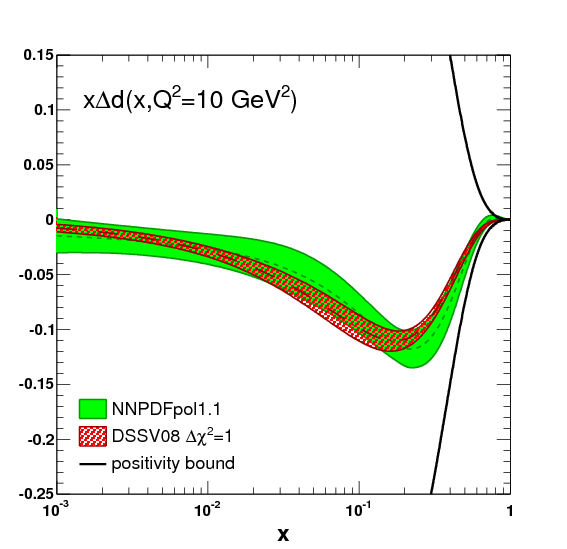} }}\\%
    \subfloat[]{{\includegraphics[width=6.5cm]{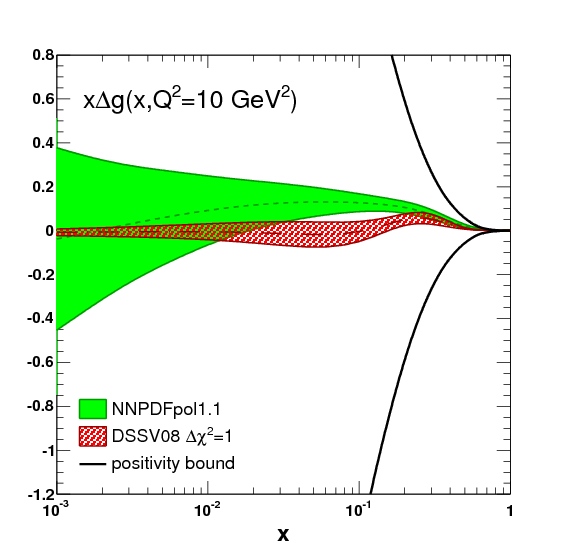} }}%
    \caption[Helicity PDFs]{\textbf{Helicity PDFs} - Current results for the helicity distributions in the proton for up quark (a), down quark (b), and gluon (c) contributions \cite{ref:NNPDF}. The red curves from de Florian, Sassot, Stratmann, and Vogelsang (DSSV) presented here are older and have been superseded by a new set of curves found in Ref. \cite{ref:DSSVupdate}.}%
    \label{fig:helicity}%
\end{figure}

The helicity flip involved with the transversity distribution is not allowed in scattering processes, meaning observable processes must be chiral-even. Therefore, the transversity distribution may only be accessed through channels that couple $h_1$$\left(x\right)$ to another chiral-odd distribution. This may be accomplished by coupling $h_1$$\left(x\right)$ to itself, as in a double spin asymmetry. Otherwise, a chiral-odd fragmentation function is sought which connects the initial and final state, and results in a chiral-even observable in the term of a single spin asymmetry $A_{UT}$. One such function is the Collins fragmentation function \cite{ref:CollinsFF}.

\begin{figure}
    \centering
    \subfloat[]{{\includegraphics[width=7.75cm]{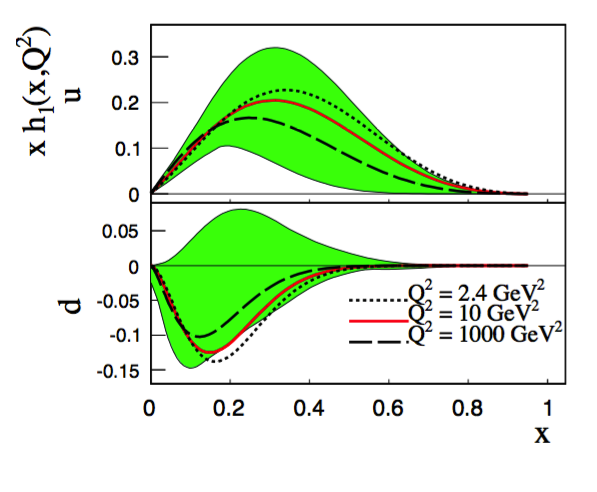} }}%
    \subfloat[]{{\includegraphics[width=6cm]{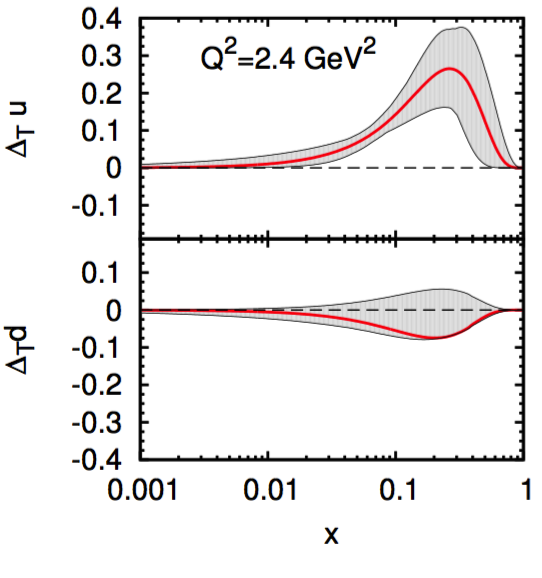} }}%
    \caption[Transversity PDFs]{\textbf{Transversity PDFs} - Current results for the transversity distributions in the proton for up and down quark contributions from (a) Kang \textit{et al.} \cite{ref:KangTransversity} and (b) Anselmino \textit{et al.} \cite{ref:Anselmino2015}.}%
    \label{fig:transversity}%
\end{figure}

As noted, the Collins FF, $\Delta^N D_{\pi/q^\uparrow}$$\left(z,j_T\right)$, describes the probability of a transversely polarized quark fragmenting to a pion which carries a certain fraction of the primordial quark momentum $z$, and a final state transverse momentum, $j_T$, defined with respect to the fragmenting quark momentum. Because of this transverse momentum dependence, the Collins function describes the azimuthal distribution of the final state pions which result from transversely polarized quarks. When $\Delta^N D_{\pi/q^\uparrow}$$\left(z,j_T\right)$ is convoluted with $h_1\left(x\right)$, the spin dependent azimuthal modulation of final state hadrons, $A_{UT}^{\sin\left(\phi\right)}$, the so-called Collins asymmetry, may be observed.

The total cross section for SIDIS contains all of the asymmetry terms that may be extracted \cite{ref:BacchettaSIDIS}. However, the full cross section contains terms that are not relevant for this discussion. Focusing only on the terms that are associated with a transversely polarized proton target (denoted by $\mathbf{S}_\perp$) and an unpolarized beam, the cross section looks like:

\begin{multline}
\label{eq:sidisxsec}
\frac{d\sigma}{dx dy d\psi dz d\phi_h dP_{h\perp}^2} \propto \lvert\mathbf{S}_\perp\rvert \Bigg[A_{UT}^{\sin\left(\phi_H-\phi_S\right)}\sin\left(\phi_H-\phi_S\right) + \varepsilon A_{UT}^{\sin\left(\phi_H+\phi_S\right)}\sin\left(\phi_H+\phi_S\right) \\
+ \varepsilon A_{UT}^{\sin\left(3\phi_H-\phi_S\right)}\sin\left(3\phi_H-\phi_S\right) + \sqrt{2\varepsilon\left(1+\varepsilon\right)} A_{UT}^{\sin\left(\phi_S\right)}\sin\left(\phi_S\right) \\ + \sqrt{2\varepsilon\left(1+\varepsilon\right)} A_{UT}^{\sin\left(2\phi_H-\phi_S\right)}\sin\left(2\phi_H-\phi_S\right) \Bigg] + ...
\end{multline}

\noindent where $\varepsilon$ is the ratio of longitudinal to transverse photon flux \cite{ref:BacchettaSIDIS}. The first two terms are the most relevant ones for the discussion in this thesis, as they have a direct dependence upon transversity and Sivers PDFs and the respective fragmentation functions. The remaining terms contain mixed dependencies on the leading twist TMD PDFs found in Table \ref{table:tmdPDFs} as well as higher twist terms. These asymmetry terms may be studied, however they will not be discussed as we move forward.

At present, SIDIS and $e^+e^-$ annihilation experiments have contributed the majority of $A_{UT}^{\sin\left(\phi\right)}$ results used in extractions of $h_1$$\left(x\right)$ and $\Delta^N D_{\pi/q^\uparrow}$$\left(z,j_T\right)$ \cite{ref:Anselmino2015}. In SIDIS, Collins asymmetries are extracted by measuring the $A_{UT}^{\sin\left(\phi_H+\phi_S\right)}$ asymmetry, which is the coefficient of the $\sin$$\left(\phi_H+\phi_S\right)$ term in Equation \ref{eq:sidisxsec}. In this formalism $\phi_S$ is the angle between the proton spin vector and the lepton scattering plane and $\phi_H$ is the angle between the lepton scattering plane and the outgoing hadron momentum vector \cite{ref:TrentoConvention}. Note that the SIDIS Collins asymmetries are a convolution of the transversity and Collins functions, which will contrast the $e^+e^-$ asymmetries that will be discussed later.
 
\begin{figure}
    \centering
    \subfloat[]{\includegraphics[trim={0 0 0 9cm},clip,width=7cm]{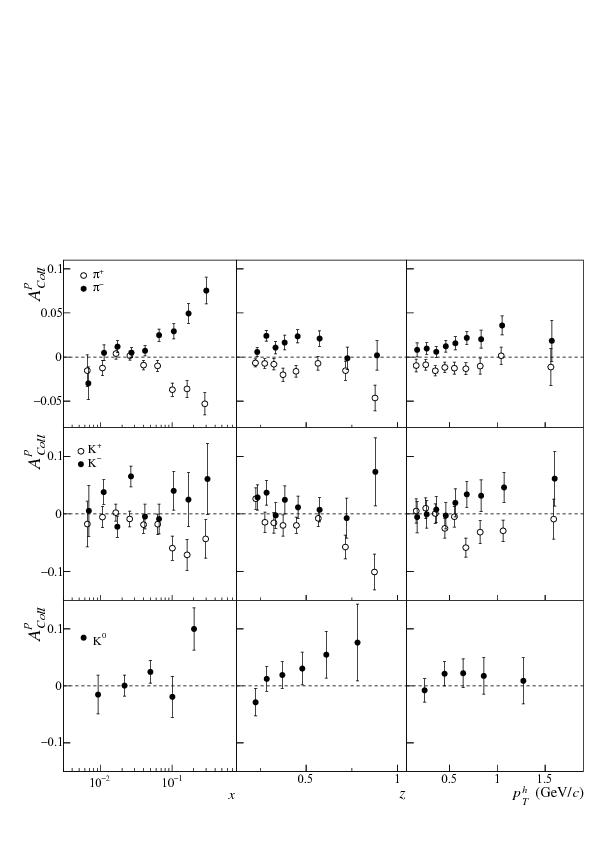} \label{SIDIS_COMPASS}} %
    \subfloat[]{\includegraphics[width=7.75cm]{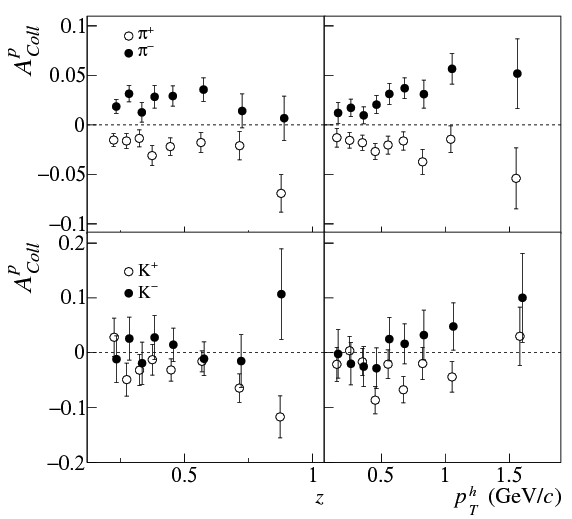} \label{SIDIS_COMPASS_X}}\\ %
    \subfloat[]{\includegraphics[width=7.25cm]{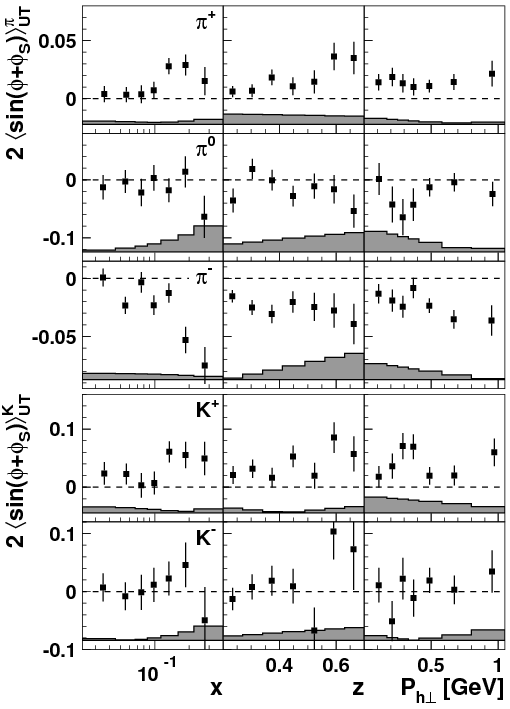} \label{SIDIS_HERMES} }%
    
    \caption[SIDIS Collins Asymmetries]{\textbf{SIDIS Collins Asymmetries} - Recent SIDIS results for the Collins asymmetry from the COMPASS \cite{ref:CompassAsymmetries} (a)/(b) and HERMES \cite{ref:HermesCollins} (c) collaborations. In addition to using outgoing charged pions, HERMES also analyzes the asymmetry for outgoing charged kaons and neutral pions. Similarly, COMPASS computes the asymmetry using outgoing charged and neutral kaons. The asymmetries in (b) show results from (a), only for the range $x>0.032$.}%
    \label{fig:CollinsSIDIS}%
\end{figure}

Figure \ref{fig:CollinsSIDIS} shows recent Collins asymmetries from proton targets from the COMPASS \cite{ref:CompassAsymmetries} and HERMES \cite{ref:HermesCollins} collaborations. In Figures \ref{SIDIS_COMPASS} and \ref{SIDIS_HERMES} the asymmetry is plotted as a function of the quark momentum fraction $x$, the hadron momentum fraction $z$, and the transverse momentum of the outgoing hadron. While the HERMES data are not as statistically significant as the COMPASS data, both results show significant asymmetries at high $x$ for the charged pion channel. The charged kaons show hints of non-zero asymmetries, but the statistical precision is not high enough to draw solid conclusions. HERMES and COMPASS find the neutral pion and kaon asymmetries, respectively, are consistent with zero. The asymmetries plotted against the other kinematic variables integrate over the whole range of $x$, including the small asymmetries at low x. In an effort to enhance the shape of the nonzero asymmetries, Figure \ref{SIDIS_COMPASS_X} shows the COMPASS asymmetries of Figure \ref{SIDIS_COMPASS} for the range $x>0.032$. In this range, the asymmetry shows no strong $z$ dependence, but does seem to be linearly increasing with $p_T^h$ for the charged pions.

In $e^+e^-$ processes, the annihilation process results in two outgoing quarks which fragment and hadronize, but their polarization is not known. Therefore asymmetries may not be extracted using the same formalism used in SIDIS, rather two hadron correlations can be measured two ways \cite{ref:Anselmino2015}:

\begin{itemize}
\item The scattering plane is formed by placing the incoming $e^+e^-$ pair in the same plane as one outgoing hadron, where the outgoing hadron momentum defines the $\hat{n}$-axis (Figure \ref{phi0_scat}). In this case, the angle $\phi_0$ is the angle between the scattering plane and other outgoing hadron momentum, measured around $\hat{n}$. Then, the extracted asymmetry, $A_0$, is the coefficient of the $\cos$$\left(2\phi_0\right)$ term in the cross section.
\item The scattering plane is formed by placing the incoming $e^+e^-$ pair and outgoing $q\bar{q}$ pair in the same plane (Figure \ref{phi12_scat}). Here, the jet thrust axis, which identifies the outgoing quark momentum direction, defines the $\hat{n}$-axis. For this method, the hadronic angles, $\phi_1$ and $\phi_2$, are measured between the scattering plane and each of the two outgoing hadron momenta about the $\hat{n}$ direction. The extracted asymmetry, $A_{12}$, is the coefficient of the $\cos$$\left(\phi_1+\phi_2\right)$ term in the cross section.
\end{itemize}

\begin{figure}
    \centering
    \subfloat[]{\includegraphics[width=7.7cm]{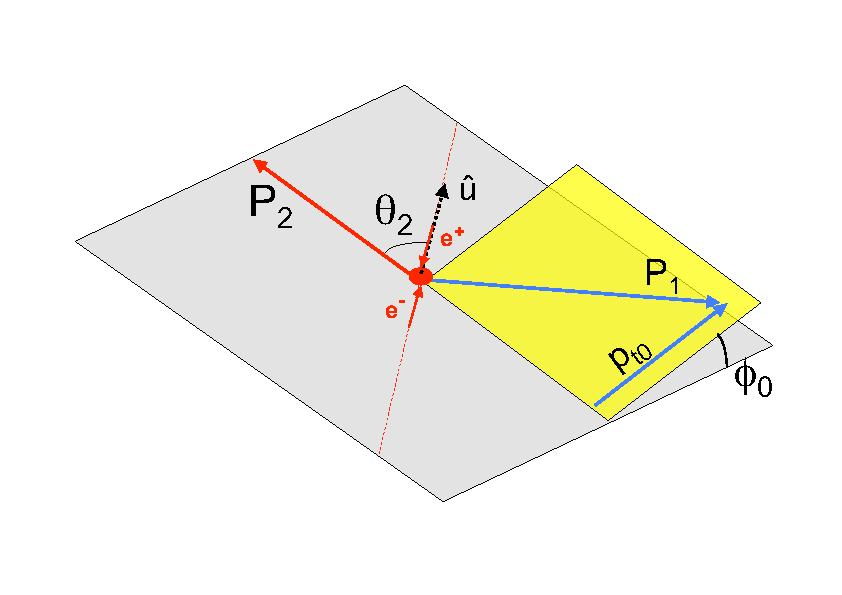} \label{phi0_scat}}%
    \subfloat[]{\includegraphics[width=7.7cm]{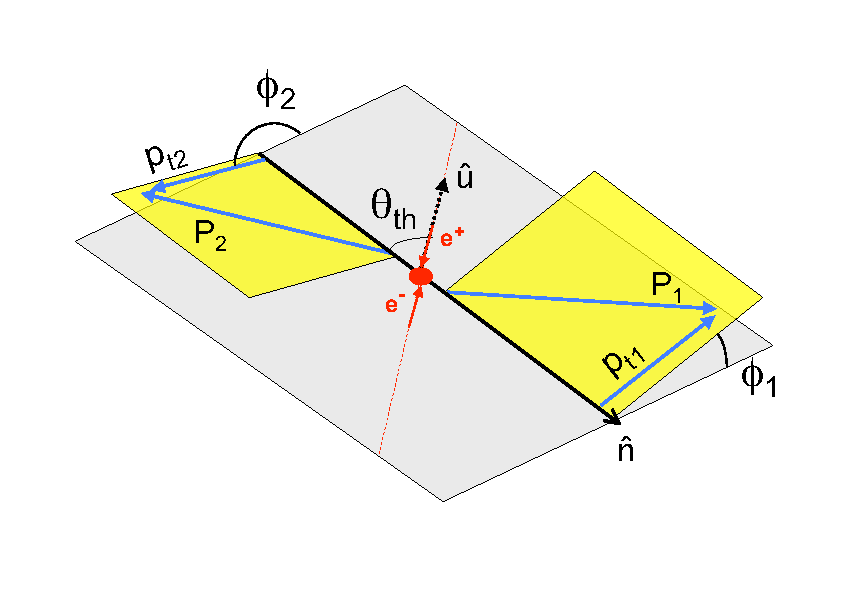} \label{phi12_scat}}%
    \caption[$e^+e^-$ Analysis Reference Frames]{\textbf{$e^+e^-$ Analysis Reference Frames} - The two scattering reference frames used in $e^+e^-$ analyses to extract the $A_0$ (a) and $A_{12}$ (b) asymmetries \cite{ref:BaBarCollins}. In both reference frames, the unit vector $\hat{u}$ points along the direction of the $e^+e^-$ beam.}
    \label{fig:AnnihilationScattering}%
\end{figure}

Figure \ref{fig:AnnihilationCollins} shows recent Collins results for identified pion pairs from the BaBar \cite{ref:BaBarCollins} and BELLE \cite{ref:BelleCollins} collaborations. These asymmetries are plotted in a two dimensional way as functions of the fractional pion momenta $z_1$ and $z_2$. The asymmetries from each experiment tell a similar tale, that is there is a stronger dependence of the asymmetry on $z_2$ as $z_1$ increases. Because the asymmetry here relies only on Collins FFs, this strong dependence on the fractional energies means that the Collins function grows in magnitude with both $z_1$ and $z_2$.

\begin{figure}
    \centering
    \subfloat[]{\includegraphics[width=7.25cm]{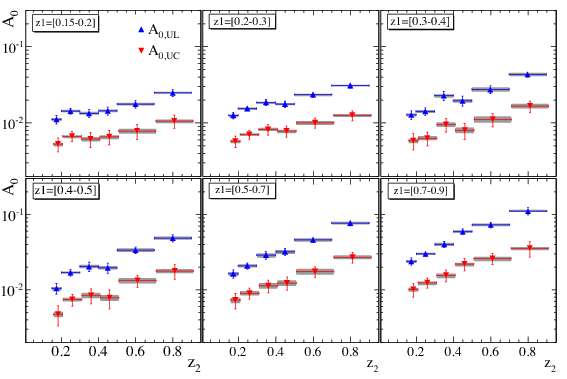} }%
    \subfloat[]{\includegraphics[width=7.25cm]{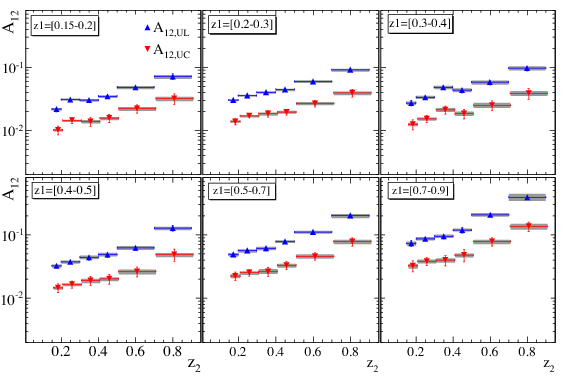} }\\%
    \subfloat[]{\includegraphics[width=7.25cm]{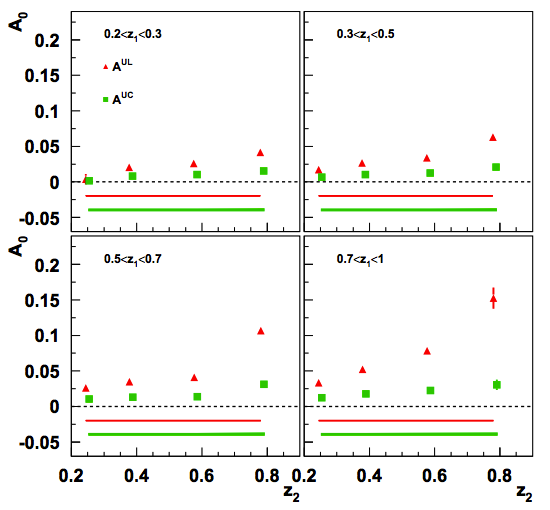} } %
    \subfloat[]{\includegraphics[width=7.25cm]{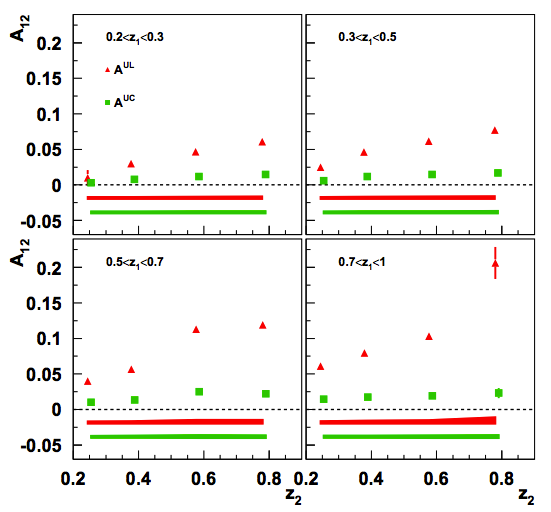} }%
    \caption[$e^+e^-$ Collins Asymmetries]{\textbf{$e^+e^-$ Collins Asymmetries} - Recent Collins results from the BaBar \cite{ref:BaBarCollins} (a)/(b) and BELLE \cite{ref:BelleCollins} (c)/(d) collaborations. Asymmetries are extracted using the two reference frames, $A_0$ and $A_{12}$.}
    \label{fig:AnnihilationCollins}%
\end{figure}

The real beauty of the $e^+e^-$ channel is that the asymmetries do not depend on quark PDFs, only fragmentation functions, which can be extracted via theoretical fits to the data. Current results on the Collins FFs using recent data from BELLE and BaBar are shown in Figure \ref{CollinsFn_Anselmino} for favored and disfavored fragmentation at two different values of $Q^2$ \cite{ref:Anselmino2015}. Favored fragmentation means the pion is formed from a fragmenting quark that appears in its valence quarks (i.e. $u\rightarrow\pi^+$), while disfavored fragmentation means the pion is formed from a fragmenting quark that does not appear in its valence quarks (i.e. $u\rightarrow\pi^-$). The disfavored fragmentation function turns out to be negative, with the interpretation that pions formed from disfavored fragmentation have the opposite azimuthal distribution to those coming from favored fragmentation.

\figuremacroW{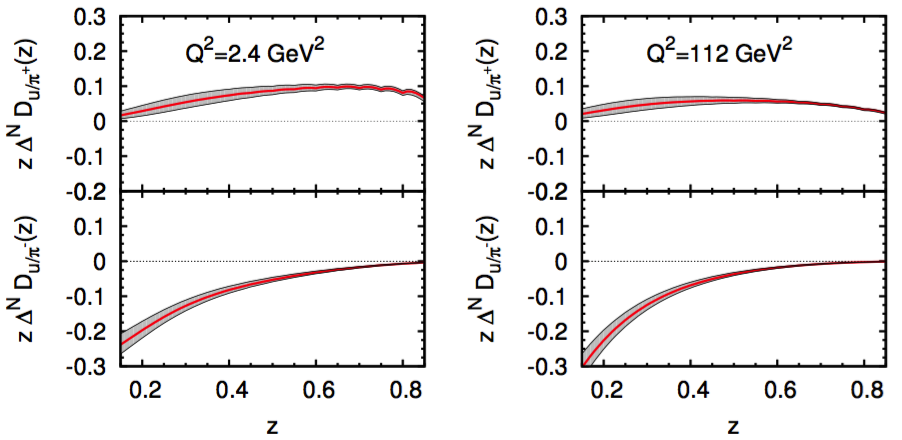}{Extracted Collins Fragmentation Function}{Current extraction results for the Collins fragmentation function for different values of $Q^2$ \cite{ref:Anselmino2015}. The top plots are for favored fragmentation and the bottom plots are for disfavored fragmentation.}{0.85} 

It is also possible to construct asymmetries in $p^\uparrow +p$ collisions that are sensitive to the convolution of the transversity and Collins functions. One example is the spin-dependent azimuthal distribution of hadrons around the axis of an associated jet. The kinematics work out differently for this channel, and the asymmetry $A_{UT}^{\sin\left(\phi\right)}$ is modulated by a $\sin\left(\phi_S-\phi_H\right)$ term \cite{ref:DAlesio} which is not the same modulation as in Equation \ref{eq:sidisxsec} for the SIDIS experiments. An exploratory analysis using the 2006 transversely polarized $p+p$ data taken at STAR revealed a statistically limited, but tantalizing asymmetry, shown in Figure \ref{StarCollinsPrelim_2006}, that seemed relatively constant as a function of the pion momentum fraction $z$. However, this result does seem to show a significant separation between the charged asymmetries for $\pi^+$ and $\pi^-$ \cite{ref:StarCollinsPrelim}. Furthermore, it seems that increased statistics could enhance the significance upon reducing the error bars. This is supported by the prediction in Figure \ref{DAlesio_Collins}, where the quark contribution to the Collins asymmetry is maximized to the Soffer inequality: $h_1\left(x\right) \leq \frac{1}{2}\Big(f\left(x\right)+\Delta f\left(x\right)\Big)$. Taking this at face value, there could be asymmetries upwards of 5\% at $p_{T,jet}$$=10$ GeV/c and at mid-rapidity ($\lvert\eta\rvert < 1$). The coverage coordinate $\eta$, or pseudorapidity, is defined as:

\begin{equation}
\eta = -\ln\left[\tan\left(\frac{\theta}{2}\right)\right]
\label{eq:rapidity}
\end{equation}

\noindent where $\theta$ is the scattering angle of the jet (see Section \ref{sec:stardetector} for further details).

\figuremacroW{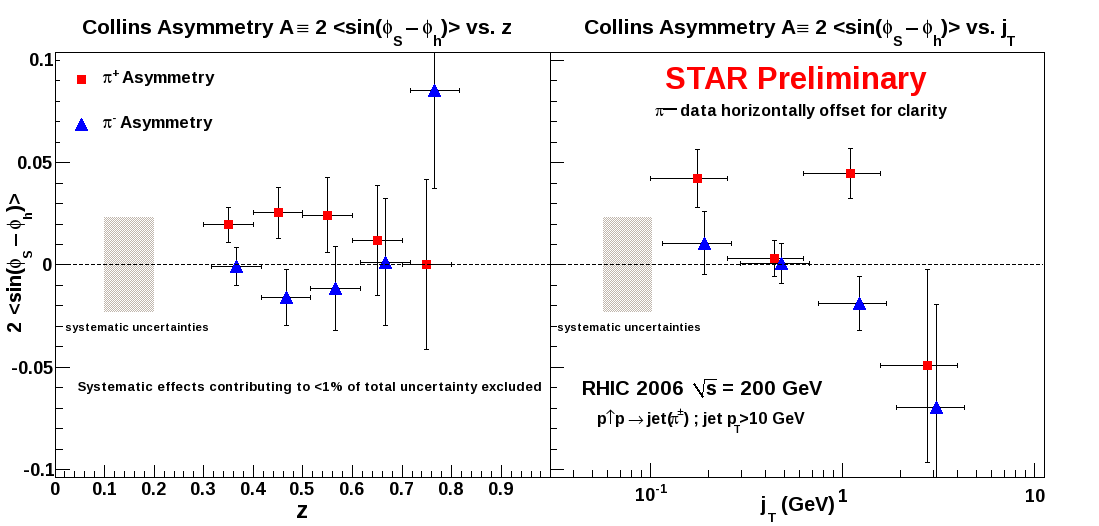}{Preliminary STAR Collins Asymmetry}{Preliminary results from 2006 $\sqrt{s}=200$ GeV STAR data for charged pions in jets \cite{ref:StarCollinsPrelim}. The result is statistically limited, but hints at a separation of charges and possible statistical significance for increased statistics.}{1} 

\figuremacroW{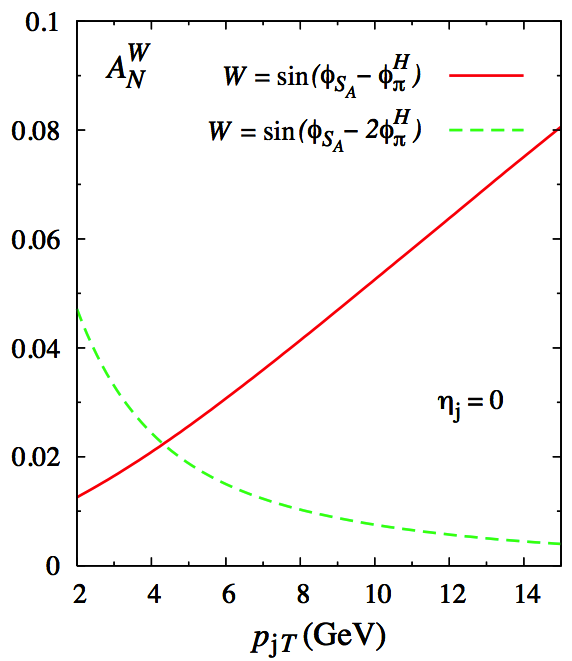}{Maximized Quark Collins Asymmetry Prediction}{The potential size of the Collins asymmetry for the $p^\uparrow +p\rightarrow jet+\pi^+ +X$ process at $\sqrt{s}=200$ GeV if the transversity distribution is maximized to the Soffer bound \cite{ref:DAlesio}.}{0.4} 

\subsection{Asymmetry Sensitive to the Sivers Function}
The Sivers distribution, $f_{1T}^\perp$, describes the connection between the intrinsic parton $k_T$ and the proton's spin in a transversely polarized proton \cite{ref:SiversFunction}. More specifically, $f_{1T}^\perp$ describes an overall preferred direction of quark and gluon movement inside of a transversely polarized proton, or a left-right asymmetry of partonic momentum with respect to a plane which spans the proton's momentum and polarization directions. This partonic asymmetry in the proton, or the Sivers effect, shows up as a left-right asymmetry in the final state hadrons or jets. In contrast to the transversity distribution, the Sivers function is chiral-even meaning it is not necessary to couple it to another distribution and it does exist for gluons. 

The SIDIS experiments have contributed the most results for the Sivers asymmetry, where azimuthal asymmetry of outgoing hadron tracks is measured. To do this, the angle between the spin vector ($\phi_S$) and outgoing hadron ($\phi_H$) is measured, and the coefficient of the $\sin$$\left(\phi_H-\phi_S\right)$ modulation from Equation \ref{eq:sidisxsec} is extracted as the Sivers asymmetry \cite{ref:AnselminoSivers2010}. Figure \ref{fig:SiversSIDIS} shows Sivers asymmetry results from the COMPASS \cite{ref:CompassAsymmetries} and HERMES \cite{ref:HermesSivers} collaborations. The results agree qualitatively where there is a statistically significant asymmetry for positively charged hadrons, but nothing significant for neutral and negatively charged hadrons. DIS experiments cannot easily access gluon contributions, so these asymmetries represent results for the Sivers effect for quarks.

\begin{figure}
    \centering
    \subfloat[]{\includegraphics[trim={0 0 0 9cm},clip,width=7.75cm]{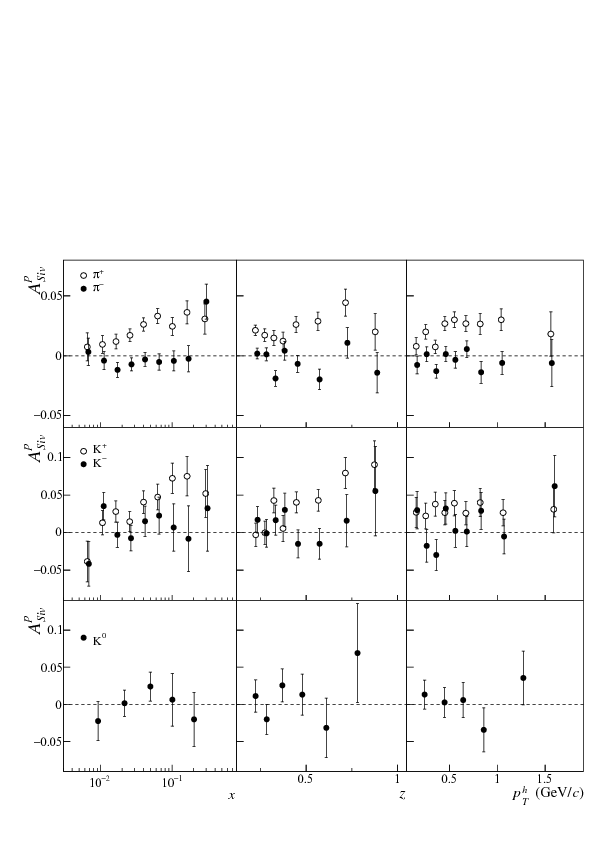} }%
    \subfloat[]{\includegraphics[width=7cm]{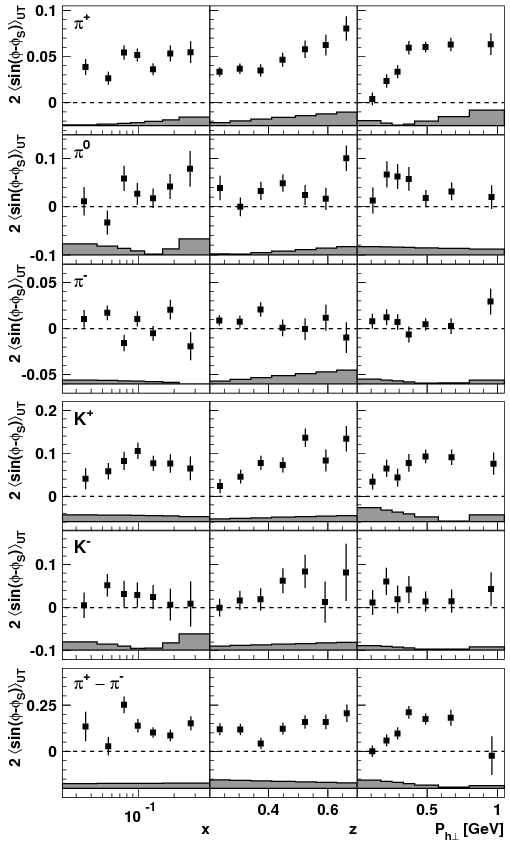} }%
    \caption[Sivers Asymmetries from SIDIS]{\textbf{Sivers Asymmetries from SIDIS} - Sivers asymmetry results from the COMPASS \cite{ref:CompassAsymmetries} (a) and HERMES \cite{ref:HermesSivers} (b) collaborations as a function of $x$, $z$, and $p_T^h$. In both cases there is a clear and statistically significant signal for positively charged hadrons, yet nothing for neutral and negatively charged hadrons.}
    \label{fig:SiversSIDIS}%
\end{figure}

Using SIDIS data from the HERMES and COMPASS collaborations, the Sivers distribution may be extracted from the asymmetry results. These functions are shown for both valence and sea quark contributions in Figure \ref{Anselmino_SiversDistribution} \cite{ref:AnselminoSivers2010}. From these, it is clear that the valence up and down quark distributions are of roughly the same magnitude but opposite in sign. Since outgoing hadrons are tagged in the analyses, unpolarized fragmentation functions describe the probability of a parton fragmenting to the detected particle\footnote{These fragmentation functions are in the TMD framework, and therefore depend upon the final state transverse momentum of the hadrons, $p_{h\perp}$. This dependence has not been measured, thus an ansatz of a Gaussian distribution is used to describe the $p_{h\perp}$ dependence of the unpolarized fragmentation functions in Ref. \cite{ref:AnselminoSivers2010}.}. Therefore, considering favored and disfavored fragmentation functions as given in Figure \ref{Anselmino_UnpolarizedFF} \cite{ref:AnselminoSivers2009}, it is easy to naively compute the asymmetry for $\pi^+$ and $\pi^-$ particles and show how the asymmetry for $\pi^-$ particles should be much smaller than the $\pi^+$ asymmetry.

\figuremacroW{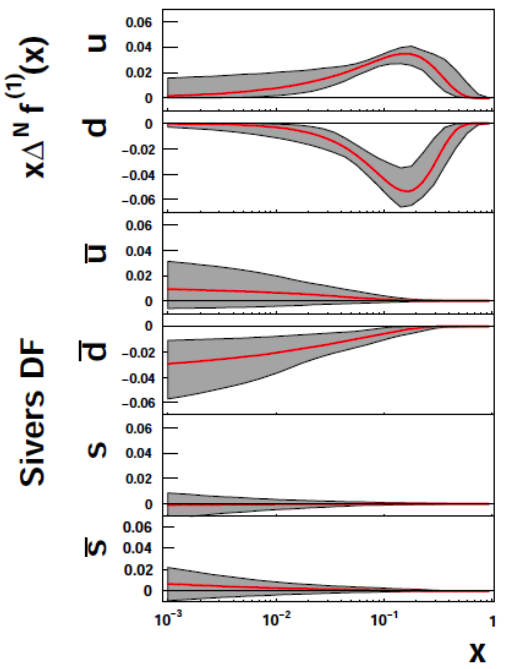}{Extracted Sivers Distributions}{Sivers distributions extracted for valence and sea quark contributions by using the asymmetries measured by the HERMES and COMPASS collaborations \cite{ref:AnselminoSivers2010}.}{0.4} 

\figuremacroW{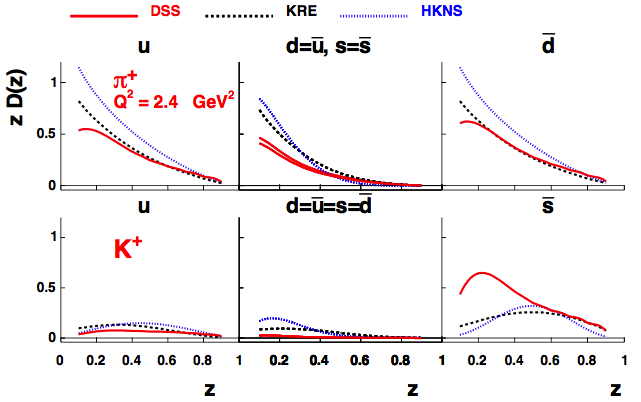}{Unpolarized Fragmentation Functions}{Unpolarized fragmentation functions using three different parameterizations \cite{ref:AnselminoSivers2009}. The left column shows the valence favored fragmentation functions, the center column shows the disfavored fragmentation functions, and the right column shows the favored fragmentation functions from the sea. The top is for $\pi^+$ particles and the bottom for $K^+$ particles, although similar curves result for negatively charged particles.}{0.6} 

In transversely polarized proton collisions, the single spin inclusive jet asymmetry is sensitive to the initial state twist-3 quark-gluon correlators. These correlators, described by the Efremov-Teryaev-Qiu-Sterman (ETQS) function \cite{ref:RHIC_QCDplan,ref:Efremov,ref:QiuSterman}, are related to the leading twist TMD Sivers function probed by the SIDIS asymmetries in Figure \ref{fig:SiversSIDIS}. The inclusive jet asymmetry, also called $A_N$, is extracted as the coefficient of the $\sin\left(\phi_S\right)$ term, where $\phi_S$ is the angle between the proton spin vector and the jet. 

\figuremacroW{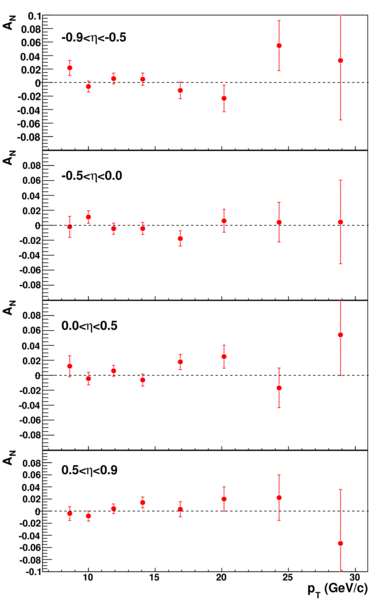}{STAR Inclusive Jet Asymmetry}{The single spin inclusive jet asymmetry, as extracted by the STAR collaboration in four bins of pseudorapidity ($\eta$) \cite{ref:StarSivers2006}.}{0.5}

The STAR collaboration has extracted and published the inclusive jet asymmetry in the past using $\sqrt{s}=200$ GeV proton collisions, and the results are shown in Figure \ref{StarSiversPRD} \cite{ref:StarSivers2006}. The asymmetries are presented in four bins of the pseudorapidity, $\eta$, with positive $\eta$ meaning forward scattering and negative $\eta$ meaning backward scattering with respect to the polarized beam. For the inclusive jet asymmetry analysis in Ref. \cite{ref:StarSivers2006} the collider beams are treated as polarized independently, so the asymmetry is computed for each beam separately and the results are combined, thus $\eta$ is defined with respect to whichever beam direction is under analysis. As $\eta$ grows large, different values of $x$ are being accessed, with higher $\eta$ accessing larger values of $x$. Clearly, for all $\eta$ bins the asymmetry is consistent with zero. 

Since there is only a single hard scale ($p_{T,jet}$) for inclusive jet asymmetry measurements in $pp$, the twist-3 factorization is required to describe the asymmetries theoretically. In contrast, SIDIS asymmetries have two hard scales (Q and $p_T^h$) thus they may be described by the TMD factorization, as shown above. Using the twist-3 factorization, the inclusive jet asymmetries are constrained by the data from Figure \ref{StarSiversPRD}. This result is shown in Figure \ref{Kanazawa_TwistThreeStar} \cite{ref:KanazawaTwist3}, and it is clear that with the STAR data there is no significant inclusive jet asymmetry expected in mid-rapidity $\sqrt{s}=200$ GeV $pp$ collisions.

\figuremacroW{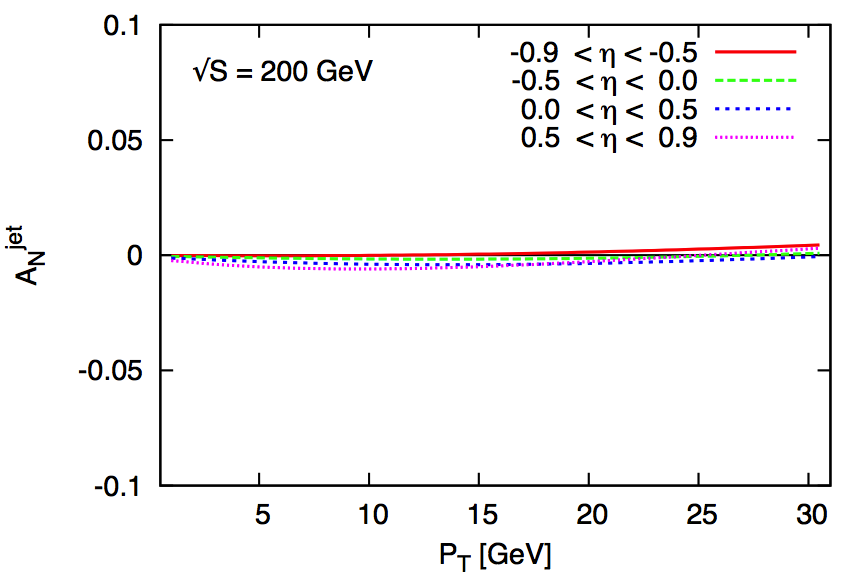}{Twist-3 Sivers Prediction}{Theoretical constraint placed on the Sivers asymmetry using the twist-3 factorization and STAR data shown in Figure \ref{StarSiversPRD} \cite{ref:KanazawaTwist3}.}{0.65}

\chapter{RHIC and The STAR Detector} 

\ifpdf
    \graphicspath{{ch2_RhicAndStarDetector/figures/PNG/}{ch2_RhicAndStarDetector/figures/PDF/}{ch2_RhicAndStarDetector/figures/}}
\else
    \graphicspath{{ch2_RhicAndStarDetector/figures/TIFF/}{ch2_RhicAndStarDetector/figures/}}
\fi

The proton collisions that provided the data presented in this thesis were generated by the Relativistic Heavy Ion Collider (RHIC), located at Brookhaven National Laboratory (BNL) in Upton, New York on Long Island. The collisions were detected by the Solenoidal Tracker at RHIC (STAR). This chapter will discuss the experimental components of the RHIC complex and STAR detector used during the collection of the data presented in this thesis.

\section{The Relativistic Heavy Ion Collider}
In addition to the heavy ions which are in the accelerator's name, RHIC is also the world's only polarized proton collider. It is split into two 3834 m long rings which carry particles in opposite directions and can collide particles with center of mass energies between 50 and 500 GeV. The rings are arbitrarily designated ``blue beam'' and ``yellow beam'', labels which are convenient to use when looking at results from each beam. Looking down on the RHIC ring from above, the blue beam travels clockwise and the yellow beam travels counterclockwise. There are six interaction points (IP) along the ring where the accelerated beams cross and collide creating opportunities for physics. These are located at 12, 2, 4, 6, 8, and 10 o'clock, as shown in Figure \ref{RHIC}. At present, only STAR and PHENIX (IP 6 and 8, respectively) are active experiments, and these are the two IPs which have spin rotators that allow for colliding protons with either longitudinal or transverse polarization, providing opportunity for multiple physics analyses related to the proton's spin. The proton beam polarization is monitored at IP 12 by the proton-carbon (pC) targets and hydrogen gas jet (H-jet) polarimeters and will be discussed further later in this section.

\figuremacroW{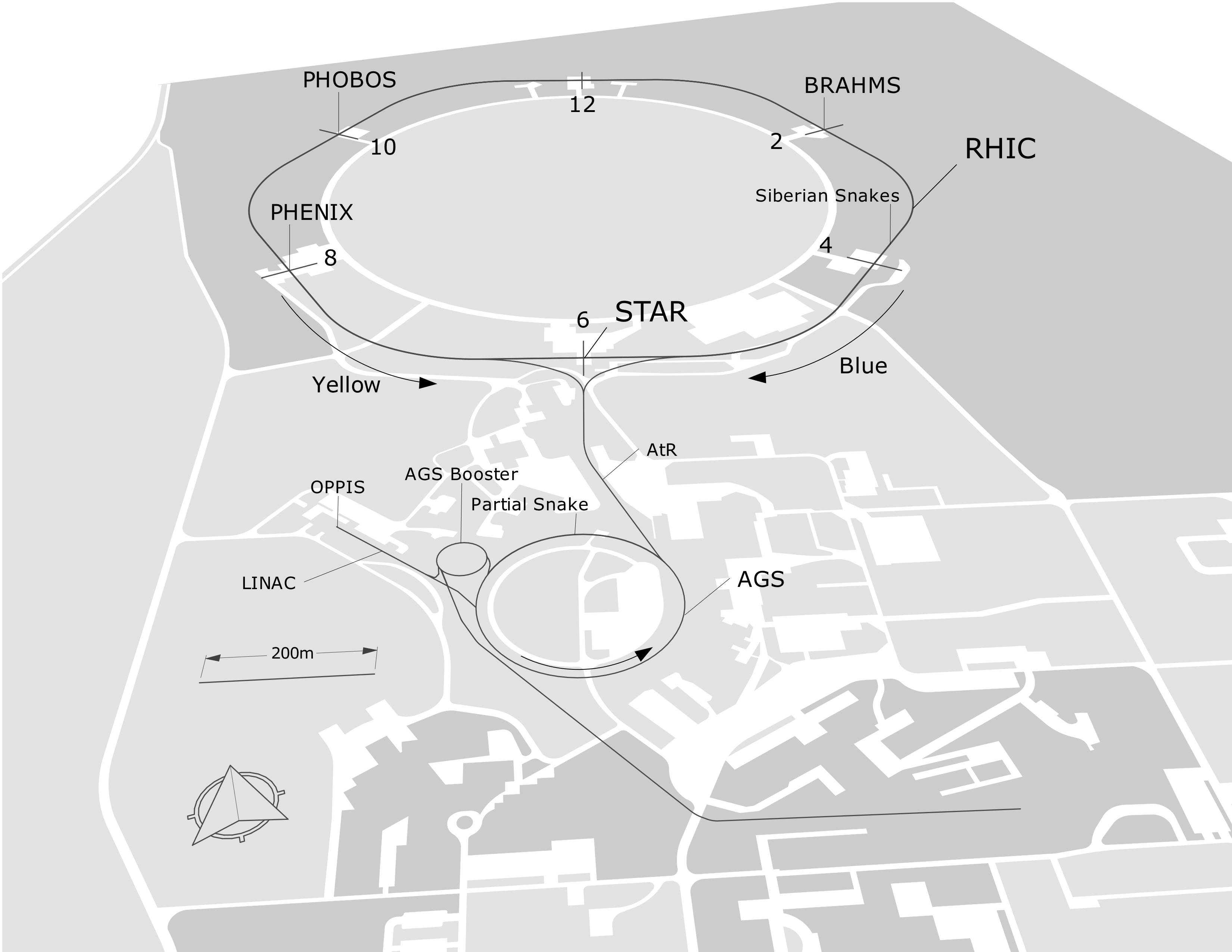}{RHIC Complex Layout}{The layout of the RHIC accelerator and surrounding complex, including the location of the interaction points, Siberian snakes, and pC/H-jet polarimeters.}{0.8}

\subsection{Protons on a Collision Course}
Protons take quite a journey before they finally collide in RHIC. It all begins at the optically pumped polarized $H^-$ source (OPPIS) \cite{ref:OPPIS}, where 300 $\mu s$ pulses of a 0.5 mA current produces 35 keV transversely polarized $H^-$ ions with upwards of 85\% polarization. The bunches produced by OPPIS are accelerated from 35 keV to 200 MeV by a radio frequency quadrupole (RFQ) and a linear accelerator. Only about half of the original ions produced by OPPIS are accelerated by the linac, resulting in a bunch with about $4\times10^{11}$ polarized protons. Once the $H^-$ ions are accelerated to 200 MeV by the linac, the electrons are stripped off, and the remaining protons are injected into the booster ring where they are accelerated to about 1.5 GeV. After the booster, the bunch is handed off to the alternating gradient synchrotron (AGS) where it is accelerated further to 24 GeV and injected into RHIC for acceleration to either 100 or 250 GeV per beam depending on the experimental running period. This process is repeated for each bunch produced by OPPIS.

\subsection{Beam Bunches, Spin Patterns, and Rotators}
Once the pulsed bunches are accelerated by the AGS, they are ready to be transferred to the RHIC rings. There are 360 radio frequency (RF) cavities (or buckets) in the RHIC ring which may be filled with bunches of polarized protons. During running, only 120 of these buckets are filled with bunches, with two empty buckets between each bunch which are often ignored when numbering or counting the filled buckets. Once the beams are accelerated to the appropriate energy they must be ``cogged'', meaning the position on the ring where the first bunches in each beam collide is set. For 2012, the beams were cogged so that the first bunches in each beam collided with each other at IP 2 and IP 8. In each beam the last nine buckets are intentionally left empty as an ``abort gap'' which is useful for beam background studies. Once the beams are cogged, there will always be crossings at each IP where protons from the blue(yellow) beams collide with the yellow(blue) abort gaps. When the beams are accelerated and cogged, the beams are then maintained as a ``store'' or ``fill'' for physics collisions. 

In each beam, a particular spin pattern is set throughout the buckets that tells the direction of the spin of the protons in each bunch, and repeats after every eight buckets. For example, the repeating pattern in each beam could be ($-+-++-+-$) for blue and ($--++--++$) for yellow, resulting in all possible spin directions in each beam colliding with each other. For 2012, there were four such pattern sets arbitrarily named $P1-P4$. Since all bunches are not identical, there are variations in shape and average polarization, changing the spin pattern with each fill helps to randomize these bunch characteristics among different spin states which will reduce systematic effects that could result from sampling one spin state more than another. These spin patterns are also where knowing the beam cogging becomes important. If it is known where each bunch from each ring is colliding, then it is known what the polarization is at that interaction point, which is necessary for spin sorting in the analysis code. 

The preceding discussion about spin patterns works for either longitudinal (helicity) or transverse polarization. The equilibrium state in RHIC is for the protons to be transversely polarized, thus when a longitudinal polarization is desired the spin rotators are used. The rotators are positioned just up and just downstream at both STAR and PHENIX experiments. The rotators are ramped with the beam and rotate spin to longitudinal direction for collisions and then back to transverse. This allows for the different experiments to set their own spin orientation independently of each other for as long as desired.

\subsection{Siberian Snakes}
One requirement for studying spin properties of the proton is to maintain the polarization throughout the acceleration process and throughout the store in RHIC so that the final collision is between protons of a particular spin orientation. Unfortunately, there are depolarization resonances caused by imperfections in the bending and focusing magnets that will reduce the polarization of the beam \cite{ref:MeiBaiPaper}. To combat the polarization loss, Siberian snakes are installed in both the AGS and RHIC rings \cite{ref:RhicConfig}. In the RHIC ring, there are two snakes installed directly across from each other near the 4 o'clock and 10 o'clock positions. These are so called ``full'' snakes, flipping the spin completely by $180\degree$ in a horizontal plane at each location and thus leaving the spin invariant at each interaction point. In the AGS where energies are lower, the depolarization is not as strong meaning the snakes can be ``partial'' snakes and rotating the spin by less than $180\degree$ at each passing. These partial snakes are not the same strength, with one operating at 5.9\% and another operating at 15\%-20\% \cite{ref:MeiBaiPaper}. This configuration is enough to maintain the polarization during the AGS acceleration. 

\subsection{Beam Polarization Measurements}
The beam polarization in RHIC is measured two ways, with a proton-carbon (pC) Coulomb Nuclear Interference (CNI) polarimeter \cite{ref:pCPolarimeter} and a hydrogen jet (H-jet) polarimeter \cite{ref:HJetPolarimeter}. The pC polarimeter provides relative polarization measurements on a fill-by-fill basis, and the output is calibrated to obtain the absolute polarization using the output from the H-jet polarimeter. The pC polarimeters are fast detectors and therefore may be used to monitor and extract the time dependence of the beam polarization throughout the fill. Fits to the polarization time dependence can then be used to weight each event by the correct beam polarization. These values will be used in the asymmetry analysis, and discussed further in Chapter \ref{chapter:AsymmetryAnalysis}.

\section{The STAR Detector}
\label{sec:stardetector}
The detector at the heart of the analysis presented in this thesis is the Solenoidal Tracker At RHIC (STAR) \cite{ref:StarOverviewNIM}, located at IP 6 on the RHIC ring. The large acceptance of STAR, covering $2\pi$ in azimuth and $-1 < \eta < 2$, makes it ideal for reconstructing high multiplicity heavy ion collisions and jet events. A cutaway cross section depiction of the detector is shown in Figure \ref{stardetector}, which shows the coverage, overall size, and the definition of the STAR coordinate system. The coordinate system is left-handed with the z-axis lying along the beam line, with $z=0$ defined as the center of the interaction region, the y-axis points towards the top of the detector and the x-axis points out of the page. The blue beam travels west along the $+z$ direction, and the yellow beam travels east along the $-z$ direction. With these definitions, the coverage coordinate $\eta$ has the same definition as in Equation \ref{eq:rapidity}:

\begin{equation}
\eta = -\ln\left[\tan\left(\frac{\theta}{2}\right)\right]
\end{equation}

\noindent where $\theta$ is the angle between the $+z$ axis and the detected momentum vector.

While STAR is a single detector, it is built up from a set of detector subsystems all working in unison. For this analysis, the relevant subsystems are the time projection chamber and solenoidal magnet, the barrel electromagnetic calorimeter, the endcap electromagnetic calorimeter, and the vertex position detector. Each of these subsystems will be discussed in detail in the following sections.

\figuremacroW{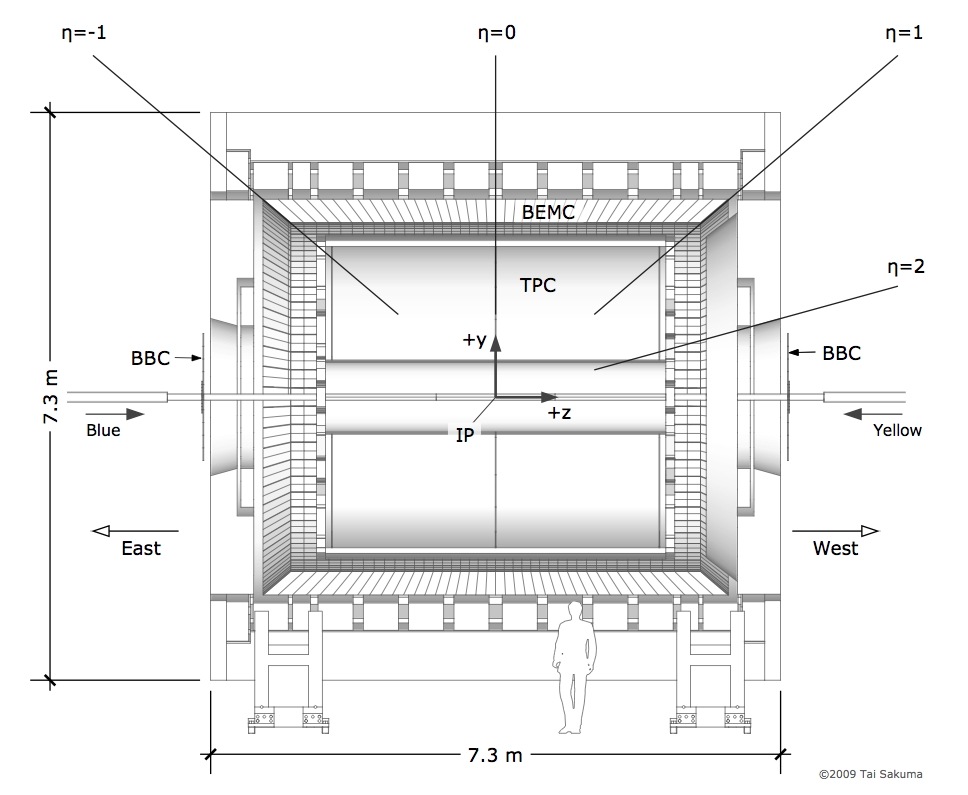}{STAR Detector}{The cutaway cross section of STAR shows the coverage of each subsystem, as well as the overall size of the detector in relation to the human depiction at the bottom \cite{ref:sakuma}.}{1}

\subsection{Time Projection Chamber and Solenoidal Magnet}
The time projection chamber (TPC) \cite{ref:tpcNIM} is at the center of STAR. It is a cylindrical detector which surrounds the interaction region, with diameter 4 meters and total length 4.2 meters, and covers the range $\lvert\eta\rvert < 1.3$ and $\Delta\phi = 2\pi$. It is used for charged particle tracking and momentum reconstruction as well as measuring particle	 energy loss (dE/dx) which will be used in particle identification. Figure \ref{tpcSchematic} shows a diagram of the TPC, including the interior.

\figuremacroW{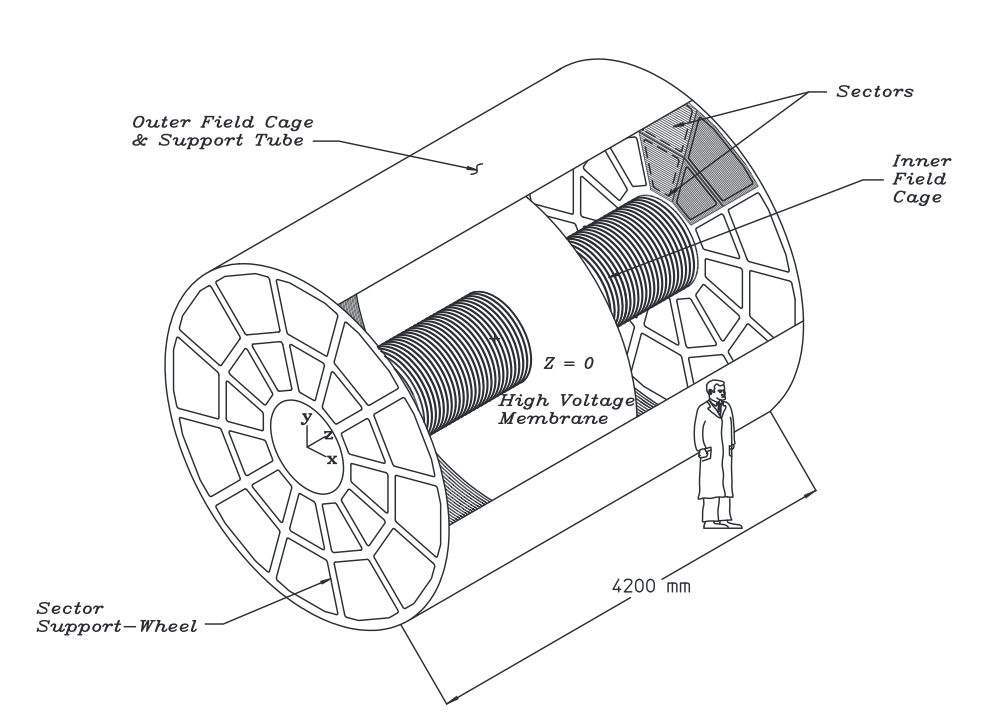}{Time Projection Chamber}{A schematic of the TPC which shows the central membrane at $z=0$, the end pad planes, and the inner and outer field cage \cite{ref:tpcNIM}.}{1}

The TPC is filled with a 90\% argon and 10\% methane gas mixture (commonly called P10) held at 2 mbar above atmospheric pressure, which is continuously being recirculated and resupplied \cite{ref:tpcGasNIM}. When a charged particle moves through the gas it leaves a trail of ionized electrons in its wake. The electrons drift along the z-axis towards the ends of the TPC in a constant electric field. The central membrane, at $z=0$ in Figure \ref{tpcSchematic}, is a cathode held at 28 kV, and each end of the TPC is an anode held at ground. The inner and outer field cages bind the gas volume at the inner ($r=100$ cm) and outer radius ($r=400$ cm) of the TPC. The anode planes at each end also serve as the readout for the drifting electrons, thus developing a picture of the original charged track. Using the start time of the collision and the drift velocity of electrons, the original z-axis position of the tracks can be determined and the collision vertex reconstructed by looking to where the tracks all point back to.

As charged particles move through the volume, they are influenced by a uniform magnetic field held at 0.5 T along the z-axis supplied by the solenoidal magnet \cite{ref:starMagnetNIM}. The magnetic field is necessary for reconstructing the track momentum and assigning the correct charge sign based on the direction of curvature. The magnetic field is aligned with the electric field to help minimize distortions to the track due to electrons spreading in the transverse or longitudinal direction.

The energy lost ($dE/dx$) by each track in the gas is also measured by the TPC. When used in conjunction with the track momentum, $dE/dx$ is a very useful tool to identify the track particle species. In a given medium, the characteristic energy loss for different particle species may be theoretically calculated \cite{ref:PdgReview}, and have different dependencies as a function of the particle momentum. Therefore, plotting the track energy loss as a function of the track momentum, the $dE/dx$ curves separate themselves by particle species. Using these curves, the tracks can be identified on the momentum range ${\sim} 0.1$ GeV/c $\leq p \leq 10$ GeV/c \cite{ref:ShaoParticleId}.

\subsection{Barrel Electromagnetic Calorimeter}
The barrel electromagnetic calorimeter (BEMC) \cite{ref:bemcNIM} surrounds the TPC, sitting inside the solenoid, and measures energy deposited by electromagnetically interacting particles, mainly electrons and photons. The BEMC provides $-1<\eta<1$ coverage, and the same $\Delta\phi=2\pi$  coverage as the TPC. In contrast to the TPC which operates with a slow $5.45$ cm/$\mu$s drift velocity and ${\sim}40$ $\mu$s readout time, the BEMC boasts a very fast readout response rate, able to keep up with the ${\sim} 9.35$ MHz collision rate of RHIC.

The BEMC subsystem is again a collection of detectors all working in unison. There are 4800 total calorimeter ``towers'' in the BEMC, each covering an area $\Delta\eta\times\Delta\phi=0.05\times0.05$. While the towers are all individual calorimeters, the actual construction of the BEMC is made up of 120 modules consisting of 40 towers each. These modules consist of two towers in $\phi$ and 20 towers in $\eta$ thus covering an area of $\Delta\eta\times\Delta\phi=1.0\times0.1$. There are 60 such modules for $\eta>0$, and 60 modules for $\eta<0$. As $\eta$ increases, each tower is angled more towards the interaction region at the center of the TPC. Figure \ref{bemcTowerProj} shows this projective nature of the towers.

\figuremacroW{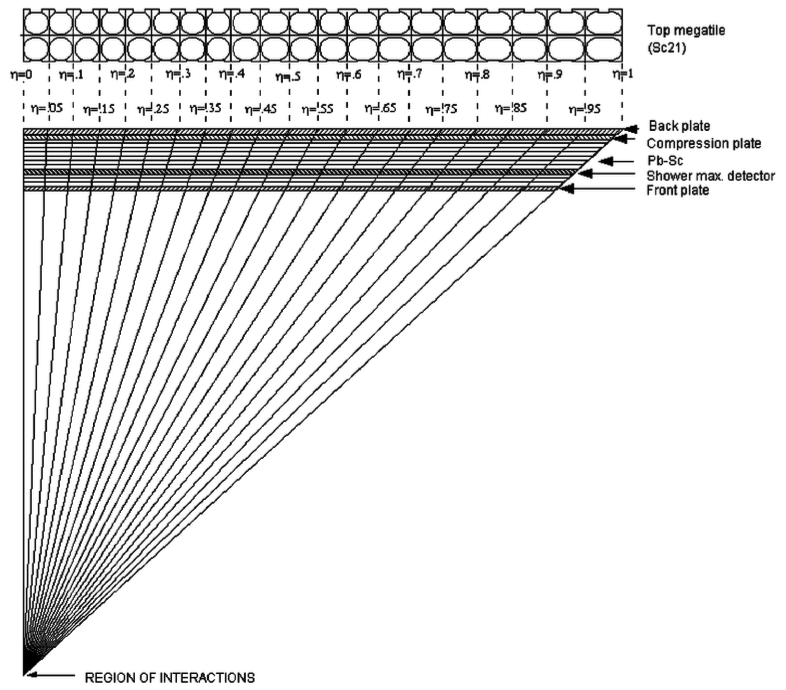}{BEMC Tower Layout}{Depiction of how the towers look in half of the BEMC including the $\eta$ coverage of each tower as well as how all towers project back to the interaction region. A schematic of the top megatile, Sc21, is shown at the top \cite{ref:bemcNIM}. This schematic of the megatile should not be confused with any physical location.}{0.9}

The interior of each module is constructed of a lead and scintillator stack. The scintillator layers in each module are constructed as a so-called ``megatile'', each containing 40 optically separated tiles. The optically separated tiles define the different towers in each megatile, as depicted in Figure \ref{bemcTowerProj}. The modules consist of 20 layers of 5 mm thick lead plates, alternating with 21 scintillating layers, for a total of about 20 radiation lengths at $\eta=0$. 19 of the scintillating layers are also 5 mm thick, but the first two in each module are 6 mm thick and represent the preshower detector. This configuration is sufficient to stop and contain a 60 GeV electromagnetic shower. There is also a shower maximum detector (SMD) buried at ${\sim}5$ radiation lengths (at $\eta=0$) which is used for more specific analyses involving high resolutions of photons, for example from $\pi^0$ decays. An end view of the final construction of a module interior is shown in Figure \ref{bemcModule}.

\figuremacroW{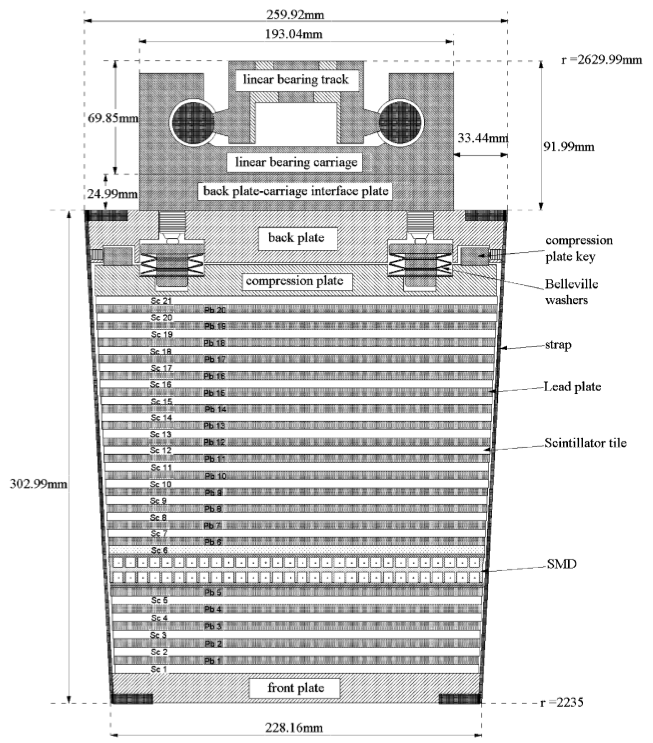}{BEMC Module Construction}{End view of the final interior construction of a BEMC module. The first two scintillating layers are thicker for the preshower detector. The SMD is buried at about five radiation lengths from the front plate at $\eta=0$ \cite{ref:bemcNIM}.}{0.8}

Since the BEMC is a fast detector, it is well suited as a triggering detector that will decide if slower subsystems (such as TPC) should read out data or simply move to the next event. Because the tiles in each megatile are optically separate, information from the towers may be read out individually or as the sum of a group. Using the transverse energy ($E_T$) as the triggering threshold, events may be triggered by single towers or a group of towers above the set threshold. This option is quite useful for analyses which seek events with high energy tracks such as W boson analyses, as well as those which seek events containing a collection of towers like a jet. The triggers and thresholds for this analysis will be discussed further in Chapter \ref{chapter:DataSelection}.

\subsection{Endcap Electromagnetic Calorimeter}
The endcap electromagnetic calorimeter (EEMC) is complementary to the BEMC, pushing STAR's calorimeter coverage to the more forward region \cite{ref:eemcNIM}. The EEMC is an annulus installed on the west end of STAR that covers $1.086 < \eta < 2$ (with a small gap between the barrel and endcap detectors) and $\Delta\phi = 2\pi$. Like the BEMC, the EEMC uses a lead and scintillator stack, with the scintillating layers made into megatiles which contain optically separated tiles which define the towers that project back towards the interaction region. Similar to the BEMC tower design, the EEMC towers are also about 20 radiation lengths deep and contain thicker preshower layers as well as an SMD buried in each tower at about 5 radiation lengths. To help aid in differentiating between electrons and charged hadrons, dedicated postshower scintillator layers are added to the EEMC. The layout and tower features are depicted in Figure \ref{eemcLayout}. 
 
\figuremacroW{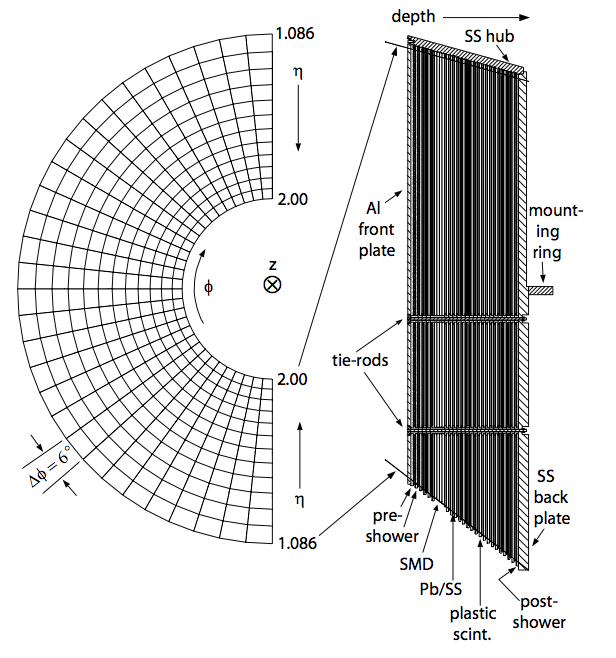}{EEMC Layout and Construction}{On the left, the layout of the towers for half of the EEMC is shown. On the right is how the towers are constructed including the preshower, SMD, and postshower detectors \cite{ref:eemcNIM}.}{0.8}

The EEMC is also a fast detector and useful for triggering in conjunction with the BEMC. Like the BEMC, single towers or groups of towers may be used to define triggered events. For this analysis, the EEMC will only be used in a triggering capacity and in cases where jets overlap with the BEMC. Pure EEMC jets will be cut out of the data sample later on, as midrapidity jets are desired for this analysis.

\subsection{Vertex Position Detector}
The vertex position detector (VPD) \cite{ref:vpdNIM} is a pair of timing detectors which sit outside of the nominal STAR volume that includes the previously discussed subsystems. Each of the detector systems is an array of 19 cylinders, all housing a layer of lead and scintillator, and a photomultiplier tube (PMT). They are symmetric detectors, mounted directly on the beamline 5.7 m to the east and west of the center of STAR. Being so close to the beamline, the VPD covers a very far forward region of pseudorapidity of $4.24\leq\eta\leq5.1$. The layout of each housing array is shown in Figure \ref{vpdFront}.

\figuremacroW{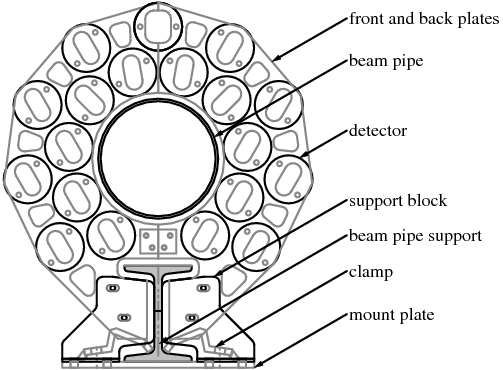}{VPD Front View}{Schematic front view of the vertex position detector, showing all 19 of the individual detector positions in the full housing. The housing splits down the middle to clamp directly around the beamline \cite{ref:vpdNIM}.}{0.75}

The VPD was designed and constructed to detect far forward particles produced from the primary collision. In proton-proton collisions the hits in the VPD originate from charged pions and $\pi^0$ decay photons. These particles are traveling at or near the speed of light, and arrive at the detectors nearly instantaneously, providing a fast trigger, vertex reconstruction and serving as the start time counter for the time-of-flight (TOF) detector. For this analysis the triggering and vertex information from the VPD will be used.

The z-axis vertex position may be reconstructed simply by using the timing information in each of the VPDs. Specifically, the vertex position may be calculated by

\begin{equation}
\label{eq:vpdVtx}
z_{VPD} = c\frac{T_{east}-T_{west}}{2}
\end{equation}

where $T_{east}$ and $T_{west}$ are the times when photons reached the east and west VPD, respectively. The speed of light gives the link between time and distance for the vertex since the photons are emitted from the same position. In proton-proton collisions, the timing resolution of a single VPD channel is ${\sim} 150$ ps \cite{ref:vpdNIM}, which corresponds with a ${\sim} 1$ cm primary vertex resolution along the z-axis.

Like the calorimeters, the VPD is also a fast detector and therefore used as a trigger detector. The VPD trigger was developed by selecting events that satisfy a timing cut ensuring coincidence between the east and west VPDs. This trigger is a so-called ``minimum bias'' trigger, as the events that are selected have to pass the requirement of coincidence thus introducing a small bias into the triggered sample. This is a trigger which will contribute to the results presented later, and the triggering requirements and data implementation will be discussed further in Chapter \ref{chapter:DataSelection}. 

\subsection{Time of Flight}
The time of flight (TOF) \cite{ref:tof} provides timing information within each event, vastly improving the particle identification capabilities of STAR. The TOF also sits inside the solenoidal magnet, and is sandwiched between the TPC and BEMC detectors, providing the same coverage as the BEMC ($\Delta\phi=2\pi$ and $-1<\eta<1$). The TOF and the VPD work in conjunction, with the VPD providing the start time of the event and the TOF acting as the stopwatch for each track in the event. 

The TOF is constructed from 120 trays that immediately surround the TPC. Within each tray there are 32 multi-gap resistive plate chamber (MRPC) modules, for a system with a total of 3840 detectors. The MRPC modules are gaseous detectors that contain several layers of glass planes that are separated by gaps of the gas. The top and bottom of each module contain electrodes which apply an electric field, so when particles pass through the gas filled layers and cause electron avalanches, the emitted electrons are guided toward the six readout pads on the outside of the electrodes. This construction provides a high degree of granularity for events with a high multiplicity of tracks, yet causes minimal interference on incident particles, so the performance of the BEMC is not hindered. The module construction is shown in figure \ref{tofModule}.

\figuremacroW{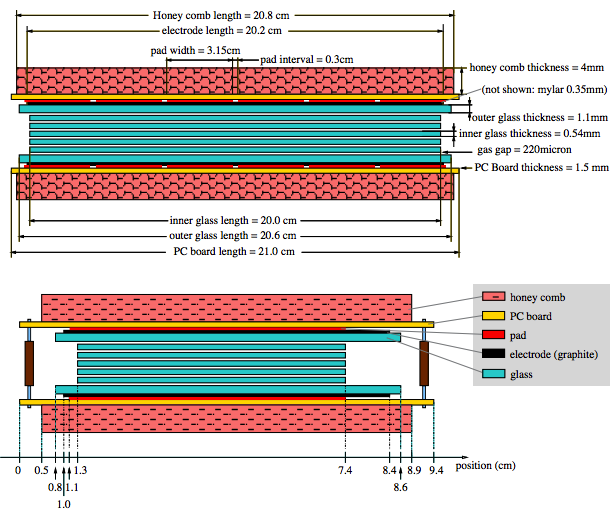}{TOF MRPC Module}{Side views of an MRPC module, for the long (top) and short (bottom) sides. The color codes on the bottom show the different detector components discussed in the text \cite{ref:tof}}{0.9}

The timing information is incredibly useful for particle identification. Using the TOF, the velocity of each track can easily be calculated via

\begin{equation}
\label{eq:tofBeta}
\beta=\frac{s}{c\left(t_{TOF}-t_{VPD}\right)}=\frac{s}{c\Delta t}
\end{equation}

where $s$ is the path length of the track calculated from the TPC. Using the velocity, the mass is calculated by

\begin{equation}
\label{eq:tofMass}
m^2=p^2\left(\left(1/\beta\right)^2-1\right)
\end{equation}

where, again, the momentum ($p$) is given by the TPC. The TOF was used along with dE/dx from the TPC to estimate kaon and proton contaminations to the pion samples. These will be presented in Chapter \ref{ch:Systematics}.



\chapter{Data Selection and Jet Reconstruction} 
\label{chapter:DataSelection}

\ifpdf
    \graphicspath{{ch3_DataSelection/figures/PNG/}{ch3_DataSelection/figures/PDF/}{ch3_DataSelection/figures/}}
\else
    \graphicspath{{ch3_DataSelection/figures/EPS/}{ch3_DataSelection/figures/}}
\fi

\section{Data Collection}
Between February 6, 2012 and March 12, 2012 RHIC collided transversely polarized protons at $\sqrt{s}=200$ GeV. During this period RHIC delivered an integrated luminosity of $36.1$ $pb^{-1}$ to STAR at an average beam polarization of $59$\%. Data are collected in a series of ``runs'' in the control room, which vary in length and number of events captured. Ideal runs average about 30 minutes in length, and collect on the order of 1.5-2 million events.

\section{Triggering}
\label{section:triggering}
It is not feasible to collect and record all of the collisions sampled by the STAR detector.  Instead, \textit{triggers} are implemented to collect only the events which are likely to contain the observables of interest. For example, for the analysis presented here higher $p_T$ jets are desired since they access more quark dominated partonic scattering than the low $p_T$ jets which are dominated by gluons that carry no transversity. Thus, a trigger which enhances statistics for higher $p_T$ jets is sought after.

The triggering system at STAR relies on the subsystems which are able to read out data very fast. The electromagnetic calorimeters read data out fast enough to keep up with the collision rate ($\sim$9.35 MHz) at RHIC. This is ideal to be able to capture events from every collision and pass data to an algorithm that decides whether the event satisfies the trigger conditions. If the event is a good triggered event, data from the much slower tracking detectors are also written out, otherwise it is thrown away. 

For this analysis, events were analyzed if they satisfied conditions for two kinds of triggers: jet patch (JP) and minimum bias (MB). To form the JP trigger, the BEMC towers are split up into 18 $\Delta\eta\times\Delta\phi=1\times 1$ ``jet patches'', each containing 400 towers. There are 6 patches on each of the east and west ends of the barrel, and 6 which overlap the other jet patches across $\eta=0$. Similarly, the EEMC is split up into 6 jet patches which also cover $\Delta\eta\times\Delta\phi=1\times 1$. Finally, there are 6 jet patches that span the gap between the BEMC and EEMC detectors which line up with the physical patch locations in each detector but carry asymmetric coverage of $\Delta\eta\times\Delta\phi=0.6\times 1$ in the west half of the BEMC and $0.4\times 1$ in the outer part of the EEMC for the same total $1\times 1$ coverage as all other jet patches. For each jet patch in every event, ADC values are summed and passed to a Data Storage and Manipulation (DSM) tree to apply the thresholds \cite{ref:StarTrigger}. There are three JP triggers with increasing thresholds as outlined in Table \ref{table:JetPatchTriggers}, with the DSM ADC threshold being converted to an approximate transverse energy threshold (in units of GeV) by the simple relation in Eq. \ref{eq:dsmThreshold}. To guard against false positive trigger fires, every event is also run through a trigger simulator during offline analysis which applies thresholds and decides whether or not the event should have fired the hardware. Events can then be checked to see if both the hardware and software triggers fired during analysis.

\begin{equation}
\label{eq:dsmThreshold}
E_T = 0.236\times\left(ADC-5\right)
\end{equation}

\begin{table}[hb]
\begin{center}
\begin{tabular}{ c c c c c }
& & \multicolumn{2}{c}{Thresholds} & \\ \cline{3-4}
Name & Offline ID & DSM ADC & $E_T$ & Avg. Prescale\\
\hline
JP0 & 370601 & 20 & 3.5 GeV & 141.35\\
JP1 & 370611 & 28 & 5.4 GeV & 2.5\\
JP2 & 370621 & 36 & 7.3 GeV & 1.0\\
\hline
\end{tabular}
\end{center}
\caption[Jet Patch Trigger Parameters]{\textbf{Jet Patch Trigger Parameters} - Thresholds applied to candidate jet patch (JP) triggered events in the DSM trees, and applied prescales averaged over the run.}
\label{table:JetPatchTriggers}
\end{table}

The final rate of a given trigger may be set using a prescale, which is designed to whittle down the amount of events that are written out and stored. Only one event is kept from the prescaled number, for instance a prescale of 10 would mean one in every ten candidate events is triggered. Notice in Table \ref{table:JetPatchTriggers} that for higher threshold triggers where jets are produced at a lower frequency than low threshold jets, the prescale is 1 so we keep every triggered event.

The minimum bias trigger is formed by requiring a coincidence of particles, enforced by a timing cut, in the VPDs positioned close to the beam pipe on the east and west ends of the detector. This trigger condition introduces a small bias into the data sample since the event is only kept if there is a coincidence, instead of randomly selecting events based on the prescale as is the case for the zero bias trigger. Table \ref{table:MinBiasTriggers} outlines the minimum bias triggers used in this analysis.

\begin{table}[h]
\begin{center}
\begin{tabular}{ c c c c }
Name & Offline ID & Trigger Condition & Avg. Prescale\\
\hline
\multirow{2}{*}{VPDMB} & 370001  & \multirow{2}{*}{East/West VPD Concidence}  & 16280.5 \\ & 370011 & &105.8\\ 
\hline
\end{tabular}
\end{center}
\caption[Minimum Bias Trigger Parameters]{\textbf{Minimum Bias Trigger Parameters} - Triggering requirement for minimum bias, which is purely a hardware trigger. Note the very high average prescale value here.}
\label{table:MinBiasTriggers}
\end{table}

The varying prescales cause each trigger to sample a different number of events. The constraints placed on the triggers are chosen so that they sample a different kinematic region. These constraints also imply that each trigger records a different luminosity. Table \ref{table:trigLumi} gives the number of events and amount of luminosity recorded by each trigger that will be used for this analysis.

\begin{table}[hb]
\begin{center}
\begin{tabular}{ c c c c }
Name & Offline ID & Luminosity [$pb^{-1}$] & Recorded Events [M] \\
\hline
VPDMB & 370001 & 0.000 & 4.571\\
VPDMB & 370011 & 0.029 & 734.853\\
JP0 & 370601 & 0.195 & 53.835\\
JP1 & 370611 & 9.143 & 180.587\\
JP2 & 370621 & 22.879 & 65.170\\
\hline
\end{tabular}
\end{center}
\caption[Events and Luminosity Recorded by Triggers]{\textbf{Events and Luminosity Recorded by Triggers} - The number of events and luminosity recorded during data collection by the triggers used for analysis.}
\label{table:trigLumi}
\end{table}

\section{Event-Level Quality Analysis}
\label{section:eventQA}
During running, problems may arise due to detector or trigger malfunctions that render a run unusable. The majority of these runs are flagged as bad during running and excluded from analysis, but some problems are unknown during running allowing bad runs to enter the data sample. Before starting an analysis for physics results, all runs should be checked for proper quality.

One method to ensure the quality of the data is to plot the average reconstructed detector response (i.e. track $p_T$, tower $E_T$, vertex position, etc.) as a function of the run number. This can be split up further by plotting for each trigger that may be used later in the analysis. For this top layer quality check no cuts were placed on the data, the only requirement is the event satisfy the various triggering conditions. The physics is same for each run, so the average value is expected to be constant for all runs. Runs which vary from the average value are subject to further investigation.

Figure \ref{BEMC_Et_BeforeQA} is an example of just one of these plots. It shows the average BEMC tower $E_T$ versus run for JP0, JP1, JP2, and VPDMB triggers. Note that results from additional triggers are shown that were dropped for the final analysis. This plot makes it very clear that several runs do not fit the general trend. These obviously bad runs are immediately thrown away as a first pass.

\figuremacroW{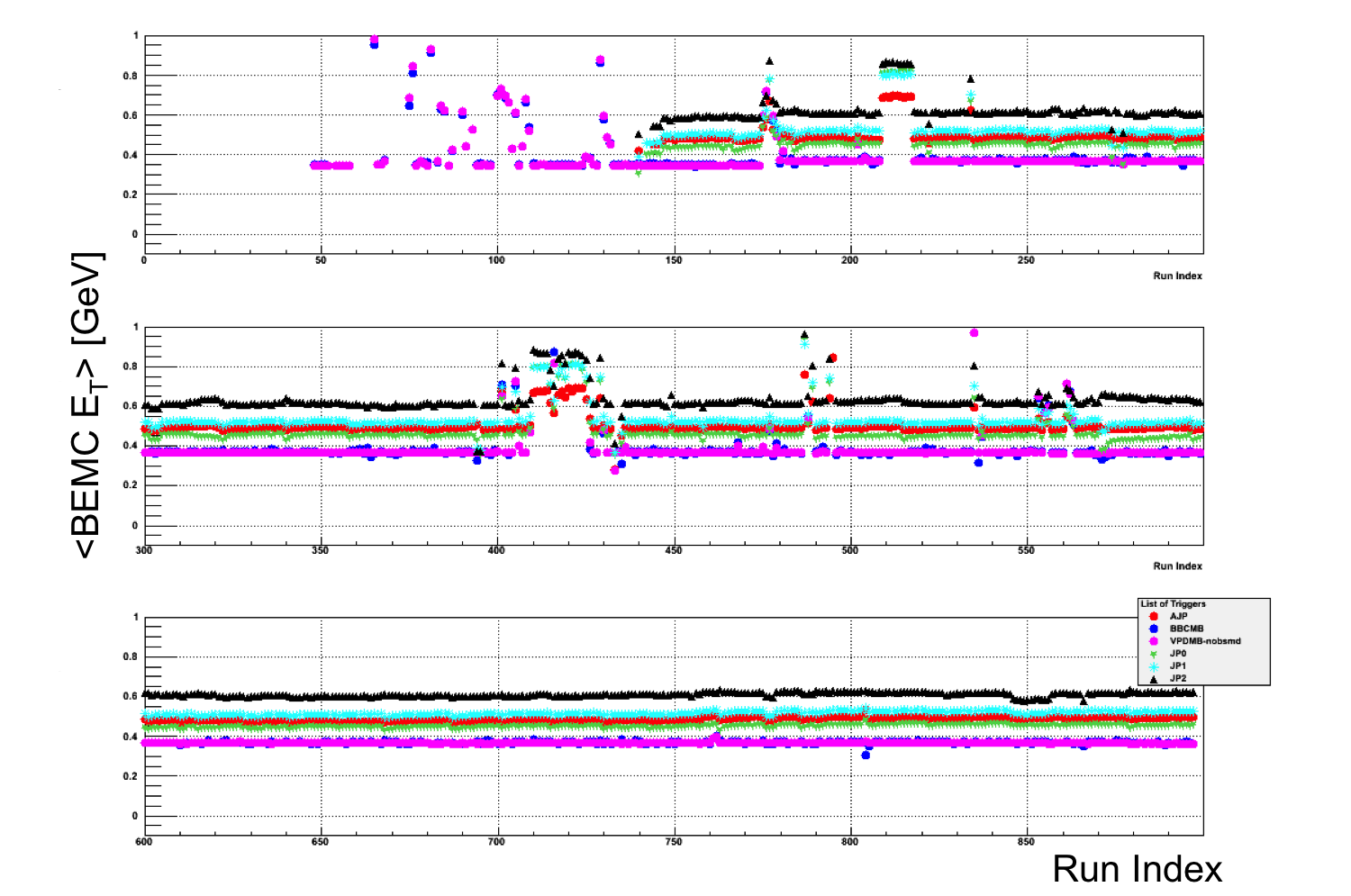}{Average BEMC Tower $E_T$ vs. Run Index}{The average BEMC $E_T$ before removing any runs due to anomalies or inconsistencies. ``Run index'' is a label used to number the runs starting from zero and does not reflect the actual run number.}{1.0}

The runs which are not visible outliers need to be picked out using a more systematic mathematical approach. To do this, the average value plots (with visible outliers removed) for each trigger are fit with a constant polynomial as a function of the ``run index'' to extract the mean ($\mu$) and RMS ($\sigma$). With these parameters it is a simple exercise to pick out those runs which have values that lie outside the interval $\mu \pm 2\sigma$. Each run that falls outside this interval is investigated using the STAR shift log and an array of detector trigger output plots that are produced during real-time running. The decision to keep or discard the run is made based on these notes and plots.

This exercise is repeated for each reconstructed value and for all triggers in the plots until all flagged runs have been reviewed and removed or kept based upon a lack of symptoms to diagnose a true problem. As a result, 320 runs were removed from the original sample because of detector issues, or because the triggers had not yet been properly commissioned as noted in the first 140 run index entries where several triggers have no proper reconstructed values.

\figuremacroW{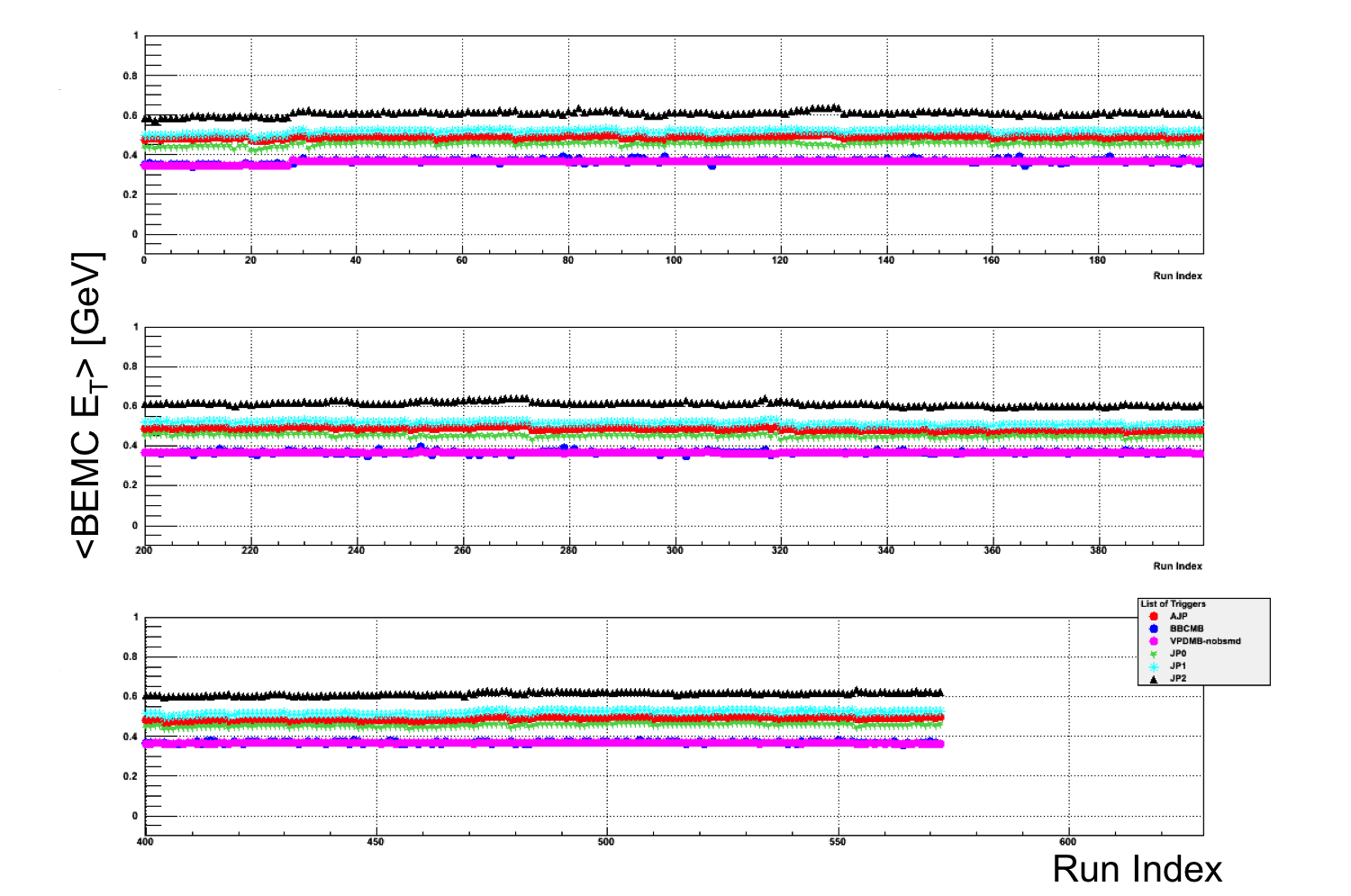}{Average BEMC Tower $E_T$ vs. Run Index}{The average BEMC $E_T$ after removing problematic runs.}{1.0}

Figure \ref{BEMC_Et_AfterQA} shows the result after removing all flagged runs. As a result of the quality analysis checks the average BEMC tower $E_T$ is essentially constant across all runs, which is also true for the other reconstructed values, yielding confidence in the health of the detectors for this sample of runs. 

\section{Jet Reconstruction}
The proton beams delivered to STAR are at sufficiently high energies that partons are elastically scattered from each other and ejected from the parent protons. Because of confinement, these ejected partons quickly fragment and hadronize into jets of colorless particles. Studying properties of the jets reveals information about the primordial partons involved in the hard scattering.

To study the properties of jets, they must be reconstructed from the raw tracks in the TPC and energy deposits in the calorimeters. There are multiple jet algorithms that can be applied as a part of the FastJet package in C$++$ \cite{ref:FastJet}. From these, the anti-$k_T$ jet reconstruction algorithm \cite{ref:AntiKtJets} is used for this analysis.

\subsection{The Anti-$k_T$ Jet Algorithm}
The anti-$k_T$ jet algorithm is a recombination scheme which relies on a single user input jet cone radius parameter R, and the distance between particles and the beam line to define jets. Two distances are calculated in this algorithm between entities in the event. Using the cone radius (R), inverse track momentum ($k_T$), rapidity ($y$) and azimuthal angle ($\phi$), the distance between two tracks is labeled $d_{ij}$ (Eq. \ref{eq:d_ij}). Using only the inverse track momentum, the distance between a track and the beam line is labeled $d_{iB}$ (Eq. \ref{eq:d_iB}).

\begin{gather}
\label{eq:d_ij}
d_{ij} = min\left(\frac{1}{k_{Ti}^2},\frac{1}{k_{Tj}^2}\right)\times\frac{\Delta_{ij}^2}{R^2}\\
\label{eq:d_iB}
d_{iB} = \frac{1}{k_{Ti}^2}\\
\Delta_{ij}^2=\left(y_i-y_j\right)^2+\left(\phi_i-\phi_j\right)^2
\end{gather}

If $d_{ij}$ is the smallest of $d_{ij}$ and $d_{iB}$, the entities $i$ and $j$ are combined into a single entity, and the iteration continues. Otherwise, if $d_{iB}$ is the smallest then $i$ is called a jet and removed from the list of entities. This method continues until there are no entities left in the event. 

In the end, jets are formed from the particles that were clustered together as a result of the minimum distance being $d_{ij}$. The properties of each jet are determined by the E-scheme recombination, meaning the four momentum vectors of all particles in the jet are summed. The resultant four vector is used to define the kinematic properties of the jet. Because this algorithm relies on distances between entities and not on a cone-style seed particle to define jets it is collinear and infrared safe, two important points to be a theoretically favorable algorithm.

Collinear safety means that the jet reconstruction algorithm results in the same output of jets in the presence of collinear splitting of the fragmenting parton. Gluon emission of a fragmenting parton would be an example of this. In this case, a collinear safe algorithm returns the same jets for the two cases:

\begin{itemize}
\item When all of the the energy is carried by the fragmenting parton and deposited into a single detector component
\item When the energy is spread out over the parton plus the emitted gluon, and thus deposited over several detector components
\end{itemize}

\noindent This is an issue with a cone-style algorithm which relies on high energy seed particles to help define the jet axis and begin the jet-finding process \cite{ref:CDFcone}. This is demonstrated in Figures \ref{collinear} and \ref{collinear2}. The anti-$k_T$ jet algorithm is not sensitive to these collinear splittings.

\figuremacroW{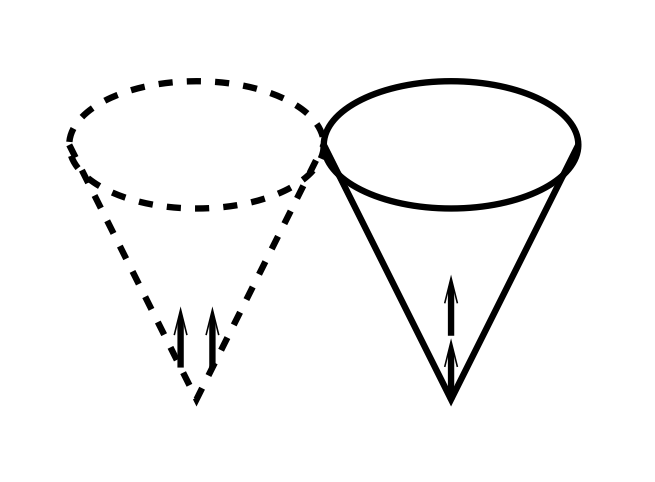}{Collinear Safety Issue}{Multiple low energy particles (left) would fail to produce a seed in a cone-style algorithm, whereas a single high energy particle would (right) \cite{ref:CDFcone}. This issue is resolved with the anti-$k_T$ algorithm where all particles are grouped by their distances from each other and the beam, where the same jet would be produced regardless of energy deposit in these examples.}{0.5} 

\figuremacroW{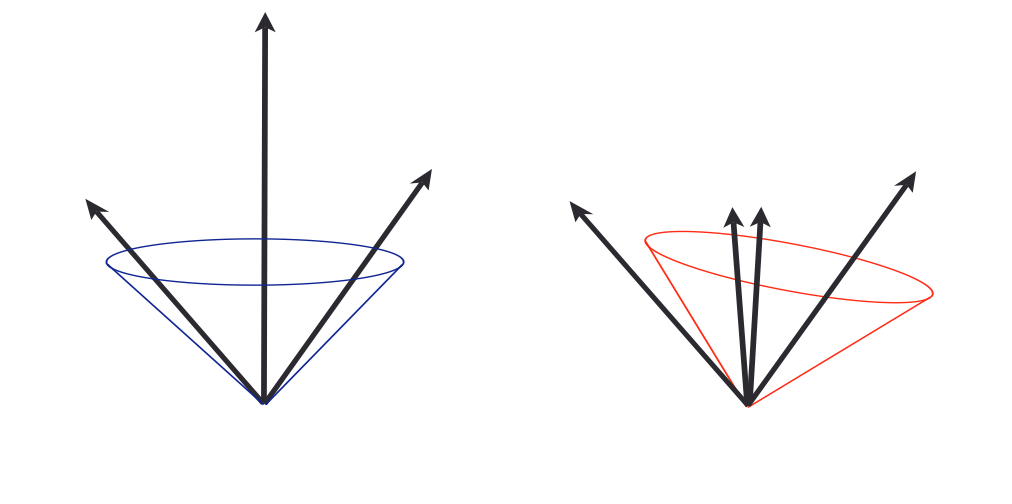}{Collinear Safety Issue}{Jets in a cone-style algorithm are sensitive to particle ordering. The correct jet (left) would not be formed without the high energy seed in the middle. If this seed were rather two lower energy particles (right), the jet would take on a different shape \cite{ref:CDFcone}.}{0.5}

In the context of jet reconstruction, infrared safety refers to how the jets are reconstructed in the presence of soft radiation, or very low energy particles. Cone-style algorithms which form jets around seed tracks may merge two independent jets if a low energy particle exists in the vicinity of two seeds, as demonstrated in Figure \ref{infrared}. Soft radiation does not adversely affect the anti-$k_T$ algorithm since all particles in the event are treated equally until all have been grouped into a jet. Thus jets are formed the same with or without soft radiation.

\figuremacroW{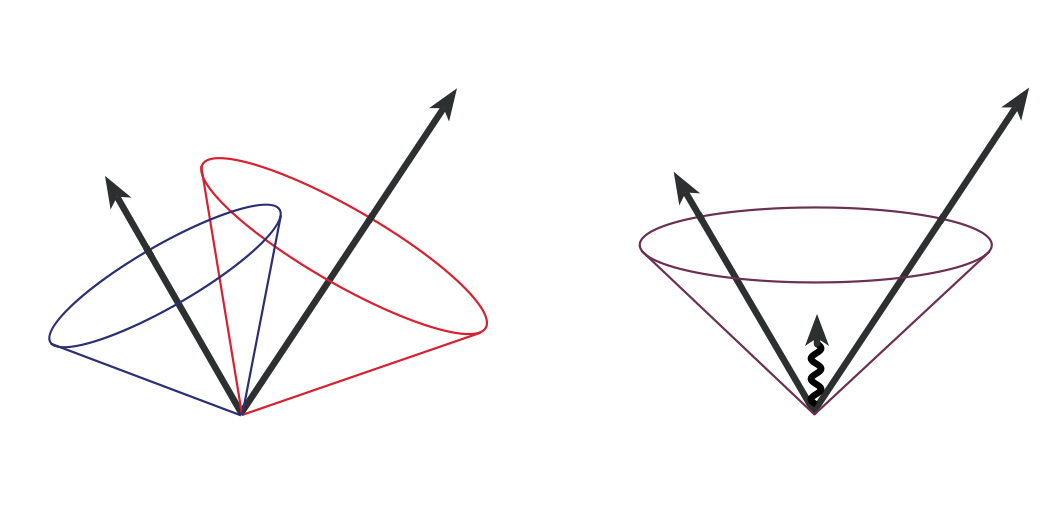}{Infrared Safety Issue}{Two jets correctly formed around two seed tracks (left) may be formed differently in the presence of soft radiation (right) when using a cone-style algorithm. This low energy particle could cause the jets to have enough in common that they are merged rather than saved as two independent jets \cite{ref:CDFcone}. For the anti-$k_T$ algorithm, the measured distance between particles is what associates particles with jets. Thus, for this case, the anti-$k_T$ algorithm would cause one of the initial two particles to absorb the soft radiation still resulting in two jets.}{0.5}

\subsection{Reconstruction Parameters}
As mentioned in the previous section, the anti-$k_T$ algorithm requires a single input parameter, R, that describes the cone radius of each jet so the algorithm knows how big the cone should be to group tracks. For this analysis, the reconstructed jet radius is set to $0.6$. 

In addition to the radius parameter, the cuts summarized in Table \ref{table:JetFinderCuts} are placed on the tracks and towers used in jet reconstruction. To reduce the chance of the reconstructed tracks being ``split tracks'', or just part of a real track, it is required that the number of hits registered on the TPC padrows is greater than 51\% of the possible hits for that track, where a hit is the crossing of a TPC pad. To reduce background and pile-up tracks, a distance of closest approach (DCA) between the closest TPC hit on the track to the primary reconstructed vertex is applied depending on the track $p_T$. The DCA must be within 2 cm for track $p_T$ < 0.5 GeV/c and within 1 cm for track $p_T$ > 1.5 GeV/c. For track $p_T$ values between these cutoff limits, the DCA must satisfy a linearly decreasing cut. The last hit in the TPC padrows can be a maximum of 125 cm from the center to avoid historical issues with the outer pads.

The offline status for each tower must be good, or set to the value 1. This means that during data collection the tower was acting properly and did not have any issues. The offline tower status values are calculated from real data on a RHIC fill-by-fill basis at the end of each run. To ensure the hits in the towers are true physics hits, and not from the pedestal, the difference between the tower ADC and pedestal should be greater than four, and it should also be greater than three times the pedestal width. Finally, to prevent double counting of tracks pointing to towers, a subtraction scheme is applied. This means, after a track is used, 100\% of its energy is subtracted from the matching tower energy so it is not used again. If this causes the tower energy to be negative, then the energy for that particular tower is set to zero.

\begin{table}[h]
\begin{center}
\begin{tabular}{ |c | c| }
\hline
\textbf{Cut} & \textbf{Value}\\
\hline
Track $N_{hits}/N_{possible}$ & > 0.51\\
\hline
Distance of Closest Approach (DCA) to vertex & Track $p_T$ dependent\\
\hline
Track $p_T$ & 0.2 GeV/c < $p_T$ < 200 GeV/c\\
\hline
Track $\eta$ & $-2.5 < \eta < 2.5$\\
\hline
Radius of last track fit point & R > 125 cm\\
\hline
Track ``flag'' & > 0\\
\hline
Tower Subtraction Scheme & 100\% \\
\hline
Offline EMC tower status & 1\\
\hline
EMC tower $ADC-pedestal$ & > 4\\
\hline
EMC tower $ADC-pedestal$ & $>3*RMS_{pedestal}$\\
\hline
EMC tower $E_T$ & > $0.2$ GeV\\
\hline
Jet $p_T$ & 5 GeV/c < $p_T$ < 200 GeV/c\\
\hline
\end{tabular}
\end{center}
\caption[Anti-$k_T$ Jet-Finding Cuts]{\textbf{Anti-$k_T$ Jet-Finding Cuts} - Jet reconstruction cuts applied to the data as the jets are identified.}
\label{table:JetFinderCuts}
\end{table}

Running the anti-$k_T$ algorithm over each run of raw data with these additional cuts results in files that contain information about the jet properties as well as the tracks and towers they contain. Another round of quality checks should be performed as jets can bring out detector defects more clearly than the raw tracks alone.

\section{Jet-Level Quality Analysis}
\label{section:JetQA}
When looking at jets, there are further plots beyond those used for event level quality analysis that can be used to check the quality of the jet sample. These include the number of jets per reconstructed interaction vertex, number of towers per jet, and number of tracks per jet. These can be useful when trying to identify calorimeter towers which are hot, meaning they are registering hits that cannot be attributed to a real signal. Hot towers can be a problem in a jet analysis as several jets could be reconstructed about a hot tower that do not have true tracks pointing to it.
 
 \figuremacroW{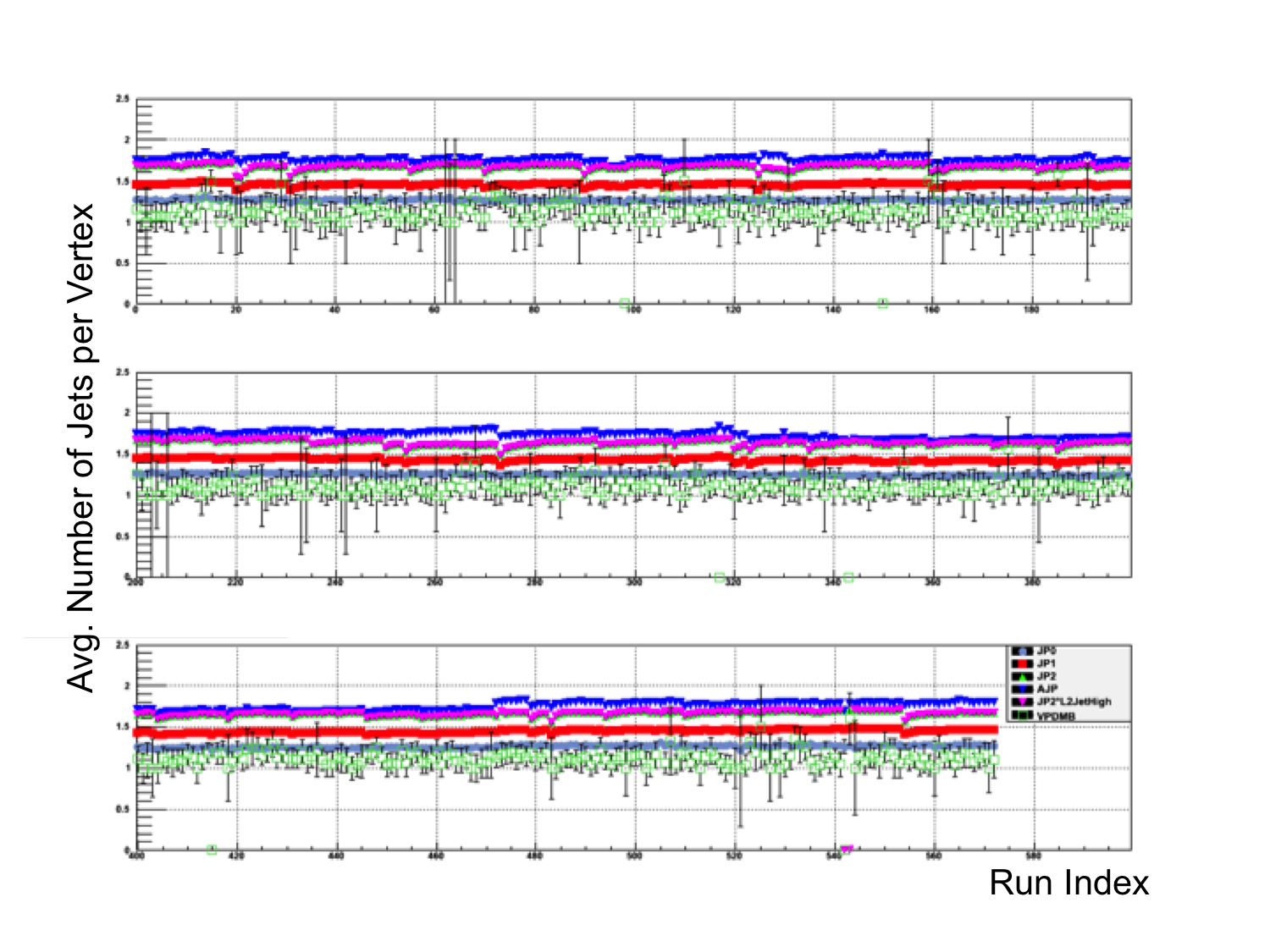}{Number of Jets per Vertex vs. Run Index}{The average number of jets per reconstructed vertex for each analysis trigger before removing any runs due to anomalies or inconsistencies.}{1.0}
 
Figure \ref{NumJets_BeforeQA} is a plot of the number of jets per reconstructed vertex before any runs were removed from the full sample of jets. In total, 70 runs were removed based on the same method outlined in Section \ref{section:eventQA}. The run-by-run values are already essentially constant as a result of a thorough event level quality check, leaving the visually bad values for further investigation. As a result of all quality checks there are a total of 503 good physics runs as shown in Figure \ref{NumJets_AfterQA}. However, 21 of these runs have missing polarization measurements in the database and are therefore dropped since polarization will be a necessity when performing the asymmetry analysis presented in this thesis.

 \figuremacroW{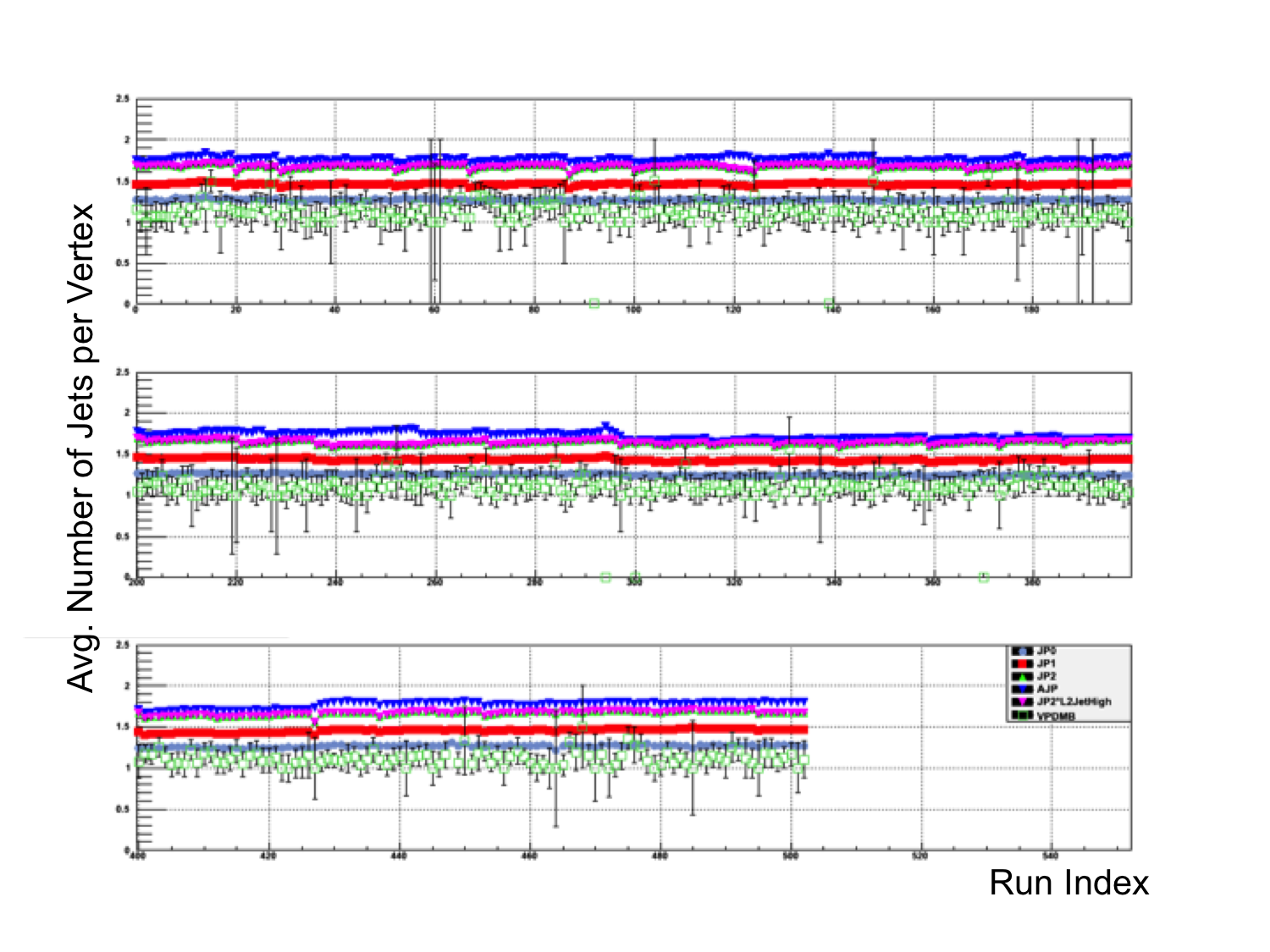}{Number of Jets per Vertex vs. Run Index}{The average number of jets per reconstructed vertex for each analysis trigger after the 70 flagged runs were investigated for problems and subsequently removed. }{1.0}

Of all the runs removed in both jet and event level quality analyses, 196 of them were removed because of apparent hot towers in the calorimeters. It was concerning that such a large part of the sample showed similar behavior, so further investigation was pursued. This investigation led to a new tower-by-tower calibration of the BEMC, which will be discussed at length in Chapter \ref{chapter:BemcCalibration}. As a result of this calibration and subsequent analysis of these removed runs, 140 runs were added back into the runlist, however 21 of these did not have updated database entries which include pedestal values for the BEMC towers and thus were deemed unreliable. Therefore, the final analysis run list includes 601 runs. These run numbers are documented in the appendix for completeness.



\chapter{Calibration of the Barrel Electromagnetic Calorimeter} 
\label{chapter:BemcCalibration}

\ifpdf
    \graphicspath{{ch4_BemcCalibration/figures/PNG/}{ch4_BemcCalibration/figures/PDF/}{ch4_BemcCalibration/figures/}}
\else
    \graphicspath{{ch4_BemcCalibration/figures/EPS/}{ch4_BemcCalibration/figures/}}
\fi

It was revealed in Section \ref{section:JetQA} that 196 runs were removed from the final run list as a result of event and jet level quality analysis checks due to towers which exhibited traits of being hot in the BEMC. This is a considerable amount of the initial data sample. The origin of this problem should be understood, including whether these runs really have hot towers or if there is a more fundamental underlying issue.

\section{Towards a New Calibration}
The first step towards gaining deeper insight into this problem is to identify those towers which are problematic. This is a simple exercise of counting the number of tower hits with transverse energy above a low tower threshold, such as 2 GeV, in each of the 4800 BEMC towers. This is demonstrated in Figure \ref{HotTowerSpectrum} where it is clear that there are several towers registering a lot of hits. 

The next step is to pick a criterion to identify the towers which are registering more hits than surrounding towers. This should be done by calculating the mean number of hits across all towers, and then selecting the ones which have many more hits than this average. However, the towers which could dramatically skew the average up need to be thrown away and excluded from the average counts calculation. This is done by identifying the towers which carry more than 20\% of the total counts in the histogram, where this threshold is set so that only the huge spikes in tower hits are above it. These towers are added to a list of bad ones to investigate, and they are excluded from the average value calculation. The average is calculated using the remaining towers in the histogram, and those which have 10 times more hits than the average are also added to the list of bad ones.

\figuremacroW{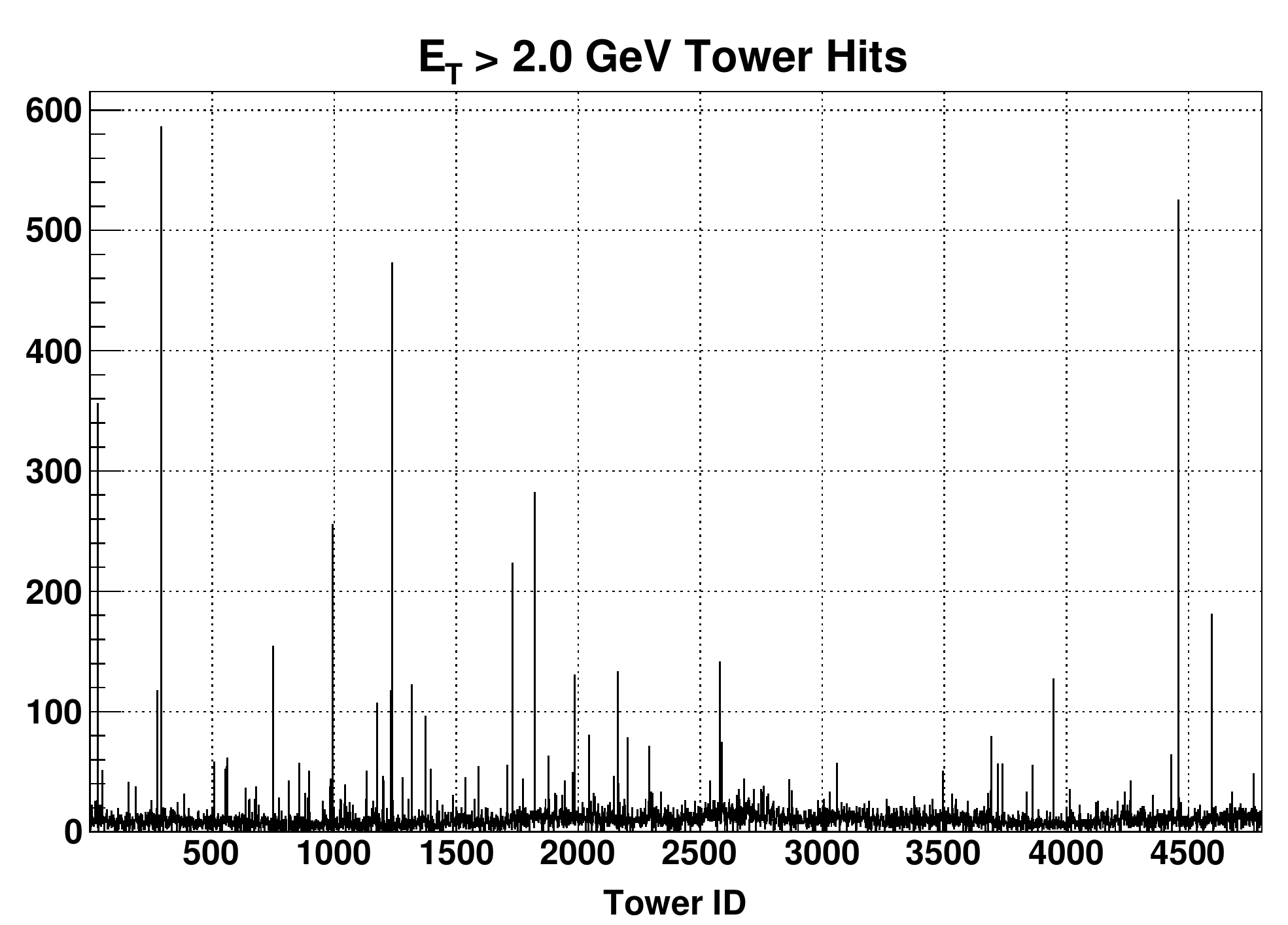}{BEMC Hot Tower Spectrum}{Number of hits in each BEMC tower that carry a transverse energy greater than 2 GeV. The spikes which have many more counts than the average number of hits are identified as hot tower candidates.}{0.75}

With these bad towers identified, they are analyzed further. The most promising method to bring problems to the foreground is to compare the ADC spectrum from a tower in the list of bad ones to a tower which wasn't tagged as bad. To compare, though, the towers should be at the same value of $\eta$ but a different value of $\phi$ so the physics is the same assuming rotational invariance. Also, both towers should have the same gain constant ($C_{gain}$) so they are making an identical conversion between ADC counts and energy. ADC values are converted to energy by the relationship:

\begin{equation}
\label{eq:GainDef}
E=\left(ADC-pedestal\right)*C_{gain}
\end{equation}

An example ADC comparison is shown in Figure \ref{SpectrumCompare}, with the bad tower spectrum given in blue and the good tower spectrum in green. The spectra are normalized to their respective total integral and are shifted by their respective pedestal value so the pedestal peaks are at zero (the red line at zero is for reference) for direct comparison. It is clear that the pedestal peaks line up well, but the physics slope is different for each tower. Both towers have the same gain constant, so the normalized spectra should look identical. Since they do not have the same slope, it is believed that the identified bad tower is not actually hot, rather there is an issue with its gain calibration constant.

\figuremacroW{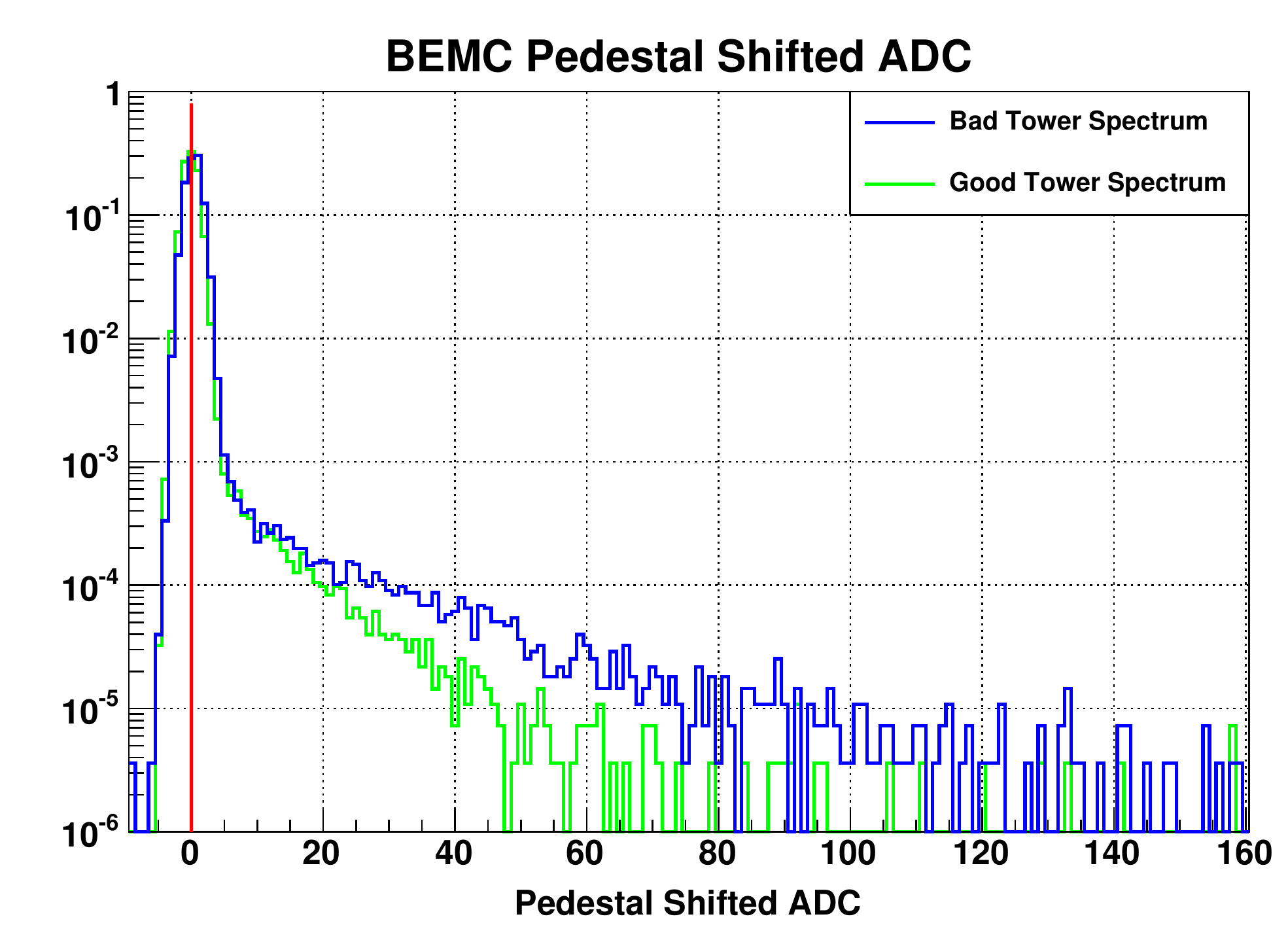}{BEMC Tower Spectrum Comparison}{Pedestal shifted ADC spectra for an identified bad, hot tower spectrum (blue) and for a good tower spectrum (green) that is in the same $\eta$ ring, but at a different value of $\phi$. The good and bad tower both have the same gain constant value ($C_{gain}$), thus the deviation in the physics slope is unexpected and shows that the gain constants should be updated.}{0.75}

Before 2012, the most recent gain calibration of the BEMC was completed using data collected in 2009. The tower efficiency is assumed to decrease by approximately $1.5\%$ each year, so the calibration constants should be updated to take this efficiency into account.

\section{Calibration Algorithm: The Big Picture}
The best case scenario for a calibration of the BEMC would have numerous light, charged particles whose mass may be safely ignored at high energies, such as electrons, detected in each tower. The momentum will be known well from the TPC, so the ratio of energy to momentum (E/p) can be calculated by measuring the energy deposited in the tower from each electron. For light particles any deviation in this ratio from one tells how much the tower ADC should be adjusted by a calibration constant to be properly calculating the energy.

Unfortunately, STAR does not see this many electrons so other avenues must be pursued. One such method is to use minimum ionizing particles (MIPs), which are particles that only deposit a minimum amount of their energy into a calorimeter tower when they pass through and are very plentiful. In fact, there are enough MIPs that they can be used to calculate a relative calibration constant for each tower. These are relative constants because they do not directly use the energy of the MIPs, rather they are calculated using a characteristic ADC value ($ADC_{MIP}$), and \textit{corrected} afterward using groups of electrons. Equation \ref{eq:relCalConst} gives the relationship to calculate the relative calibration constant for each tower \cite{ref:2009calibReport}. In this relationship, $\eta$ is the tower pseudorapidty and $\theta$ is the tower polar angle. An in depth treatment of the MIP analysis will be given in the next section.

\begin{equation}
\label{eq:relCalConst}
C_{relative}=\frac{0.264}{ADC_{MIP}}\frac{1+0.056\eta^2}{\sin\theta}
\end{equation}

The final step of the calibration is to take the relative calibration constants which are not directly dependent upon the tower energy, and adjust them based upon the measured tower energy for identified electrons. As previously stated, the ratio E/p should be one for electrons. Any deviation in the ratio from one tells how much the relative calibration constants should be adjusted, and once adjusted they are the final absolute gain constants needed for ADC to energy conversion. Since there are not enough electrons for a tower-by-tower analysis, they must be grouped effectively to be statistically useful and represent the physics well. Therefore they must be grouped at the same value of $\eta$, which means they may be grouped into an entire $2\pi$ $\eta$ ring that includes 120 towers. Or they may be grouped into so-called crate slices which are eight towers at the same value of $\eta$ and in the same electronics crate. Crate slices give a better resolution to the calibration constants, as there are 15 crate slices for each $\eta$ ring. Once grouped, the E/p spectrum will be a Gaussian spread about one, and may be fit to extract the mean. Equation \ref{eq:absoluteCalConst} shows how this mean is used to adjust the relative calibration constants to attain the absolute gain constants. More specifics of the electron analysis will be given in a later section.

\begin{equation}
\label{eq:absoluteCalConst}
C_{absolute}=\frac{C_{relative}}{\left<\frac{E}{p}\right>}
\end{equation}
\section{Relative Calibration with Minimum Ionizing Particles}
The ultimate goal of the MIP relative calibration is to fill histograms with the pedestal subtracted ADC value for each identified MIP. These histograms are then fit to extract the characteristic MIP value and apply Equation \ref{eq:relCalConst} to each tower. 

\subsection{Developing the MIP Spectra}
Several cuts are applied to tracks in each event to identify them as MIPs. Table \ref{table:mipcuts} outlines these cuts. Some of them apply to a $3\times 3$ cluster, where the central tower is the one that is being analyzed to identify MIPs and fill the ADC histograms. The neighboring towers are used for isolation cuts on the central tower, so there are no large energy deposits in the surrounding towers. MIPs do not deposit enough energy in the calorimeter towers to cause a trigger to fire, so no triggering requirement is enforced for this part of the calibration analysis.

\begin{table}[htp]
\begin{center}
\begin{tabular}{|c|c|}
\hline
\textbf{Track Cut} & \textbf{Value} \\
\hline
Vertex rank & $>10^6$ \\
\hline
Vertex z-position & -30 cm $ \leq v_z \leq$ 30 cm\\
\hline
Number of tracks per central tower & 1 \\
\hline
Track momentum & $p>1$ GeV/c\\
\hline
Tower ID & $ID_{enter}=ID_{exit}$\\
\hline
Neighboring ADC in 3x3 cluster & $(ADC-pedestal)<2\times RMS_{pedestal}$\\
\hline
Central tower ADC & $(ADC-pedestal)>1.5\times RMS_{pedestal}$\\
\hline
Triggering requirement & None\\
\hline
\end{tabular}
\end{center}
\caption[MIP Identification Criteria]{\textbf{MIP Identification Criteria} - MIP identification cuts applied to tracks event-by-event.}
\label{table:mipcuts}
\end{table}

An example tower MIP spectrum is given in Figure \ref{MipSpectrum} which results after all cuts have been applied. It is clear that there is a defined peak around $ADC\approx 16$, this is the peak that will be fitted to extract the characteristic ADC value for this tower. There are also two bins close to zero that have several counts. These bins are pedestal counts from the original ADC spectrum that passed all applied cuts and snuck into the MIP spectrum. These counts will be ignored when fitting the spectrum for the characteristic ADC value.

\figuremacroW{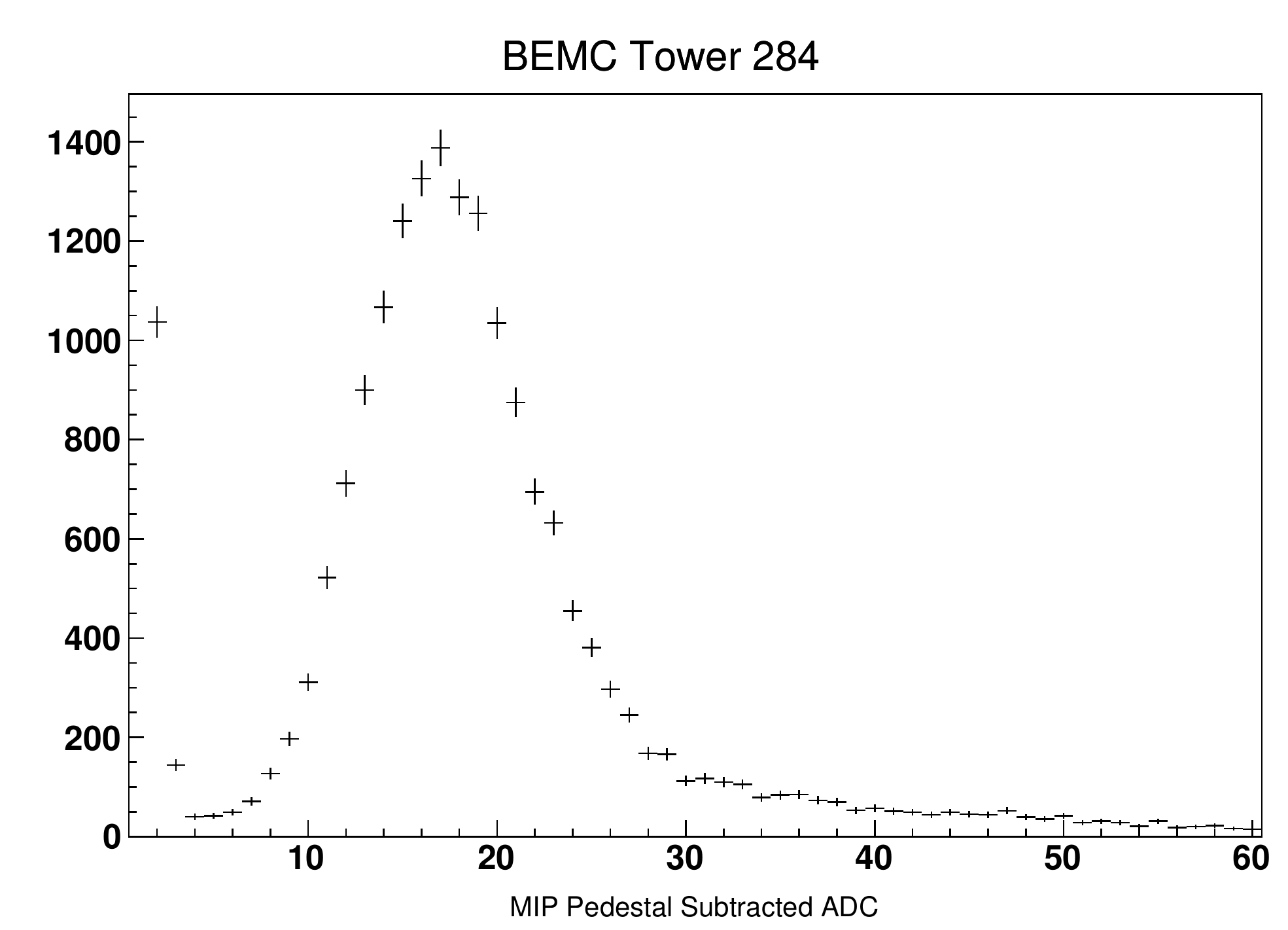}{MIP ADC Spectrum}{Resulting MIP ADC spectrum after all cuts from Table \ref{table:mipcuts} have been applied.}{0.75}

\subsection{Fitting the MIP Spectra}
With the MIP spectrum for each calorimeter tower in hand, it is necessary to identify the characteristic ADC value to be used in the relative gain constant calculation. At this point there are two viable methods to extract the ADC value: 

\begin{enumerate}
\item Fit the spectrum across a narrow range so the peak is very well known, and use the peak value as the ADC value.
\item Develop a fit which describes a very broad range very well, and calculate the mean of this fit across the range of the histogram as the ADC value.
\end{enumerate}

Unfortunately, the first method ignores the underlying physics in the spectrum, such as backgrounds which behave in an unknown way. These backgrounds may shift the peak around, yet the characteristic ADC will remain constant. In other words, the peak is not always the best characterization of the ADC spectrum. However, the second method does take these backgrounds into account by fitting a more broad range to characterize the physics of the MIPs. Then using this fit as a template to compute the mean across the whole range of the histogram - while appropriately ignoring pedestal counts - describes the characteristic ADC value very well even with the unknown backgrounds. It turns out there is very little difference between calculating the mean over the range of the fit versus calculating the mean over the range of the histogram using the fit as a template yielding confidence in the method.

With the method understood, a fitting function which best fits the data is the final piece of the puzzle to complete the analysis. The MIP spectra are not exactly Gaussian as the peak is asymmetric, which is clear in Figure \ref{MipSpectrum}. In fact, the MIP spectra should contain both Gaussian statistics from the photons in the photo tubes, as well as Landau statistics from MIPs passing through the calorimeters. The Landau distribution has an asymmetric peak with a long high energy tail, as with these MIP spectra \cite{ref:PdgReview,ref:LeoBook}. The fit should contain both pieces for the best fit possible to a wide area. 

It turns out that the best fit to the data results from the product of Gaussian and Landau distributions, with five free parameters. These parameters are an overall vertical scale parameter, Gaussian mean and $\sigma$ values, and Landau most probable value and $\sigma$ parameters. These parameters are initialized in ROOT \cite{ref:ROOT} using the mean and width of the MIP distribution, as well as the height of the bin with the most counts not including the residual pedestal bins. This initialization does not fix any of the parameters, rather it gives ROOT a good starting point for adjusting the parameters for the best possible fit to the data. An example fit is shown drawn in blue in Figure \ref{MipFit}, with the calculated mean drawn as a vertical red line.

\figuremacroW{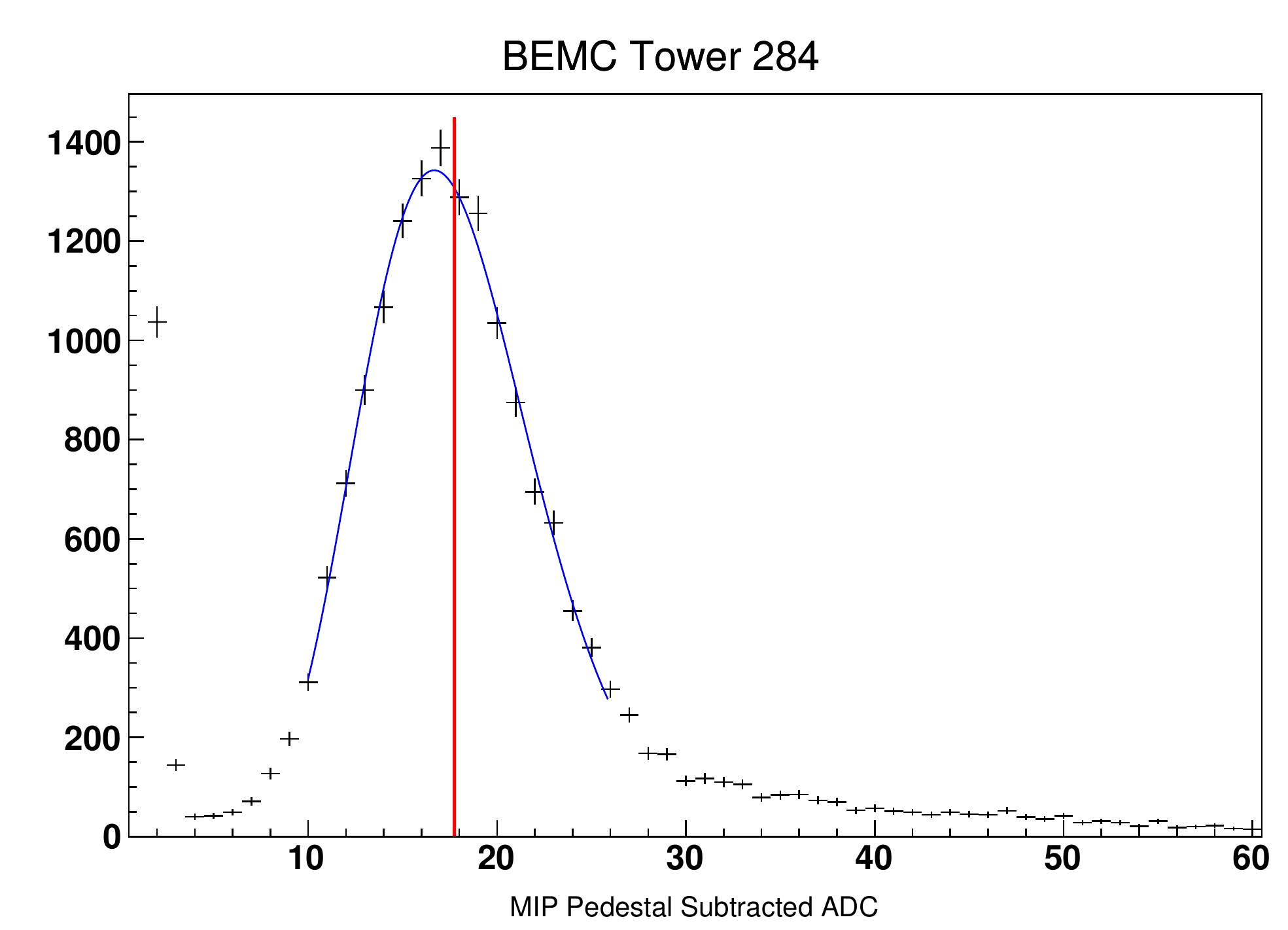}{MIP Spectrum Fit}{Fit applied to the MIP spectrum is drawn in blue. The red line gives the mean, which is used to calculate the MIP relative calibration constant in Equation \ref{eq:relCalConst}.}{0.75}

The same fit is applied to the MIP spectra from all 4800 calorimeter towers. After fitting, each spectrum and fit are reviewed one-by-one to ensure every tower has a good fit and MIP spectrum. Those towers with a bad fit or spectrum are flagged as bad, and have a zero calibration constant into the database, and are given a bad ``status''. This provides an extra layer of security against using bad towers in data analysis, as the gain statuses are checked in analysis code, and bad status towers are skipped. The means and statuses from each tower are stored and passed on to the electron calibration algorithm.

\section{Absolute Calibration with Electrons}
The final gain calibration constants are obtained by shifting the relative constants according to Equation \ref{eq:absoluteCalConst}, with the electrons grouped into rings and crate slices to maximize statistics. First the same procedure of developing the E/p spectra and fitting them must be applied before correcting the relative constants.

\subsection{Developing the E/p Spectra}
\label{section:EoverPspectra}
The same sample of data is analyzed with a different set of cuts applied to pick out electrons instead of MIPs, as outlined in Table \ref{table:electronCuts}. Some cuts again apply to a $3\times 3$ cluster formed around a central tower, which is the one under analysis. The surrounding towers are known as neighboring towers. The tower energy is calculated using the MIP relative calibration constants in conjunction with Equation \ref{eq:GainDef}, used both in the cuts and in filling the E/p spectra.

\begin{table}[htp]
\begin{center}
\begin{tabular}{|c|c|}
\hline
\textbf{Track Cut} & \textbf{Value}\\
\hline
Vertex rank & $10^6$\\
\hline
Vertex z-position & -60 cm $\leq v_z \leq$ 60 cm\\
\hline
Track momentum & 1.5 GeV/c $< p <$ 10 GeV/c\\
\hline
Track dE/dx & $3.5(10^{-6}) <$ dE/dx $< 5.0(10^{-6})$\\
\hline
Track $n_\sigma\left(\pi\right)$ & \textbf{Exclude} $-1< n_\sigma\left(\pi\right) < 2.5$\\
\hline
Track $n_\sigma\left(e\right)$ & $-1< n_\sigma\left(e\right) < 2$\\
\hline
$\Delta R$ between track and tower & $\Delta R \leq 0.02$ \\
\hline
Offline tower status & 1\\
\hline
MIP status & 1\\
\hline
Track number of TPC hits & $\geq 10$\\
\hline
Tower ID & $ID_{enter} = ID_{exit}$\\
\hline
Number of tracks in central tower & 1\\
\hline
Central tower ADC & $(ADC-pedestal)>2.5\times RMS_{pedestal}$\\
\hline
Number of neighboring tracks & 0\\
\hline
Maximum cluster $E_T$ & Must reside in central tower\\
\hline
Triggering requirement & See section \ref{sec:electronTriggering}\\	
\hline
\end{tabular}
\end{center}
\caption[Electron Identification Criteria]{\textbf{Electron Identification Criteria} - Electron identification cuts applied to tracks event-by-event.}
\label{table:electronCuts}
\end{table}

In these cuts the so-called fiducial cut is defined by a $\Delta R$ term. This is defined as the distance between the center of the tower and where the track hits the tower:

\begin{equation}
\Delta R =\sqrt{\left(\eta_{track}-\eta_{tower}\right)^2+\left(\phi_{track}-\phi_{tower}\right)^2}
\label{eq:towerTrackDr}
\end{equation}

\noindent The cuts reference the values $n_\sigma\left(\pi\right)$ and $n_\sigma\left(e\right)$, defined as the $\log\left(\frac{dE}{dx}\right)$ distribution divided by the expected mean and $\frac{dE}{dx}$ resolution for pions and electrons \cite{ref:ShaoParticleId}. These are used as particle identification cuts, where the $n_\sigma\left(e\right)$ cut is used for identifying tracks as electrons in conjunction with the true $\frac{dE}{dx}$ value cut. For those tracks which are truly pions misidentified as electrons and satisfy the electron cuts show up as backgrounds in the E/p spectra. The $n_\sigma\left(\pi\right)$ cut rejects pion tracks and helps cut down on backgrounds that will add to the augmentation of the E/p peak.

\subsection{Triggering for the E/p Spectra}
\label{sec:electronTriggering}
In addition to the data cuts applied to each track or tower, there are triggering requirements that must be met before the electron's information is added to the E/p spectra. Smaller sized ROOT trees are created from the original data files, and only information of interest is stored. This includes the triggers that are of interest for this analysis, which are outlined in Table \ref{table:BhtTriggers}. These are called ``high tower'' triggers, meaning that only one calorimeter tower in the barrel must be above a given threshold for the trigger to fire. In the trees, triggering information is stored if it fired either in the hardware or in the offline simulator. Otherwise nothing is stored to help keep the size of the files down.

\begin{table}[htb]
\begin{center}
\begin{tabular}{ c c c c c }
& & \multicolumn{2}{c}{Thresholds} & \\ \cline{3-4}
Name & Offline ID & DSM ADC & $E_T$ & Avg. Prescale\\
\hline
BHT0*VPDMB & 370501 & 11 & 1.4 GeV & 7.387\\
BHT1*VPDMB & 370511 & 15 & 2.4 GeV & 1.003\\
BHT2 & 370531 & 18 & 3.1 GeV & 1.0\\
BHT2*BBCMB & 370521 & 18 & 3.1 GeV & 1.0\\
BHT2*BBCMB & 370522 & 18 & 3.1 GeV & 1.0\\
\hline
\end{tabular}
\end{center}
\caption[High Tower Trigger Parameters]{\textbf{High Tower Trigger Parameters} - Thresholds applied to candidate barrel high tower (BHT) triggered events applied in the DSM trees. The prescales averaged over all runs is also given.}
\label{table:BhtTriggers}
\end{table}

In any triggered sample, a great concern is how the triggers are affecting the final result. One such trigger effect is demonstrated in Figure \ref{EoverP_vsP_HT}, where a clear momentum dependence of E/p is observed, and the spectrum is broken in the vicinity of the trigger threshold for tracks that satisfy the BHT2 trigger. It would be for the best if this dependence could be avoided, for which there are two options.

\figuremacroW{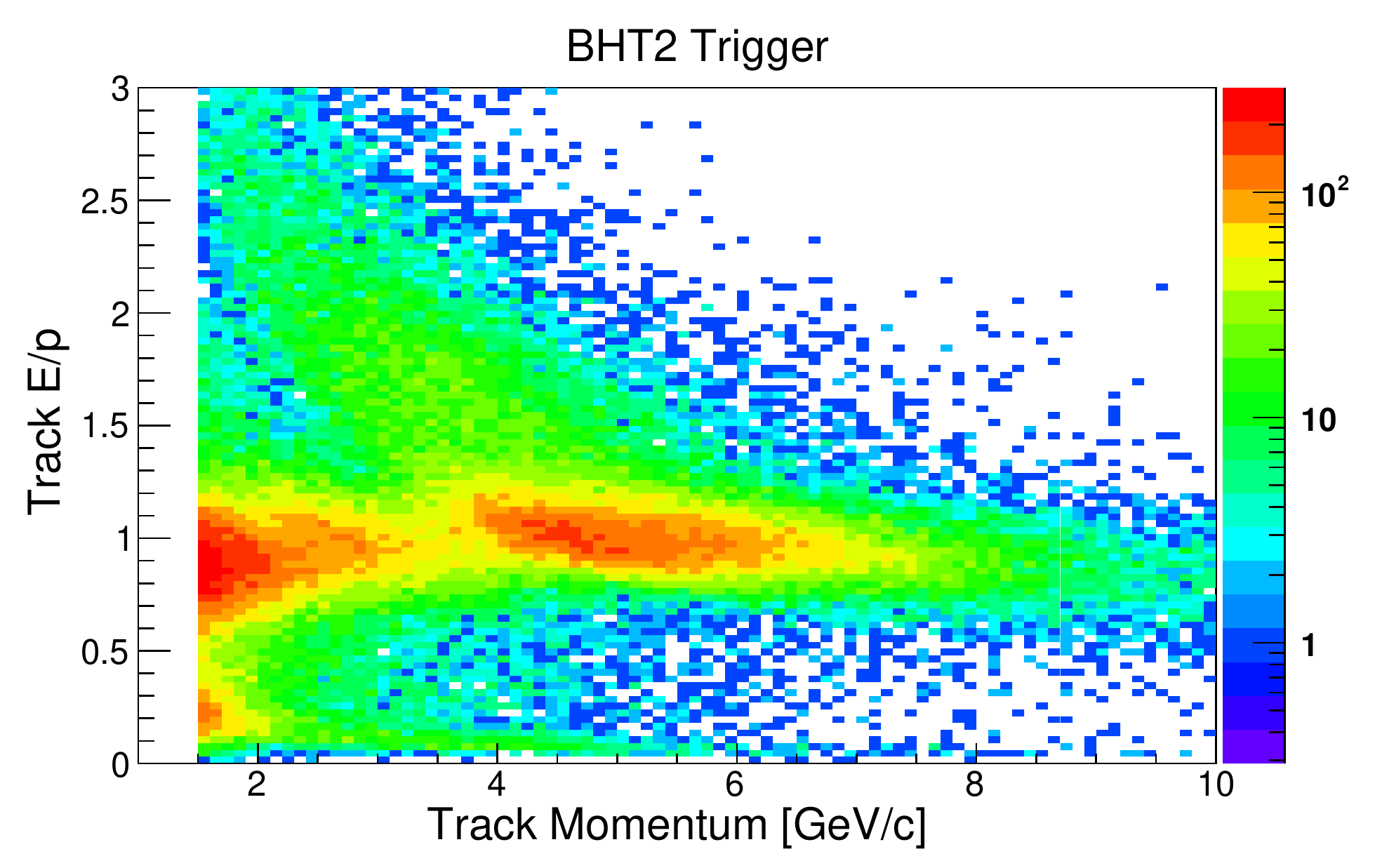}{BHT2 E/p vs. Track Momentum}{This shows the momentum dependence of E/p in the BHT2 triggered events. Also, it is clear to see where the trigger ``turns on'' around 3.5 GeV/c.}{0.75}

First, the BHT2 trigger could be used, so long as the other two lower BHT triggers did not fire, at high enough momentum that it is certain the events are no longer being influenced by the trigger threshold. Then, for the lower momentum tracks, an ``unbiased'' triggered sample may be used. For this, events are chosen that are triggered by ``something else''. The way the ROOT trees were set up, this means the event should satisfy one of two conditions:

\begin{enumerate}
\item No high tower triggers were stored in the event. This means none of them fired in the hardware or software, and thus the event is assumed to be triggered on something else, such as a jet patch.
\item The triggering information was stored, but the hardware trigger did not fire, which means the software trigger did fire. This removes the obvious hardware trigger biases and momentum dependence around the threshold found in Figure \ref{EoverP_vsP_HT}.
\end{enumerate}

Applying this so-called unbiased triggering condition, the result is shown in Figure \ref{EoverP_vsP_Unbiased}. The momentum dependence present in the high tower spectrum is now gone, so the conditions seem to work as planned. The only remaining decision is how to apply the momentum condition. Looking at small momentum slices of E/p separately for both triggering scenarios in Figures \ref{EoverP_HTProj} and \ref{EoverP_UnbiasedProj}, it is clear that the two samples give consistent peaks if the BHT2 events are used above 5 GeV/c in momentum, and the unbiased events are used below this mark. This is important since the statistics from the two triggered samples will be combined together, so the peaks should not be shifted drastically from each other.

\figuremacroW{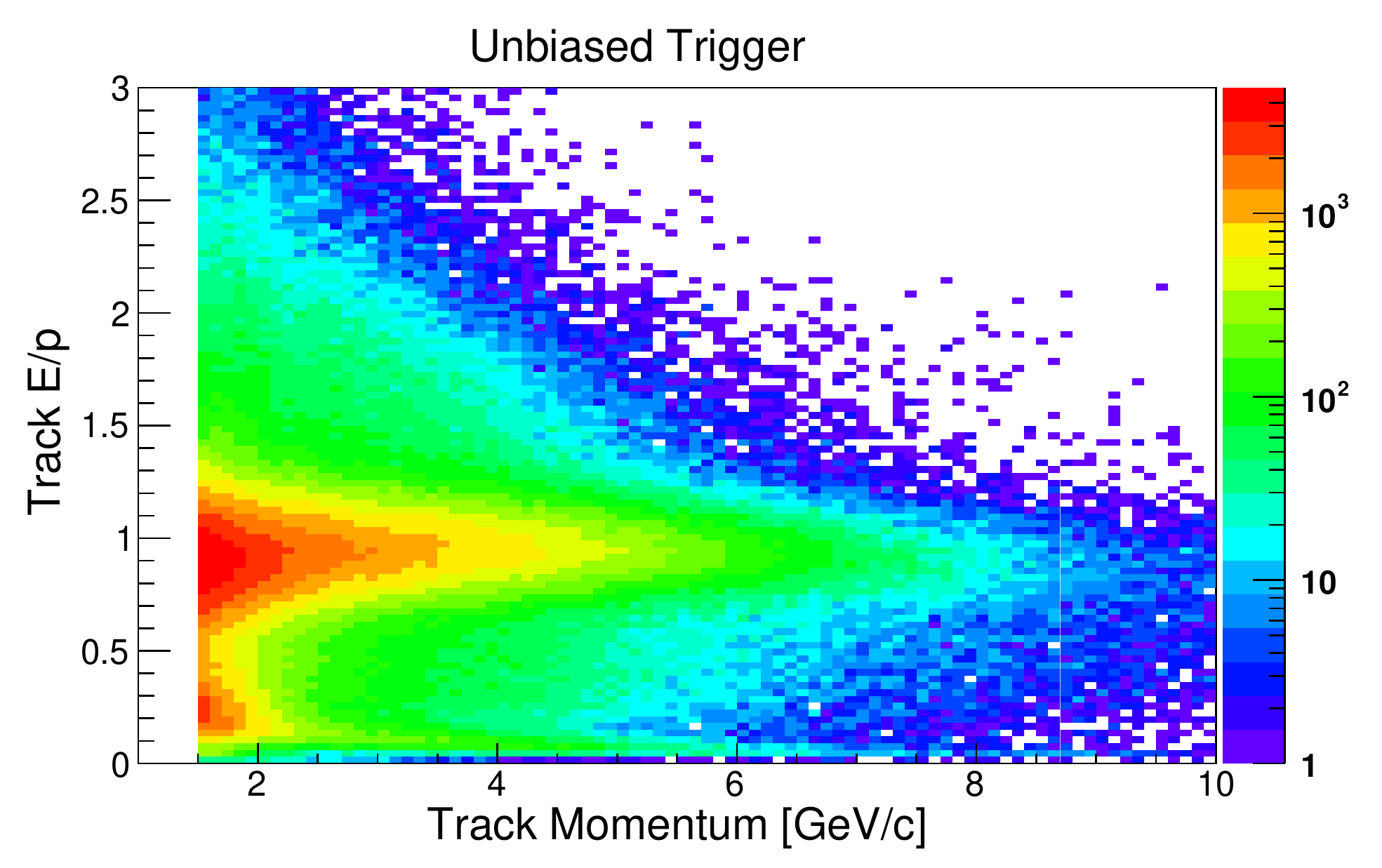}{Unbiased E/p vs. Track Momentum}{The momentum dependence is removed if we look at the so-called unbiased trigger. The backgrounds at low E/p are much bigger in this triggering setup, though.}{0.75}

The unbiased triggering scheme will be applied for track momenta less than 5 GeV/c. Above this momentum, the slices develop a background shoulder at low E/p below the peak. These increased backgrounds will be difficult to incorporate into the fit, and could shift the peak. Looking at the BHT2 momentum slices, the backgrounds at low E/p and threshold effects at high E/p are gone above 5 GeV/c, where this triggering scheme will be applied. Even though the statistics are obviously less in the BHT2 scheme than in the unbiased scheme, a more accurate and clean peak is worth the statistical loss.

\figuremacroW{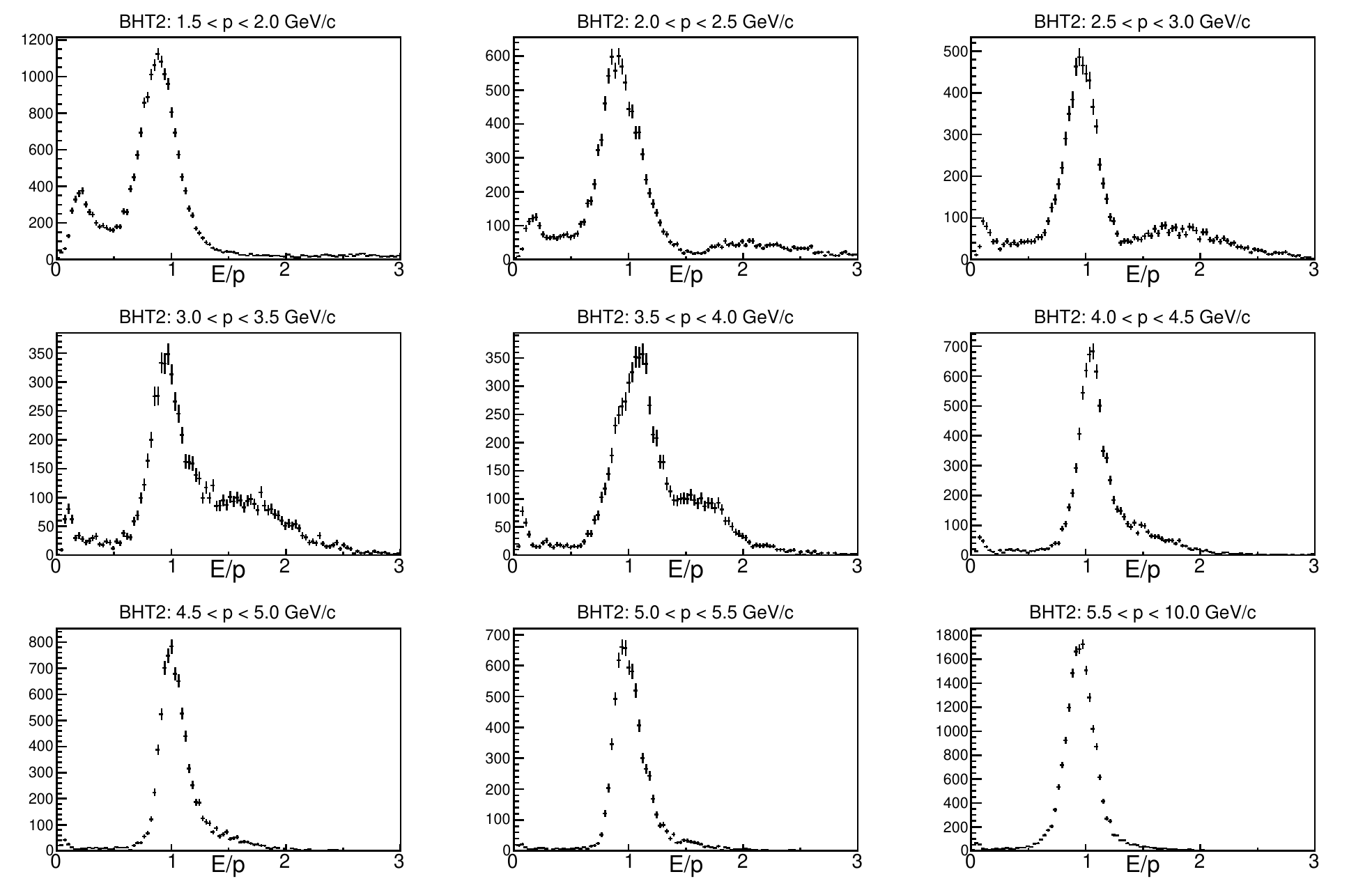}{BHT2 E/p Momentum (p) Slices}{Above 5 GeV/c, the threshold effects should be gone, and the shoulder at higher E/p has disappeared.}{1}

\figuremacroW{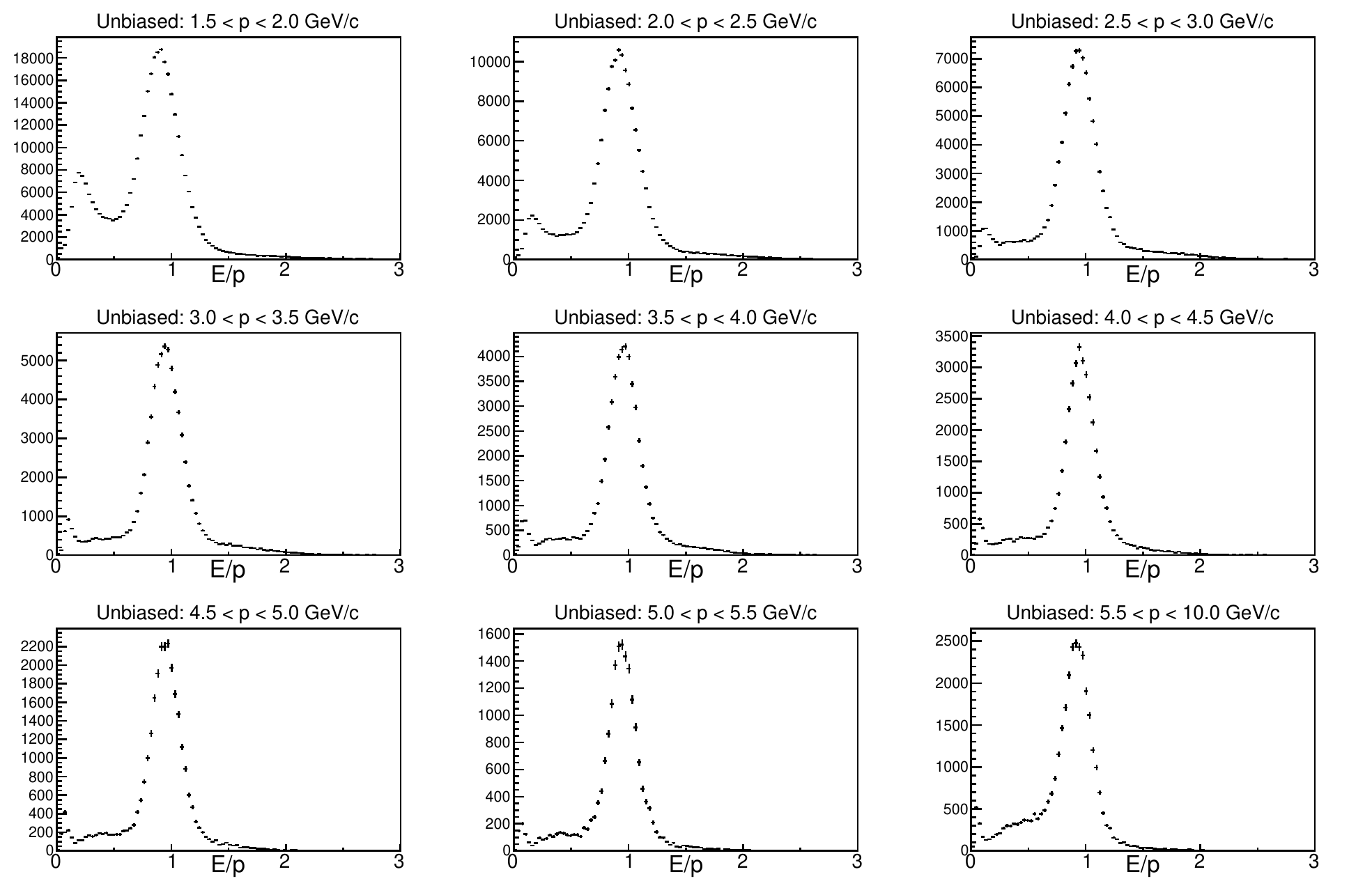}{Unbiased E/p Momentum (p) Slices}{Below 5 GeV/c the peaks are consistent with the peaks in the BHT2 spectra when that momentum is above 5 GeV/c. In these momentum slices, an odd background shoulder below the peak is developing for momentum above 5 GeV/c, which will be difficult to characterize.}{1}

Summing it up, the following two triggered samples will be used for the calibration:

\begin{enumerate}
\item If the track momentum is less than 5 GeV/c, the unbiased trigger condition will be applied. This means that either the event was triggered by something other than a high tower trigger, or any high tower trigger fired in the simulator but not in the hardware to avoid those biases.
\item If the track momentum is greater than 5 GeV/c, the BHT2 trigger will be applied. This means that the BHT0 and BHT1 triggers didn't fire in either the hardware or software.
\end{enumerate}

The results of these triggering conditions are shown in Figure \ref{RingEoverPSpectra} for a midrapidity ring. The statistics for the unbiased scheme are much greater than those in the BHT2 plot, but the backgrounds are also larger. The BHT2 plot shows a much cleaner and exact peak, which will be easier to fit and extract the mean E/p value.

\figuremacroW{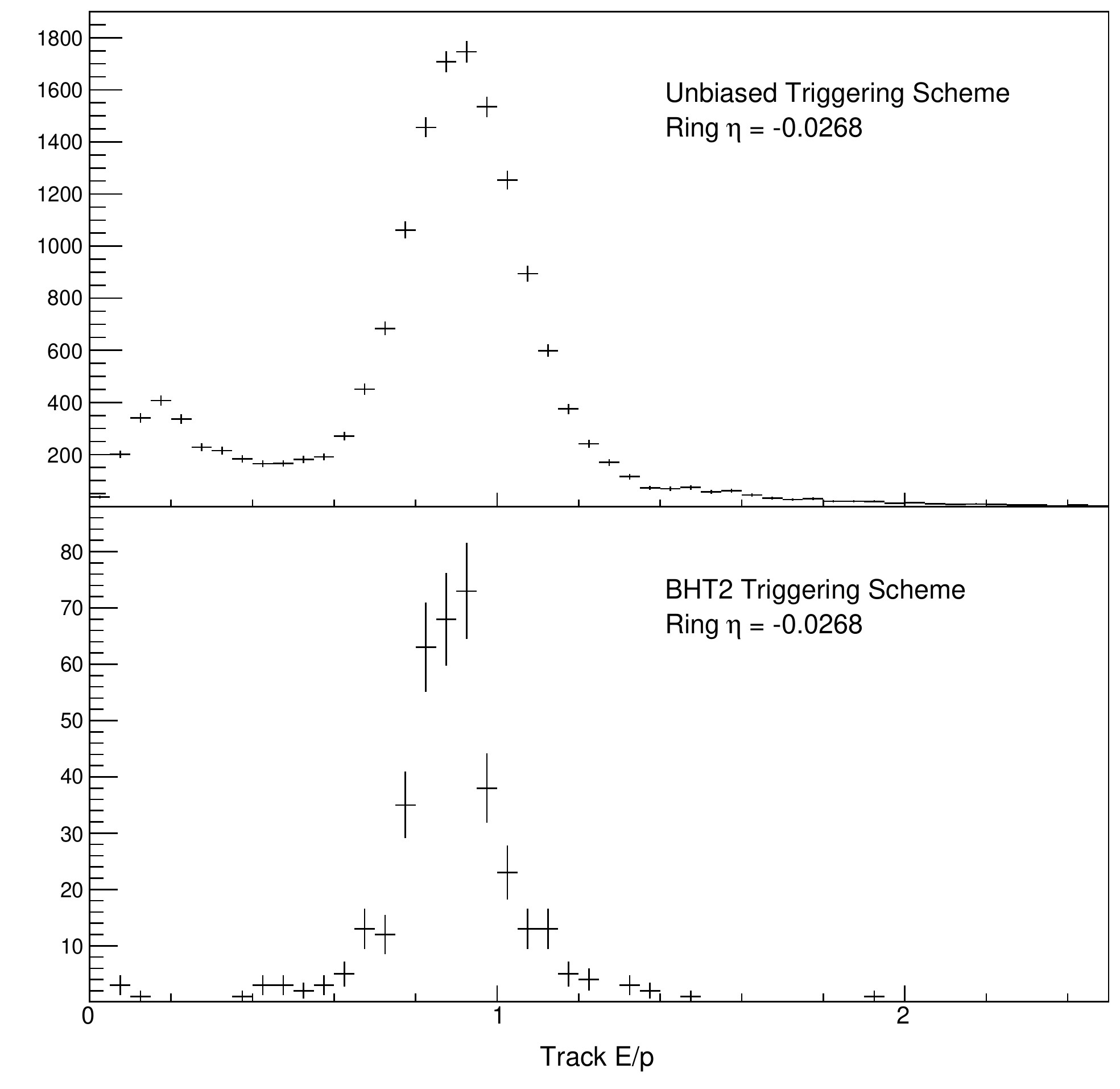}{Sample Ring E/p Spectrum}{E/p spectra for the different triggering scenarios once all cuts have been applied, shown here for a midrapidity ring. It is clear that the unbiased spectrum (top) has way more statistics than the BHT2 spectrum (bottom), but exhibits a much more prominent background that will need to be fit correctly. }{0.75}

An unfortunate circumstance of applying the unbiased trigger condition is that it is not possible to know if events were truly triggered by something other than a high tower trigger, since only high tower trigger information was stored. It is assumed, though, that every event that does not satisfy any high tower trigger is likely triggered by something else. Recreating the ROOT trees with additional trigger information vastly increases the tree file size, so rather than try to find space to store the larger, updated trees, a quick test was completed using the original source files. The test is looking at all events from a subsample of runs, still applying the same cuts listed in Table \ref{table:electronCuts}, and counting the number of times a high tower trigger fires when nothing else fires, a type of event that could sneak into the unbiased sample that should be thrown away. This will give an estimate to how big of an effect errors in the unbiased triggering sample would be, and whether or not a tree overhaul is truly necessary. The test showed that 2.5\% of the time a high tower trigger fires by itself. This is quite small, and will be folded into the systematic error assigned to the calibration constants.

The final comparison is to understand how the mean in the unbiased triggered sample compares to the high tower sample ring-by-ring. If the two scenarios show a dramatic difference for a particular ring or set of rings, then the statistics should not be combined together for those and the best triggering scheme should be chosen. This will be shown in the next section when the fitting procedure is discussed.

\subsection{Fitting the E/p Spectra}
Much like with the MIP analysis, the fit for the E/p spectra must characterize both the peak and the backgrounds. But in this case, since the calorimeter completely stops the electrons the only statistical counting is photons in the photo tubes which follow Poisson statistics, and approaches Gaussian statistics for large sample sizes. Thus, in this analysis the mean of the Gaussian piece of the fit is the value which will be extracted to augment the relative calibration constants. Also, the backgrounds are clear outside of the peak, and can be characterized with a background specific fit, which contrasts the MIP fit logic. 

The E/p fitting procedure is a three step process:

\begin{enumerate}
\item Fit the backgrounds with across a broad range of E/p.
\item Fit the Gaussian peak, not including the background shape.
\item Use the first two fits as seeds to a combined fit that fits both the backgrounds and Gaussian peak at the same time. The mean of the Gaussian part of this final fit is used as the $\left<E/p\right>$ value in Equation \ref{eq:absoluteCalConst}.
\end{enumerate}

In 2009, the backgrounds were fit with a $1^{st}$ order polynomial, which worked out well for those statistics. With the higher statistical precision in the 2012 run, this was no longer adequate. Several functions were tested, with the best results coming from an exponential fit to the backgrounds. This, and the other pieces of the fit, are applied to the spectra in Figure \ref{RingEoverPSpectra} and are shown in Figure \ref{RingEoverPFit}. The nice feature of this background fit is that it works whether the backgrounds are present, as with the unbiased spectrum, or not present as in the BHT2 spectrum. It follows the backgrounds well, and leads to an overall fit that follows the data very well.

\figuremacroW{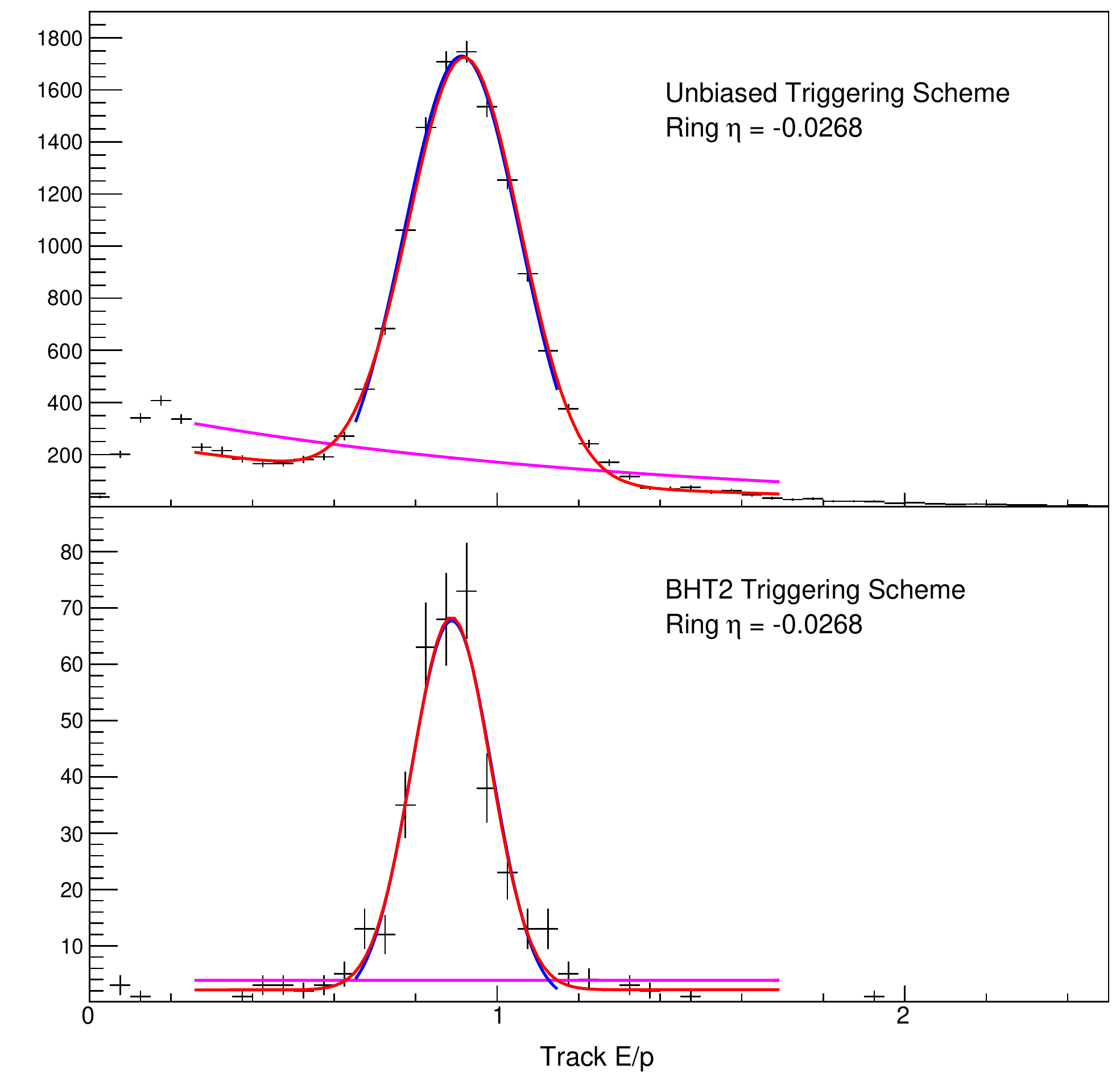}{Sample Ring E/p Fits}{Example fits applied to the previously shown E/p spectra. The pink curve is the exponential fit applied to the backgrounds, and the blue curve is the Gaussian-only fit applied to the peak. The combination of these two, the final fit, is shown as the red curve.}{0.75}

With the fitting procedure finalized, the means from the two triggering schemes may be compared. The fit is applied to each ring E/p spectrum for both unbiased and BHT2 triggering, and the Gaussian mean is extracted. These are shown in Figure \ref{RingFitMeans}, where it is clear that the BHT2 and unbiased agree at midrapidity. As $\lvert\eta\rvert$ approaches 1, the mean values begin to strongly disagree. In the rings where the means disagree, the outer most 7 rings on both ends of the barrel, the BHT2 means seem to have the most stable values. Therefore, only the BHT2 E/p spectra will be used on the outer 7 rings on each end. For the remaining rings at midrapidity, the statistics from the unbiased and BHT2 triggering schemes will be combined.

\figuremacroW{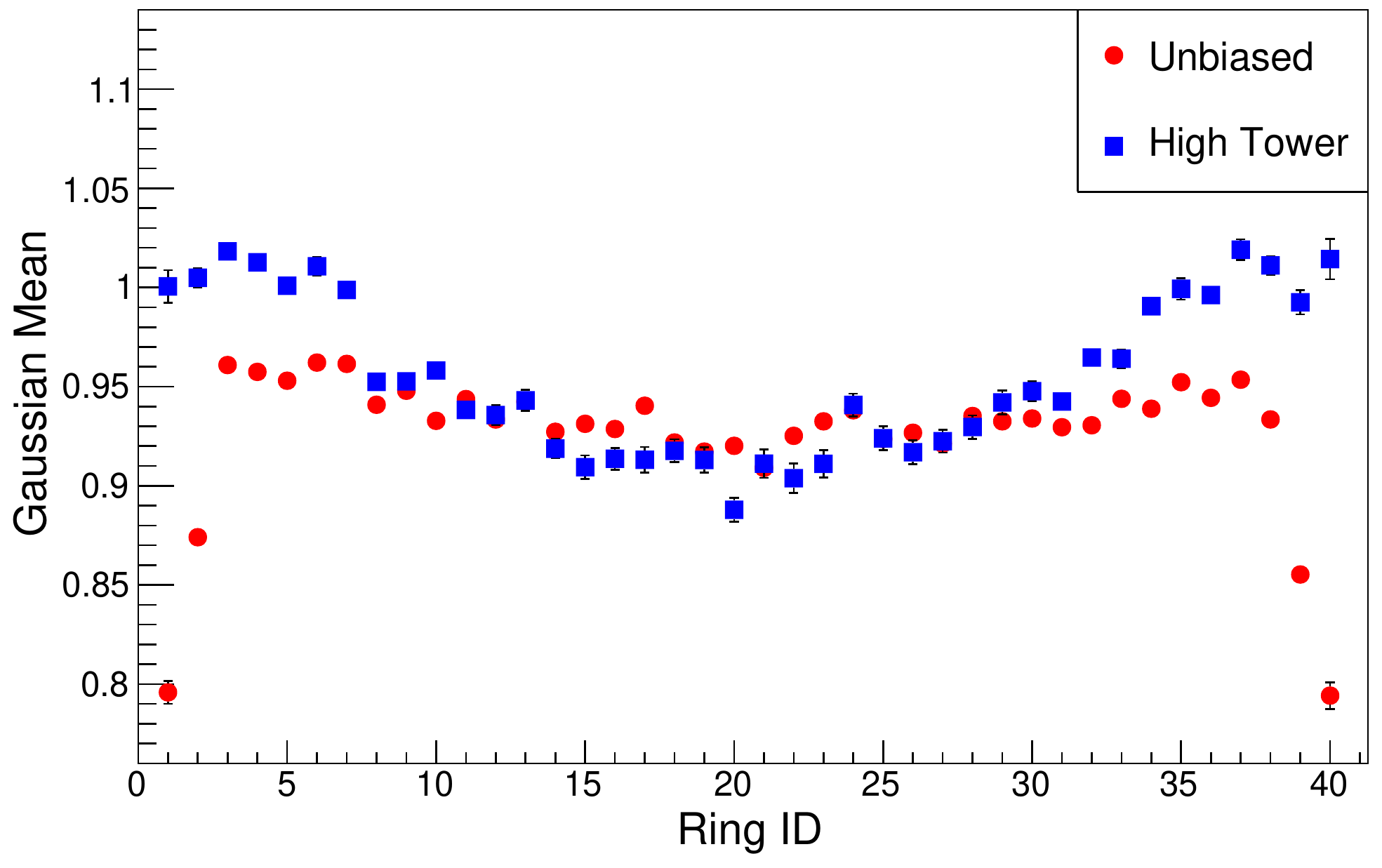}{Triggering Fit Mean Comparison}{Comparison between the unbiased fit means and the BHT2 fit means. The ``Ring ID'' value is indexed so that it increases as $\eta$ goes from $-1$ to $+1$. The highest values of $\lvert\eta\rvert$, where the mean values diverge, are known as the ``outer rings'' of the BEMC.}{0.75}

The final fitting procedure is applied and example rings are shown in Figure \ref{CombinedRingEoverPFit}, one where the statistics are combined, and one where they are not on an outermost $\eta$ ring. At midrapidity, the statistics are sufficient to split further into crate slices, where the resolution of the calibration is much more fine as the relative constants will be divided by several different $\left<E/p\right>$ values from the same ring. On the outer rings, the statistics are not good enough for this, and the relative constants from each tower in the ring will be divided by the same $\left<E/p\right>$.

\figuremacroW{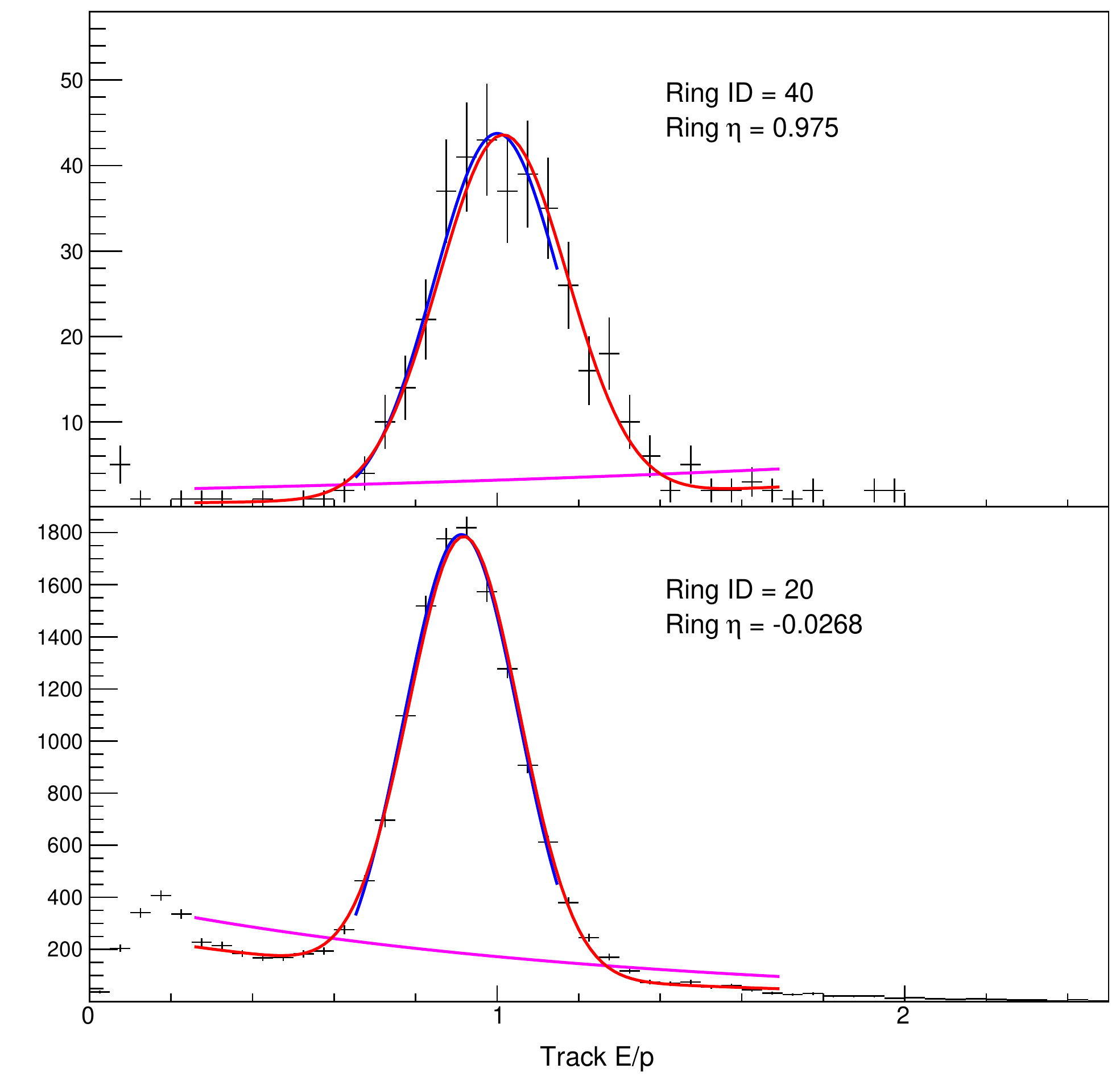}{Final Triggering Fit Sample}{The final triggering scheme and fits applied to outer ring (top) and midrapidity (bottom) towers.}{0.75}

\section{Comparing Gain Constant Results}
With the calibration completed for the 2012 data, it should be compared to the 2009 calibration at the same beam energy to see if the gains fit the expectations of $4.5\%$ yearly tower efficiency degradation. The percent difference is plotted in Figure \ref{GainPercent}, and it is obvious from the mean that the 2009 calibration constants are larger than the updated 2012 gains. This is incorrect, as smaller gains means smaller energy measured in the calorimeters. The 2012 constants should be larger than the 2009 for decreasing tower efficiency. 

\figuremacroW{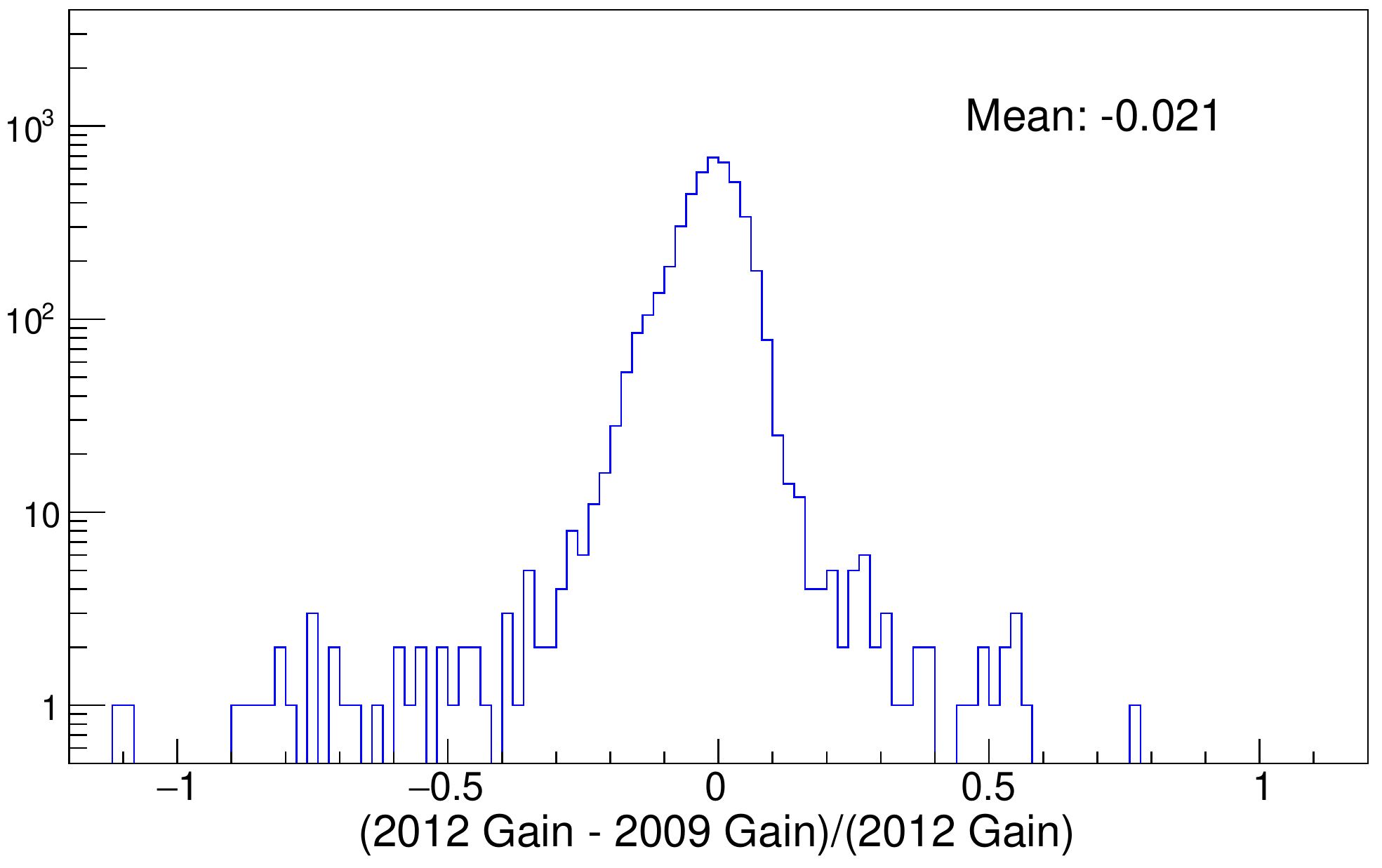}{Percent Change Between Calibrations}{This shows the size of the change between the 2012 and 2009 gain calibration constants.}{0.75}

The explanation for the decrease in the constants is easy to understand, yet troubling at the same time. In the 2009 calibration algorithm, the triggering was vastly different. In that analysis, only high tower triggers were used, and tracks were rejected from analysis if they pointed to the tower which fired the high tower. This is problematic because then the E/p values are biased low, forcing the $\left<E/p\right>$ to also be too small. If the mean is small, then the absolute gain constants calculated with Equation \ref{eq:absoluteCalConst} will be too large. 

Figure \ref{RingFitMeans2009}, shows the unbiased and high tower fit means extracted when the 2012 data and cuts are used, but the 2009 high tower rejection algorithm is restored. It is immediately apparent that there is a difference between the unbiased and high tower triggering Gaussian means, with the high tower means being much lower than the unbiased means. This clearly shows why the 2009 gains are larger than the 2012 gains, because the triggering logic biased the $\langle E/p\rangle$ values too low.

\figuremacroW{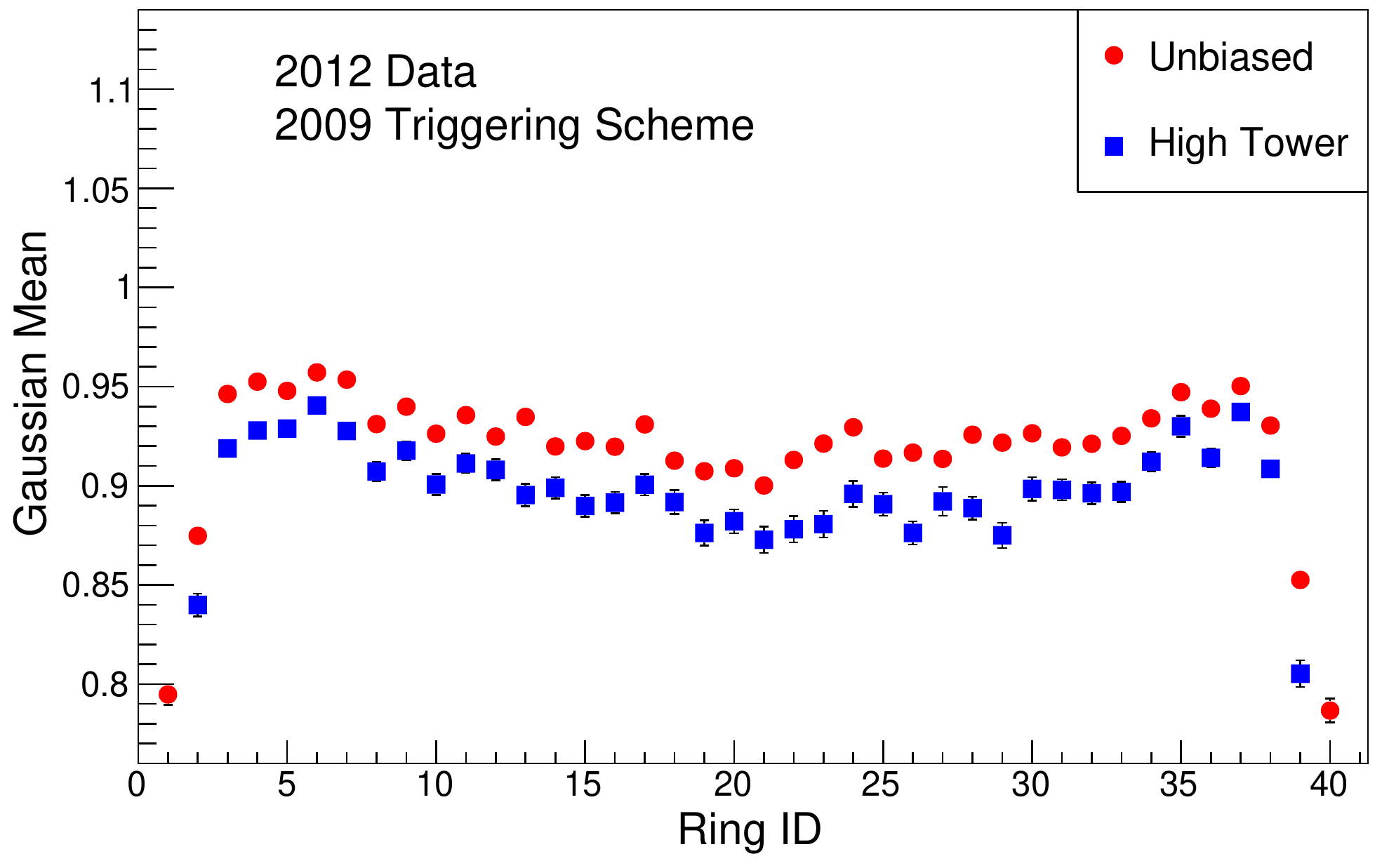}{2009 Triggering E/p Fit Means}{The unbiased and high tower Gaussian means from the 2012 data, but restoring the 2009 triggering algorithm.}{0.75}

\section{Post-Calibration Quality Analysis}
With such a small change in the calibration constants, the differences in the energies will be minimal. Therefore the issue with the hot towers before are not solved. Regardless, another round of quality analysis with updated calibration constants was performed using the 196 runs removed during the first round of data quality checks. The procedure is the same that was outlined in Section \ref{section:eventQA}, and the results are shown in Figure \ref{BEMC_Et}. In the end 147 runs were passed on to the jet level quality analysis, with the majority of the runs removed being obvious outliers.

\figuremacroW{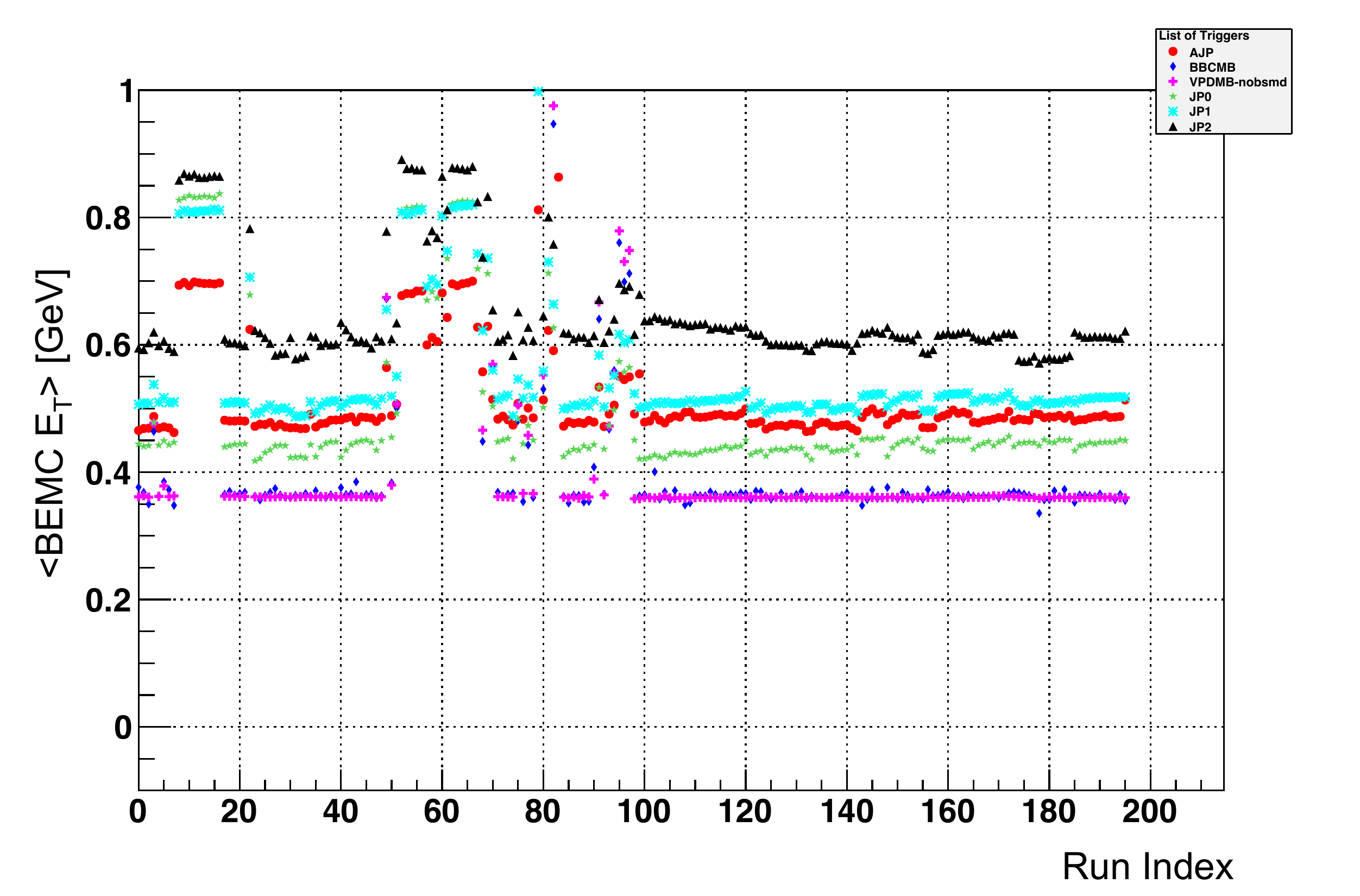}{Average BEMC $E_T$ with New Gains}{Event level quality analysis results using the 196 runs removed previously because of assumed hot towers.}{1}

The jet level quality analysis follows the procedure given in Section \ref{section:JetQA}, and the results are given in Figure \ref{TowerEtQAPlot}. Only seven runs were removed as a result of this QA, the obvious outliers. This adds 140 runs to the final list that will be used in the asymmetry analysis. The bumps in tower $E_T$ for the JP2 trigger look bad, but aren't so far outside the mean. Other triggers do not exhibit this behavior, so these bumps are assumed to be okay.

It is likely the initial round of QA was too strict on what runs were removed, at both the event and jet levels. With the wild outliers removed, this sample of runs has a stable average value, yielding confidence in the data quality.

\figuremacroW{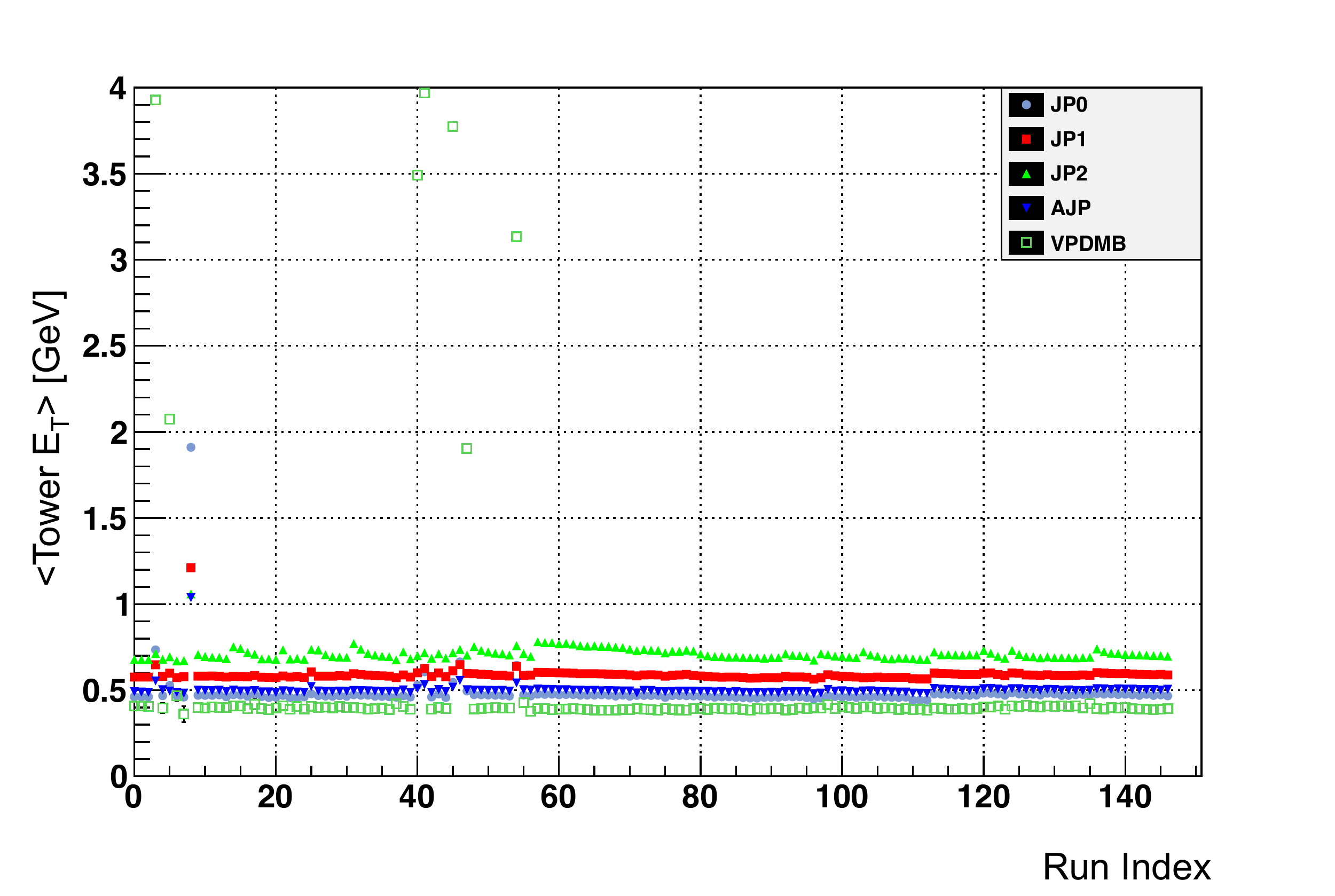}{Average Tower $E_T$ with New Gains}{Jet level quality analysis results using the updated calibration constants. Seven runs are removed here.}{1}

\section{Systematic Uncertainties}
The systematic uncertainties assigned to the final gain calibration constants are for the most part data-driven calculations. The goal of these simple analyses is to look for any changes or biases in the constants which may arise from the algorithm or from hardware fluctuations or imperfections. All of the data that were used to calculate the final constants will again be used to investigate these potential biases.

\subsection{Global Vs. Primary Tracks}
At STAR there are two types of tracks stored within our final file framework: global and primary tracks. Primary tracks are those which satisfy a matching condition to an interaction vertex, and global tracks are all tracks in the detector in that saved event which could arise from a collision, backgrounds, or pile-up. 

To create the calibration specific ROOT trees, all of the primary tracks which are matched back to a global track are saved, therefore excluding all global tracks which do not have any matching primary track and likely arise from background or pileup. This became cause for concern during the calibration, with the fear being that excluding the signal from so many global tracks would introduce a bias that would result in an overall increase in average track momentum. Returning to the original data files which were used to produce the calibration ROOT trees, studying this potential bias is a simple analysis. Using the original data, all tracks (global and primary) are stored, and can be examined. 

The analysis applies all of the cuts that are used in the final calibration, except triggering conditions, and was run two times over a subsample of 100 runs of the final data sample. The first pass was completed by looking at primary tracks which are matched to global tracks, as in the final result, and then the second pass used all of the global tracks. Figure \ref{TrackMomentum_GlobalPrimary} shows a comparison of the momentum distributions, which are normalized to their total integral, for the matched primary tracks and all global tracks. Looking at the shape and mean of each distribution, there seems to be no bias in the matched primary track distribution. To confirm, Figure \ref{TrackEoverP_GlobalPrimary_LogY} shows a comparison of the E/p distributions for matched primary and all global tracks. It is clear that the global tracks give a much higher background, as one would expect since they can originate from background events, but the two distributions line up well just below E/p = 1 and beyond. 

\figuremacroW{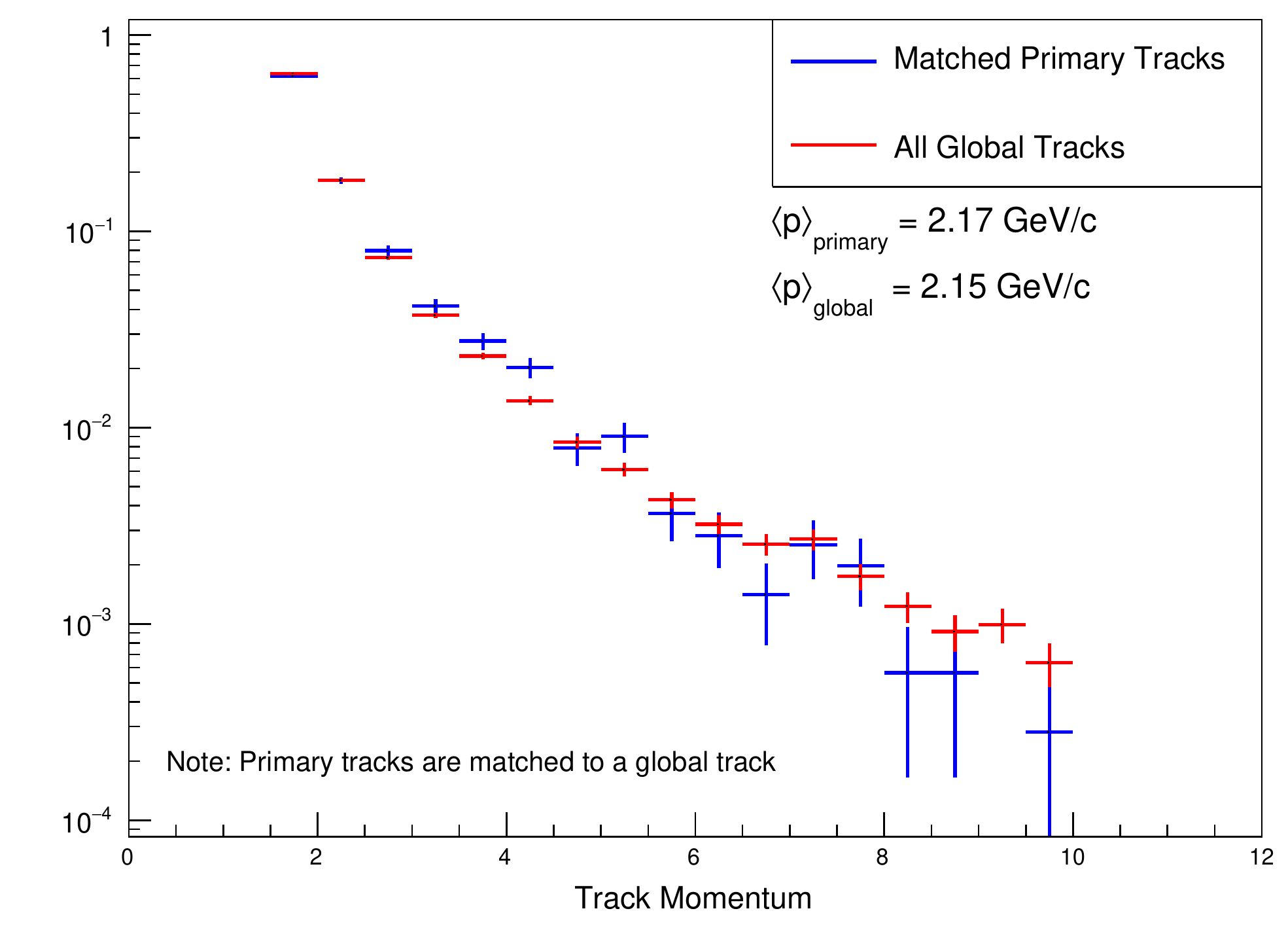}{Global vs. Matched Primary Track Momentum}{Comparison of the normalized momentum spectra for matched primary tracks and all global tracks to study any potential momentum bias. Both spectra give a very similar mean and shape overall, there seems to be no bias with using the matched primary tracks.}{0.75}

\figuremacroW{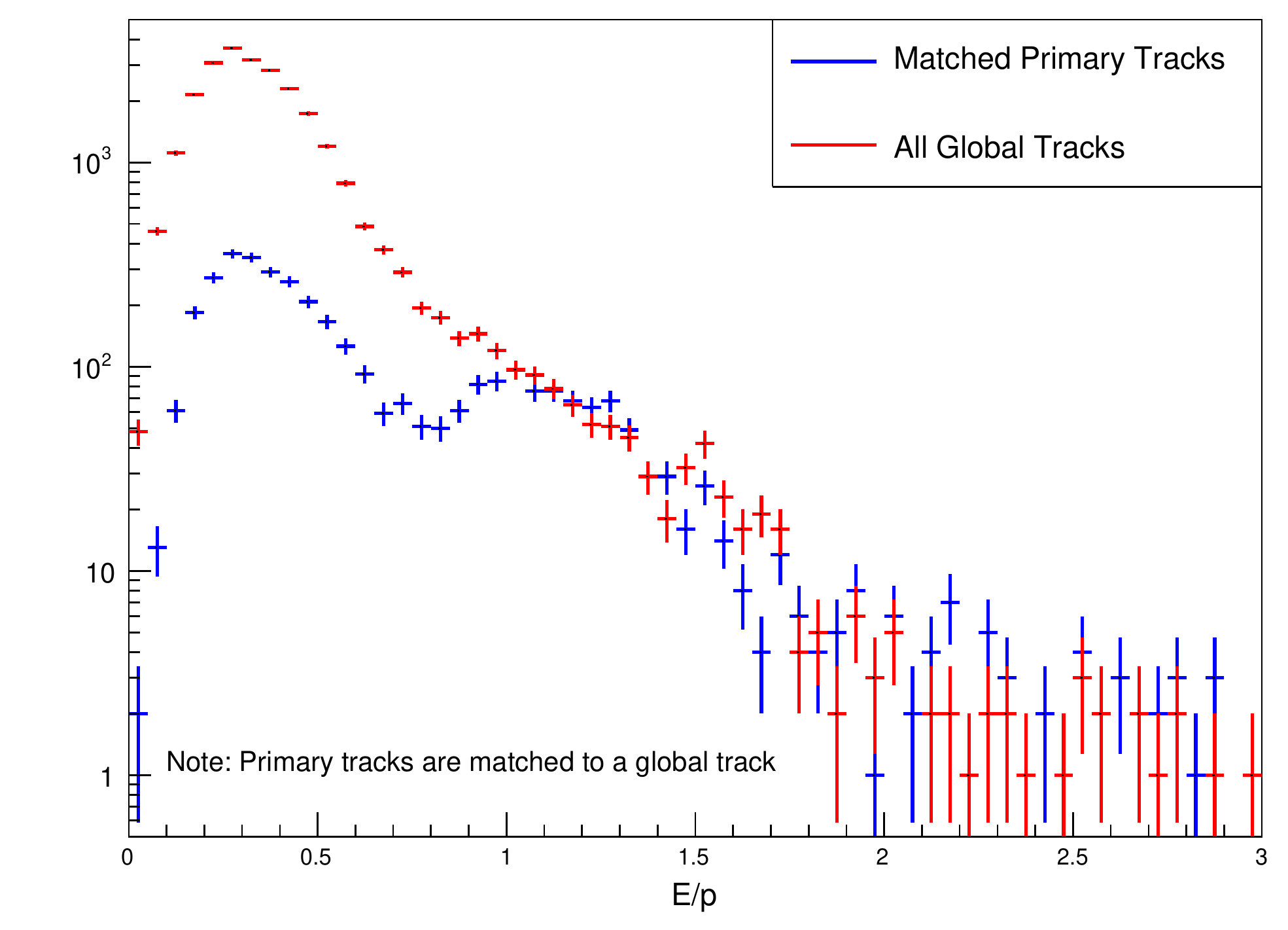}{Global vs. Matched Primary Track E/p}{Comparison of E/p distributions for matched primary tracks and all global tracks to study any potential momentum bias. The additional global tracks are clearly concentrated in the background region.}{0.75}

The outcome of this study is that the additional global tracks distribute themselves evenly in momentum, and excluding them does not cause any shift in average momentum. However, the additional global tracks do show up as background counts in the E/p spectra, but the spectra agree well around E/p = 1 and beyond. This evidence supports the conclusion that there is no benefit for changing the calibration to use all of the global tracks, and there will be no systematic error assigned.

\subsection{$e^+$/$e^-$ Differences}
Any differences in the gain constants found by splitting the final sample into subsamples of $e^+$ and $e^-$ would show what effect the TPC is having on the calibration. Quantifying any difference is a simple matter of reproducing the entire calibration procedure for two different subsamples, one each representing $e^+$ and $e^-$. The results of this analysis are summarized in Figure \ref{fig:GainPercentDiff} where the percent difference between the subsample results and the combined sample final results is plotted.

\begin{figure}[htb]
    \centering
    \subfloat[]{\includegraphics[width=0.45\textwidth]{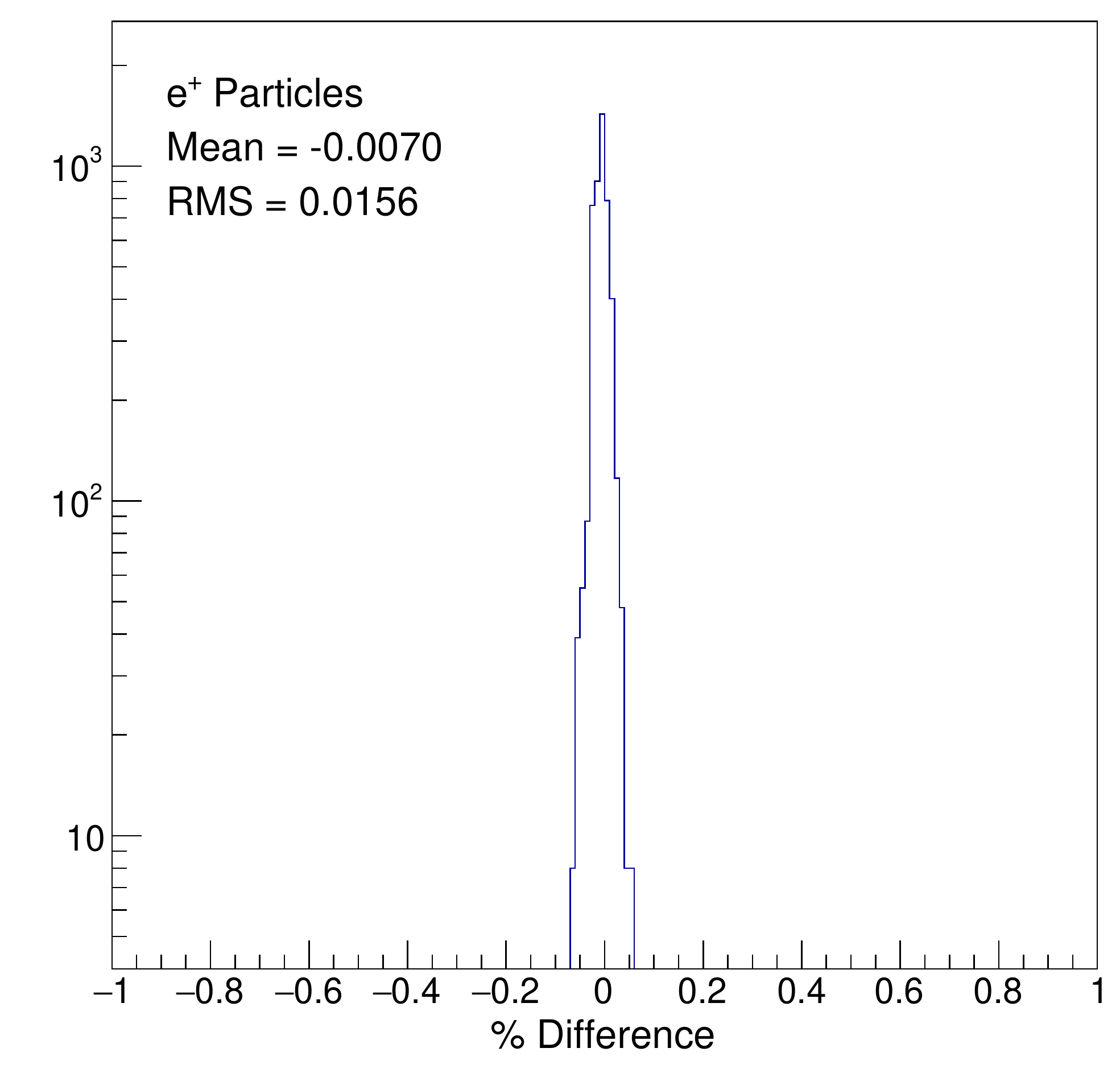} \label{GainPercentDiffPlus}} %
    \subfloat[]{\includegraphics[width=0.45\textwidth]{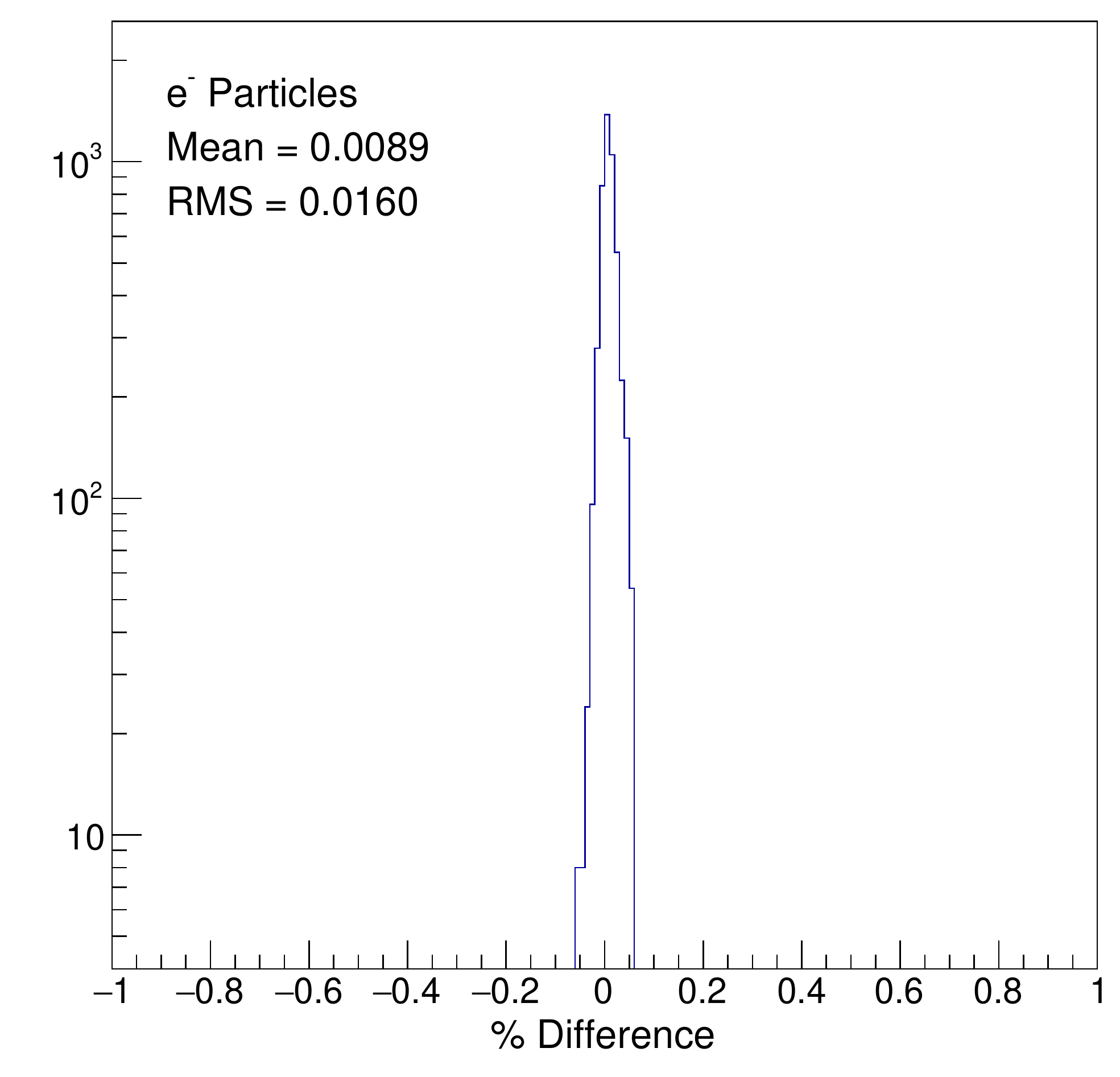} \label{GainPercentDiffMinus}}%
    \caption[$e^\pm$ Particle Gain Comparison]{\textbf{$e^\pm$ Paricle Gain Comparison} - Percent difference comparison of gain constants between the total combined sample of final result gains and subsamples of (a) $e^+$ and (b) $e^-$ particles.}%
    \label{fig:GainPercentDiff}%
\end{figure}

The mean of each distribution in Figure \ref{fig:GainPercentDiff} is offset from zero, but both cases carry a width large enough to encompass a zero percent difference. Since the distributions are offset in different directions, it seems there is some bias introduced by the TPC. To cover any bias, the largest of the two mean values, 0.9\%, will be assigned as a conservative systematic error.

\subsection{Time Dependence}

To maximize statistics for the final sample, the calibration must integrate over all runs that were collected during the 2012 200 GeV proton collision period. This covers several weeks worth of running, which means one single set of calibration constants must describe the calorimeter for that entire period. Over this period, the calorimeter hardware may change, and these changes will be mostly averaged out in the final calibration constants. Unfortunately we cannot have multiple calibrations, but we can look at how the calibration constants change over time.

One way to understand changes over time is to plot the average calibration constant for each run. This is not a difficult exercise, but statistics again are a problem. To gain enough statistics to plot a calibration constant for each run, we must integrate over the statistics in each tower in the calorimeter. Following the calibration procedure all over again, and integrating statistics over all towers, the average gain constant for each run is shown in Figure \ref{TimeDep_AllRuns}. There is a clear time dependence that has a negative slope, so the calibration constant is decreasing as the run progresses. 

\figuremacroW{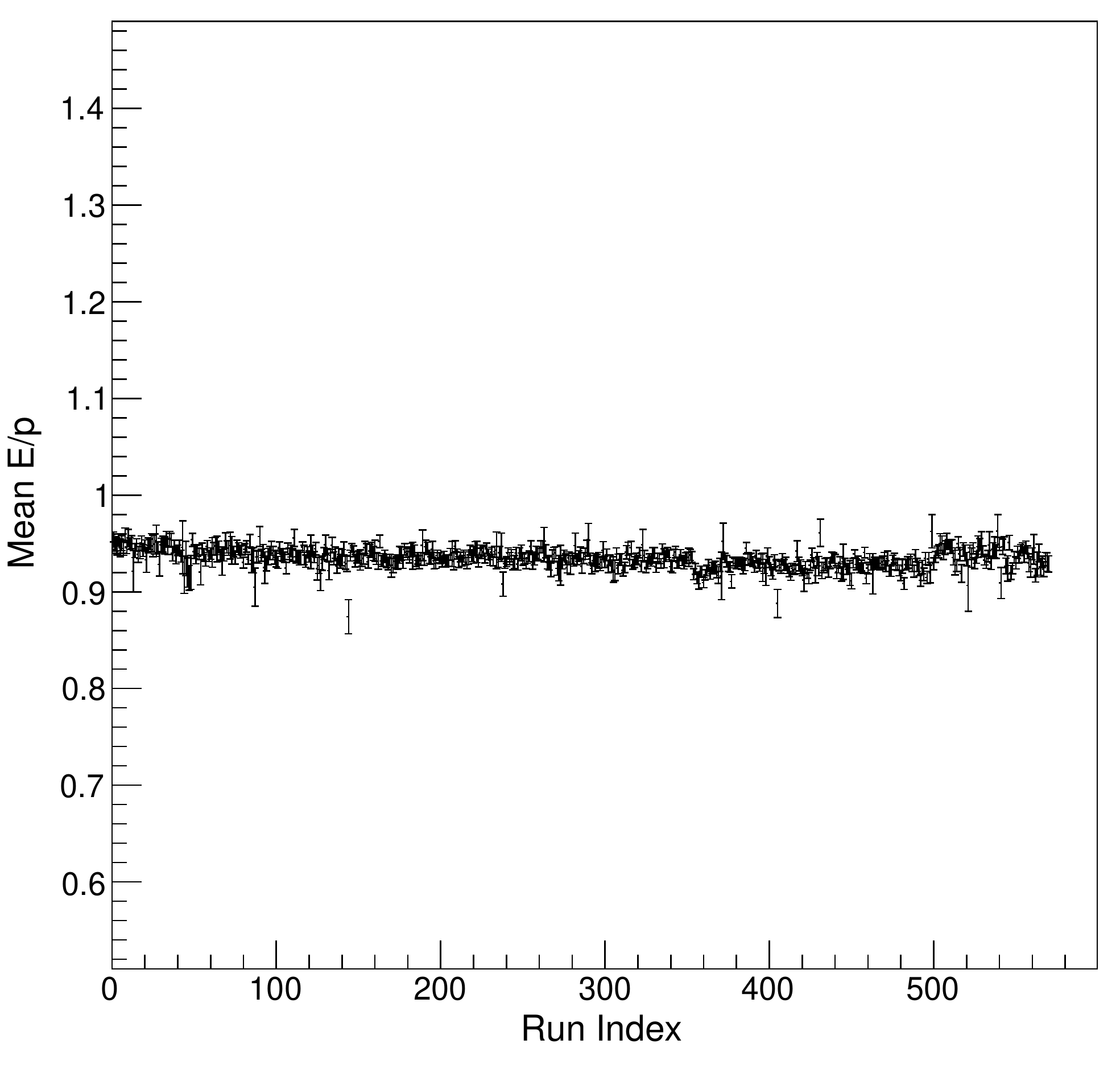}{Calibration Time Dependence}{Gain calibration constant from each run using statistics integrated over all towers. Note that runs which produce less than 100 electrons were removed as they gave nonsensical answers, and the run index is simply a number given to each run for easy plotting.}{0.75}

The systematic error for this time dependence will be two-fold. First, we will assign an error which describes the difference between the initial value ($0.9416$) and the mean of the time dependence ($0.9342$). This is a simple percent difference calculation:

\begin{equation}
\frac{0.9416-0.9342}{0.9342}\times 100\% = 0.7921\%
\end{equation}

\noindent The second piece of the error we will apply pertains to the global shift down in the mean E/p value that occurs around run index 350 and ends around index 500. This shift down is not currently understood, there is no documented reason this shift should happen. Therefore we account for it in the systematic errors by comparing the mean of the approximately 150 ``drop down'' runs ($0.9267$) to the mean of the whole range of runs. Again, it is a simple percent difference calculation:

\begin{equation}
\frac{0.9342-0.9267}{0.9342}\times 100\% = 0.8028\%
\end{equation}

\noindent The final error is these two results added in quadrature, giving a final systematic error to account for the time dependence of the calibration of 1.13\%.

\subsection{$\Delta R$ Dependence}
During the course of the calibration, one correction is applied to the data as the E/p spectra are filled. Each electron candidate is corrected for energy leakage that may result from being close to the edge of the tower. This correction is a pure simulation correction, where the fraction of leakage is plotted as a function of $\Delta R$ between the track and the center of the tower. This simulation curve is calculated for each $\eta$ ring, assuming rotational invariance. Naturally the correction goes to 1 as $\Delta R$ goes to zero, and shrinks from there. Thus, for each electron candidate, we calculate the $\Delta R$ value, evaluate the correction function at that value, and then divide the tower energy for that track by the correction value resulting in the final energy used for the track in the E/p spectra.

A curiosity arises, since we are applying a $\Delta R$ dependent correction, whether or not there is a $\Delta R$ dependence of the mean E/p values in the calibration. To study this we again combine all of the statistics from all towers, and plot the E/p spectra in bins of the $\Delta R$ value. These E/p spectra are subject to the same fitting procedure used in the final calibration, and the mean E/p values are plotted as a function of $\Delta R$. The results are shown in Figure \ref{fig:DeltaRMeans}, plotted for inner and outer towers. Take note here that the ``outer'' definition means that statistics were grouped from the outer two $\eta$ rings on either end of the detector. This is because the last two rings have the highest backgrounds, and are subject to the harshest entrance angles of tracks, and thus give the largest energy correction.

\begin{figure}[htb]
    \centering
    \subfloat[]{\includegraphics[width=0.45\textwidth]{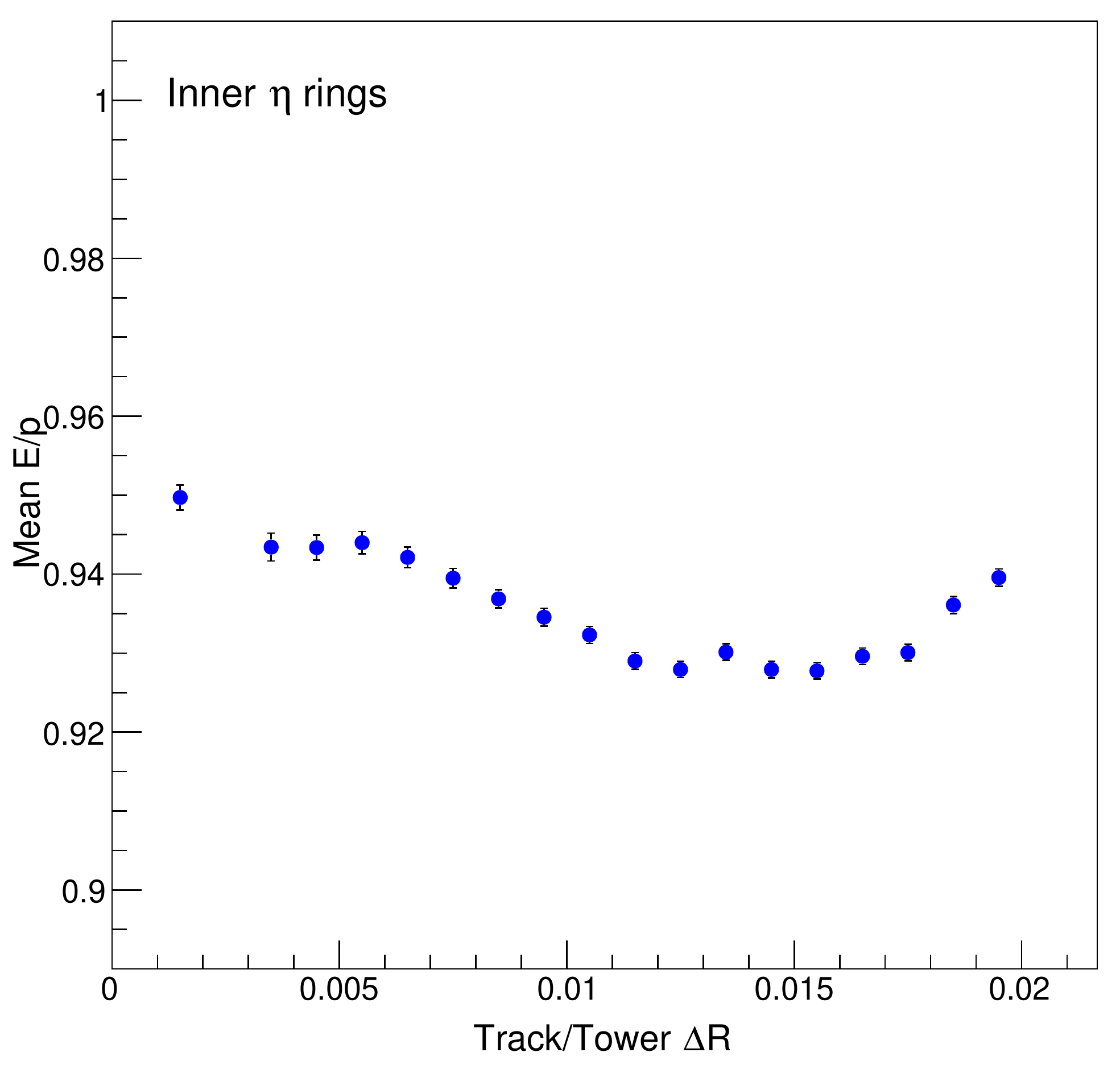} \label{TowerDrMeansIn}} %
    \subfloat[]{\includegraphics[width=0.45\textwidth]{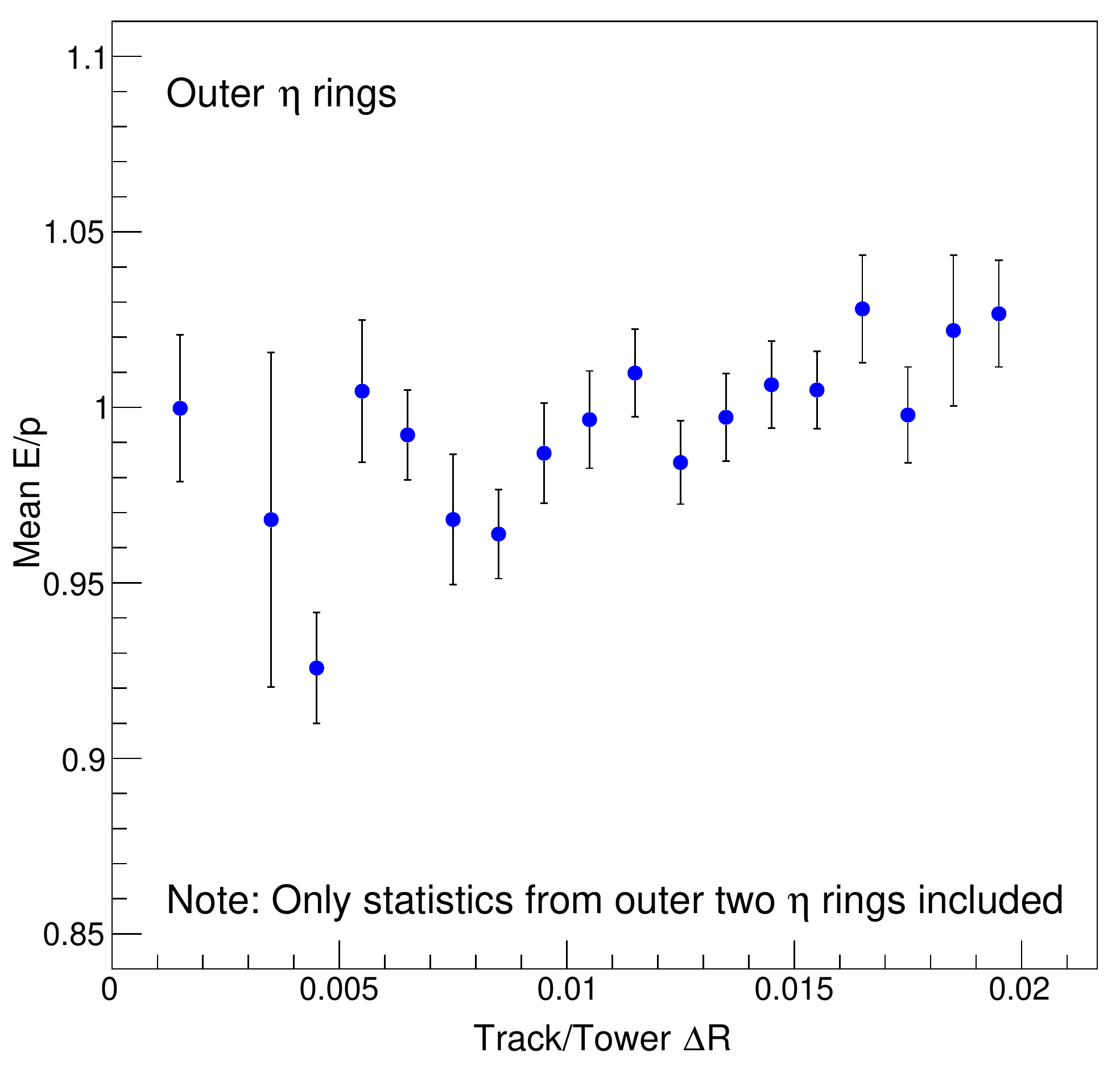} \label{TowerDrMeansOut}}%
    \caption[Mean E/p vs. $\Delta R$]{\textbf{Mean E/p vs. $\Delta R$} - Mean E/p values extracted as a function of $\Delta R$ for the (a) inner 36 towers and the (b) outer two towers on each end.}%
    \label{fig:DeltaRMeans}%
\end{figure}

The mean E/p values as a function of $\Delta R$ do show a spread in values for both the inner and the outer rings. The applied systematic will be the RMS of the spread in the mean E/p values, which is plotted in Figure \ref{fig:DeltaRSpread}. A consequence of plotting the values this way is that we will have to apply a systematic error to the inner 36 $\eta$ ring towers, and then a systematic for the outer two $\eta$ ring towers on either end of the barrel. Reading directly from the plots, the inner towers have an error of 0.67\% and the outer towers have an error of 2.45\%.

\begin{figure}[htb]
    \centering
    \subfloat[]{\includegraphics[width=0.45\textwidth]{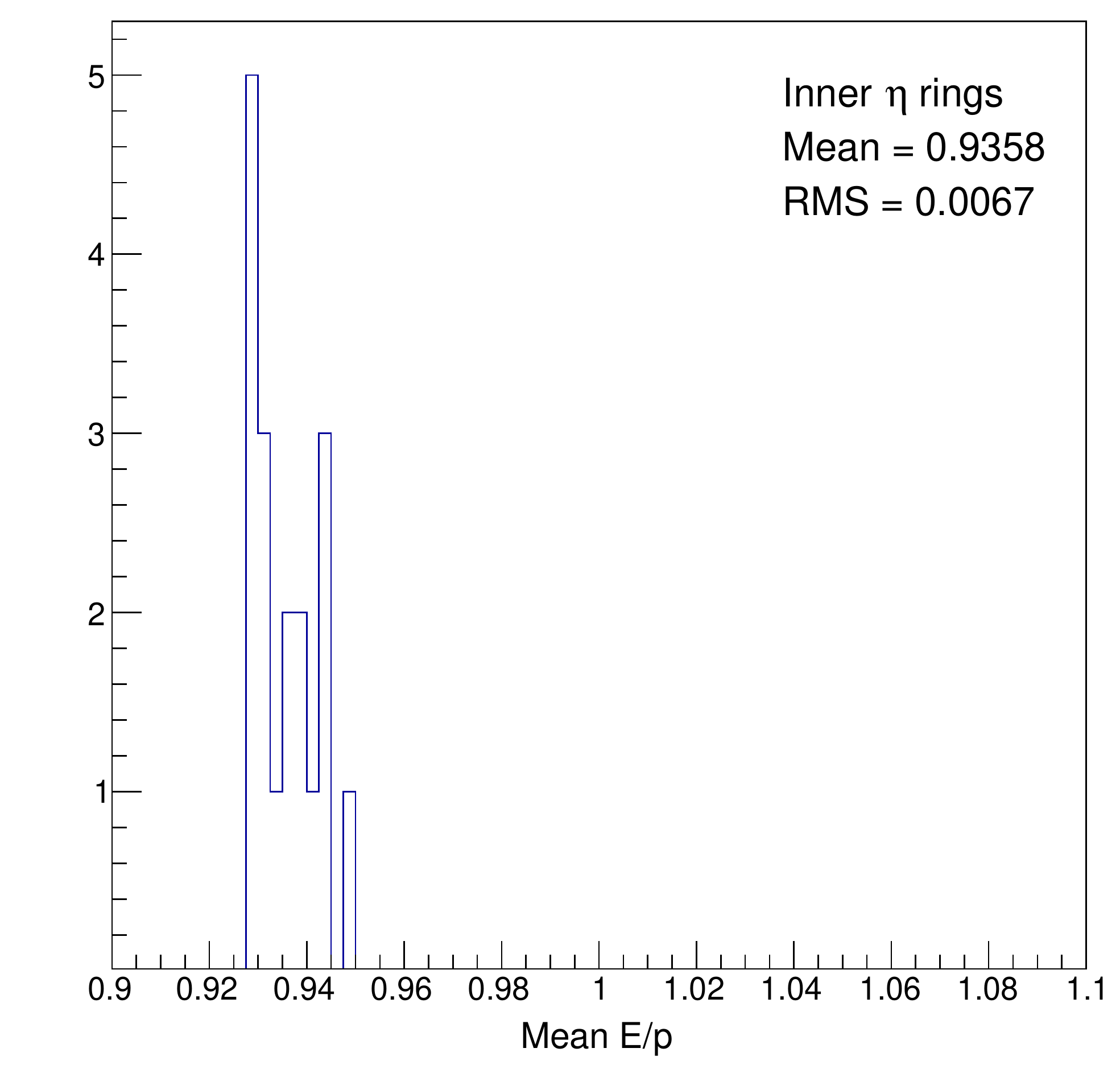} \label{TowerDrSpreadIn}} %
    \subfloat[]{\includegraphics[width=0.45\textwidth]{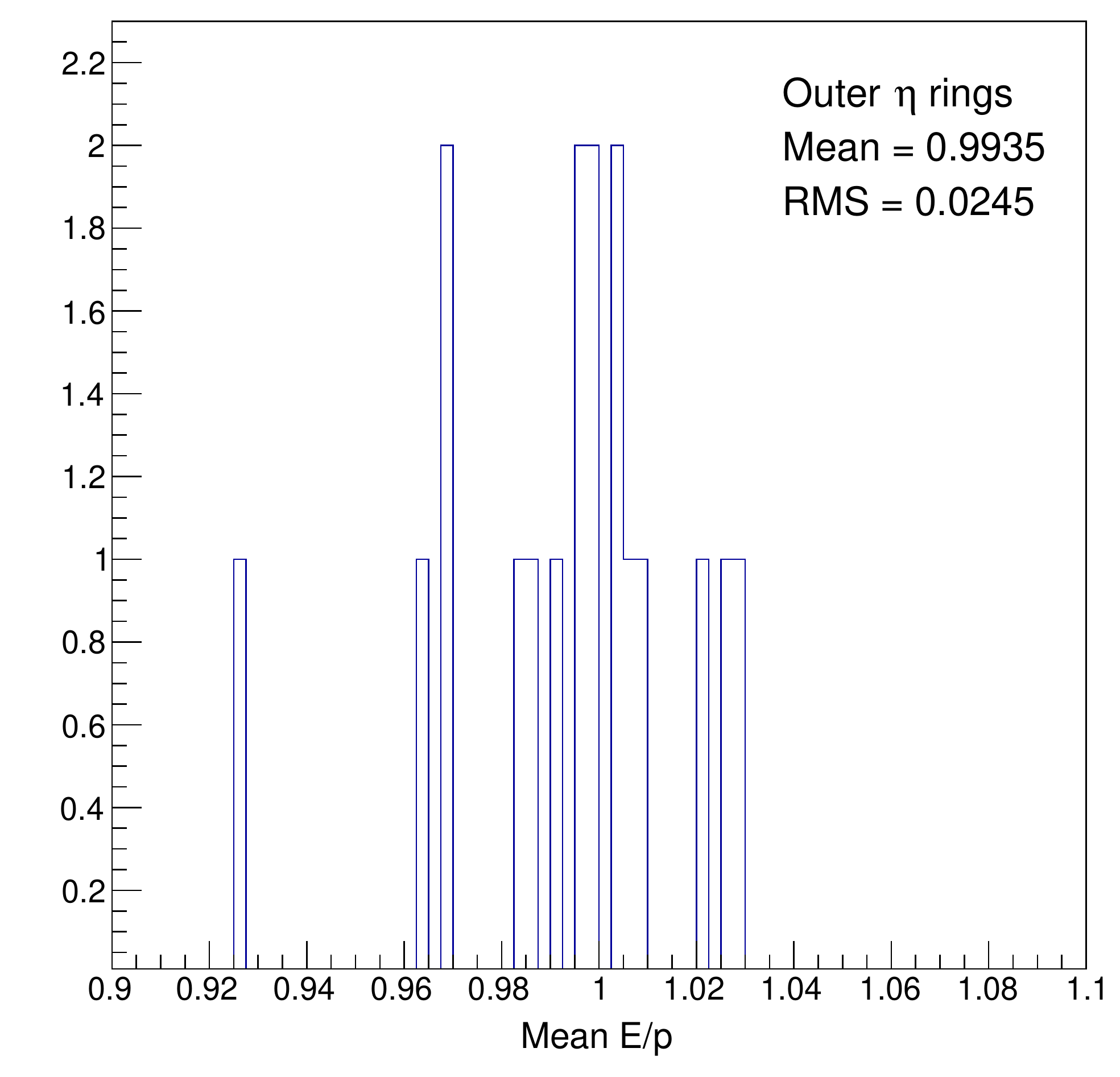} \label{TowerDrSpreadOut}}%
    \caption[Mean E/p vs. $\Delta R$ Spread]{\textbf{Mean E/p vs. $\Delta R$ Spread} - The spread in the mean E/p values from Figure \ref{fig:DeltaRMeans}, the RMS will be assigned as the systematic error.}%
    \label{fig:DeltaRSpread}%
\end{figure}

\subsection{Simulation and Edge Effects}
Along the same lines as the $\Delta R$ systematic error is an error associated with the energy leakage correction. This error, however, is purely associated with the simulation that was used. As the simulation leakage curves have not been updated for the 2012 analysis, we simply use the number which was previously calculated. There will be a 1.8\% systematic error assigned for this correction.

\subsection{Trigger Bias}
Due to how the analysis ROOT trees are created, this calibration has a rather unorthodox triggering scheme. When we apply the unbiased triggering scheme, an unfortunate truth is that some events will be a high tower only event without any other trigger firing. In fact, a simple analysis revealed that 2.5\% of the time an event slips into the unbiased sample where only a high tower trigger fired.

We can calculate the bias that the 2.5\% high tower events imparts on the unbiased spectrum. First define ratios of E/p, $R_{HT}$ and $R_{unb}$ as the high tower and unbiased E/p values, respectively. Note that this calculation is for events below the unbiased 5 GeV track momentum threshold. In this region, we also define $R_{ideal}=R_{unb}$, meaning the unbiased E/p value is the ideal value in the unbiased momentum region. Using these definitions, and the fraction of HT events in the unbiased sample we can write down the measured E/p ratio in the unbiased region:

\begin{equation}
R_{measured} = 0.975\times R_{unb} + 0.025\times R_{HT}
\end{equation}

\noindent Using this definition and the percent error equation, we can calculate the percent error between the measured value and the ideal value. This is given by:

\begin{equation}
\Delta R_{measured} = \frac{R_{ideal} - R_{measured}}{R_{ideal}} = 0.025\times\left(1-\frac{R_{HT}}{R_{unb}}\right)
\end{equation}

The values for the ratios $R_{HT}$ and $R_{unb}$ may be calculated from the analysis ROOT trees by plotting the E/p spectra and following the same fitting procedure used in the calibration algorithm. This yields $R_{HT}=0.975$ and $R_{unb}=0.934$, thus giving an error magnitude of 0.11\%, which is very small compared to some of the other systematic errors but will be folded in for completeness. 

\subsection{Crate Dependence}
During data collection, the barrel hardware has specifically set values including timing and voltages for each tower. These values may have crate-to-crate variations, which would lead to a variation in E/p among the different crates. This can be investigated easily by splitting up the final electron sample results into 30 different crate histograms. Then the mean E/p may be extracted for each crate by applying the same fitting procedure used in the analysis. The results for each crate are shown in Figure \ref{CrateMeans}, where it is clear there is quite a bit of variation in the values. The spread of the values will be used to assign a systematic error. Figure \ref{CrateMeanSpread} shows the spread of the mean E/p values from each crate and also fit with a Gaussian curve. The width of the Gaussian curve, or 1.7\%, will be assigned as the systematic error due to crate variations.

\begin{figure}[htb]
    \centering
    \subfloat[]{\includegraphics[width=0.45\textwidth]{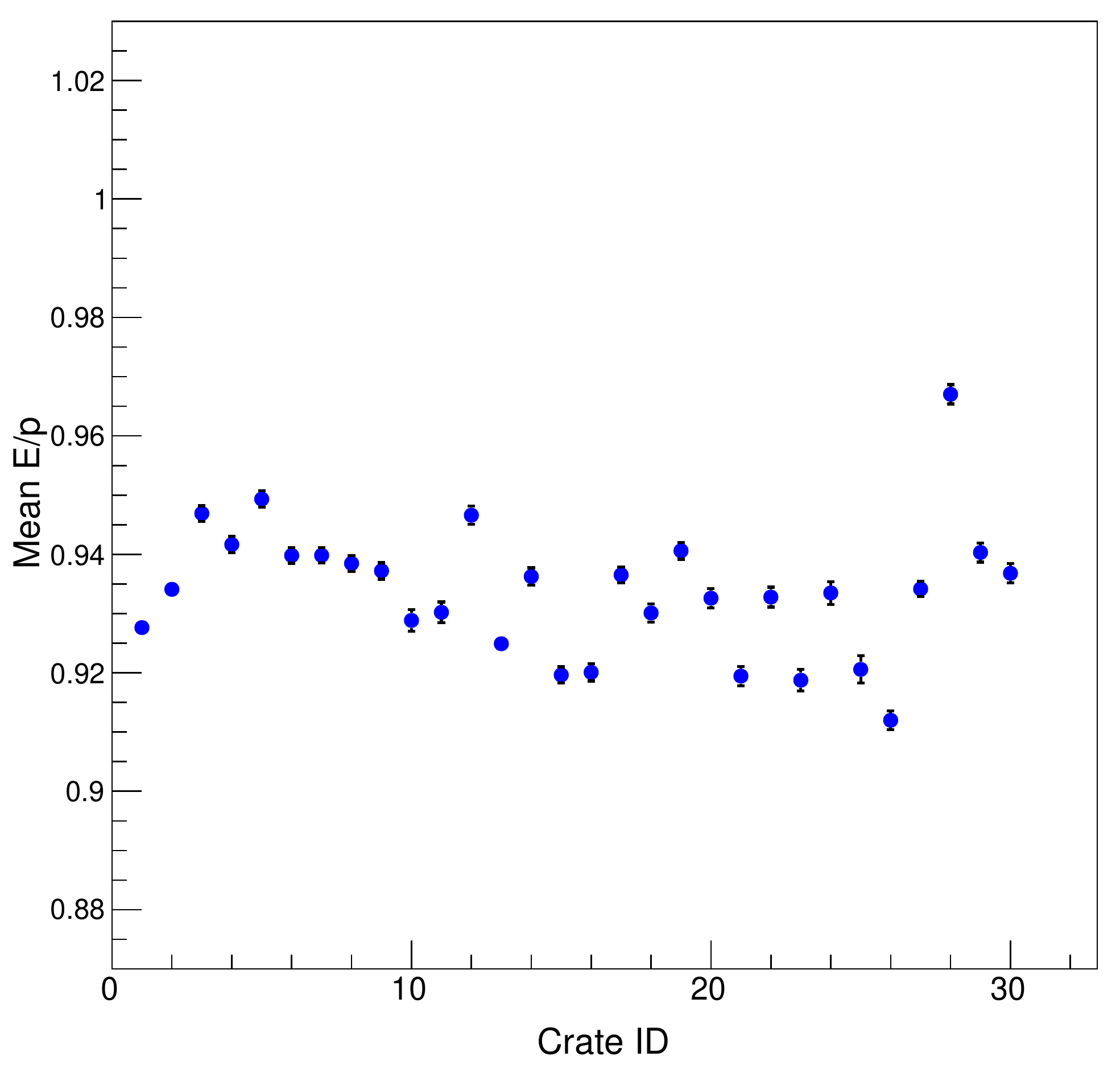} \label{CrateMeans}} %
    \subfloat[]{\includegraphics[width=0.45\textwidth]{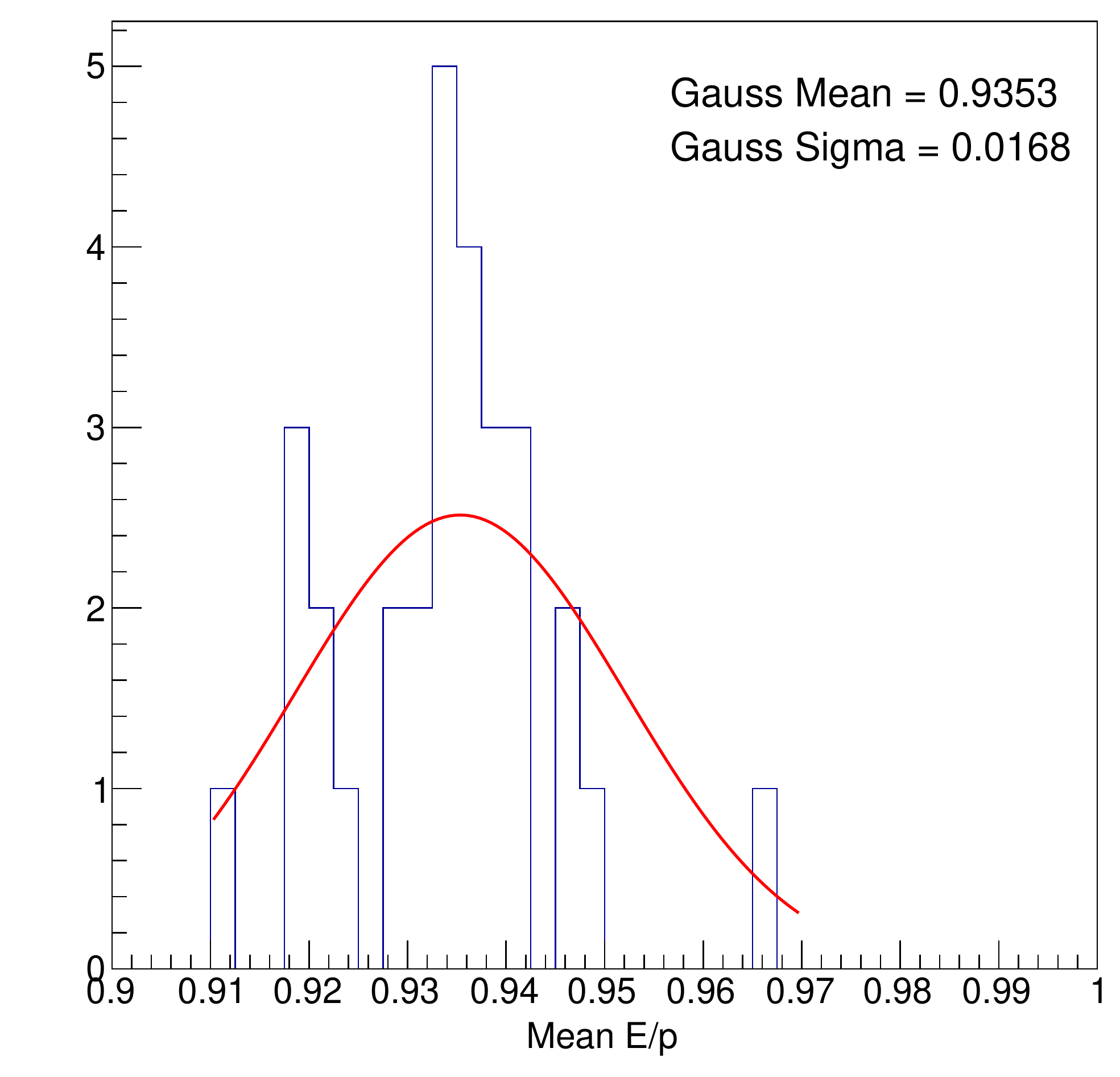} \label{CrateMeanSpread}}%
    \caption[Mean E/p vs. Crate]{\textbf{Mean E/p vs. Crate} - (a) The mean E/p values from each crate and (b) the spread in those values and Gaussian fit used to assign a systematic error.}%
    \label{fig:DeltaRSpread}%
\end{figure}

\subsection{$\eta$ Dependence}
Integrating statistics over all $\eta$ rings is one way the final results are extracted, specifically for the outer seven rings on either end of the barrel. We also rely on integrating over towers in the same ring and within the same crate (crate slice) for the inner rings. We should therefore have a look to see if there is any strong dependence of the mean E/p values over all rings. Like the other analyses, the statistics were combined for each ring and then the mean E/p value extracted by applying the same fit as in the calibration analysis. 

The results for all rings are shown in Figure \ref{RingEoverP_Combined}, where it is clear the only differences are between the inner and the outer designations that showed up as a result of the high tower and unbiased triggering samples in (see Figure \ref{RingFitMeans}). Because the inner and outer regions are treated separately, this jump between the samples is not considered as a systematic error. Looking locally at the inner and the outer regions, it is clear that there are a few fluctuations but it is not clear if these are because of the hardware gains set during running. Because of this unknown, we choose to assign no systematic error because of an $\eta$ dependence.

\figuremacroW{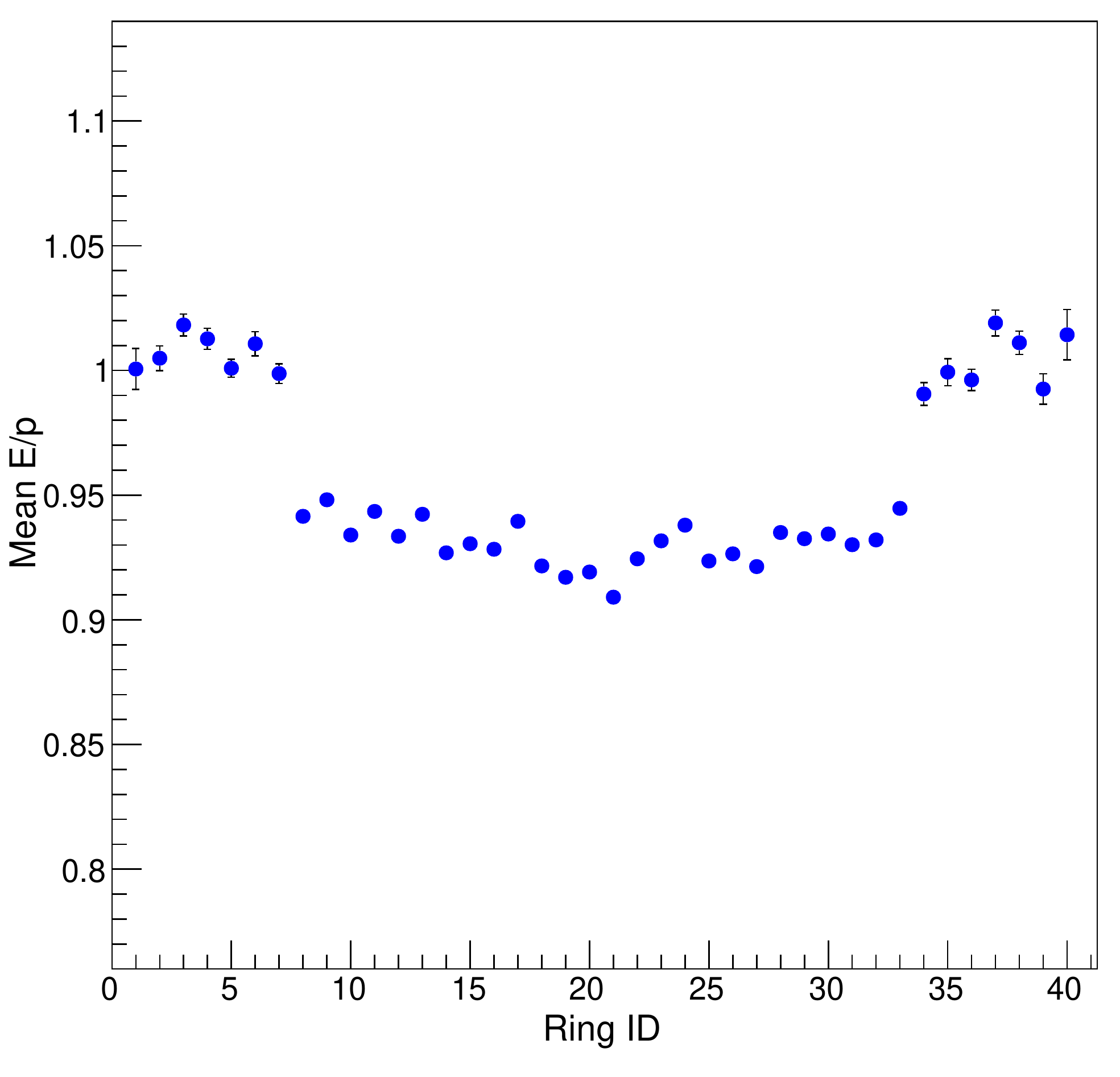}{Mean E/p vs. $\eta$ Ring}{Mean E/p values extracted for each ring and plotted to look for any dependence as a function of $\eta$.}{0.75}

\subsection{Rate Dependence}
The final systematic error we considered is determining if there is a dependence of the mean E/p value on the collision rate. The backgrounds and pile-up in STAR decrease from the beginning of a RHIC fill until the beam is finally dumped, as a result of decreased bunch size as the protons collide with each other. Any dependence on the collision rate would be a measure of how the backgrounds are affecting the mean E/p values, but since we took care to characterize the backgrounds well in our fit, there should be no dependence. 

To measure this dependence, the statistics are integrated over all towers for each run number and the E/p values are extracted for each run. These are the same data points we used previously to determine the time dependence. In this case, rather than plotting versus the run number, we plot as a function of the beam-beam counter (BBC) rate. The BBC at STAR is a two-detector system which wrap around the beam pipe on the east and west sides of the interaction region, like the VPD. The BBC serves multiple purposes, but for this calibration systematic error analysis, the main purpose is to measure the luminosity of each run by measuring coincidence of particles emitted in a collision. The output from the BBC is the rate of coincidence, which is what the mean E/p values are plotted against in Figure \ref{AllRunsEoverP_BBCrate}. Fitting this distribution with a first order polynomial results in a slope which is of the order $10^{-9}$, and giving the same $\chi^2$ value as a flat constant fit. Therefore, there is no dependence on the collision rate, and no systematic error will be assigned.

\figuremacroW{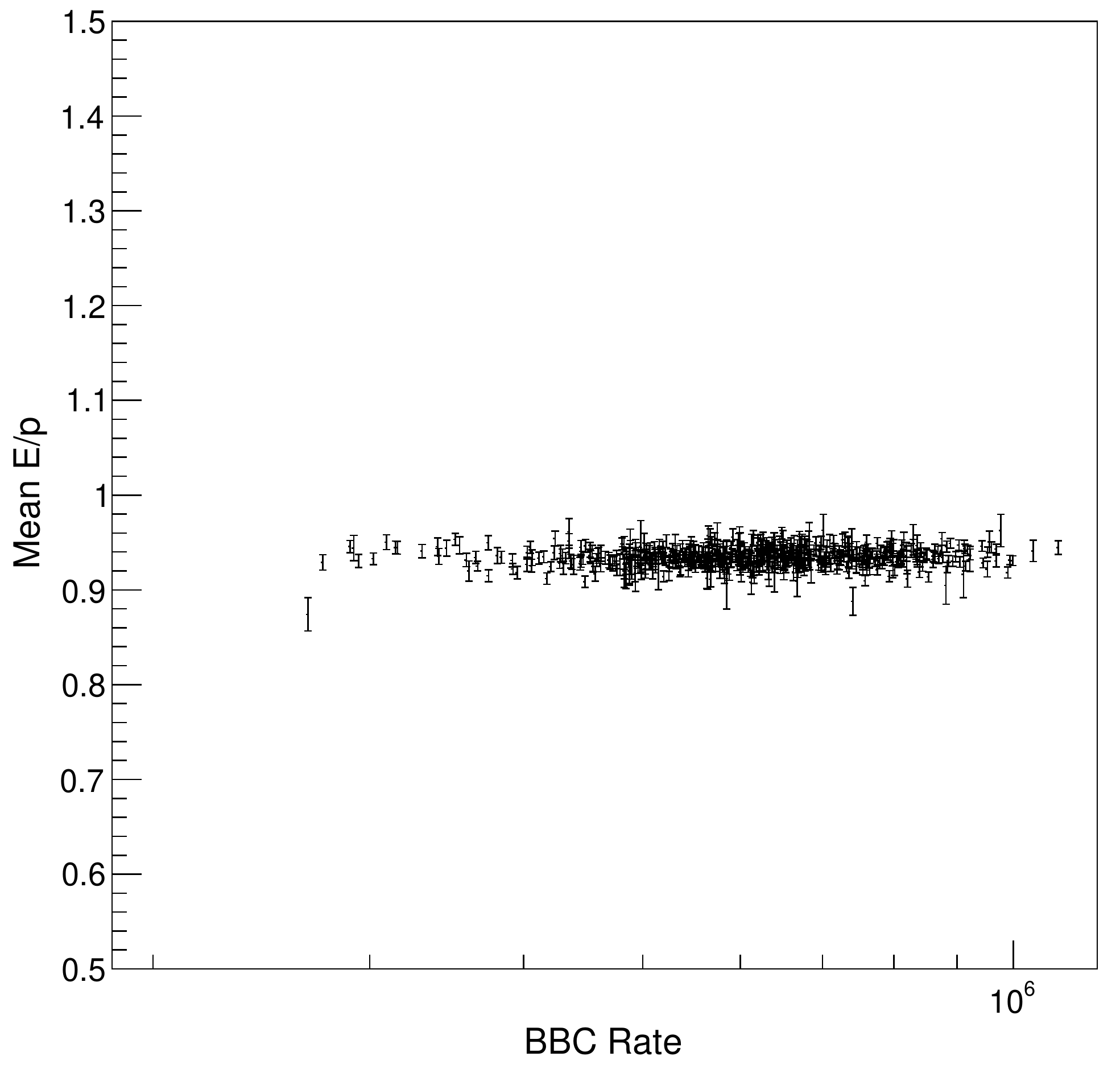}{Mean E/p vs. BBC Rate}{Mean E/p values integrated over all tower statistics and extracted for each run and plotted as a function of the BBC rate.}{0.75}

\subsection{Table of Systematic Errors}
All of the errors calculated and discussed for the BEMC calibration are summarized in Table \ref{table:CalibSystErrors}. Because of the $\Delta R$ systematic error there are columns for the inner 36 rings for the outer two rings on either end of the barrel. The bottom row shows the total of each column added in quadrature.

\begin{table}[htp]
\begin{center}
\begin{tabular}{|c|c|c|}
\hline
\textbf{Error Name} & \textbf{Inner Ring Value} & \textbf{Outer Ring Value}\\
\hline
Global/Primary Tracks & 0 & 0 \\
\hline
$e^+$/$e^-$ Difference & 0.9\% & 0.9\% \\
\hline
Time Dependence & 1.13\%  & 1.13\% \\
\hline
$\Delta R$ Dependence & 0.67\% & 2.45\% \\
\hline
Edge Effects and Simulation & 1.8\% & 1.8\%\\
\hline
Trigger Bias & 0.11\% & 0.11\% \\
\hline
Crate Dependence & 1.7\% & 1.7\% \\
\hline
Rate Dependence & 0 & 0 \\
\hline
$\eta$ Dependence & 0 & 0\\
\hline
\textbf{Total (quadrature)} & \textbf{2.95\%} & \textbf{3.77\%}\\
\hline
\end{tabular}
\end{center}
\caption[Calibration Systematic Errors]{\textbf{Calibration Systematic Errors} - Summary of systematic errors calculated for the calibration of the 2012 200 GeV proton data.}
\label{table:CalibSystErrors}
\end{table}

\newcommand{\lum}{\mathcal{L}}

\chapter{Asymmetry Analysis} 
\label{chapter:AsymmetryAnalysis}

\ifpdf
    \graphicspath{{ch5_AsymmetryAnalysis/figures/PNG/}{ch5_AsymmetryAnalysis/figures/PDF/}{ch5_AsymmetryAnalysis/figures/}}
\else
    \graphicspath{{ch5_AsymmetryAnalysis/figures/EPS/}{ch5_AsymmetryAnalysis/figures/}}
\fi

\section{Analysis Cuts}
With the QA and BEMC calibration complete, the stage is set to complete the asymmetry analysis. The first step is to select the desirable data events applicable to the various asymmetry moments of interest.

\subsection{Event Level Cuts}
A quick check of the event as a whole is applied before analyzing the jets and tracks it contains. These basic cuts are:

\begin{itemize}
\item The polarization must be transverse
\item The polarization must have a valid entry in the STAR database, and not masked out as unusable
\item Collisions from abort gap events are excluded from analysis
\end{itemize}

The last nine bunches in each RHIC beam are unfilled, and when the beams are injected into the accelerator they are offset with respect to each other. In this way, protons from one beam meet the empty abort gap in the other beam at STAR. Since the bunches from one beam are unfilled, collisions seen in the detector must be due to background. Therefore the collision rate in the abort gaps are used to monitor the amount and time dependence of the background. Any event failing any of these cuts is immediately thrown away and excluded from analysis.

\subsection{Vertex Cuts}
The position along the beamline where the hard collision occurs is called the vertex. In each event the STAR vertex finder returns a collection of possible vertices and their associated ranking. The highest ranked vertex is the one most likely to be associated with the hard collision in the event, as opposed to a pile up vertex arising from a previous events. Each TPC track is assigned to a vertex and and all tracks included in a jet must point back to the same vertex. For this analysis only the highest ranked, or ``best'', reconstructed vertex is used and is required to have:

\begin{itemize}
\item A vertex rank greater than $10^6$
\item A $z$ position within $60$ cm of the center of the TPC
\end{itemize}

The ranking cut means the vertex has at least two tracks associated with it, and these tracks are matched to the BEMC or EEMC detectors. Rankings less than this value do exist, but these vertices only have one matched track and suffer badly from pile-up track contamination. The $z$ position of the vertex is restricted to minimize uncertainties in the track momentum and particle identification. Tracks which are outside of this cut are generally too short to have lots of hits on the TPC pads, decreasing the ability to accurately reconstruct the momentum. These short tracks obviously spend less time in the detector, and therefore do not deposit as much energy as the longer tracks, thus making particle identification which relies on dE/dx less trustworthy.

\subsection{Jet Cuts}
The cuts placed on the tracks and towers that go into the anti-$k_T$ jet algorithm are given in Table \ref{table:JetFinderCuts}. The jets which come out of the anti-$k_T$ algorithm are then subject to several cuts to choose the ones which will be analyzed:

\begin{itemize}
\item Jet $R_{t}$ < 0.95
\item $\sum p_{T_{track}}^{charged} > 0.5$ GeV/c
\item No tracks with $p_{T}>30$ GeV/c
\item $\lvert\eta_{jet}^{detector}-0.1\rvert < 0.8$
\item $\lvert\eta_{jet}\rvert < 1$
\end{itemize}

Neutral energy background jets are removed via the first two cuts. Because they do not originate from the collision vertex, beam backgrounds are generally not reconstructed as tracks in the TPC. However, since the calorimeters detect photons from decays of $\pi^0$ and $\eta$ particles that do not leave tracks in the TPC, it is not possible to know how much of the calorimeter response is due to backgrounds. Since jet reconstruction algorithms are very good at picking out clusters of energy in the calorimeter, if the fraction of the jet energy is coming from the calorimeters ($R_t$) is greater than 95\% it is very likely there is a large background contribution to that jet. For the same reason, it is good to have at least one track in the jet, hence the $\sum p_{T_{track}}^{charged} > 0.5$ GeV/c cut.

Tracks reconstructed with $p_T$ > 30 GeV/c are not likely to arise from the collision vertex. To see why, consider the collision of two protons, each carrying momentum of 100 GeV/c. On average the colliding partons carry less than 30\% of the total momentum of the proton. The probability of two 30 GeV/c partons transferring enough momentum to produce a 30 GeV/c track is very low. These high momentum tracks also are very straight in the detector, making it a challenge to identify their charge. Most often these tracks are cosmic rays that enter the detector and pass close enough to the vertex to be included in the event, which are not analysis events. For these reasons, jets are removed if they contain tracks above the 30 GeV/c threshold.

The central part of each jet is required to reside completely within the fiducial area of the detector. To enforce this, the $\eta_{jet}^{detector}$ cut is applied, where $\eta_{jet}^{detector}$ is the pseudorapidity of the jet measured from $z=0$, regardless of where the reconstructed vertex is. The cut is larger on one side than the other because the endcap calorimeter is present, and can detect parts of the jet. The endcap is only on one end, though, so the other side of the detector carries a smaller fiducial cut so the jet can be fully contained and detected. The $\eta_{jet}$ cut is applied to pick out the physics of interest. $\lvert\eta_{jet}\rvert$ < 1 is commonly known as ``midrapidity'' among the theoretical community where many predictions and fits are published, and is the choice here to conform to the theoretical preference.

\subsection{Pion Identification Cuts}
For the analysis described in this thesis it is necessary to identify the charged pion tracks within the jets.  A track is flagged as a charged pion if:

\begin{itemize}
\item $\Delta R > 0.05$
\item Number of TPC hits used for track reconstruction (nHitsFit) > 20
\item $-1<n_{\sigma}(\pi) < 2.5$
\end{itemize}

$\Delta R$ is defined as the distance between the jet axis and a track in the jet:

\begin{equation}
\label{eq:trackJetDR}
\Delta R = \sqrt{\left(\phi_{jet}-\phi_{track}\right)^2+\left(\eta_{jet} - \eta_{track}\right)}
\end{equation}

\noindent and it is used to cut out tracks that are very close to the jet axis. In an upcoming section it will be shown that the asymmetry, in part, relies on measuring the angular displacement of pions around the jet axis. The transverse momentum of the pion, relative to the jet axis ($j_T$), defines this angular displacement.  As the pion shrinks closer to the jet axis, the angular displacement becomes increasingly difficult to resolve. To maximize both the sampled physics and resolution this cut value has been strategically chosen based upon simulations which will be discussed in full detail later. The jets are still used if tracks fail this cut, but the tracks that fail are not included in this analysis.

Higher quality tracks result from using a higher number of ``fit points''. Fit points are those TPC hits used when reconstructing the tracks. Tracks with lower number of hits are low quality and shorter tracks which do not deposit much energy for particle ID. Tracks are identified as pions if they satisfy the $n_\sigma(\pi)$ cut, defined in Section \ref{section:EoverPspectra}. The $n_\sigma(\pi)$ distribution is shown in Figure \ref{nSigmaPiPlusCut} for all tracks carrying positive charge in jets that satisfy the JP1 triggering condition (see Section \ref{section:trigconditions}). The green vertical lines indicate the accepted $n_\sigma(pi)$ region. This plot makes the intermingling of pions, kaons, protons and electrons clear, and how tracks could be misidentified as pions with this cut. The correction applied to the final pion asymmetries due to kaon and proton dilution will be discussed in Chapter \ref{ch:Systematics}.

\figuremacroW{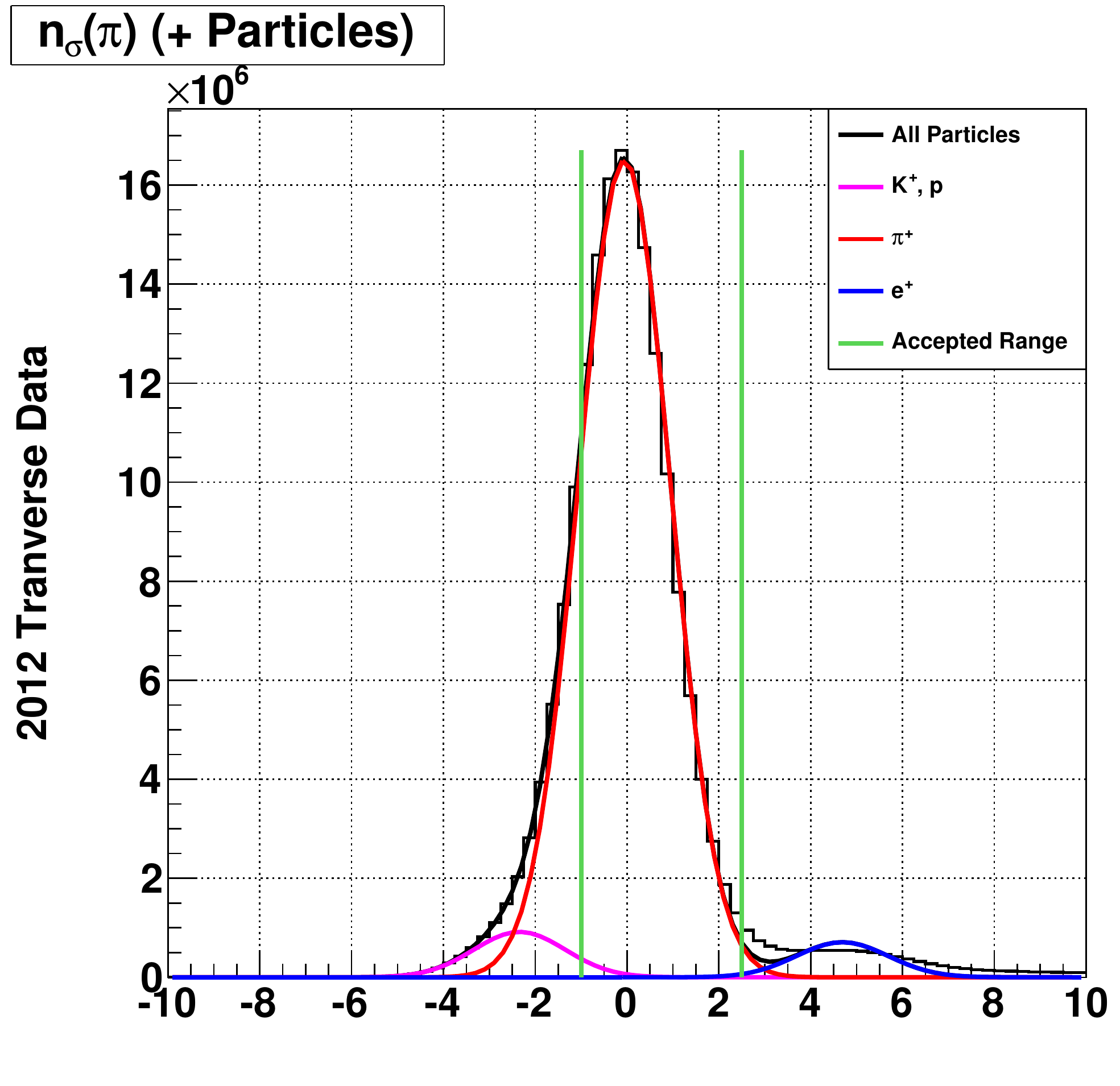}{Sample $n_\sigma(\pi)$ Distribution}{Plot of $n_\sigma(\pi)$ for all positively charged tracks that fulfill the JP1 triggering condition. Other track species are also present here and a correction will be applied to account for these contributions.}{0.75}

\section{Triggering Conditions}
\label{section:trigconditions}

The analysis triggers are the same as those outlined in Section \ref{section:triggering}. These triggers are subject to several requirements before the events are said to be triggered events. These are applied on a jet-by-jet basis:

\begin{itemize}
\item Jet patch triggers must have fired in both the hardware and in the software simulator.
\item VPDMB triggers must have fired in the hardware only and they are subject to the constraints that the $z$ position of the TPC vertex is within 30 cm of $z=0$, and the vertices found from the TPC and VPD detectors must satisfy $\lvert z_{TPC}-z_{VPD}\rvert \leq 6$ cm.
\item The jet must exceed a minimum $p_T$ value defined in Table \ref{table:minimumTrigPt}.
\item Jet patch triggered jets must geometrically match to the barrel, endcap, or overlap patch which fired the trigger.
\end{itemize}

These first two requirements have been discussed previously in Section \ref{section:triggering}. The motivation for the minimum jet $p_T$ requirement is made more clear in the JP2 triggered spectrum shown in Figure \ref{JetPtCut_JP2}. The turn over inside of the red box, which denotes rejected events, results from meeting of the natural $p_T$ spectrum which has much fewer high $p_T$ jets in it, and the enhanced triggered spectrum which selects out the higher $p_T$ jets. In this region the events are highly biased since the reconstructed $p_T$ is smeared about the turn-on value, so for all triggers, a $p_T$ value above the nominal turn-on value is chosen as the minimum to be identified as a triggered event. These values are outlined for all of the analysis triggers in Table \ref{table:minimumTrigPt}.

\figuremacroW{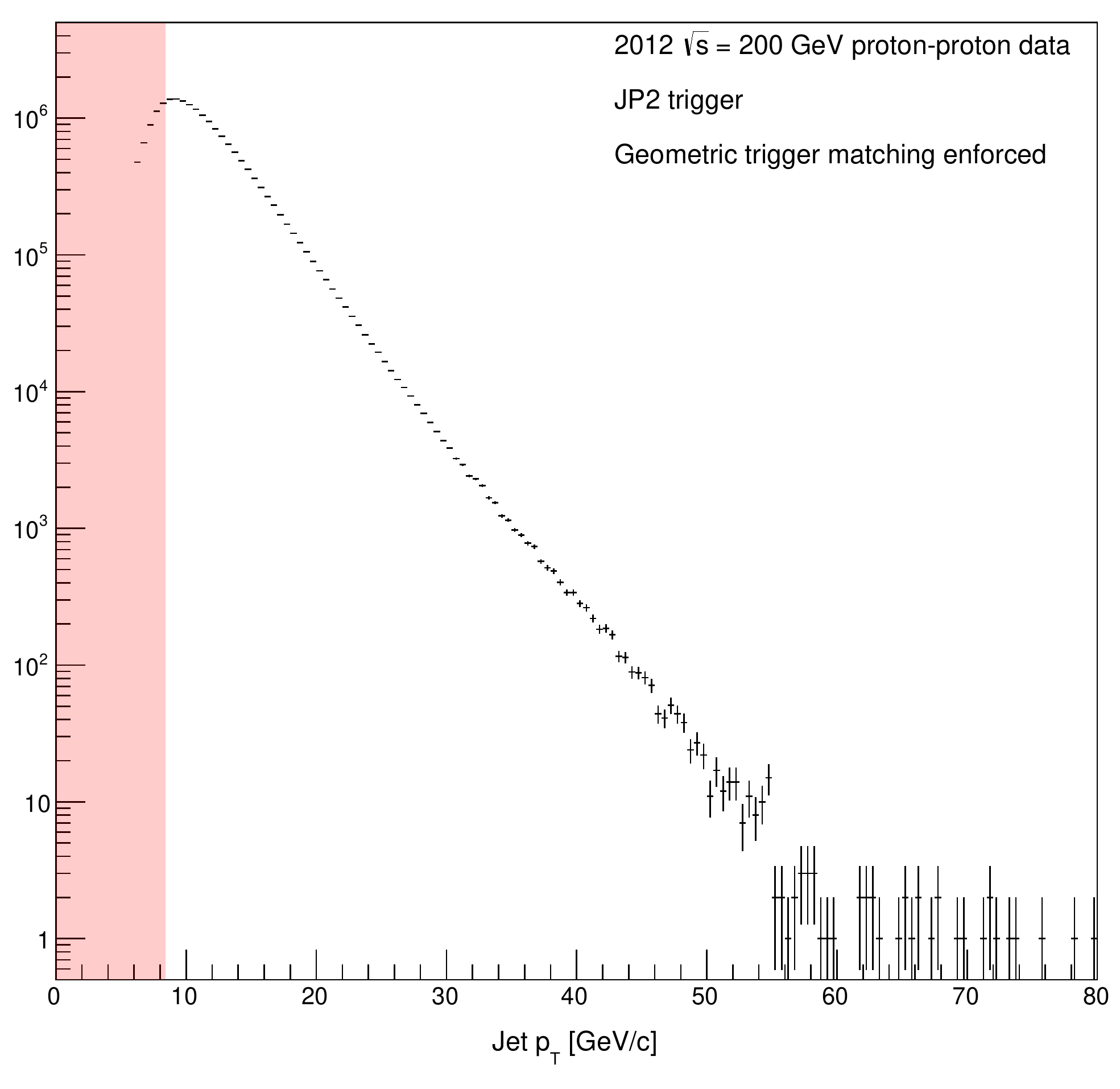}{JP2 Trigger Minimum Jet $p_T$ Cut}{JP2 triggered jet $p_T$ spectrum for jets passing all triggering requirements. Data in the red box will be ignored as analysis jets, everything outside of the box are subject to further analysis cuts.}{0.75}

\begin{table}[htp]
\begin{center}
\begin{tabular}{|c | c|}
\hline
Trigger Name & Minimum $p_T$ (GeV/c) \\
\hline
VPDMB & 5.0\\
\hline
JP0 & 5.0 \\
\hline
JP1 & 6.0\\
\hline
JP2 & 8.4\\
\hline
\end{tabular}
\end{center}
\caption[Minimum Trigger Jet $p_T$ Values]{\textbf{Minimum Trigger Jet $p_T$ Values} - Minimum jet $p_T$ values that must be met before an event will be accepted as triggered.}
\label{table:minimumTrigPt}
\end{table}%

The final requirement for jet patch triggers only is that the jet must geometrically match to the physical jet patch which fired the trigger in the detector. This, in general, removes split jets. This is a simple process for the barrel and endcap jet patches that begins with knowing which of the barrel, endcap, or overlap jet patches are above the set threshold in each event. For each of these patches the $\eta$ and $\phi$ of the jet are compared to that of the patch. If the jet fulfills the requirements $\lvert\phi_{jet}-\phi_{patch}\rvert < 0.6$ and $\lvert\eta_{jet}-\eta_{patch}\rvert < 0.6$ to any of the physical detector patches, the jet is said to be geometrically matched to a barrel, endcap, or overlap jet patch.

If all of the outlined data cuts are met, and a track is identified as a pion, the analysis yields are stored by using an ``or'' of all of the triggers. This means that if the triggering conditions are met for any trigger, then the yields are merged and not separated out by trigger. The pion yields are defined in a specific way useful for extracting asymmetries from the data in Section \ref{sec:CrossRatio}.

\section{Determination of Kinematic Observables}
\label{sec:kinematicObservables}
The asymmetries will be plotted as a function of:

\begin{itemize}
\item Jet $p_T$
\item Pion $j_T$, the transverse momentum of a pion with respect to the jet axis
\item Pion $z$, the momentum fraction of a pion in a jet
\end{itemize}

The jet $p_T$ sets the hard scale for scattering in $h_1\left(x,Q^2\right)$, with $Q^2 \sim p_T^2$, so plotting the asymmetry against it gives an understanding of how transversity behaves. The pion $z$ and $j_T$ are both dependent variables of the Collins fragmentation function, $\Delta^N\left(z,j_T\right)$, so the asymmetry as a function of these two will give insights of the fragmentation function dependence.

Calculating $z$ is simple since the total momentum of the jet, $\lvert\vec{p}_{jet}\rvert$, and pion, $\lvert\vec{p}_{\pi}\rvert$, are known from the TPC. Using these, $z$ is simply the fraction of the pion's momentum to the jet's momentum, or $z=\lvert\vec{p}_{pion}\rvert/\lvert\vec{p}_{jet}\rvert$. From this definition, it is clear that pions with a higher $z$ carry more of the momentum, and are therefore closer to the jet axis, especially for low jet track multiplicity. For high jet track multiplicity, the average $z$ value will be much lower, which is the most probable case.

Calculating $j_T$ from the data is a bit more complex since it relies on being able to measure transverse momenta of tracks with respect to the jet axis. Since the jets do not have a predefined and fixed coordinate system, a variable set of coordinate axes needs to be applied to each jet so $j_T$ can be calculated consistently. Equations \ref{eq:nls1} - \ref{eq:nls2} define an orthogonal set of axes called the NLS coordinate system. This system aligns one axis with the jet momentum and defines two other axes using the detector $\hat{z}$ axis which lies along the beamline. Using the jet momentum as the variable axis, the other two axes are always defined the same for each jet giving a consistent set of axes to measure from.

\begin{align}
\label{eq:nls1}
\vec{L} &= \vec{P}_{jet}\\
\vec{N} &= \hat{z}\times\vec{L}\\
\label{eq:nls2}
\vec{S} = \vec{N}\times\vec{L} &= \left( \hat{z}\times\vec{L} \right) \times \vec{L}
\end{align}

With a defined coordinate system for each jet, using it is simple. Given a pion within the jet, the projection of its momentum vector ($\vec{P}_{\pi}$) into the N-S plane gives the transverse momentum with respect to the L (or jet) axis, as shown in Equation \ref{eq:pionTrans}. The pion's $j_T$ value is simply the magnitude of this vector, $j_T=\lvert\vec{j}_{T}\rvert$, and thus still carries the units of momentum. 

\begin{equation}
\vec{j}_{T}=\left(\vec{P}_{\pi}\cdot\hat{N}\right)\hat{N}+\left(\vec{P}_{\pi}\cdot\hat{S}\right)\hat{S}
\label{eq:pionTrans}
\end{equation}

The unit vector $\hat{S}$ lies in the reaction plane formed by the jet axis and the beamline. The angle between this reaction plane and $j_T$ is called $\phi_H$, one part of the Collins angle that appears in the sine modulation of the Collins moment in the polarized cross section, $\phi_C=\phi_S-\phi_H$. With the coordinate system and $\vec{j}_T$ in hand, $\phi_H$ is calculated simply as the angle between $\hat{S}$ and $\vec{j}_T$ on the interval $\left[-\pi,\pi\right]$. The other angle, $\phi_S$, is discussed in the next section.

\section{Cross Ratio Asymmetry Formalism}
\label{sec:CrossRatio}
The differential cross section for the azimuthal hadron distribution in a jet may be written as \cite{ref:kangStarPrelim}:

\begin{equation}
\label{eq:ppXsec}
\frac{d\sigma}{d\eta d^2p_T dz d^2 j_T} = F_{UU}\left\{1+A_{UT}^{\sin\left(\phi_S-\phi_H\right)}\sin\left(\phi_S-\phi_H\right)\right\}
\end{equation}

\noindent where $A_{UT}^{\sin\left(\phi_S-\phi_H\right)}$ is the Collins asymmetry,  $F_{UU}$ is the unpolarized structure function and $\phi_S$ and $\phi_H$ are defined in Section \ref{sec:kinematicObservables}. There are several ways to extract the Collins asymmetry, $A_{UT}^{\sin\left(\phi_S-\phi_H\right)}$.  One way is to calculate the difference between the differential cross-sections defined in Equation \ref{eq:ppXsec}:

\begin{equation}
A=\frac{1}{P}\left(d\sigma^\uparrow-d\sigma^\downarrow\right)
\label{eq:xSecAsym}
\end{equation}

\noindent where $P$ is the polarization of the beam. Since $\phi_S$, the angle which describes the proton spin, picks up a phase shift of $\pi$ when the spin is flipped, plugging Equation \ref{eq:ppXsec} into this relationship isolates the Collins asymmetry. This works if there are sizable statistical samples of both spin states, such as at RHIC where proton bunches are polarized in spin patterns which do not favor one state over the other. 

This approach, though, requires precise understanding of the spin-dependent luminosities which will introduce a systematic error. A more appropriate method for symmetric detectors, which cancels both the detector efficiencies and spin-dependent luminosities to leading order, comes with the cross ratio formalism \cite{ref:OhlsenCR}. 

The cross-ratio exploits detector symmetry to write the spin-dependent yields such that detector efficiencies and luminosities cancel. The formalism used for the analysis in this thesis is analogous to Equation \ref{eq:xSecAsym} and is given by

\begin{equation}
\epsilon=P\times A=\frac{\sqrt{N_X^+ N_Y^-}-\sqrt{N_X^- N_Y^+}}{\sqrt{N_X^+ N_Y^-}+\sqrt{N_X^- N_Y^+}}
\label{eq:CrossRatio}
\end{equation}

Here, the yields $N_X^+$, $N_Y^-$, $N_X^-$ and $N_Y^+$ are the numbers of pions which result from initial proton spin states of up ($+$) or down ($-$) that scatter into $X$ and $Y$ halves of the detector. For a symmetric detector, splitting it into halves creates two regions where the physics is the same by rotational invariance, and therefore allowing systematic effects to cancel in Equation \ref{eq:CrossRatio}. For example, consider the case where the detector is split into left and right halves. In this case, a jet that scatters to the left from a spin up event is the same as a jet that scatters to the right from a spin down event by rotational invariance, and vice versa for jet scatter directions. In fact, for a symmetric detector, any dividing line that splits the detector into halves will work. The natural choice for the STAR coordinate system is to split the detector into upper and lower halves, which minimizes contributions from single spin asymmetries (for example, the Sivers moment) which manifest as left-right asymmetries. Therefore, we set $X=U$ and $Y=D$ in subsequent equations to reflect this choice.

The form of the detected pion yields are functions of the spin-dependent luminosity ($\lum$), the efficiency for the given detector half (I), the unpolarized cross section ($\sigma_0$), and modulations of the polarized cross section. Following this labeling convention, the functional yields are given in Equations \ref{eq:Yield1} - \ref{eq:Yield2} where only a single modulation of the polarized cross section is considered for simplicity. In this example the modulation is the Collins asymmetry, $A_{UT}^{\sin\left(\phi_S-\phi_H\right)}$, shortened to $A_C$. Also note the Collins angle, $\phi_S-\phi_H$, is shortened to $\phi_C$. Further modulations will be discussed once the framework is established.

\begin{align}
\label{eq:Yield1}
N_U^+\left(\phi_C\right)&=\lum^+I_U\left(\phi_C\right)\sigma_0\left\{1+ P\times A_C\sin\phi_C\right\}\\
N_U^-\left(\phi_C\right)&=\lum^-I_U\left( \phi_C\right)\sigma_0\left\{1 - P\times A_C\sin\phi_C\right\} \\
N_D^+\left(\phi_C\right)&= \lum^+I_D\left( \phi_C\right)\sigma_0\left\{1 - P\times A_C\sin\phi_C\right\}\\
\label{eq:Yield2}
N_D^-\left(\phi_C\right)&= \lum^-I_D\left( \phi_C\right)\sigma_0\left\{1+ P\times A_C\sin\phi_C\right\}
\end{align}

\noindent The sign of the $\sin\phi_C$ term in these yields follows from rotational invariance. $N_U^+$ and $N_D^-$ are the same sign because they represent the same physics, only with a rotation of $180^\circ$. Similarly, when the jet scatters into the opposite half of the detector from the initial proton spin ($N_U^-$ and $N_D^+$) the physics is the same if one case is rotated by $180^\circ$. The negative sign in front of the $\sin\phi_C$ term when the jet scatter and spin directions are opposite comes from a $\pi$ phase shift to $\phi_C$, which is discussed below.

The yields are easiest to understand from the basic definition of the pion cross section in the absence of any spin, $\sigma=N/\lum$. This is true for a perfectly efficient detector, but in a realistic scenario some tracks are missed. To account for this loss, an efficiency term is added so that raw pion yields would take the form $N=I\sigma\lum$. When the spin information is added, the cross section becomes a spin-dependent cross section, which is equal to an unpolarized cross section, plus terms which are the asymmetry modulations such as Collins, which accounts for all of the additional terms in these definitions.

In the functional yields the Collins angle, $\phi_C$, is defined as $\phi_C=\phi_S-\phi_H$. A depiction of $\phi_S$ and $\phi_H$ is given in Figure \ref{collinskinematics}. The angle $\phi_H$ was defined in the last section as the angle between $\hat{S}$ and $\vec{j}_T$. The angle $\phi_S$ derives from $\phi_{S,true}$, the ``true'' angle between the proton spin direction and, as defined in the STAR frame, the plane formed by the jet axis and the beamline. The quoted angle $\phi_S$ is a generalized version of this angle. For interactions with spin and jet scatter direction in the same hemisphere, $\phi_{S,true}$ is simply the angle between the jet and spin vectors, $\phi_{S,true} = \pi/2 - \phi_{Jet}$. But, this undergoes a $\pi$ rotation when the spin and jet scatter directions are in opposite hemispheres, $\phi_{S,true} = \pi/2 - \phi_{Jet} - \pi$. The sine term absorbs this phase shift and adds a negative sign in front of the asymmetry modulation term. In practice, this means that $\phi_S=+\pi/2-\phi_{Jet}$ for all jets with $\phi_{Jet}>0$ regardless of spin orientation, and by rotational invariance $\phi_S=-\pi/2-\phi_{Jet}$ for all jets with $\phi_{Jet}<0$. Since the STAR coordinate system is setup so that $\phi_{jet}>0$ in the upper half of the detector, and $\phi_{jet}<0$ in the lower half of the detector, splitting into upper and lower halves for the cross ratio is a natural choice. $\phi_S$ has the maximum limits of $\left[-\pi/2,\pi/2\right]$, and combined with the limits of $\phi_H$ the maximum range for $\phi_C$ is $\left[-3\pi/2,3\pi/2\right]$. In the analysis, $\phi_C$ is redefined to span the more natural range $\left[-\pi,\pi\right]$.

\figuremacroW{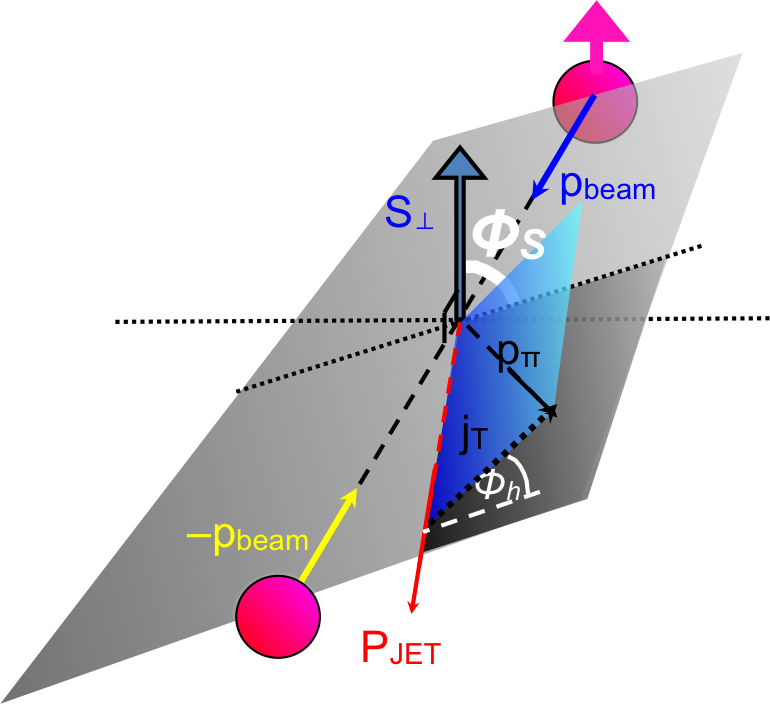}{STAR Scattering Kinematics}{Depiction of the scattering kinematics used for this analysis. The angle $\phi_S$ is the angle between the scattering plane, which intersects the beamline and the jet axis ($p_{jet}$), and the polarized proton spin ($S_\perp$). The angle $\phi_H$ is the angle between the scattering plane and $j_T$, where $j_T$ lies in the plane that intersects the pion momentum ($p_\pi$) and $p_{jet}$.}{0.85}

The result of the cross ratio is shown in Equation \ref{eq:CrossRatioResult}, where all detector efficiencies, luminosities, and cross sections cancel leaving only the asymmetry term with its sinusoidal modulation. In this case of a Collins asymmetry, where the result is a function of the Collins angle, $\phi_C$, the result must be fit with a sinusoidal polynomial $p_0+p_1\sin\phi_C$ to yield the asymmetry extraction. In this fit, $p_0$ is a cross check term which should be consistent with zero. Using the $p_0$ values extracted when calculating the asymmetry as a function of jet $p_T$, they scatter statistically according to a Gaussian distribution with a mean of $-5.08 \left(10^{-4}\right) \pm 1.43\left(10^{-4}\right)$. The $p_1$ term is the Collins asymmetry not yet corrected for the imperfect beam polarization. The beam polarization calculation and correction will be discussed shortly.

\begin{align}
\epsilon\left(\phi_C\right)&=\frac{\sqrt{N_U^+\left(\phi_C\right) N_D^-\left(\phi_C\right)}-\sqrt{N_U^-\left(\phi_C\right) N_D^+\left(\phi_C\right)}}{\sqrt{N_U^+\left(\phi_C\right) N_D^-\left(\phi_C\right)}+\sqrt{N_U^-\left(\phi_C\right) N_D^+\left(\phi_C\right)}}\nonumber \\
&=\frac{\sqrt{\sigma_0^2\lum^+\lum^-I_U\left(\phi_C\right)I_D\left(\phi_C\right)}\left(\sqrt{\left\{1+PA_C\sin\phi_C\right\}^2}-\sqrt{\left\{1-PA_C\sin\phi_C\right\}^2}\right)}{\sqrt{\sigma_0^2\lum^+\lum^-I_U\left(\phi_C\right)I_D\left(\phi_C\right)}\left(\sqrt{\left\{1+PA_C\sin\phi_C\right\}^2}+\sqrt{\left\{1-PA_C\sin\phi_C\right\}^2}\right)} \nonumber \\
&= \frac{1+PA_C\sin\phi_C-\left(1-PA_C\sin\phi_C\right)}{1+PA_C\sin\phi_C+1-PA_C\sin\phi_C} \nonumber \\
&=\frac{2PA_C\sin\phi_C}{2} \nonumber \\
&=P\times A_C\sin\phi_C
\label{eq:CrossRatioResult}
\end{align}

\subsection{Binning Considerations}
\label{sec:Binning}
The functional forms of the yields are a useful learning tool, but they are not so useful for analysis where the accessible information include track and jet observables. To extract asymmetries, two-dimensional histograms are used to count the yields. One axis holds the $\phi$ information, with twelve bins used for $\phi_C$ which translates to six bins used for describing $\phi_S$. The data are being split a lot to have four different yields, so using fewer bins is better to ensure the population of all bins.

The other axis of the histogram holds the kinematic information about jet $p_T$, $z$, and $j_T$. Tables \ref{table:jetPtBins} - \ref{table:pionJtBins} outline the kinematic bins used for the analysis. These bins were chosen to optimize the sampled physics while also ensuring the statistical power of each bin is good enough to calculate asymmetries. When the two-dimensional histograms are filled, the angle is known and the kinematic value is known, so there is a bin waiting for that value. These yields now hold all of the transverse spin asymmetry modulations, and they must now be extracted.

\begin{table}[htp]
\begin{center}
\begin{tabular}{|c | c | c|}
\hline
Bin Number & $p_T$ Low [GeV/c] & $p_T$ High [GeV/c]\\
\hline
1 & 6.0 & 7.1 \\
\hline
2 & 7.1 & 8.4 \\
\hline
3 & 8.4 & 9.9 \\
\hline
4 & 9.9 & 11.7 \\
\hline
5 & 11.7 & 13.8 \\
\hline
6 & 13.8 & 16.3\\
\hline
7 & 16.3 & 19.2 \\
\hline
8 & 19.2 & 22.7 \\
\hline
9 & 22.7 & 26.8 \\
\hline
10 & 26.8 & 31.6 \\
\hline
\end{tabular}
\end{center}
\caption[Jet $p_T$ Bins]{\textbf{Jet $p_T$ Bins} - Jet $p_T$ bins used for analysis optimized so that statistics are more evenly distributed for increasing $p_T$.}
\label{table:jetPtBins}
\end{table}%

\begin{table}[htp]
\begin{center}
\begin{tabular}{|c | c | c|}
\hline
Bin Number & $z$ Low & $z$ High \\
\hline
1 & 0.1 & 0.2 \\
\hline
2 & 0.2 & 0.3 \\
\hline
3 & 0.3 & 0.4 \\
\hline
4 & 0.4 & 0.6 \\
\hline
5 & 0.6 & 0.8 \\
\hline
\end{tabular}
\end{center}
\caption[Pion $z$ Bins]{\textbf{Pion $z$ Bins} - Pion $z$ bins used for analysis.}
\label{table:pionZBins}
\end{table}%

\begin{table}[htp]
\begin{center}
\begin{tabular}{|c | c | c|}
\hline
Bin Number & $j_T$ Low [GeV/c] & $j_T$ High [GeV/c] \\
\hline
1 & 0.05 & 0.15 \\
\hline
2 & 0.15 & 0.25 \\
\hline
3 & 0.25 & 0.375 \\
\hline
4 & 0.375 & 0.5 \\
\hline
5 & 0.5 & 1.0 \\
\hline
6 & 1.0 & 4.5 \\
\hline
\end{tabular}
\end{center}
\caption[Pion $j_T$ Bins]{\textbf{Pion $j_T$ Bins} - Pion $j_T$ bins used for analysis optimized so that statistics are roughly equalized across the bins.}
\label{table:pionJtBins}
\end{table}%

In addition to these bins, it is possible to select ranges of kinematic variables and do multidimensional plotting of the results to help map out the underlying physics. The asymmetry results that will be presented later are plotted in four different ways to highlight the physics:

\begin{itemize}
\item As a function of jet $p_T$ integrated over all $z$ and $j_T$ to highlight the behavior of the asymmetries as a function of the hard scale.
\item As a function of $z$ and $j_T$ in six ranges of jet $p_T$ (Table \ref{table:ptRanges}) to understand how the fragmentation function variables are varying with the hard scale.
\item As a function of $j_T$ in four ranges of pion $z$ (Table \ref{table:zRanges}) for jet $p_T > 9.9$ GeV/c to understand how the two fragmentation variables are dependent upon each other and disentangle their dependencies.
\end{itemize}

\begin{table}[htp]
\begin{center}
\begin{tabular}{|c | c|}
\hline
Range Number & Jet $p_T$ Range [GeV/c] \\
\hline
1 & 6.0 - 8.4\\
\hline
2 & 8.4 - 9.9\\
\hline
3 & 9.9 - 11.7\\
\hline
4 & 11.7 - 16.3\\
\hline
5 & 16.3 - 19.2\\
\hline
6 & 19.2 - 31.6\\
\hline
\end{tabular}
\end{center}
\caption[Jet $p_T$ Ranges]{\textbf{Jet $p_T$ Ranges} - Ranges of jet $p_T$ used later for plotting the $z$ and $j_T$ asymmetries.}
\label{table:ptRanges}
\end{table}%

\begin{table}[htp]
\begin{center}
\begin{tabular}{|c | c|}
\hline
Range Number & Pion $z$ Range\\
\hline
1 & 0.1 - 0.2\\
\hline
2 & 0.2 - 0.3\\
\hline
3 & 0.3 - 0.4\\
\hline
4 & 0.4 - 0.8\\
\hline
\end{tabular}
\end{center}
\caption[Pion $z$ Ranges]{\textbf{Pion $z$ Ranges} - Ranges of pion $z$ used later for plotting the $j_T$ asymmetries.}
\label{table:zRanges}
\end{table}%

\subsection{Extracting Asymmetries from Yields}
\label{sec:asymExtract}
Now Equation \ref{eq:CrossRatio} must be applied multiple times to the yields so the asymmetries may be extracted. For each kinematic bin, the cross ratio is applied to each $\phi$ bin, and the result is stored in a new histogram used for fitting. Using the same yield labeling scheme as in Section \ref{sec:CrossRatio}, the statistical error applied to the cross ratio result in each bin is given as

\begin{equation}
\label{eq:CrossRatioErr}
\delta\epsilon = \frac{\sqrt{\left(1 + \epsilon\right)^2\left(N_U^- + N_D^+\right)+ \left(1 - \epsilon\right)^2\left(N^+_U + N^-_D\right)}}{2\left(\sqrt{N^+_UN^-_D}+\sqrt{N_U^-N_D^+}\right)}
\end{equation}

Then, for each kinematic bin, the resulting cross ratio histogram is fit with a sinusoidal polynomial $p_0+p_1\sin\left(\phi_C\right)$. The value of $p_1$ is extracted from the fit and divided by the average beam polarization giving the final beam asymmetry. The statistical error from the fit and beam polarization are used to calculate the final error given to the asymmetry value. The final result of this algorithm is shown in Figure \ref{ARatio_Fit}. 

\figuremacroW{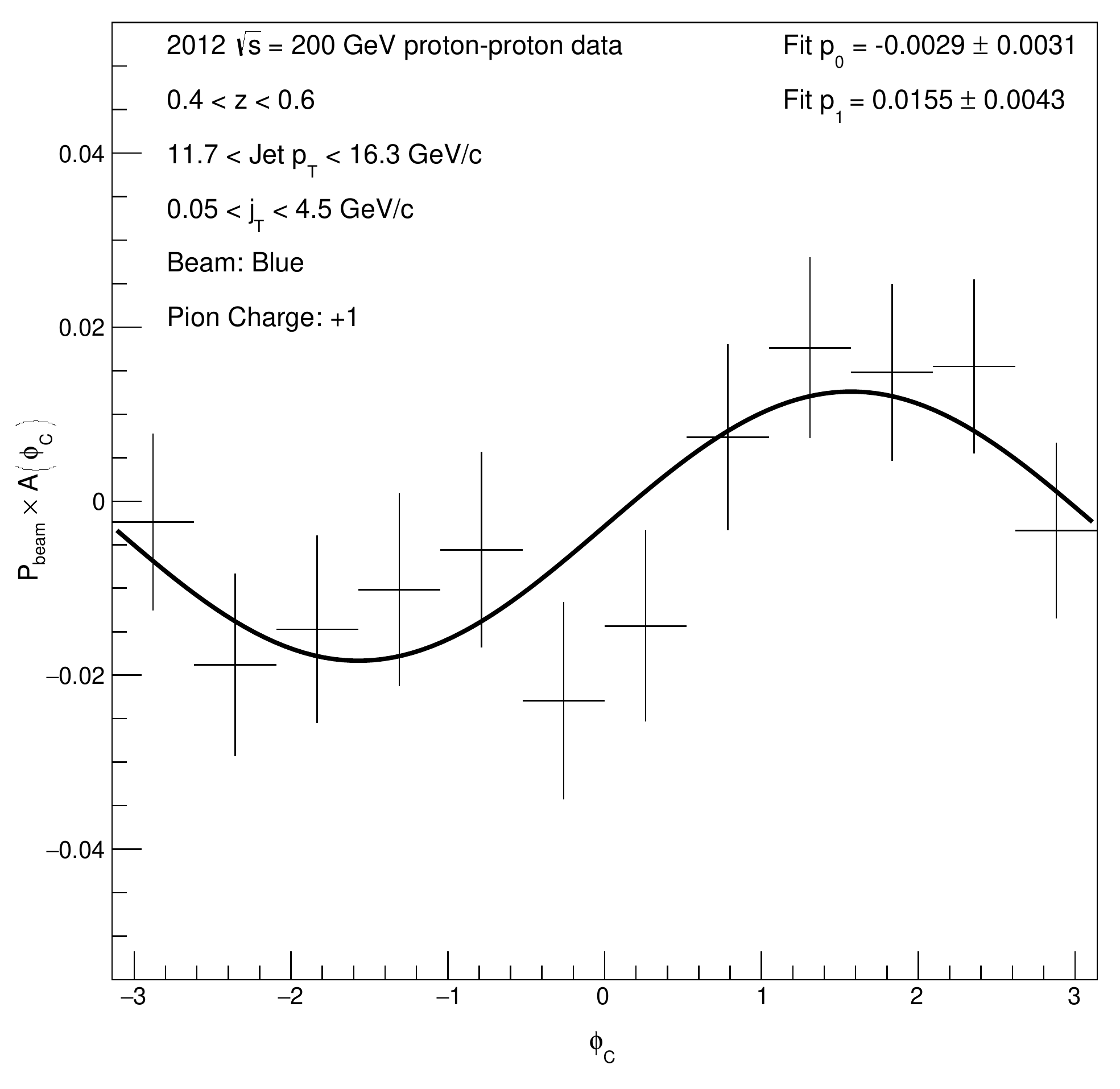}{Example Asymmetry Fit}{An example fit for a particular $z$ bin and Jet $p_T$ range. The sinusoidal behavior of the cross ratio is very apparent even without the fit.}{0.75}

This process is further complicated because of the two beams of protons that are collided. To get a final singular asymmetry value, the same algorithm is followed for each beam, blue and yellow, and then the weighted average of the two points is used for the final asymmetry value plotted for each kinematic bin.

\subsection{Polarization Determination}
The polarization needs to be known very well to correct the cross ratio asymmetries for imperfect beam conditions. During running, the polarization is measured towards the beginning of each RHIC fill so an initial value may be determined, and then monitored with several subsequent measurements spread out throughout the beam fill. Plotting these polarization measurements as a function of time, they may be fit with a straight line to extract the initial constant polarization ($P_0$) and the decay rate or slope ($dP/dt$). For each event that is analyzed, the amount of time since the initial polarization measurement ($\Delta t$) may be used to calculate the polarization for that event based on the same straight line fit used to extract the parameters:

\begin{equation}
\label{eq:EventPol}
P\left(t\right) = P_0 + \frac{dP}{dt}\Delta t
\end{equation}

The polarization values for each good analysis event are stored in a histogram so that once the analysis code has run over all data, all histograms may be combined to extract the average polarization value. This value, in turn, is used in the asymmetry extraction algorithm. Similarly the error in the polarization value is calculated event-by-event and stored so the average value may be used for the calculation of error in the final beam asymmetry value. 

Figure \ref{PolarizationVsRun} shows the average beam polarization value and its error plotted for all runs that are used in analysis. Runs which have an average polarization of zero for both blue and yellow beams are not shown on this plot. These are runs which pass all quality analysis checks but have no good events to contribute to the analysis, meaning they do not contribute to the asymmetry results. The values in Figure \ref{PolarizationVsRun} clearly show how the polarization varies within each RHIC fill, with clear jumps at the start of a new beam fill. Taking the average over all runs, the average polarization is $61.10 \pm 4.97\%$ for the blue beam and $52.73 \pm 5.92\%$ for the yellow beam. These are the polarization values and errors that are used in the calculation of the final asymmetry value for each beam. A systematic error due to the polarization is assigned to the final asymmetries and is discussed in Chapter \ref{ch:Systematics}.

\figuremacroW{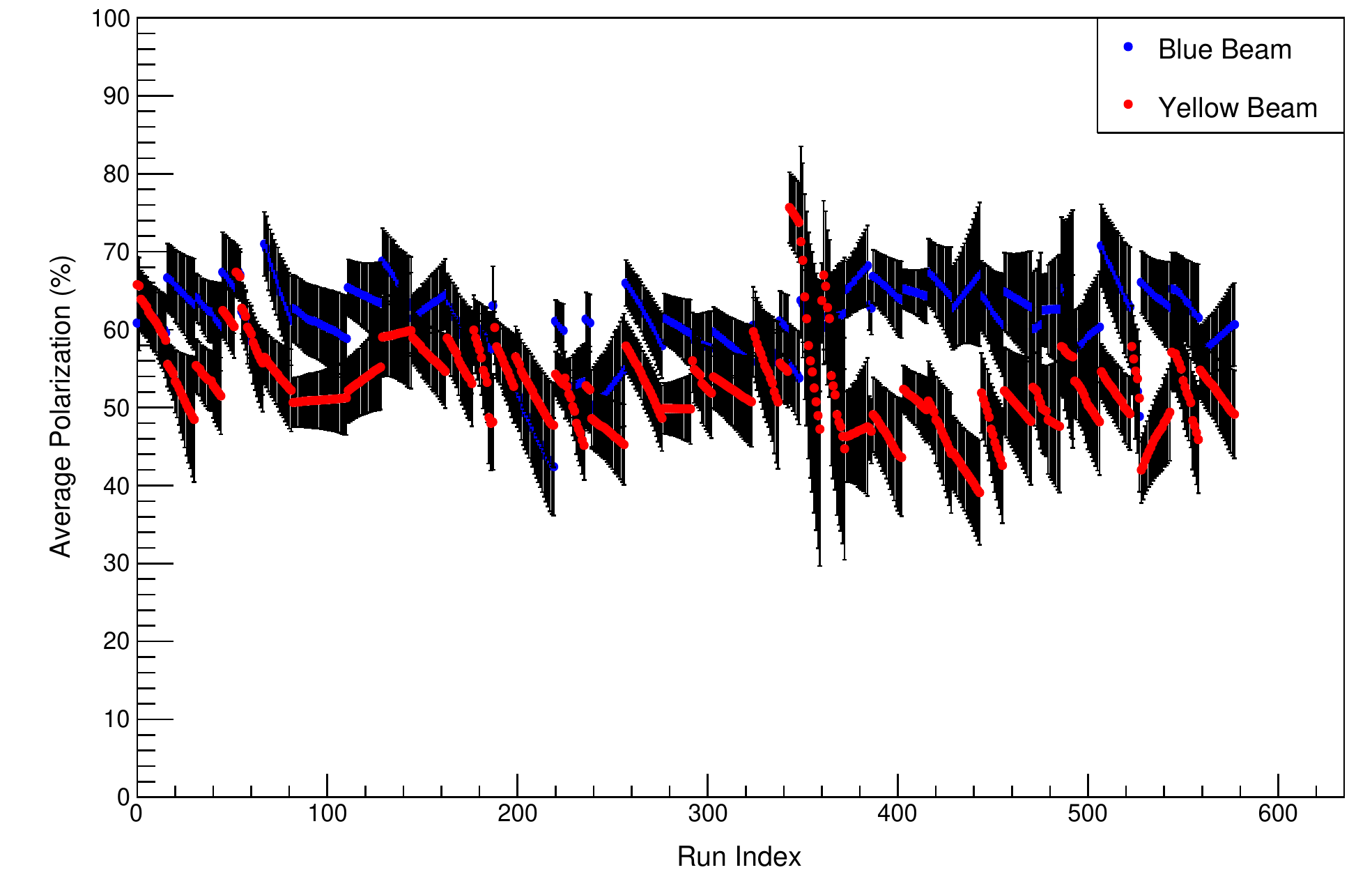}{Average Polarization vs. Run Index}{The average polarization for each run plotted as a function of the run index, which starts from zero and is not the same as the run number. Note that if the polarization for both beams is zero then the run is not shown on this plot, as it would not be included in analysis anyway.}{1}


\chapter{Corrections and Systematic Errors} 
\label{ch:Systematics}


\ifpdf
    \graphicspath{{ch6_Systematics/figures/PNG/}{ch6_Systematics/figures/PDF/}{ch6_Systematics/figures/}}
\else
    \graphicspath{{ch6_Systematics/figures/EPS/}{ch6_Systematics/figures/}}
\fi

The unfortunate truth is that using particle collisions to learn about underlying physical processes forces a reliance upon particle detectors. And in all cases these detectors, with their limited resolutions and efficiencies, systematically bias the physics result and message. As experimentalists, it is our duty to understand these systematic biases and errors so that they can be accounted for in the final result. However, trying to determine unknown biases in the data using only the detector output is challenging. Therefore, we seek a method of understanding systematic effects where the detector effects may be turned off.

For this analysis, we tune a Monte Carlo generator so that the detector response matches as well as possible in data and simulation.  With this agreement in hand we can in good faith pursue studies of the physics at multiple stages of the processes:  at the level of the partons before fragmentation (parton level), at the level of the particles after fragmentation but before detection (particle level), and at the level of the detector (detector level). Since the simulation reproduces the data well, these studies with and without the detector included gives a measure of the systematic effects imparted by the detector. This chapter is dedicated to development of the simulation sample, and measuring corrections to the data as well as systematic errors and biases that affect the final result.

\section {Simulation Methods for STAR}
STAR has implemented a standard procedure of generating QCD events with the PYTHIA Monte Carlo generator \cite{ref:Pythia6} and simulating the passage of these events through the detector geometry with GEANT3 \cite{ref:Geant}. However, this formalism does not reproduce the conditions inside the detector during data collection, such as pileup and backgrounds. The only way to reproduce these conditions is with real data, thus the PYTHIA events are embedded into real zero bias data to help the simulation sample have the best match possible to the data output. The zero bias trigger is set up to trigger randomly during any of the possible 120 bunch crossings at STAR. Since there is no detector requirement, most often the zero bias events are comprised of the pileup and background that permeate the other triggered events and may be used to encode this information into the simulation sample.

\subsection{Event Generation with PYTHIA}
PYTHIA is a robust framework used to generate complete high energy physics events with enough detail to match experimental observations. The beam particle species, beam energy, and types of processes to produce may be set to multiple different configurations to handle a wide array of experiments. For this analysis, of course, the beam species are both protons and the center of mass energy is set to 200 GeV. The physical processes in each generated event are also governed by initial input supplied by the user. This input includes the parton distribution functions used to select partons in hadronic collisions, underlying event (UE) and beam remnant contributions, fragmentation and hadronization, and initial/final state radiation to name only a few. All of these parameters may be set independently, or they may be collectively set using one of the many prepared ``tunes'' within the PYTHIA program. These prepared tunes can do a decent job of matching to STAR data and have been used out of the box exclusively in the past. However, the tune parameters are generally derived from experiments with a much higher center of mass energy (such as the Tevatron or LHC), meaning the prepared tunes need some adjusting for the best possible match to STAR data. The development of an appropriate tune for this analysis will be given later in this chapter.

\subsection{Modeling STAR with GEANT}
To make comparisons between the simulated PYTHIA events and the collected data, the detector response to the PYTHIA events must be simulated. GEANT is a framework to simulate particles passing through matter which provides the link between simulation and data. Detector specific geometry and materials may be simulated within the framework so the PYTHIA particles pass through an exact replica of the experimental setup. Therefore, the STAR geometry is updated yearly based on detector additions or extractions so that the correct amount of material is accounted for in the simulations, yielding the most accurate simulations possible.

\subsection{Embedding STAR Data}
Regardless of how well PYTHIA generates events, it can never completely reproduce the STAR detector conditions. These conditions vary depending upon the particle species in the accelerator and at what energy they are being collided. For instance, 200 GeV center of mass energy proton collisions produce, on average, lower multiplicity events than 500 GeV collisions. Higher average multiplicity leads to more pileup, or mixing of TPC hits from separately triggered events. Unfortunately, the detector response to effects of pileup and backgrounds cannot be accounted for in the pure simulation, so the PYTHIA events are embedded into the raw zero-bias DAQ file events. This embedding mixes the response from the pure simulation with the detector effects of pileup and backgrounds on an event-by-event basis, resulting in a sample that is the best recreation of the data possible.

\section{Developing a New PYTHIA Tune}
The PYTHIA Monte Carlo generator aims to describe proton-proton collisions through a mixture of theory, models like multiple parton interactions and fragmentation, and fits based on existing experimental data such as parton distribution functions, decay widths, and particle multiplicities. The vast majority of the proton-proton interaction data included in PYTHIA is taken from the Tevatron and the LHC, which include data up to center-of-mass energy of $13$ TeV, nearly two orders of magnitude greater than the center-of-mass energy of the collisions in this analysis. As a result, it is imperative to tune the available parameters in PYTHIA to data taken at $\sqrt{s}  = 200$ GeV collisions.

The ideal way to optimize PYTHIA for STAR is to compare the output from PYTHIA \& GEANT to the real STAR data. The very best case would be to produce multiple full embedding samples for each tune under investigation, and make comparisons between the embedding output and data. These comparisons could be very robust, comprising many observables from the data, and the final tune used for analysis of systematic errors would be the one which matches the best to the kinematic observables. This approach simply is not practical because the production of the embedding samples takes a long time for even a modest number of generated events.

Another approach is to generate a large amount of pure PYTHIA events at the particle level, which can be done quite fast, and compare the output to STAR minimum bias trigger data. It is important to note, though, that before the comparison is made, the STAR data must be corrected for detector inefficiencies and resolution so that it may be directly compared to pure PYTHIA particle level output. STAR has published invariant yield momentum distributions of $\pi^\pm$ particles using data collected in 2005 \cite{ref:starPion2005} and 2012 \cite{ref:starPion2012} in proton-proton collisions at $\sqrt{s}=200$ GeV. All told, these results combine to cover a momentum range of 0.3-15 GeV/c, they have been corrected for detector effects and they are ready to compare to pure PYTHIA particle level output. The combined results from the two data sets are shown in Figure \ref{DataInvYields_PiPlus} for $\pi^+$ particles. These data give a unit of measure to compare the output from each PYTHIA tune, to give a tune which matches well to STAR data. The $\pi^-$ data is published but will not be used in the discussion of tune selection as it returns the same results overall.

\figuremacroW{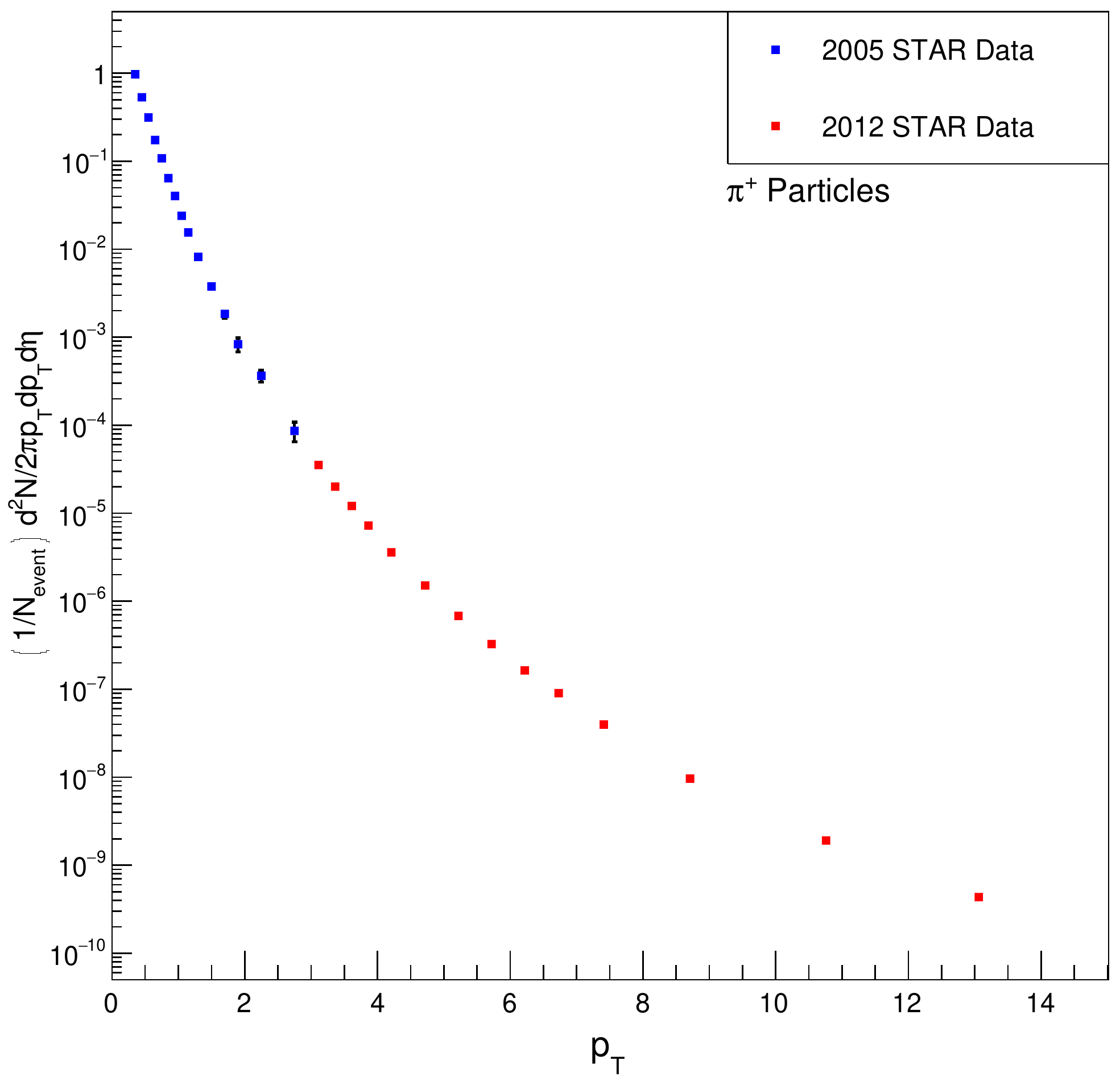}{Data Invariant Yield Momentum Distribution}{The combined 2005 and 2012 invariant yield momentum distribution for $\pi^+$ particles.}{0.75}

The charged pion invariant yields were compared to several pure PYTHIA simulation samples, each representing a different tune. This analysis utilized the Perugia 0 tune, which was used in previous $\sqrt{s}=200$ GeV analyses \cite{ref:2009ALL}, as well as the updated Perugia tunes: Perugia 2012, Perugia 2012-ueHi, and Perugia 2012-ueLo \cite{ref:PerugiaTunes}. These tunes are also compared to CDF Tune A \cite{ref:cdfTuneA1,ref:cdfTuneA2,ref:cdfTuneA3,ref:cdfTuneA4,ref:cdfTuneA5,ref:cdfTuneA6}, a much older tune which was developed for Tevatron data and used in very early STAR jet analyses. The updated Perugia 2012 tunes set the intrinsic partonic transverse momentum $k_T=1$ GeV/c by default, and allow the user to access an updated PDF set that is not standard for the Perugia 0 tune. CDF Tune A was included to see how it agreed with the low $p_T$ yields, a region where the UE contributions are large. The comparison between the data and these tunes using the parameters set to their nominal values is shown in Figure \ref{InvariantYieldRatios_PiPlusAllPt}. 

\figuremacroW{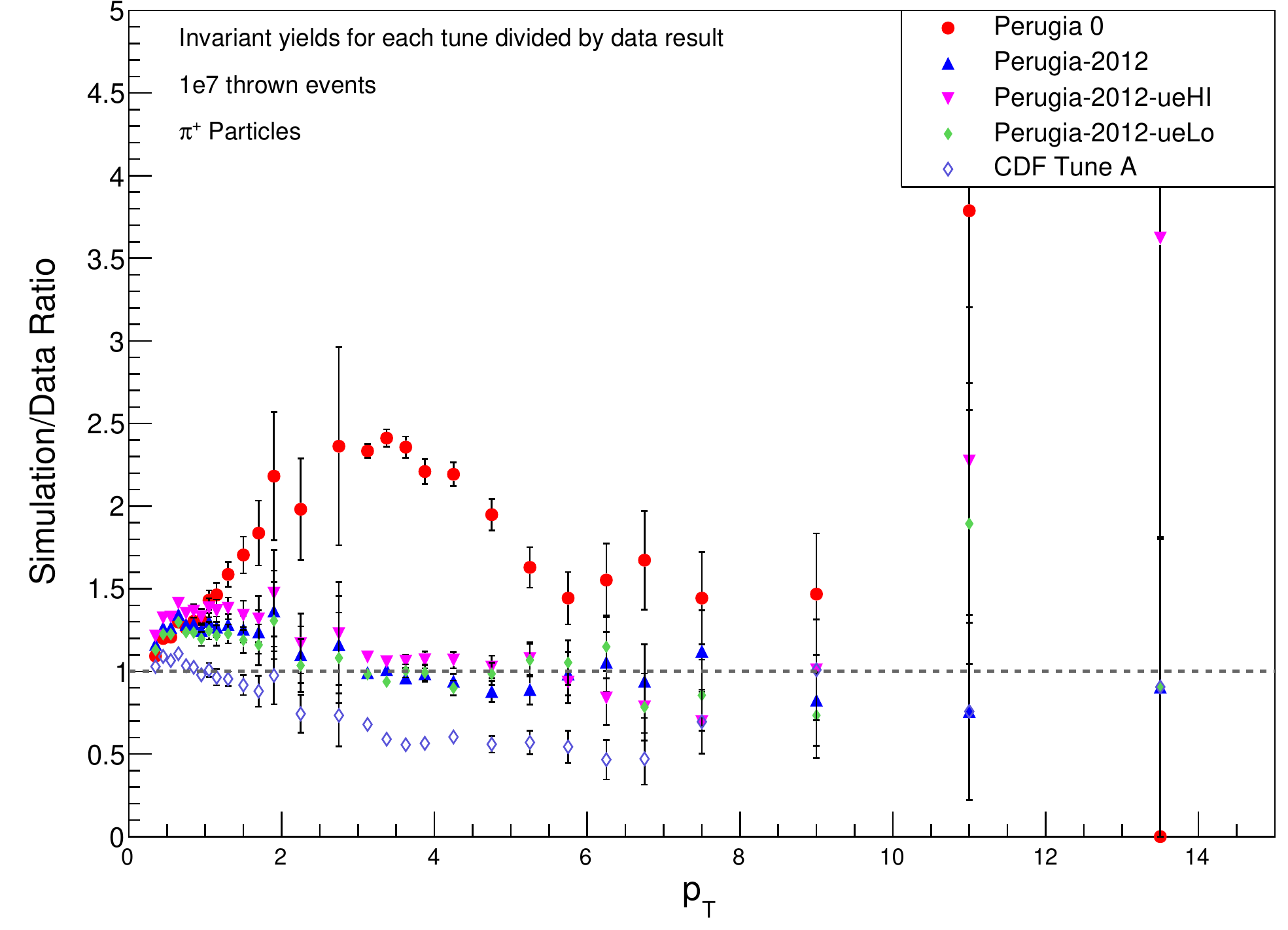}{Nominal Tunes Comparison}{Comparison between the invariant yields from data and from the various PYTHIA tunes with parameters set to their nominal values.}{0.75}

The immediate conclusion from the results in Figure \ref{InvariantYieldRatios_PiPlusAllPt} is that the Perugia 2012 tunes match the data well out of the box at higher $p_T$ and the CDF Tune A matches better at lower $p_T$. Perugia 0 gives the poorest agreement of all the tunes. Previous studies at STAR suggest that this disagreement is reduced by setting the intrinsic $k_T$ parameter to $1$ GeV/c. Setting $k_T = 1$ GeV/c in Perugia 0 results in a vastly improved comparison, shown in Figure \ref{InvariantYieldRatios_PiPlusAllPt_p0kT1}. The modified Perugia 0 tune performs on par with the Perugia 2012 tunes.

\figuremacroW{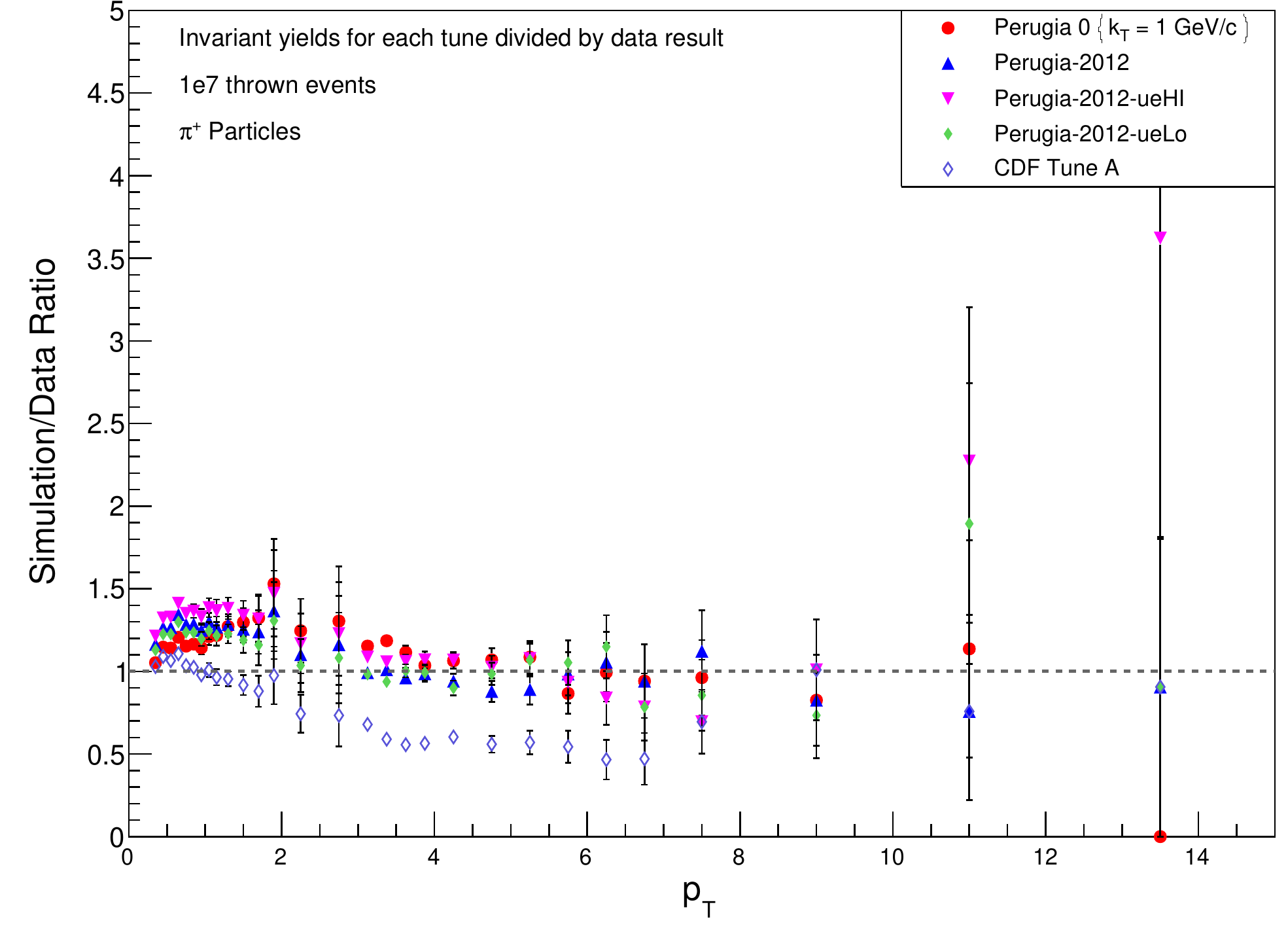}{Perugia 0 Reduced $k_T$ Ratio}{Comparison between the invariant yields from data and from the various PYTHIA tunes with parameters set to their nominal values, except for Perugia 0 where the primordial $k_T$ is set to 1 GeV/c.}{0.75}

Although the performance of all the Perugia tunes is now very similar, the Perugia 2012 tune was selected because it uses the updated CTEQ6L1 PDF set \cite{ref:CTEQ6L1} that is not available in Perugia 0. The only issue with Perugia 2012 is that it does not match the data well at lower $p_T$ where the UE contributes to the process the most. Fortunately, the UE contributions may be tuned by adjusting either the $p_{T,0}\left(s_{ref}\right)$ or $P_{90}$ exponent parameters. These two parameters are directly related by the equation 

\begin{equation}
\label{eq:pT0}
p_{T,0}^2\left(s\right) = p_{T,0}^2\left(s_{ref}\right)\left(\frac{s}{s_{ref}}\right)^{P_{90}}
\end{equation}

\noindent where $p_{T,0}\left(s\right)$ is the infrared regularization scale for multiple parton interactions. The nominal  $p_{T,0}\left(s_{ref}\right)$  and $P_{90}$ terms are tuned to the $\sqrt{s} = 7$ TeV LHC data. Trial and error tells us that the best match to the STAR data comes when the $p_{T,0}\left(s_{ref}\right)$ is increased by 10\%. However, as noted, this parameter is expected to be correctly tuned to the $\sqrt{s_{ref}}$ value, so it is best to make similar adjustments to the $P_{90}$ exponent which is not tuned to reference energy. With the reference energy being so large, the RHIC center of mass energy of 200 GeV gives quite a large lever arm to tune the energy dependence of the UE contributions, and give a good match to the STAR data. Decreasing the energy exponent is an effective increase in the $p_{T,0}\left(s_{ref}\right)$ parameter, and decreasing the exponent by 11.25\% yields the same result as increasing $p_{T,0}\left(s_{ref}\right)$ by 10\%. Both of these changes to the parameters are shown in Figure \ref{InvariantYieldRatios_PiPlusAllPt_ReducedExp}, and the effects are the same. Visually, the match to the data is equally as good whether we make adjustments to the UE or not when $p_T > 3$ GeV/c, but below this mark the changes to the tune do a world of good resulting in a much better match to the STAR data.

\figuremacroW{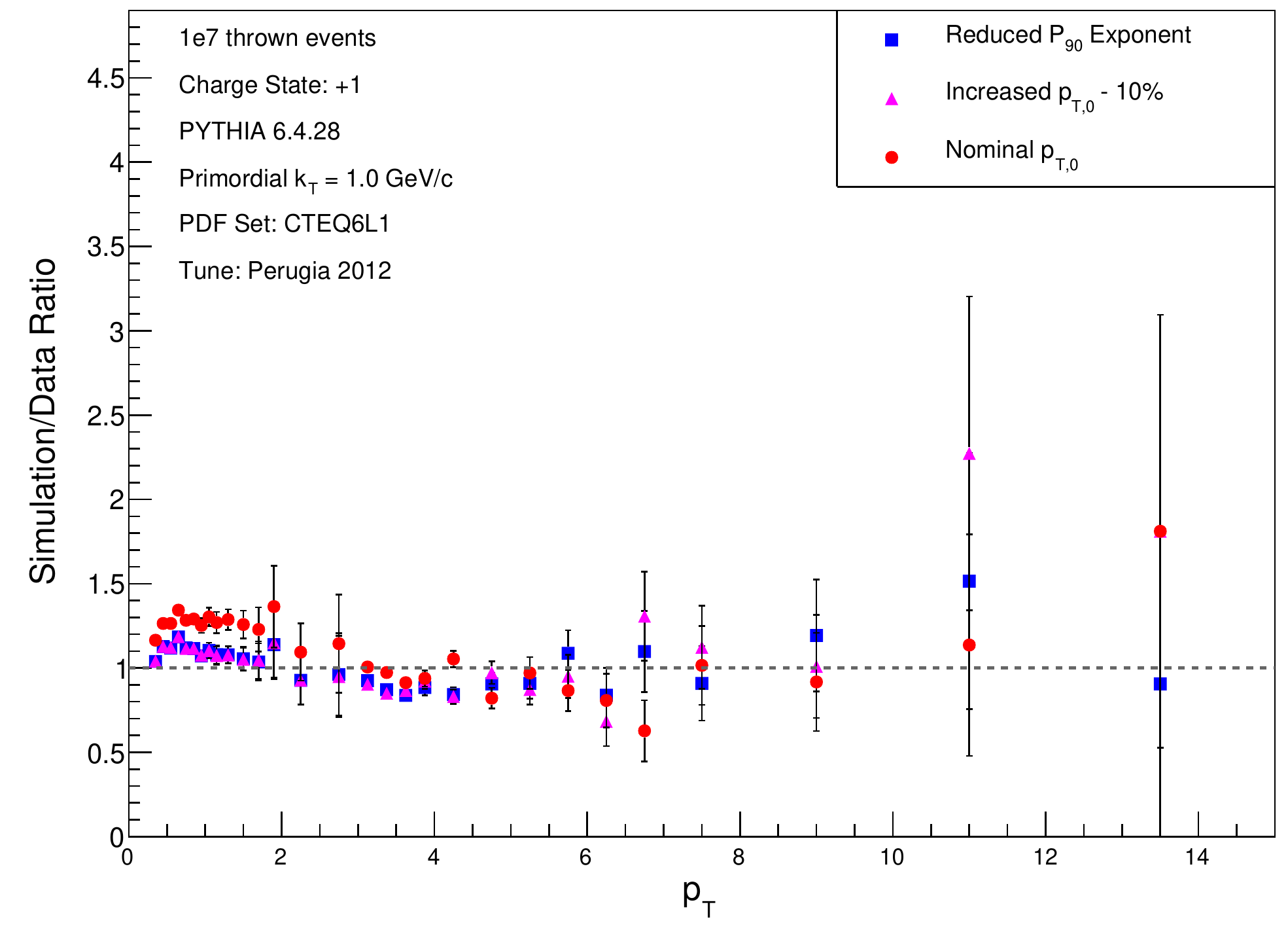}{Reduced $P_{90}$ vs. Increased $p_{T,0}\left(s_{ref}\right)$}{Comparing the effects of reducing the energy exponent vs increasing the $p_{T,0}\left(s_{ref}\right)$ parameter. The result of each change is shown compared to the original out of the box Perugia 2012 tune.}{0.75}

In summary, we choose Perugia 2012 because of the excellent ``out of the box'' agreement for $p_T > 3$ GeV/c and the updated CTEQ6L1 PDF set. Then to get good agreement at low $p_T$ where the UE contributions are strong, the energy scaling $P_{90}$ parameter (PARP(90) in PYTHIA) is reduced by 11.25\% from 0.240 to 0.213. 

\section{Embedding Sample Statistics and QA}
\label{sec:embedQA}
The number of events produced for this sample was determined by scaling the statistics used for the 2009 embedding sample. The 2009 sample was used to determine systematic errors for the preliminary version of this result. In the data we use the jet patch triggers to enhance the high $p_T$ jet sample. In the embedding we do this by producing samples of increasing partonic $p_T$ ($\hat{p}_T$), meaning the $p_T$ in the center of mass frame of the partons engaged in the hard scattering process. Generating events this way ensures that the embedding statistics are comparable to the data statistics across the entire jet $p_T$ range of the measurement. It is also possible to simply generate a minimum bias sample and use the emulated trigger simulation to select higher $p_T$ jets. This method would require a vast increase in the simulated data we would need to produce. The $\hat{p}_T$ bins we are using as well as the statistics generated and the partonic cross section for all included subprocesses for each bin are given in Table \ref{table:EmbedStats}.

\begin{table}[htp]
\begin{center}
\begin{tabular}{|c|c|c|}
\hline
$\hat{p}_T$ Bin [GeV/c] & Generated Events & Partonic Cross Section (mb)\\
\hline
2-3 & 2100295 & $9.006$\\
\hline
3-4 & 600300 & $1.462$ \\
\hline
4-5 & 600300 & $3.544\times 10^{-1}$\\
\hline
5-7 & 300289 & $1.514 \times 10^{-1}$ \\
\hline
7-9 & 300289 & $2.489\times 10^{-2}$\\
\hline
9-11 & 300289 & $5.846\times 10^{-3}$ \\
\hline
11-15 & 160295 & $2.305 \times 10^{-3}$\\
\hline
15-20 & 100302 & $3.427 \times 10^{-4}$\\
\hline
20-25 & 80293 & $4.563 \times 10^{-5}$\\
\hline
25-35 & 76303 & $9.738 \times 10^{-6}$\\
\hline
35-$\infty$ & 23307 & $5.020 \times 10^{-7}$\\
\hline
\end{tabular}
\end{center}
\caption[Embedding Statistics]{\textbf{Embedding Statistics} - Number of embedding events generated for each of the 11 $\hat{p}_T$ bins.}
\label{table:EmbedStats}
\end{table}%

With statistics split up into several $\hat{p}_T$ bins, they must be combined to have one complete sample. The analysis of the simulation is completed for each $\hat{p}_T$ bin, and then the various distributions we have filled are summed from each $\hat{p}_T$ bin and normalized with a weight that is proportional to the number of statistics within the bin to be sure all bins are represented equally. The weight used is the inverse luminosity for each bin, where the luminosity is calculated simply as $\mathcal{L}=N/\hat{\sigma}$. In this relationship $N$ is the number of events produced for each bin and $\hat{\sigma}$ is the average cross section for all included subprocesses generated by PYTHIA which is extracted during simulation production. For this analysis, the luminosity is normalized to the luminosity of the $35-\infty$ GeV/c bin so that the inverse luminosity of each bin, the weight, is greater than 1 and therefore makes for more intuitive recombined distributions. A simple test that this weighting scheme is working correctly is to plot the $\hat{p}_T$ distribution, as in Figure \ref{PartonicPt}, and check that it is smooth. Without proper representation of the luminosity from each bin, the distribution would exhibit jumps at the partonic bin edges.

\figuremacroW{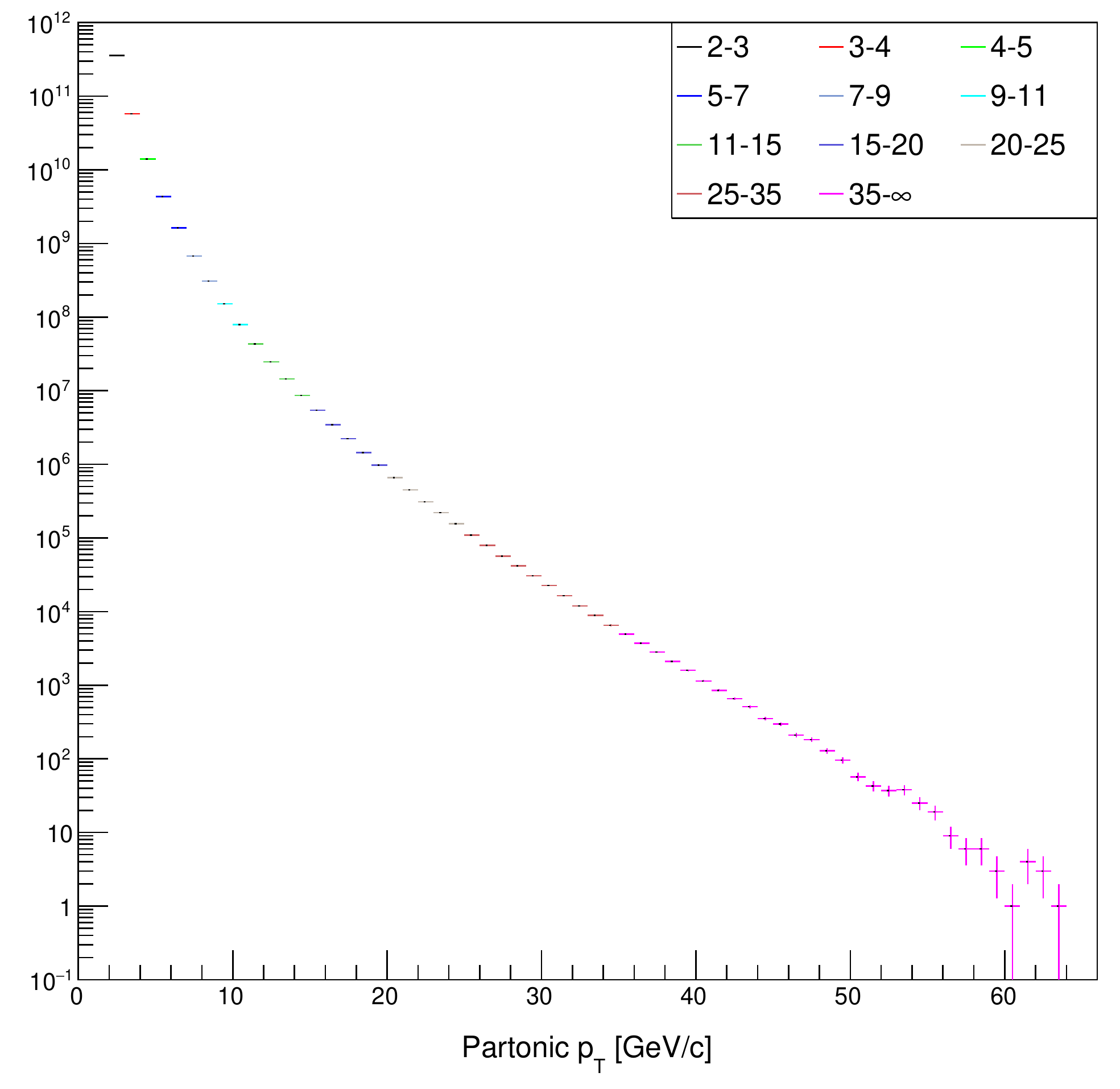}{Partonic $p_T$ Sepctrum}{The full partonic $p_T$ spectrum that combines statistics from all $\hat{p}_T$ bins. The distribution is smooth at all $\hat{p}_T$ bin edges, showing that the weighting scheme for combining bins is correct.}{0.8}

One of the input parameters to the simulation is to setup the vertex distribution that is used by PYTHIA to generate events. From the data, we get the mean position of vertex distribution on the $\left(x,y,z\right)$-axes as well as the spread in these distributions. From these input parameters, a random vertex is generated and used to generate the PYTHIA event. However, when the vertices are reconstructed at the detector level, the z-axis position distribution may not match up with the data. To ensure they align, the data distribution is divided by the simulation distribution and the result is fit with a $4^{th}$ order polynomial, with the dependent variable being the z-axis position of the vertex. Then on an event-by-event basis, a vertex weight is calculated using the output from the polynomial fit, and each given vertex z-axis position. This weighted vertex distribution should now match the data much better, as shown in Figure \ref{fig:vertexWeighting} where the unweighted and weighted distributions are compared to the data. This vertex weight is subsequently applied to every distribution that is filled in the simulation so that the vertex correction is propagated to all simulation analysis output.

\begin{figure}
    \centering
    \subfloat[]{{\includegraphics[width=7cm]{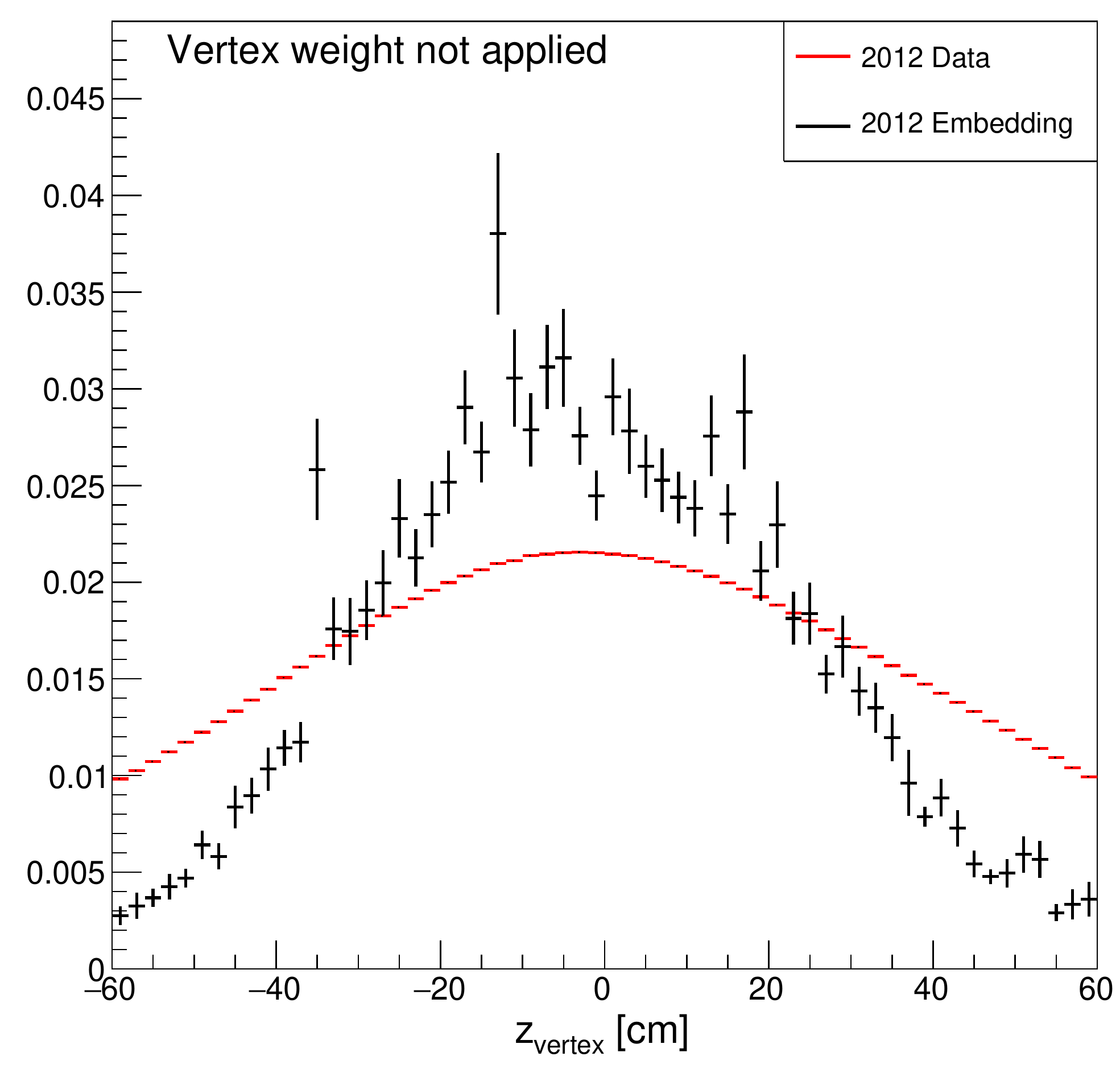} }}%
    \subfloat[]{{\includegraphics[width=7cm]{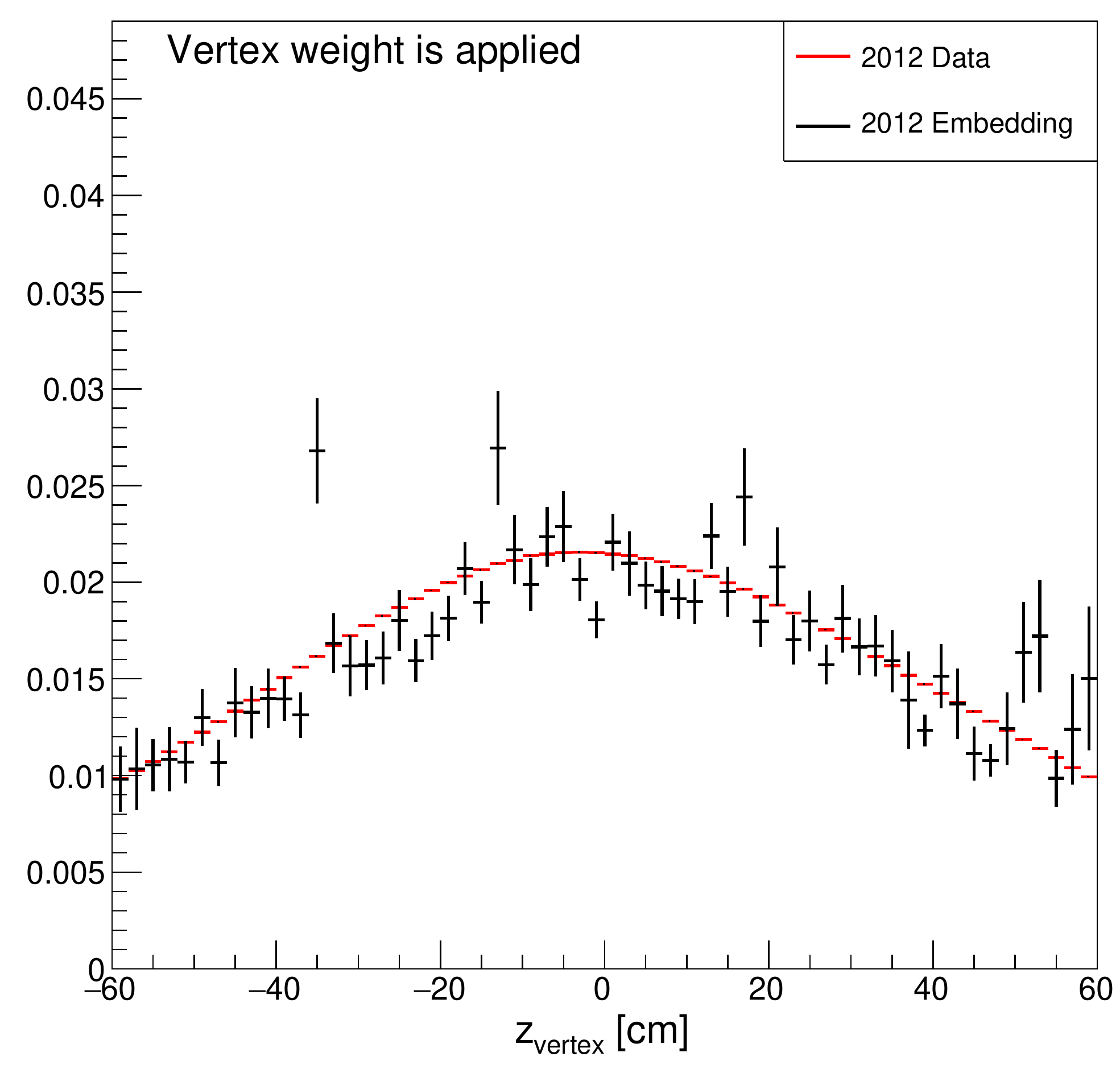} }}%
    \caption[Vertex Weight Comparison]{\textbf{Vertex Weight Comparison} - Comparison of the vertex z-axis position distributions for (a) unweighted simulation output and (b) weighted simulation output. Clearly the weighting is producing more accurate simulation distributions than the raw detector reconstructed distribution.}%
    \label{fig:vertexWeighting}%
\end{figure}

To get true comparisons to the data, the triggers must be combined to get a single, final comparison plot for all kinematic variables. Unlike in the data, the embedding does not know anything about prescales for the various triggers, meaning how often to allow a particular trigger to fire. Rather, the embedding is ``take-all'', meaning if an event satisfies a trigger, then it is assigned to that triggering algorithm. The immediate consequence of this is that triggers overlap with all triggers that are ``below'' them, e.g., if an event satisfies the JP2 criteria, then that event will also satisfy the requirements for JP1 and JP0. The prescales in the data reduce the amount of overlap between the trigger samples, but we must mock up this effect in the embedding by applying an appropriate weight and triggering scheme.

To remove the overlaps between triggers, every jet event is assigned to a single trigger category which are not allowed to overlap. For this analysis, jet events are assigned to the most restrictive trigger category starting with JP2. Specifically this means that any event which satisfies the JP2 conditions is only a JP2 event, otherwise we check if it satisfies JP1 or JP0 conditions in that order. If the jet event does not satisfy any of the jet patch conditions, then we assign it to a ``min-bias'' category. There is no hardware minimum bias condition for the embedding events, however we assume that every event in the embedding is part of the minimum bias sample. Therefore, the events which are not a jet patch event are immediately shoved into the minimum bias sample, which receives the corresponding prescale.

With the overlaps removed, we must also weight the events that enter into each distribution which will mimic the prescales that exist in the data. In the data, the triggering conditions of Section \ref{section:trigconditions} are applied to sort the data into triggered categories. If the triggering condition is true, then a corresponding triggering flag (flagJP2, flagJP1, flagJP0, and flagVPDMB) is set to true. Since the triggers are prescaled, an event is accepted if a simple ``or'' of all the trigger flags is true:

\begin{quote}
\centering
if (flagJP2 || flagJP1 || flagJP0 || flagVPDMB) accept
\end{quote}

\noindent All data distributions are filled using this triggering logic. To produce analogous prescaled plots in the embedding, we apply a weight to each event as we fill all histograms.

The prescale weight is simply the inverse of the effective prescale for the given triggering sample. The effective prescale is calculated by applying the defined trigger scheme to the data and the embedding separately, and counting the number of resulting true trigger flags. For each trigger, the number of triggered events are compared to the number of JP2 triggered events since JP2 always carries a prescale of 1. This gives a calculated prescale for both the data and the embedding, but since we want the embedding to emulate the data result, we take a ratio of the data to embedding calculated prescales to get out the effective prescale that is applied as a weight in the embedding analysis. Special care is given to the JP0 prescale, because we must consider the case where an event satisfies both the JP0 and VPDMB triggers and both events are kept. This only happens when the z-vertex satisfies the constraint $\lvert v_z \rvert < 30$ cm, as this is the constraint placed upon VPDMB triggered events. Therefore this same analysis is repeated for both $\lvert v_z \rvert < 30$ cm and $\lvert v_z \rvert > 30$ cm and the JP0 prescale is assigned in the analysis depending on the z-axis position of the vertex. This same argument could be made for the JP1 trigger prescale, but it is so small in comparison to the VPDMB prescale that the overlap effect is negligible, and we therefore ignore it. All of the calculated prescales used for the embedding weight are summarized in Table \ref{table:EmbedPrescales}.

\begin{table}[htp]
\begin{center}
\begin{tabular}{|c|c|}
\hline
Trigger & Embedding Prescale \\
\hline
JP2 & 1.00 \\
\hline
JP1 & 2.67 \\
\hline
JP0 ($\lvert v_z\rvert < 30$ cm) & 134.65 \\
\hline
JP0 ($\lvert v_z\rvert > 30$ cm) & 65.32 \\
\hline
VPDMB & 752.90 \\
\hline
\end{tabular}
\end{center}
\caption[Embedding Trigger Prescales]{\textbf{Embedding Trigger Prescales} - The prescale value used for each trigger in the embedding as a weight for each event as distributions are filled. Using these weights with the triggering algorithm allows for a direct comparison to the combined data.}
\label{table:EmbedPrescales}
\end{table}%

This trigger weighted embedding will be used for correcting kinematic effects and estimating systematic errors so it is important to understand how well these distributions match those in data.  The main kinematic variables used for asymmetries are jet $p_T$, $z$ and $j_T$. Comparisons for each of these are shown in Figures \ref{Combined_JetPt_2012emb} - \ref{Combined_PionJt_Plus_2012emb}. Only the $\pi^+$ comparisons shown for the $z$ and $j_T$ distributions, but the $\pi^-$ comparisons look equally as good. Comparisons between the data and embedding distributions were made for many types of track, tower, and jet variables, and more of them are given in the appendix. The excellent agreement in these kinematic distributions means the updated PYTHIA tune is working well, and that the trigger algorithm is being implemented correctly.

\figuremacroW{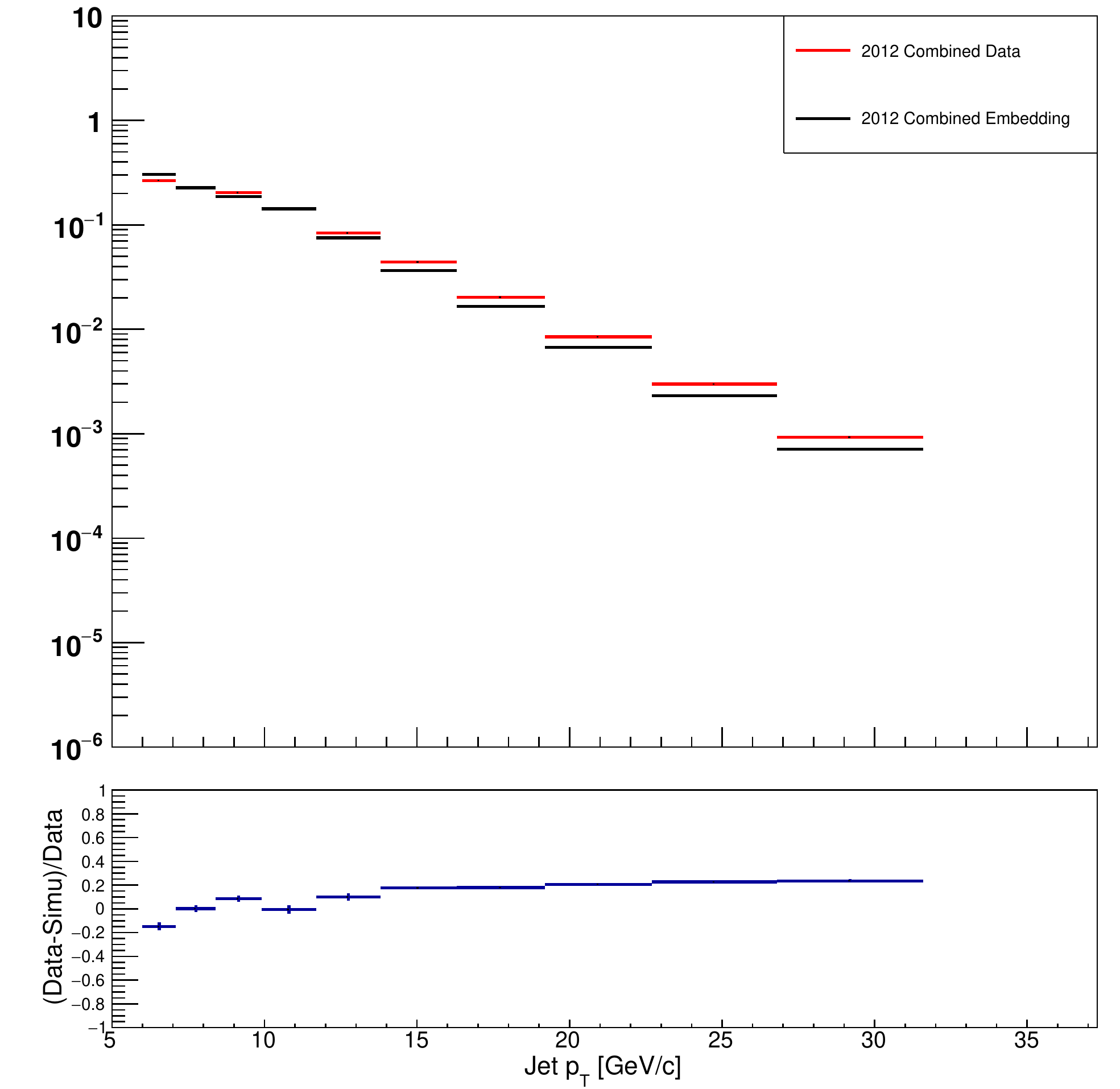}{Combined Triggers Jet $p_T$}{Comparing the jet $p_T$ distributions for the combined triggers in data and embedding.}{0.6}

\figuremacroW{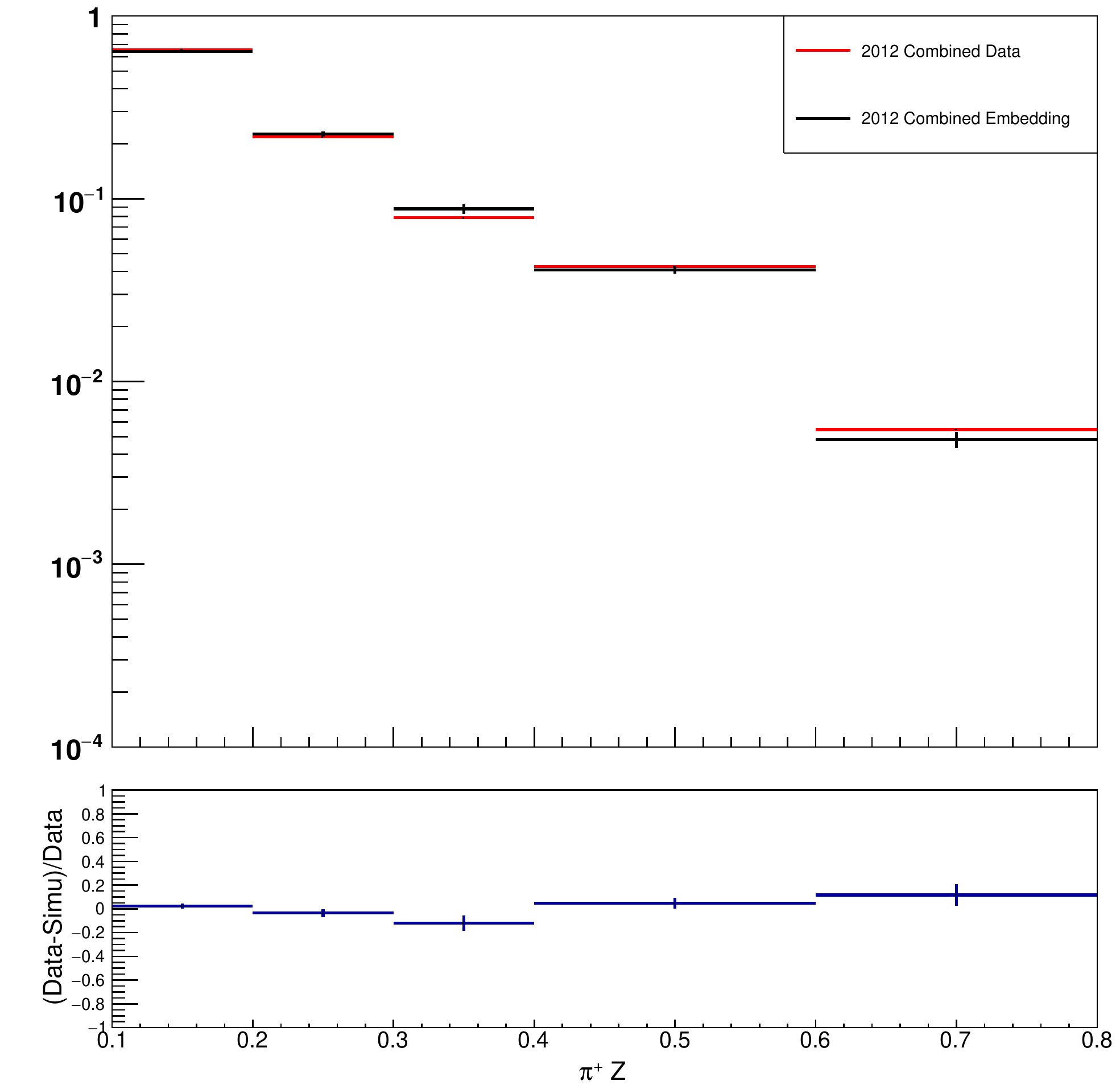}{Combined Triggers $\pi^+$ $z$}{Comparing the $\pi^+$ $z$ distributions for the combined triggers in data and embedding.}{0.6}

\figuremacroW{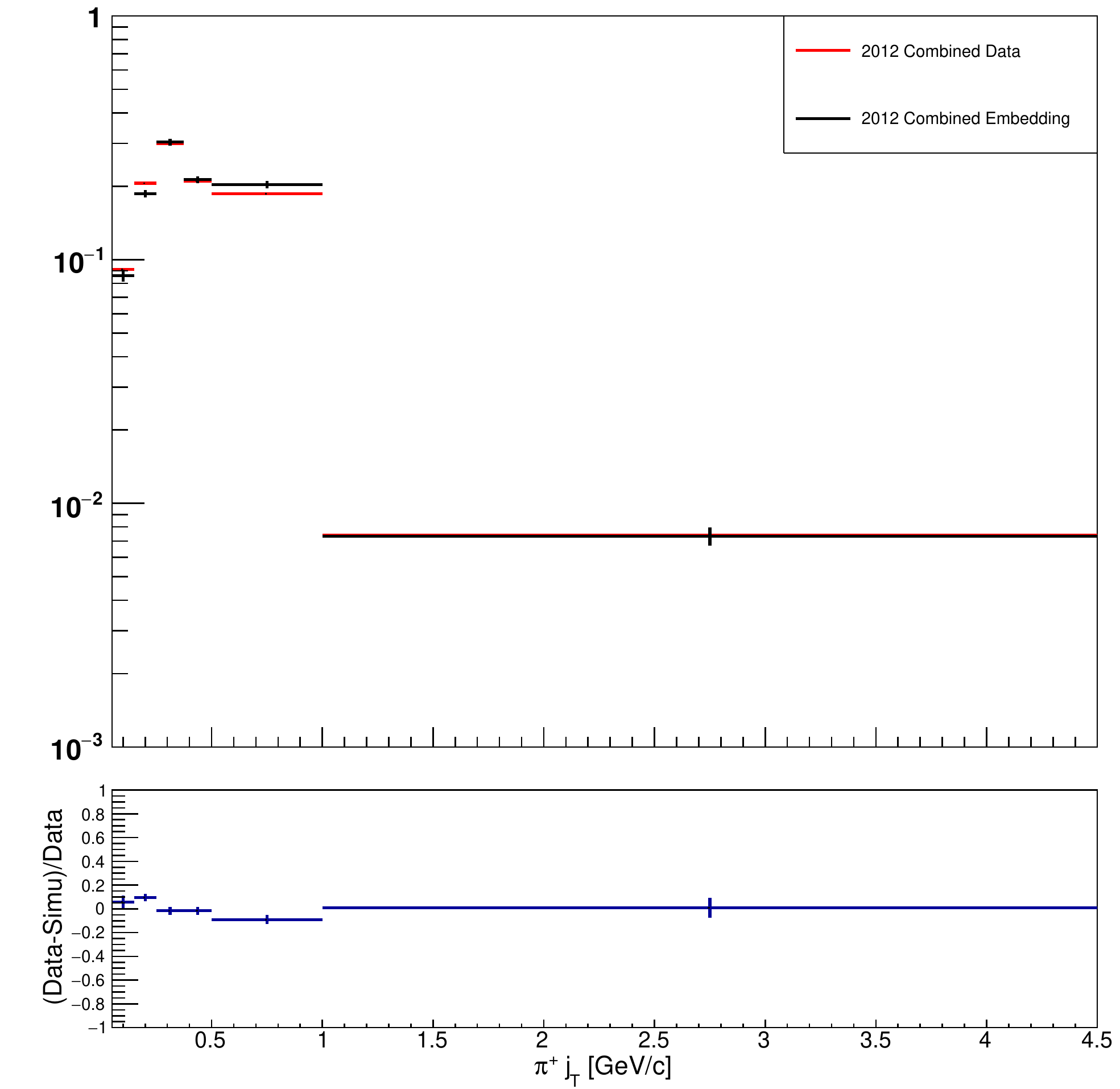}{Combined Triggers $\pi^+$ $j_T$}{Comparing the $\pi^+$ $j_T$ distributions for the combined triggers in data and embedding.}{0.6}

\section{Estimating Data Corrections}
The excellent agreement between the embedding samples and the data motivates the use of these samples to determine corrections and systematic errors for the asymmetries. In the case of a correction, the final asymmetry points may be corrected from their measured values to what the true value would be in the absence of detector and reconstruction effects. This section is dedicated to exploring the corrections that will be applied to the asymmetry results.

\subsection{Finite Bin Width Correction}
\label{sec:finiteBins}
As noted in Chapter \ref{chapter:AsymmetryAnalysis}, the yields which go into the cross ratio calculation are binned as a function of the azimuthal angle $\phi$. To have consistent bin population, there are twelve bins used for $\phi_C$ across a range of $2\pi$ and six bins for $\phi_S$ across a range of $\pi$. Using a finite number of bins introduces a dilution which will decrease the amplitude of the sine curve and the extracted asymmetry by definition. The dilution shrinks in magnitude as the number of bins increases, dropping to zero as the number approaches infinity.

This effect is easy to demonstrate by a straightforward calculation. Let us define the width of a single bin as $\Delta$, and calculate the average value of the sine within the bin. This gives:

\begin{equation}
\begin{aligned}
\langle \sin x \rangle &= \frac{1}{\Delta} \int_{x-\frac{\Delta}{2}}^{x+\frac{\Delta}{2}} \sin x^\prime dx^\prime = \frac{1}{\Delta}\left\{\cos\left(x-\frac{\Delta}{2}\right) - \cos\left(x+\frac{\Delta}{2}\right)\right\} \\ &= \frac{2}{\Delta}\sin\left(x\right) \sin\left(\frac{\Delta}{2}\right)
\end{aligned}
\end{equation}

\noindent This equation becomes a little more useful if we expand the sine term about $\Delta /2$. If we do that, and ignore terms that are higher than third order, we get:

\begin{equation}
\langle \sin x \rangle = \sin\left(x\right)\times\left\{1-\frac{\Delta^2}{24}\right\}
\end{equation}

\noindent where the term in brackets is our dilution factor, call it D:

\begin{equation}
D=1-\frac{\Delta^2}{24}
\end{equation}

\noindent Clearly the dilution goes away as the width of the bin goes to zero, and the average value of the sine within the bin is equivalent to the actual value of the sine within that bin.

The measured asymmetry is too small by the amount of D given above, so to correct it we divide the final asymmetry value by the value of D calculated from the width of the bins. The bin width is calculated simply by dividing the entire range by the total number of bins (N). Having two different ranges on the axis, we get two:

\begin{equation}
\Delta_{\phi_C} = \frac{2\pi}{N_{\phi_C}} \qquad\qquad \Delta_{\phi_S} = \frac{\pi}{N_{\phi_S}}
\end{equation}

\noindent However, since the number of bins for the inclusive jet asymmetry is halved, the width of the bins stays the same. Therefore, the dilution factor for each asymmetry is the same, but we keep two equations for completeness:

\begin{equation}
D_{\phi_C} = 1-\frac{1.645}{N_{\phi_C}^2} \qquad\qquad D_{\phi_S} = 1-\frac{0.411}{N_{\phi_S}^2}
\end{equation}

\noindent The dilution factors are equivalent given the binning scheme used in this analysis, and the value is $0.989$ which will be applied as the asymmetries are plotted. The method of implementation for this correction will be to divide both the measured asymmetry value and statistical error bar by this dilution factor to yield the true values.

\subsection{Kinematic Shifts}
\label{sec:kinShifts}
Detector resolutions and inefficiencies result in measured kinematic variables, such as track $p_T$ and tower $E_T$, being shifted on average from their true value. The issue with this bias arises when the asymmetry results are used by theorists who know nothing of our detector resolutions and biases, they work at the level of particles without detector effects. If our asymmetries are plotted at incorrect values on the x-axis then any comparisons to theoretical extractions and results from other experiments would not be accurate. To avoid these issues, it is essential to correct our kinematic values and plot them at their true values.

Luckily this is the easiest detector effect to correct using the embedding sample, although the method to do so is somewhat involved. What we are looking for is the true value of jet $p_T$, $z$, and $j_T$ within each of the detector level bins defined in Section \ref{sec:Binning}. A two dimensional histogram is the simplest way to do this. On the x-axis of the histogram, the kinematics will be binned in the detector level bins and filled with the detector level value, and on the y-axis the same kinematic variable will be plotted very finely at the particle level and then filled with the true kinematic value from the particle level. Then, it is a simple matter of reading out the average at the particle level within each detector level bin and calling that the true value of jet $p_T$, $z$, or $j_T$.

Getting to the particle level is the most difficult part of this analysis. When filling at the particle level, it is imperative to know that the particle level jet and subsequent jet particle are the same ones that we are considering at the detector level. A ``matching condition'' is applied to enforce matches between the detector and particle level. Each detector level jet that satisfies the same cuts as applied in the data is matched to the particle level if it satisfies a $\Delta R$ cut. Where the $\Delta R$ cut is the overall distance between the jets:

\begin{equation}
\Delta R_{jets} = \sqrt{\left(\eta_{detector} - \eta_{particle}\right)^2 + \left(\phi_{detector} - \phi_{particle}\right)^2} < 0.6
\end{equation}

\noindent Note that this cut is only applied to the particle level jet which has the minimum $\Delta R$ to the detector level jet, so there is only one possible matching particle level jet for each detector level jet. The value of this cut is set at 0.6 because that matches the radius of the jets which are found using the anti-$k_T$ jet finding algorithm. Once the jets are matched, we look at each track within the detector level jet and impose the same cuts on it as the data. For each track passing the cuts, again it must be matched to a particle within the particle level jet. This also requires a $\Delta R$ cut, and in this case it takes the same form only for the two tracks and enforces a different cut value:

\begin{equation}
\Delta R_{particles} = \sqrt{\left(\eta_{track} - \eta_{particle}\right)^2 + \left(\phi_{track} - \phi_{particle}\right)^2} < 0.02
\end{equation}

\noindent Again this cut is applied only to the particle level jet particle which has the minimum $\Delta R$ to the detector track, for each detector track in the jet. For each track that is matched to a particle in the particle jet, the histogram described above is filled with either the jet $p_T$, $z$, or $j_T$ using the detector and particle level values. This is shown for jet $p_T$ in Figure \ref{ptShiftThesis}, where the horizontal black bar within each detector level bin gives the true value of the jet $p_T$ within that bin and will be used when the asymmetry values are plotted. The same analysis is repeated for $z$ and $j_T$ and the true points are used to plot the asymmetry.

\figuremacroW{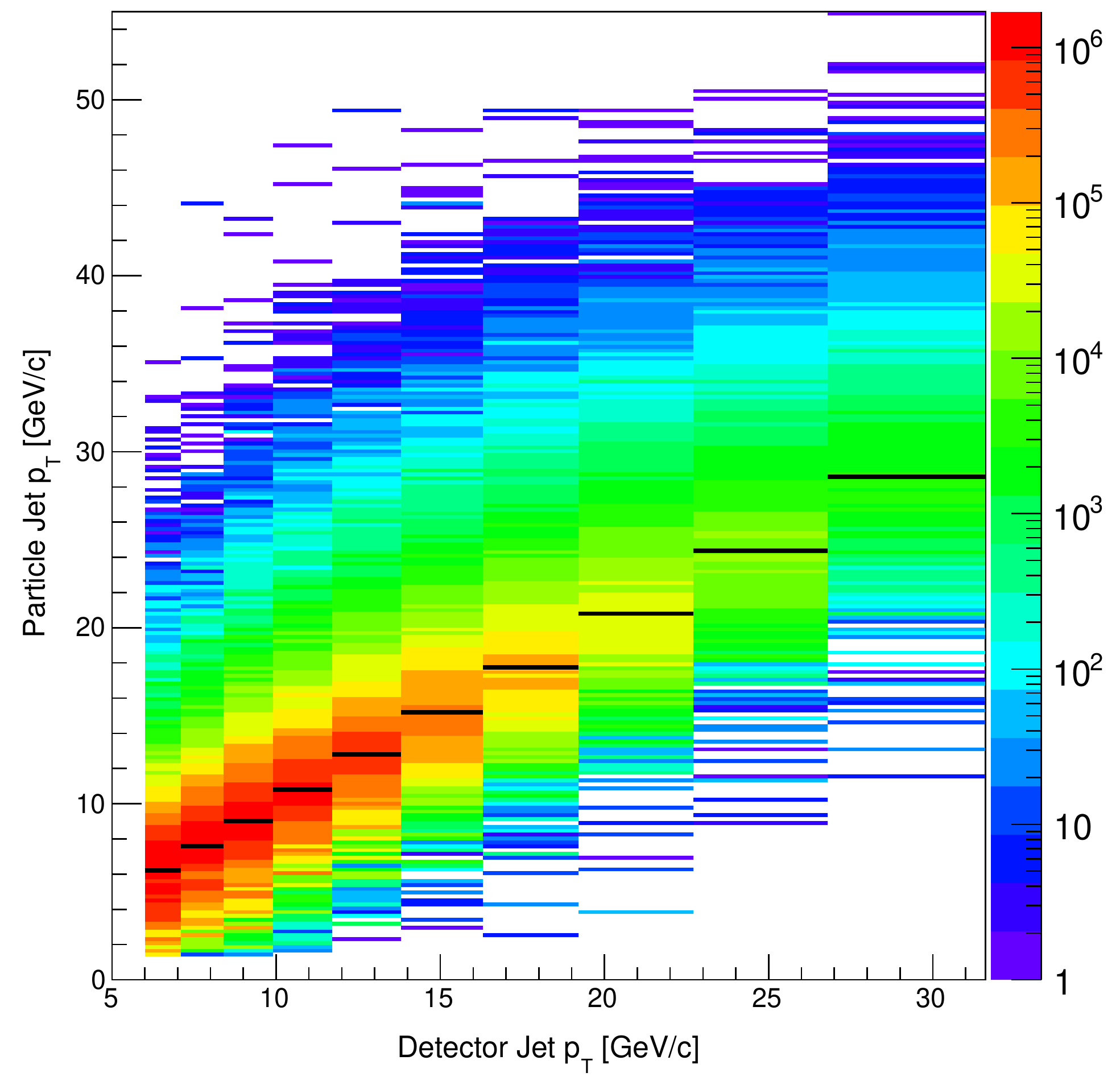}{Collins Analysis Jet $p_T$ Shift}{Calculating the true value of the jet $p_T$ for plotting the final Collins asymmetries, with the detector kinematics plotted on the x-axis and the particle level kinematics on the y-axis binned finely. The true value of the jet $p_T$ is given by the black bar in each detector level bin.}{0.8}

Note that the preceding discussion applies to the Collins asymmetries. When one considers the inclusive jet asymmetry, the analysis must change a little because there is no consideration of tracks. Therefore, to calculate the shift the same setup as in Figure \ref{ptShiftThesis} is used, only filled for all detector level jets that pass analysis cuts and excludes any track matching. The result of this very similar analysis will be used when plotting the inclusive jet asymmetry.

\subsection{$\phi_C$ Resolution}
It is very important to reconstruct kinematics correctly for the asymmetry analysis, and this is especially true for the Collins angle, $\phi_C$. Incorrectly reconstructing $\phi_C$ leads to shifts within the cross ratio which propagate to the fit that is used to extract the asymmetry. Smearing of the reconstructed $\phi_C$ about the true value leads to measuring a smaller asymmetry than the true value. We can use the embedding to estimate the effect of this smearing and correct the asymmetries back to their true values.

On the average, the simulation does a good job of reconstructing $\phi_C$ at the detector level as shown in Figure \ref{fig:phiCcompare}, with any differences appearing to be within statistical scatter. Thus, we can use the value of $\phi_C$ at the detector and particle levels to characterize the smearing. To do so, the difference in the detector and particle level $\phi_C$ value is plotted for each kinematic bin and range. This difference is fit with a triple Gaussian function as in Figure \ref{PhiCollFitEx_Plus}: a central one to characterize the peak and one each to the left and right of the central peak to characterize any residual bumps. This fit is applied individually for each kinematic bin.

\begin{figure}
    \centering
    \subfloat[]{{\includegraphics[width=7.25cm]{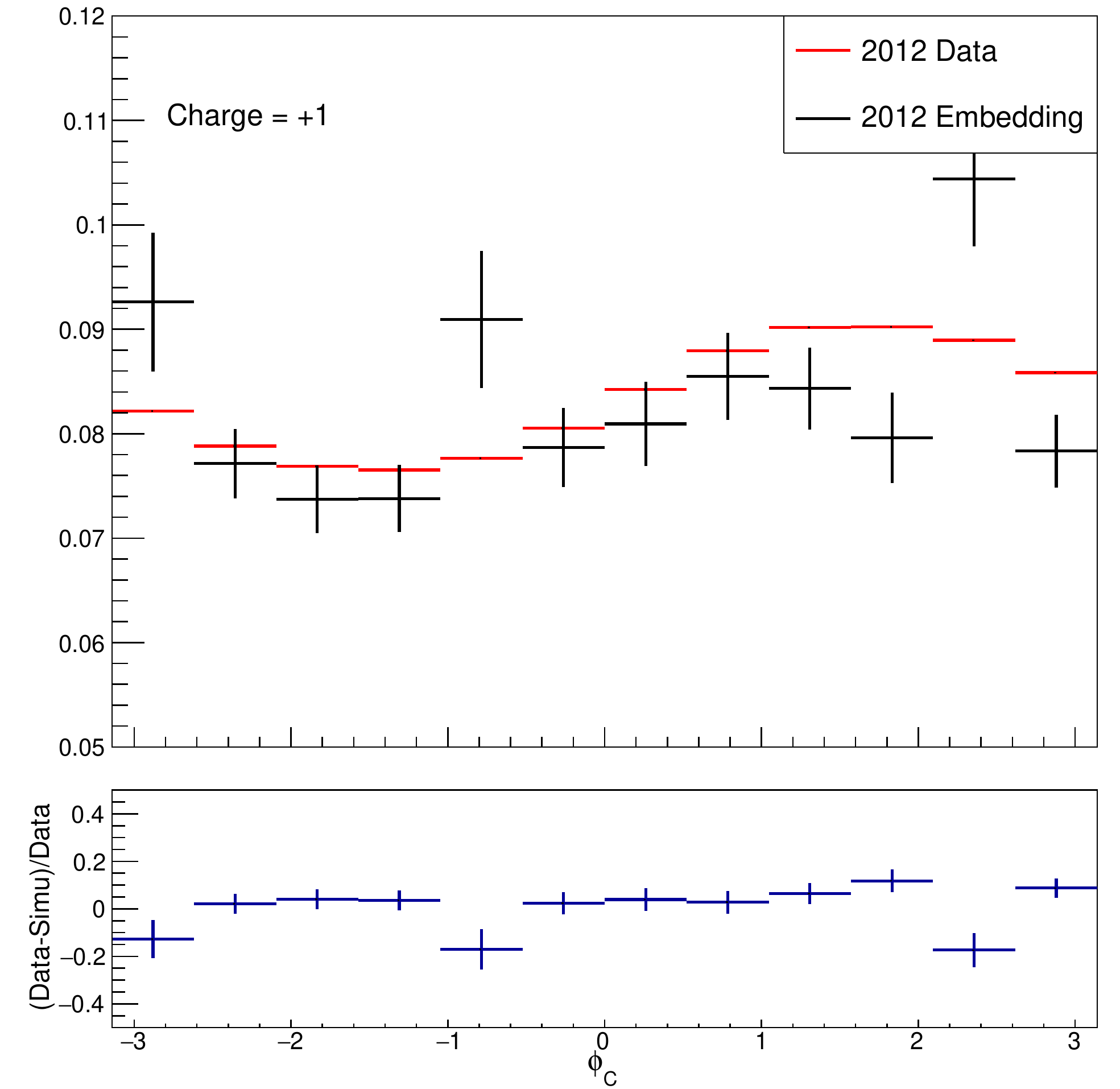} }}%
    \subfloat[]{{\includegraphics[width=7.25cm]{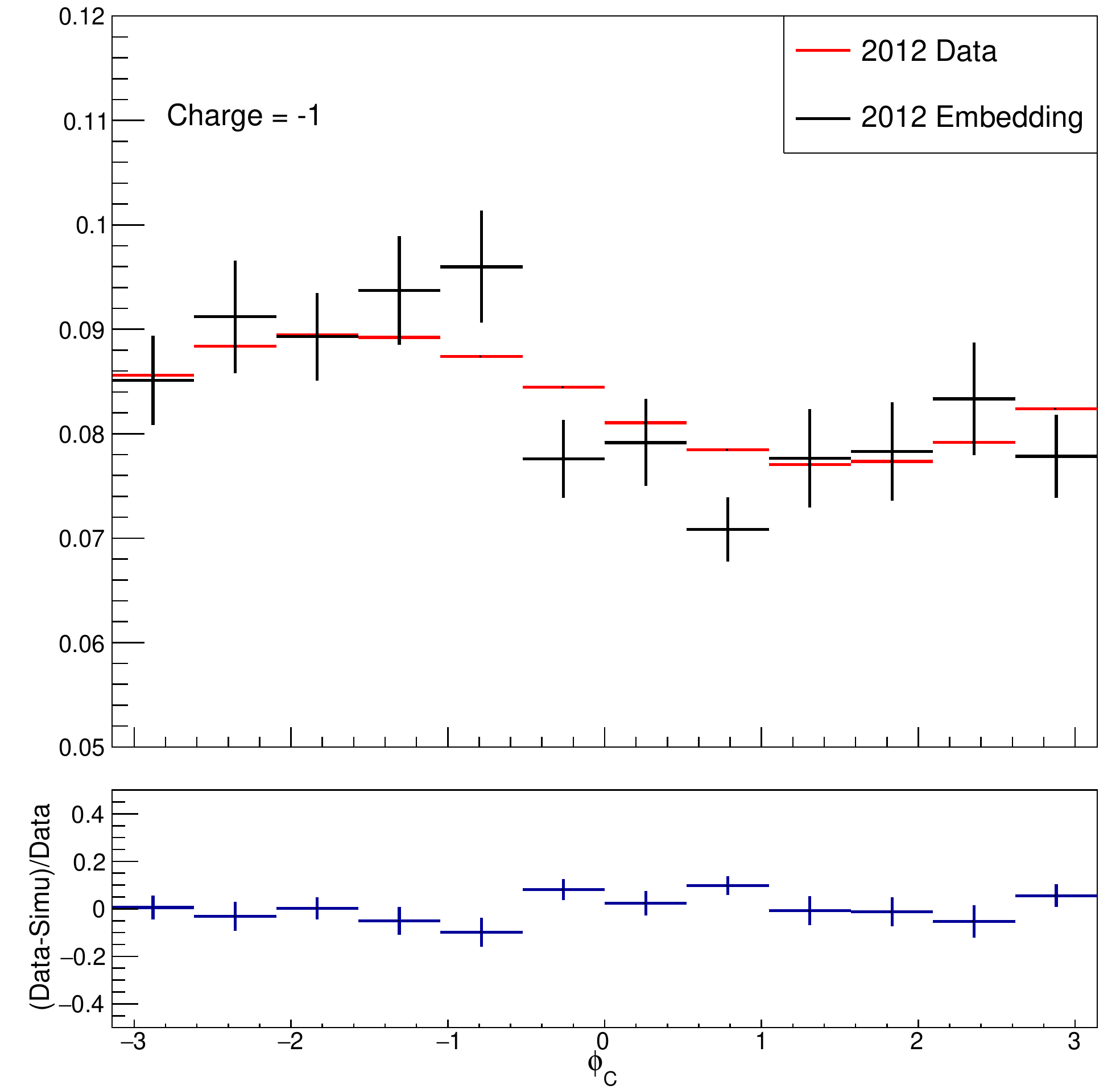} }}%
    \caption[Data/MC $\phi_C$ Comparison]{\textbf{Data/MC $\phi_C$ Comparison} - Comparison of the values of $\phi_C$ in the data and at the detector level in the simulation for (a) positively and (b) negatively charged identified pion tracks. These plots integrate over all kinematic ranges, and thus represent average values for each bin.}%
    \label{fig:phiCcompare}%
\end{figure}

\figuremacroW{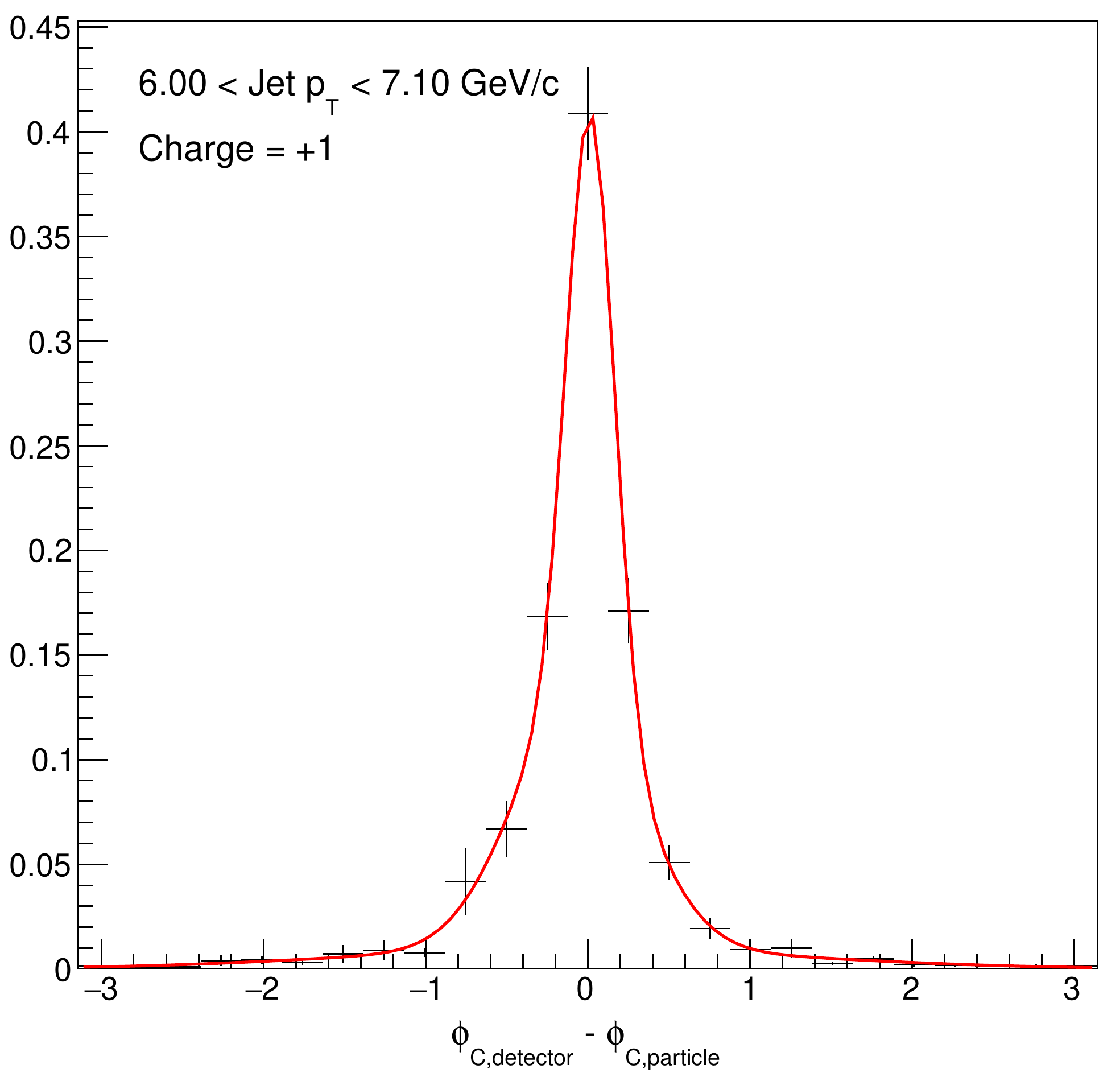}{$\phi_C$ Resolution Example Fit}{A triple Gaussian fit to the spread in detector minus particle level $\phi_C$ values.}{0.8}

The fits to the spread in $\phi_C$ values are, in turn, used to calculate the dilution. This is a two step process that begins with taking the convolution of the resulting triple Gaussian fit function with a sine function of unit amplitude on the full range of $\phi_C$, $[-\pi,\pi]$, which will result in another sine function whose amplitude reflects how the smearing affects the asymmetry result. This convolution result is then plotted as a function of $\phi_C$ and fit with another sine function to extract the amplitude. The resulting amplitude is the dilution that is imparted by the resolution in $\phi_C$. An example of this final sine fit is shown in Figure \ref{PhiCollDilutionEx_Plus}.

\figuremacroW{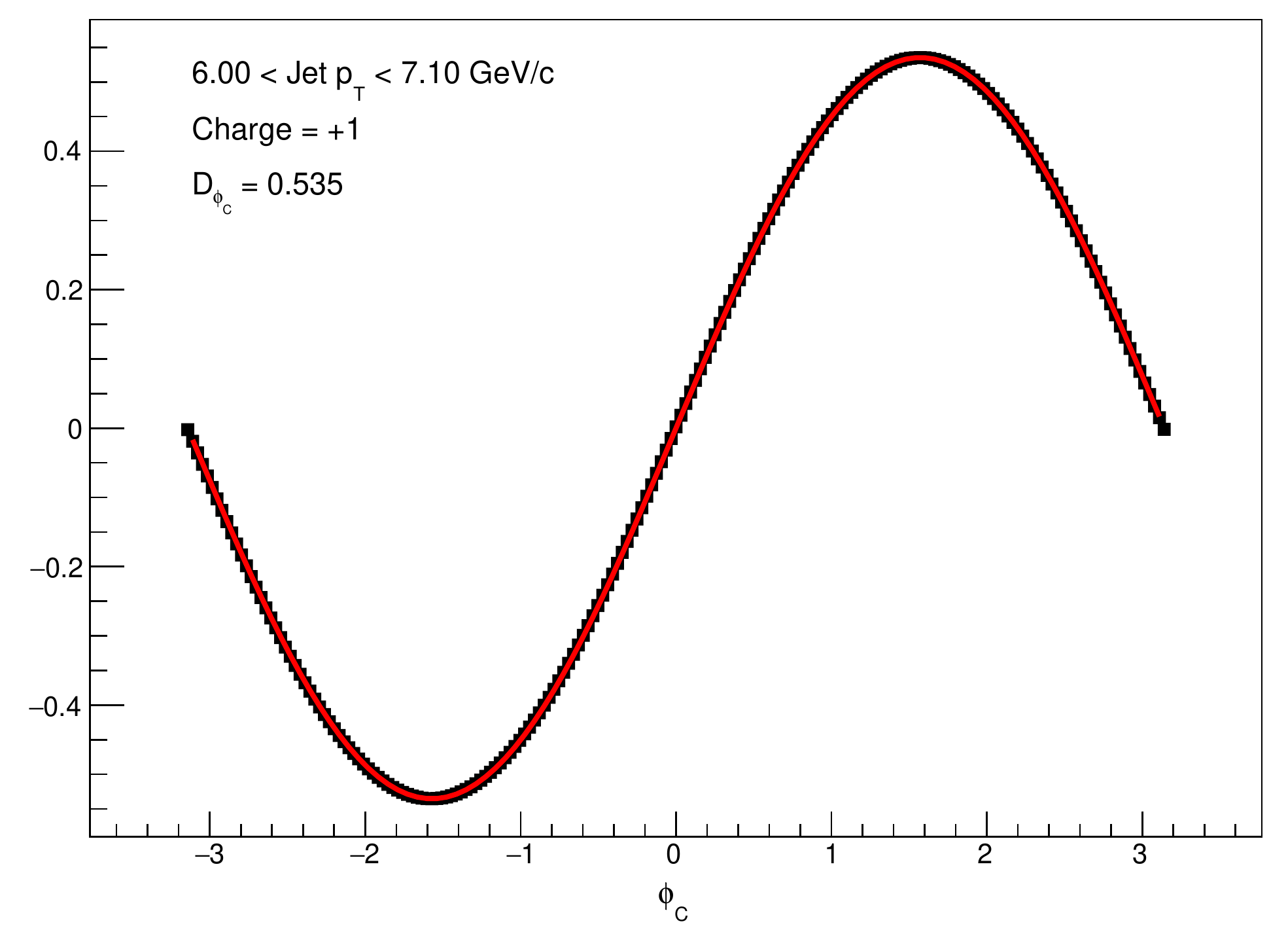}{$\phi_C$ Dilution Example Fit}{A sine function is fit to the convolution result. The amplitude of the fit is taken as the dilution parameter for that kinematic bin, in this case it is $0.535$.}{0.8}

As expected, from these results, the smearing of $\phi_C$ causes a reduction in the true asymmetry value. Therefore, the applied correction will increase the measured asymmetry. Thus, the measured asymmetry is divided by the dilution due to $\phi_C$ smearing for each kinematic bin. The bin-by-bin dilution values are given in the asymmetry result tables in the appendix.

\subsection{Pion Sample Contamination Correction}
\label{sec:pionCorrection}
When performing the Collins analysis, it is important to determine the purity of the signal particle sample. For instance, it is interesting and valid to compute the Collins asymmetry for any outgoing hadron such as $\pi^\pm$, $K^\pm$, or $p/\bar{p}$. However, due to differences in PDFs and fragmentation functions, there is good reason to believe the asymmetries will not remain the same for each species sample. For this analysis, $\pi^\pm$ are identified in jets and the asymmetry signal is calculated from the angular distribution in those jets. Therefore we must understand what the contamination to the $\pi$ sample is due to $K^\pm$, $p/\bar{p}$, and $e^\pm$ (which carry zero Collins signal) and correct the final asymmetry for it, since each contaminating particle presents some contamination to our true signal.

Two detectors are used to get a handle on the $\pi$ sample contamination. At high track momentum, fits to the TPC $n_\sigma \left(\pi\right)$ (defined in Section \ref{section:EoverPspectra}) distribution are used to identify and measure the fraction of the contaminating samples. At low track momentum, fits to the TOF $m^2$ distributions (Equation \ref{eq:tofMass}) are used to identify and measure the fraction of the contaminating samples. Above 1 GeV it is difficult to use the TOF for particle identification.

Looking at the sample $n_\sigma\left(\pi\right)$ distribution in Figure \ref{nSigmaPiPlusCut} the contamination curves for $K^+/p$ and $e^+$ show up as shoulders on the $\pi^+$ curve and are clearly protruding into the $\pi^+$ sample. This is the overlap that we want to quantify, but here is the first issue. In this sample figure, all track momenta are integrated over and then Gaussian functions are fit to the result of that. However, looking at Figure \ref{dEdxDistribution}, the overlaps are not constant with momentum, rather there are six distinct regions of track momentum outlined in Table \ref{table:hadMomRegions} that will be investigated. To begin, we fill $n_\sigma\left(\pi\right)$ distributions for each of the six track momentum range in Table \ref{table:hadMomRegions} for each kinematic bin that we have for our final asymmetry, one set each for $\pi^+$ and $\pi^-$.

\figuremacroW{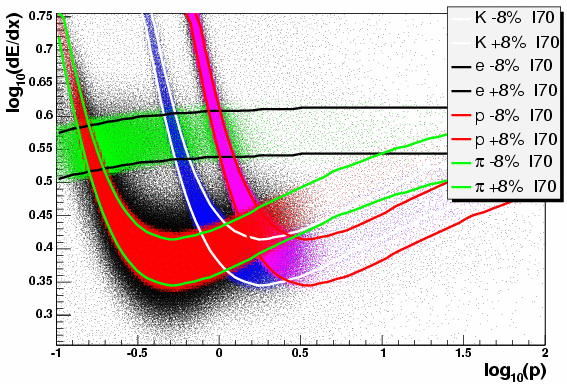}{dE/dx Distribution}{The dE/dx distribution from the TPC for all track momenta \cite{ref:ShaoParticleId}. Clearly, there are different overlaps in the curves as the track momentum changes. These changes need to be addressed in the $\pi$ contamination study.}{0.75}

\begin{table}[htp]
\begin{center}
\begin{tabular}{|c|c|c|}
\hline
Region & $p_{track}$ Range [GeV/c]\\
\hline
1 & $0.00-0.80$ \\
\hline
2 & $0.80-1.00$ \\
\hline
3 & $1.00-1.26$ \\
\hline
4 & $1.26-1.58$ \\
\hline
5 & $1.58-2.50$ \\
\hline
6 & $2.50-\infty$ \\
\hline
\end{tabular}
\end{center}
\caption[dE/dx Track Momentum Regions]{\textbf{dE/dx Track Momentum Regions} - The regions of dE/dx split up by track momentum which will be used to calculate contaminations to the $\pi$ sample.}
\label{table:hadMomRegions}
\end{table}%

With the distributions filled, we apply a fitting procedure for each track momentum bin. The final fit is the sum of four Gaussian curves, one each for $\pi$, $K$, $p$, and $e$, but to get the most accurate final fit possible we need to build it up from several different pieces. First, we need to know where the $K$, $p$, and $e$ peaks appear on the $n_\sigma\left(\pi\right)$ distribution. The simplest way to do this is to fill three two-dimensional histograms for each track momentum bin within each kinematic bin with one axis holding $n_\sigma\left(\pi\right)$ values and the other axis holding $n_\sigma\left(K\right)$, $n_\sigma\left(p\right)$, or $n_\sigma\left(e\right)$ respectively. Taking advantage of features in ROOT, we can take a TProfile along the $n_\sigma\left(\pi\right)$ axis, and fit the result with a $3^{rd}$ order polynomial. Evaluating this polynomial at $n_\sigma\left(K/p/e\right) = 0$ tells us where that flavor appears on the $n_\sigma\left(\pi\right)$ distribution. A sample of this for the lowest jet $p_T$ bin within the $1.26$ GeV/c $< p_{track} < 1.58$ GeV/c bin is shown in Figure \ref{nSigmaKaonMean}, where the result of the $3^{rd}$ order polynomial fit to the horizontal black lines is shown by the smooth, continuous black line. This fit result is used to extract where the $n_\sigma\left(K\right)$ peak shows up on the $n_\sigma\left(\pi\right)$ distribution.

\figuremacroW{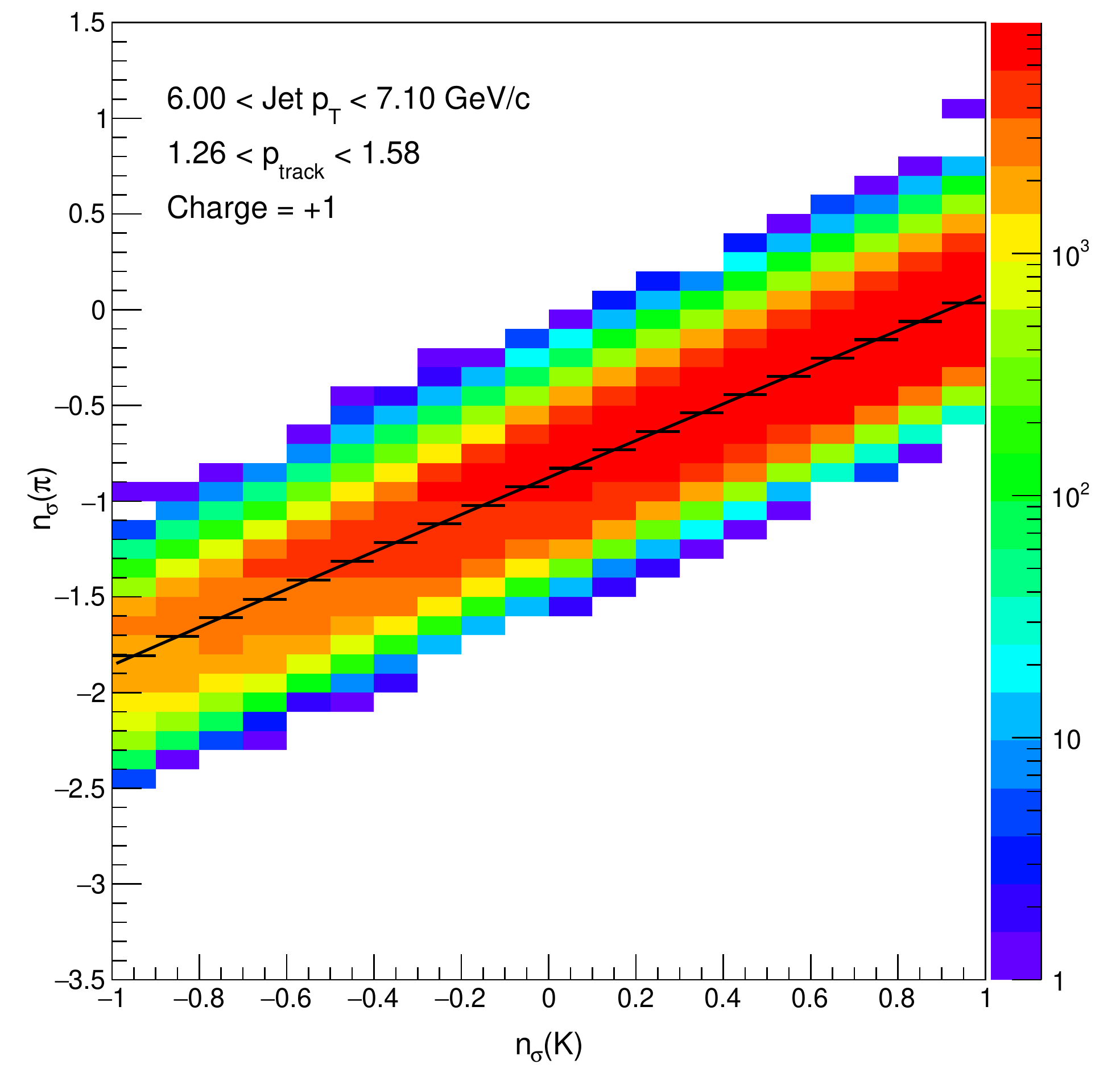}{$n_\sigma\left(K\right)$ Mean Fit}{A sample fit to extract the peak location of $n_\sigma\left(K\right)$ on the $n_\sigma\left(\pi\right)$ distribution for the lowest jet $p_T$ bin within the $1.26$ GeV/c $< p_{track} < 1.58$ GeV/c bin.}{0.8}

Having the means of all the distributions in hand, the next step is to fit each particle flavor peak with a Gaussian distribution. The output parameters from each of these fits are then used to seed the final fit with a sum of four Gaussian distributions. In this final fit we let all parameters float except for the means of the $K$, $p$, and $e$ peaks since we know where those need to be, they are fixed to what is extracted. Now we have all the information necessary to integrate the final fit over each flavor, and divide that result by the integral of the whole distribution to extract the fraction of $K$, $p$, and $e$ that enter the $\pi$ sample. For these fractions, we set two ranges of integration:

\begin{itemize}
\item $-1 < n_\sigma\left(\pi\right) < 2.5$ defined as the ``signal'' region for our asymmetries.
\item $-10 < n_\sigma\left(\pi\right) < -1 \cup 2.5 < n_\sigma\left(\pi\right) < 10$ defined as the ``background'' region for our asymmetries.
\end{itemize}

\noindent These fractions calculated in both ranges will be used later in this section to correct the measured asymmetry values. Figure \ref{nSigmaPionFitsExample} summarizes the total combined fit for the entire $n_\sigma\left(\pi\right)$ distribution for the lowest jet $p_T$ bin and the $1.26$ GeV/c $< p_{track} < 1.58$ GeV/c bin, and the final contributions to the total fit which are integrated to calculate the contamination fractions. This procedure is repeated for all track momentum bins within each kinematic bin for both charge states to get fractions for all final asymmetry points.

\figuremacroW{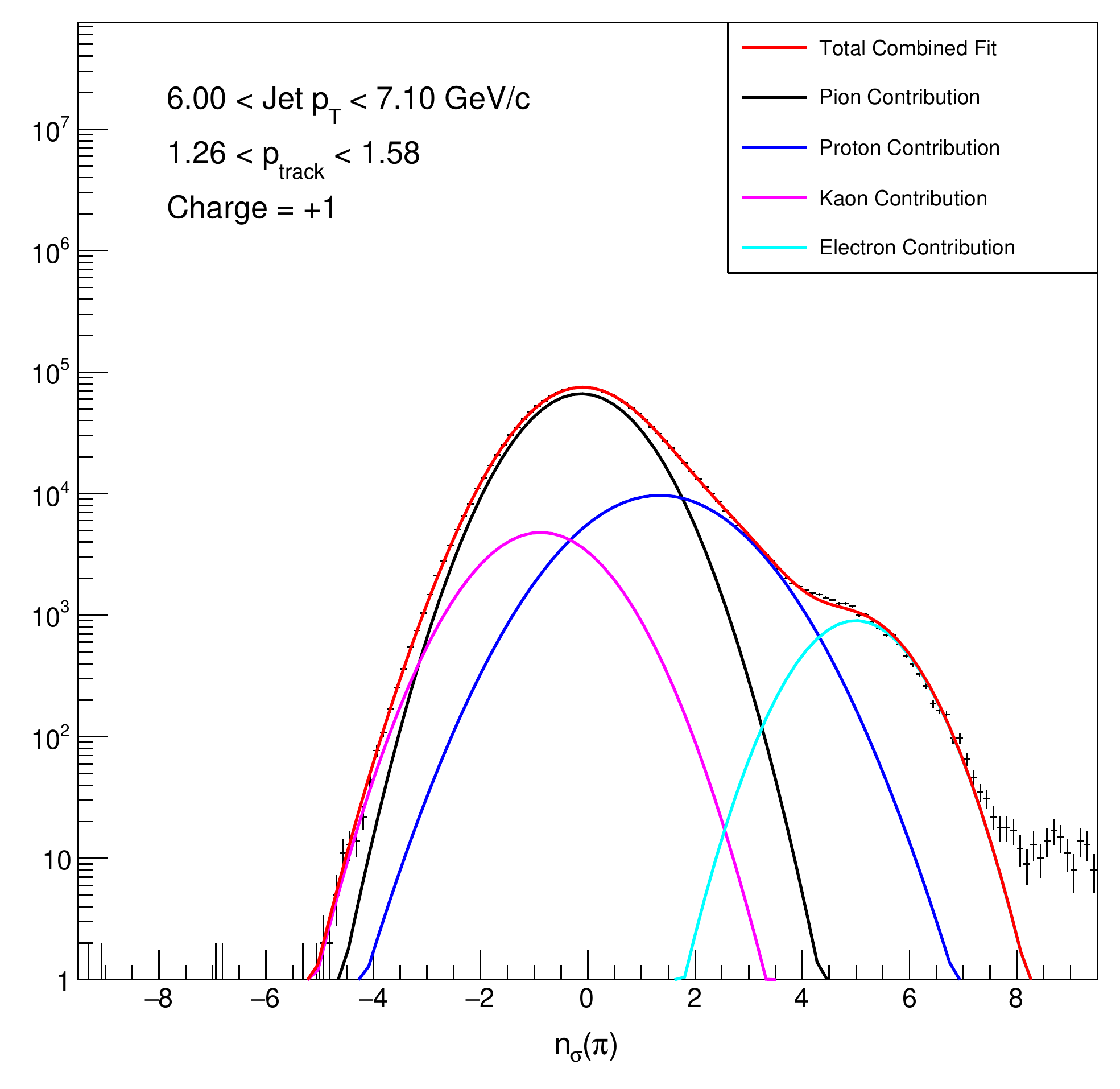}{$n_\sigma\left(\pi\right)$ Sample Fit}{A sample fit to the $n_\sigma\left(\pi\right)$ distribution, and the contribution from all particle species to the total combined fit which are used to calculate the contamination fractions.}{0.8}

Now that the dE/dx method is understood and implemented, the next task is the analysis of the $m^2$ distributions from the TOF. The mass of all particle species is understood and fixed, so there will not be movement among the peaks like we had with the dE/dx analysis. However, to maintain consistency, the $m^2$ distributions are split up into the same track momentum ranges that are outlined in Table \ref{table:hadMomRegions} for each kinematic bin, and for both charge states. It should be noted, however, since the electron mass is very small compared to the pion, it will hover around zero on the $m^2$ distribution and will be engulfed by the $\pi$ peak. Thus, in this analysis, we get out $K$ and $p$ dilutions, but not for $e$.

The fitting procedure for the $m^2$ distributions is not the same as what we used previously for the dE/dx analysis. Since the mass is a known quantity, we can set the value for all particle species in the $m^2$ fit, and jump straight to the final fit. It turns out that a Voigt profile, the result of the convolution of Gaussian and Lorentzian distributions, is the best fit for the $m^2$ distributions, and the total fit in this case is a sum of three of them, one each for $\pi$, $K$, and $p$. Precursory fits are not required to achieve a good final fit, we can simply begin with the sum of three Voigt profiles as shown in Figure \ref{MassSquaredFitExample} for the lowest jet $p_T$ bin and the $1.00$ GeV/c $< p_{track} < 1.26$ GeV/c bin. Another difference from the previous dE/dx analysis is how the fractions are extracted. The $m^2$ analysis is performed within the confines of the two $n_\sigma\left(\pi\right)$ ranges above for signal and background regions, making the distributions totally closed. In this case, we can use ROOT to our advantage by defining the fit to return the fraction as an output parameter and removing that much work to be done later by hand. Since the $n_\sigma\left(\pi\right)$ distributions are not closed, we cannot implement this method there.

\figuremacroW{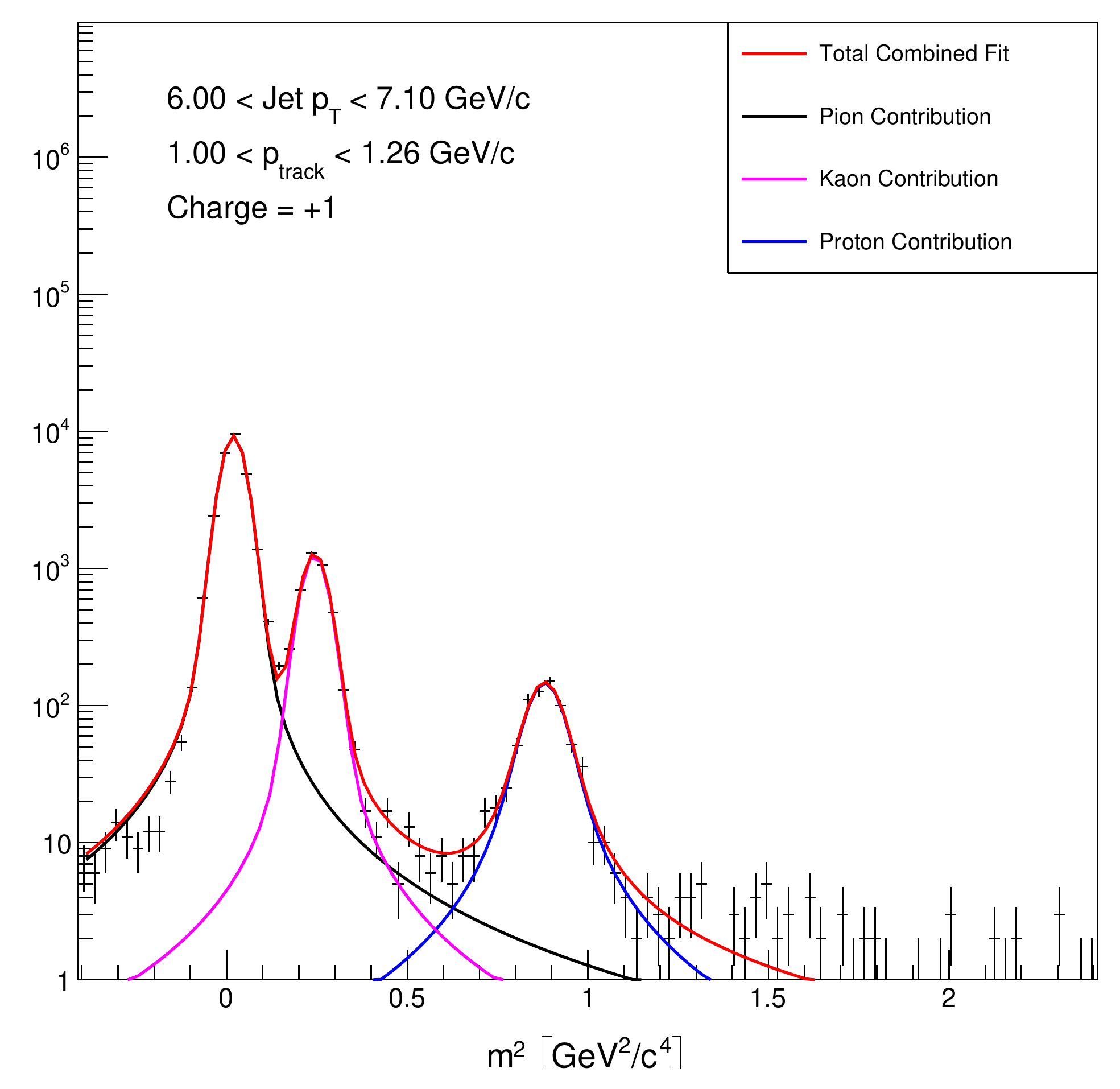}{$m^2$ Sample Fit}{A sample fit to the $m^2$ distribution, and the contribution from all particle species to the total combined fit.}{0.8}

With the fractions for our two different regions calculated, but still separated by track momentum, we need to combine the results to get the final fractions for each kinematic bin. Beyond the $1.00$ GeV/c $< p_{track} < 1.26$ GeV/c track momentum bin the TOF resolution begins to severely degrade, and in this bin the $K$ and $\pi$ $n_\sigma\left(\pi\right)$ peaks overlap leading to unreliable pion fractions. Therefore, we choose to use the fractions from the TOF $m^2$ analysis in the lowest three track momentum bins, and the fractions from the dE/dx analysis from the highest three track momentum bins. We take a weighted average over all six bins as the final fraction for each kinematic bin, using the number of events from the dE/dx analysis as the weight for all six bins. This is because the $m^2$ analysis uses only the VPDMB trigger since it contains the timing information for the TOF, whereas the dE/dx analysis uses all triggered samples combined because there is no difference between triggers. For equal weighting, the statistics from the dE/dx results will be used as the weight. The results for the signal and background fractions for positively charged tracks are shown in Figure \ref{AllFracPtExample}, note that the electron fractions are not plotted, they are all on the order of $10^{-2}$ or smaller and would not show up.

\figuremacroW{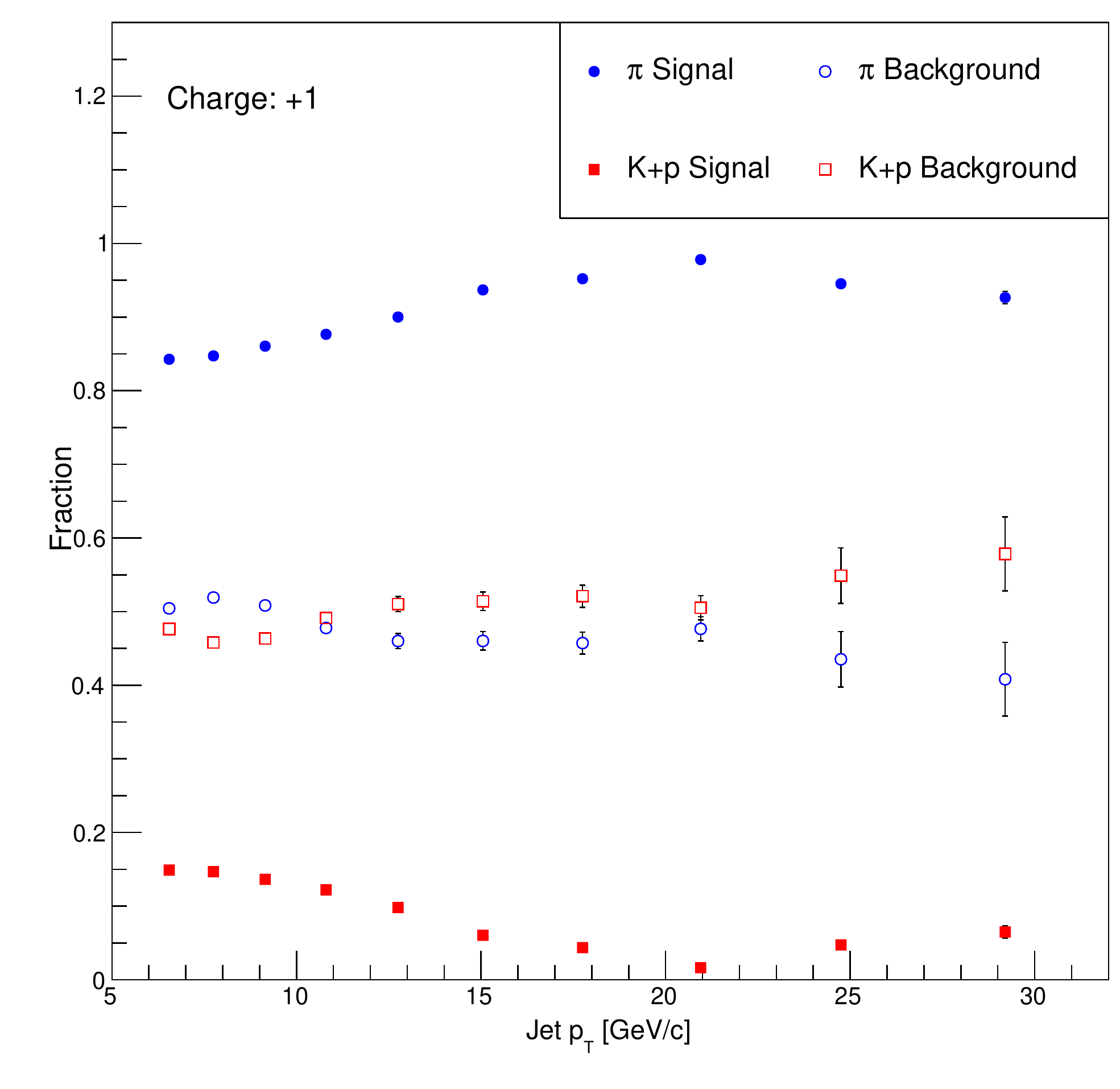}{$\pi$ Sample Particle Fractions}{The signal and background fractions for $K+p$ and $\pi$.}{0.8}

Using these fractions, the method for calculating the correction is straightforward. First, write down expressions for the measured asymmetries in each of the two regions defined above (signal and background) that include contributions from combined $K+p$ and from $\pi$:

\begin{equation}
\begin{aligned}
A_{meas}^{sig} &= A_\pi^{true}f_\pi^{sig} + A_{kp}^{true}f_{kp}^{sig} \\
A_{meas}^{bk} &=  A_\pi^{true}f_\pi^{bk} + A_{kp}^{true}f_{kp}^{bk}
\end{aligned}
\end{equation}

\noindent Note that since $e^\pm$ do not come from the quark, they cannot carry any asymmetry. Therefore there is no $A_e$ term in this expression, however the $e^\pm$ contribute to the denominator of the $\pi$ and $K+p$ fractions ($f$). What we want in the end is $A_\pi^{true}$, so solving the system for this variable, we have:

\begin{equation}
A_\pi^{true} = \frac{A_{meas}^{sig}f_{kp}^{bk} - A_{meas}^{bk}f_{kp}^{sig}}{f_\pi^{sig}f_{kp}^{bk} - f_\pi^{bk}f_{kp}^{sig}}
\label{eq:contamCorrection}
\end{equation}

\noindent In this equation $A_{meas}^{sig}$ is the Collins asymmetry measured within our cut region of $-1 < n_\sigma\left(\pi\right) 2.5$, and $A_{meas}^{bk}$ is the Collins asymmetry measured in the background region outside of the cut. Assuming that the background region is made up of contaminating particles, which can only be $K$ and $p$, then we measure the Collins asymmetry outside of the $n_\sigma\left(\pi\right)$ cut, but we choose particles which satisfy $-2 < n_\sigma\left(K\right) < 3$ or $-2 < n_\sigma\left(p\right) < 4$ to try and only select contaminating particles rather than additional $\pi$. The same analysis is applied to this background asymmetry as is applied to the signal asymmetry.

On a kinematic bin-by-bin basis, the asymmetry is corrected using Equation \ref{eq:contamCorrection} and the corrected values are given in the tables in the Appendix. The error for the correction is also calculated by propagating the error through Equation \ref{eq:contamCorrection}. The statistical error will increase for each asymmetry point, but this increase is still less than the sizable systematic error that would be applied if the contamination was treated as an error instead of a correction.

\section{Estimating Systematic Errors}
Any systematic effects which cannot be accounted for by correcting the final asymmetries must be assigned an error. This section is dedicated to understanding and quantifying the systematic errors.

\subsection{Kinematic Shift Errors}
\label{sec:kinShiftError}
For the kinematic shifts that were presented in Section \ref{sec:kinShifts}, we assign a systematic error that reflects the uncertainty in the calculation of these shifts. The calculation of the error is handled differently for jet $p_T$ and $z$ than it is for $j_T$, because the latter has no reliance upon the jet itself, only upon the identified pion track momentum. The errors are calculated for each kinematic variable using these equations:

\begin{align}
\delta p_{T,true} &= \sqrt{\delta p^2_{T,track-loss} + \delta p^2_{T,stat.} + p^2_{T,jet,BEMC} + \delta p^2_{T,jet,tracks}} \label{eq:kinErr1} \\
\delta z_{true} &= \sqrt{\delta z^2_{track-loss} + \delta z^2_{stat.} + \delta p^2_{jet,BEMC} + \delta p^2_{jet,tracks} + \delta p^2_{\pi}} \label{eq:kinErr2} \\
\delta j_{T,true} &= \sqrt{\delta j^2_{T,track-loss} + \delta j^2_{T,stat.} + \delta j^2_{T,tracks}} \label{eq:kinErr3}
\end{align}

\noindent Note that these expressions have several terms in common. In the following discussion, unless otherwise noted, jet $p_T$ will be used as the example, but the proper substitutions should be made to calculate the error in the other kinematic shifts. For instance, when calculating the error in $z$, one should use $p_{jet}$ rather than $p_{T,jet}$, a detail that could easily get lost in the notation above.

Two sources of error contribute to all of the cases. The first contribution,\\ $\delta p_{T,track-loss}$, accounts for discrepancies between the embedding and the data for the TPC track reconstruction efficiency. This is calculated by removing $4\%$ of the tracks from each event and running the jet finding algorithm again, where $4\%$ is a conservative estimate of the tracking efficiency based on the analysis in Ref. \cite{ref:tpcTrackLoss}. Using these jets with tracks removed, and comparing to jets with no tracks removed we get a measure of how accurately the embedding recreates the TPC efficiency. To account for any differences in TPC efficiency that affect the kinematic shifts, a systematic error is included which is the difference in the shifts using both tracking efficiencies. The second contribution, $\delta p_{T,stat.}$, encapsulates the statistical error due to a finite embedding sample size. It is calculated for each detector bin in the correlation matrices (for example see Figure \ref{ptShiftThesis}) by

\begin{equation}
\delta p_{T,stat.} = \sqrt{\frac{\sum\limits_{i=1}^{N_y}w^2_i\left(y_i-\bar{y}\right)^2}{\sum\limits_{j=1}^{N_y}w^2_j}}
\end{equation}

\noindent In this calculation $y_i$ is the value of each particle level bin along the y-axis in Figure \ref{ptShiftThesis} and $\bar{y}$ is the weighted average along the y-axis. The weight, $w_i$, is the number of counts in that bin (also the same used in the weighted average calculation), and the sums run over the total number ($N_y$) of bins along the y-axis.

For the jet $p_T$ and $z$ kinematics which clearly rely upon jet finding, there are errors in the jet momentum scale arising from the BEMC towers efficiency, energy calibration uncertainty and the TPC track momentum uncertainty. The error due to the barrel is easily calculated from two pieces: the uncertainty on the efficiency set to $1\%$, and the uncertainty on the gain calibration which is combined between inner and outer $\eta$ rings to be a conservative value of $3.8\%$. Using these two pieces, and the neutral fraction of energy in the jet ($R_t$) measured by BEMC, the error in the jet $p_T$ imparted by the BEMC is 

\begin{equation}
\delta p_{T,jet,BEMC} = \langle p_T\rangle\times R_t\times\sqrt{\delta^2_{gain} + \delta^2_{eff}}
\end{equation}

The error imparted by the track uncertainty is a bit more involved, and is coming from the charged portion of the jets: ($1-R_t$). There is an error which is due to the uncertainty in the track momentum, which is set at $1\%$, calculated simply as $\delta p_{T,tracks}=\langle p_T \rangle\times\left(1-R_t\right)\times0.01$. There is another piece that encapsulates how well the simulation emulates hadronic interactions from charged hadrons and long lived neutral particles (like $n$, $\bar{n}$, $K^0_S$ and $K^0_L$) in the BEMC towers. The analysis in Ref. \cite{ref:starBemcTrkError} determined there is a $\delta f_{E_{had}}= 0.09$ systematic error associated with how well the hadronic effects are emulated. Having this value in hand, we simply need to multiply it by the charged hadronic portion of the jet $p_T$ and apply the appropriate scaling parameters to get the error in the jet momentum scale due to charged tracks in the BEMC:

\begin{equation}
\label{eq:bemcTrkUncert}
\delta p_{T,jet,BEMC_{tracks}} = \langle p_T\rangle \left(1-R_t\right)\delta f_{E_{had}} \left\{\frac{f_{E_{dep}}\left(S_{neutral}-\epsilon_{track} f_{track_{dep}}\right)}{\epsilon_{track}}\right\}
\end{equation}

\noindent Here the charged component of the jet is isolated by the product $\langle p_T\rangle \left(1-R_t\right)$. The expression in the brackets is the appropriate scaling factor that must be applied. We account for the inefficiency of the TPC detection of these tracks by dividing by the efficiency of the TPC ($\epsilon_{track}$) which is set to $\epsilon_{track}=81\%$. Next we account for the average fraction of hadronic energy deposited in the BEMC ($f_{E_{dep}} = 0.3$). Finally, long lived neutral hadrons which are not detected by the TPC, but do deposit energy into the BEMC, are accounted for. This is called the neutral hadron scale-up, $S_{neutral}$, and it is set to $1/0.86=1.163$ \cite{ref:starBemcTrkError}. $S_{neutral}$ is offset by the fraction of energy deposited by hadrons into a single isolated tower, $f_{track_{dep}} = 0.5$, which is likely a conservative value since we use the $100\%$ energy subtraction scheme for BEMC towers in our jet finding algorithm. Therefore, the appropriate factor is $\left(S_{neutral}-f_{track_{dep}}\epsilon_{track}\right)$, where the $\epsilon_{track}$ factor offsets the value in the denominator of Equation \ref{eq:bemcTrkUncert}. 

The total track uncertainty for the jets is simply these results added in quadrature:

\begin{equation}
\delta p_{T,jet,tracks} = \sqrt{\delta p_{T,tracks}^2 + \delta p_{T,jet,BEMC_{tracks}}^2}
\end{equation}

\noindent Then the final result for the uncertainty in the jet momentum scale is $\delta p_{T,jet,tracks}$ and $\delta p_{T,jet,BEMC}$ added in quadrature. For the jet $p_T$ shift error, this is all we need, and the result may be calculated by following Equation \ref{eq:kinErr1}.

In addition to the contributions that have been discussed thus far, there is one more error that contributes to the total systematic in the $z$ shift. This is a contribution from the uncertainty in the pion momentum, which goes into the final calculation of $z$. This is simply calculated using the $1\%$ track momentum uncertainty as $\delta p_\pi = \langle p_\pi\rangle\times 0.01$. With this, the error in the $z$ shift may be calculated by plugging all relevant information into Equation \ref{eq:kinErr2}.

Similarly to $z$, the error for $j_T$ contains a contribution from the track momentum uncertainty because, after all, $j_T$ is the transverse momentum of the pion with respect to the jet axis. This error is calculated easily using the $1\%$ track momentum uncertainty again as $\delta j_{T,tracks}=\langle j_T\rangle\times0.01$. Now the calculation of the error for the $j_T$ shift may completed by filling in all values in Equation \ref{eq:kinErr3}. Note that all of the errors in the kinematic shifts are calculated on a kinematic bin-by-bin basis and are reported in the tables in the Appendix.

\subsection{Trigger Bias}
\label{sec:triggerBias}
Applying a triggering algorithm to the data as it is collected is an effective way to enhance higher $p_T$ jets that would otherwise get swallowed by the vast amount of low $p_T$ jets that are much more readily produced. However, the trigger also tends to bias the sample by enhancing the number of quark jets. This effect is due to the fact that gluon jets fragment more broadly than quark jets and therefore some of the energy from the gluon jet is more likely to fall outside of the jet patch. This means gluon jets are less likely to cause a trigger, especially at lower values of $p_T$. Because we cannot tag quark or gluon jets in the data, the only way to measure this effect is to use the simulation where we know the particle species that are in each event. 

The size of the trigger bias is determined by calculating the fraction of events that originate from a quark at the detector level, where we apply the trigger, and taking the ratio with the same fraction calculated at the particle level where trigger and detector effects are removed. 

Starting at the detector level and applying all analysis level cuts to jets, we match detector jets to particle level jets using the same matching outlined in Section \ref{sec:kinShifts}. Then we match the particle level jet on back to the level of partons before hadronization, applying the same matching condition as for detector to particle level. In the simulation, we have access to the partons which were a part of the hard scattering and thus the partons that form the jets which are called the ``mother'' partons. Having this we can match the parton jet to the mother partons, using a $\Delta R$ calculation. To be considered a match, we look for the mother parton which has the minimum $\Delta R$ with respect to the parton jet, and it must satisfy $\Delta R < 0.3$ which is smaller than the jet matching condition because the mother which created the parton jet should be very close to that jet. Once the mother parton species is known, distributions of jet $p_T$, $j_T$ and $z$ are filled on a track-by-track basis, ignoring charge sign, for events that satisfy the triggering condition. The distributions are filled separately for quark originated events and gluon originated events.

The same analysis is repeated for all of the particle level jets, without consideration of the detector level jets, for all jets that satisfy analogous cuts to the analysis. The particle jets are matched to the parton level jets, and the parton jets are matched to the mother partons, and distributions are filled on a particle-by-particle basis within the particle jet if the particle is a pion and satisfies other analysis cuts. These distributions are also filled separately for quark and gluon originated events.

What we are left with is raw counts in each kinematic bin for quark and gluon originated events, hence we have the number of quark events and number of gluon events in each bin. So, letting $q$ and $g$ denote quark and gluon events, respectively, we define the quark fractions at the detector and particle level as

\begin{equation}
\label{eq:quarkFrac}
f_{det}^{quark} = \frac{q_{det}}{q_{det}+g_{det}} \qquad\qquad f_{particle}^{quark} = \frac{q_{particle}}{q_{particle}+g_{particle}}
\end{equation}

\noindent Calculating these fractions on a kinematic bin-by-bin basis allows us to calculate the ratio of the detector to particle level fractions, $R^{quark} = f_{det}^{quark}/f_{particle}^{quark}$. This ratio gives a measure of the bias imparted by the trigger at the detector level. As a function of jet $p_T$, the results for the detector and particle level quark and gluon fractions are plotted in Figure \ref{ProcessFractionPt}, where the gluon fraction is calculated the same way as in Equation \ref{eq:quarkFrac} only with the number of gluons in the numerator. Within the statistical precision, there is an overall trend that shows enhancement of quark jets at the detector level relative to the particle level, and a suppression of gluon jets at the detector level for $p_T>10$ GeV/c.

\figuremacroW{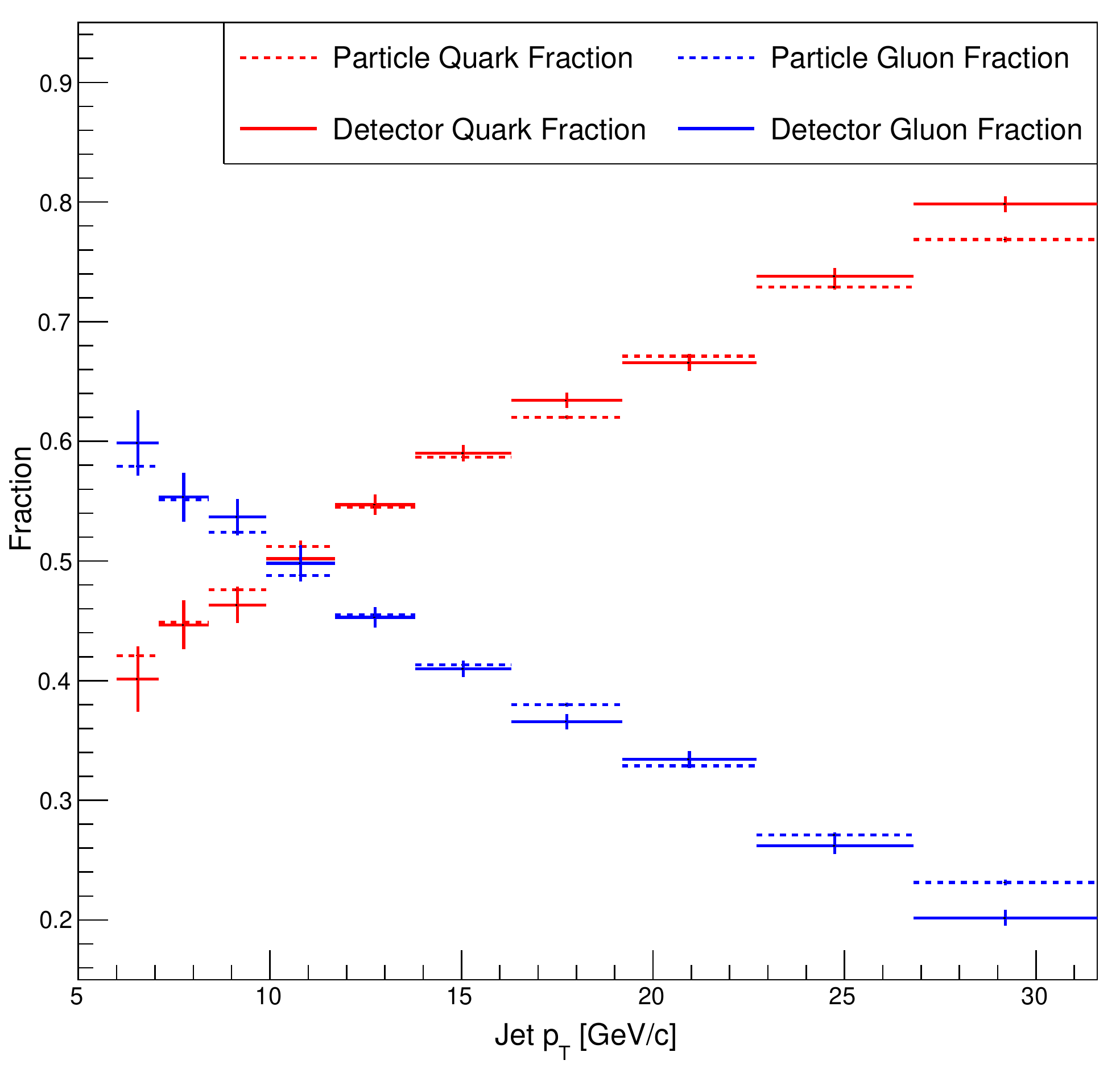}{Quark and Gluon Fractions vs. Jet $p_T$ }{Detector and particle level quark and gluon fractions. These fractions go into the ratio, $R^{quark}$ to calculate the systematic error due to trigger bias.}{0.8}

To get a handle on the size of the bias imparted by the trigger, $R^{quark}$ is plotted in Figure \ref{QuarkRatioPt}. There is not a clear dependence here upon the jet $p_T$, all values seem to be oscillating about and consistent with $1$ within their statistical errors. Because of this lack of structure, the approach to assigning an error for this bias is to combine all of the statistics for all jet $p_T$ bins and assign a flat rate error for all bins. To do this, all statistics are combined to get a single detector and particle level fraction, and thus a single quark ratio for jet $p_T$. The result of this is a quark ratio of $1.0834$, meaning an error of $8.34\%$ will be assigned for each jet $p_T$ bin. 

\figuremacroW{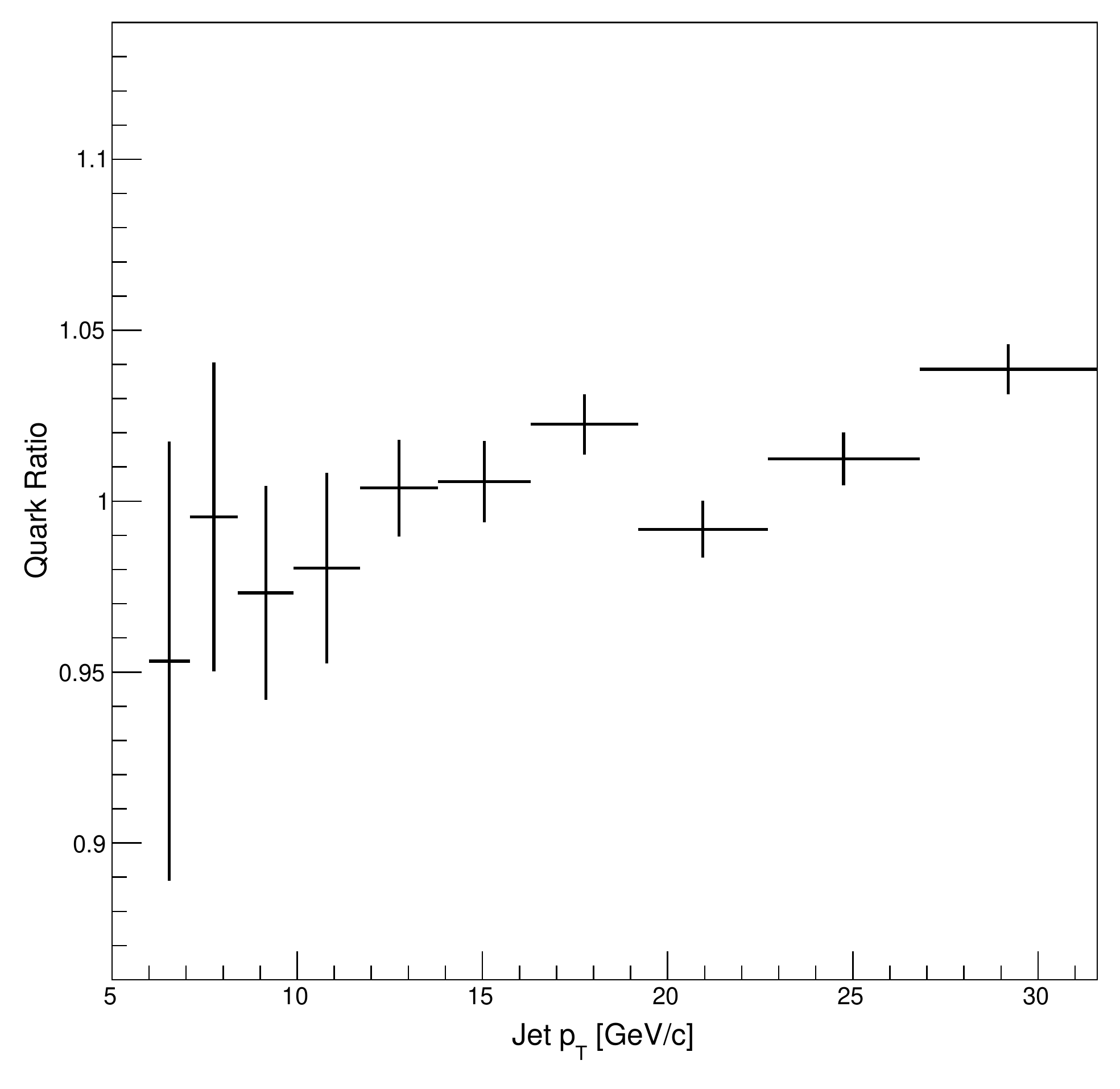}{$R^{quark}$ vs. Jet $p_T$}{The ratio of the detector to particle level quark fractions as a function of jet $p_T$ that can be found as the red lines in Figure \ref{ProcessFractionPt}.}{0.8}

Figures \ref{QuarkRatioZ_PtBins} - \ref{QuarkRatioJt_zBins} are the analogous plots to Figure \ref{QuarkRatioPt}. Since the lack of structure propagates throughout all kinematic ranges and bins, the errors are assigned in the same way. Tables \ref{table:ptRangeTB} - \ref{table:zRangeTB} of the systematic errors for the jet $p_T$ and pion $z$ ranges are given below. Note that $z$ and $j_T$ are plotted in ranges of jet $p_T$ but these use the same statistics, so identical errors will be assigned for both plotting arrangements.

\figuremacroW{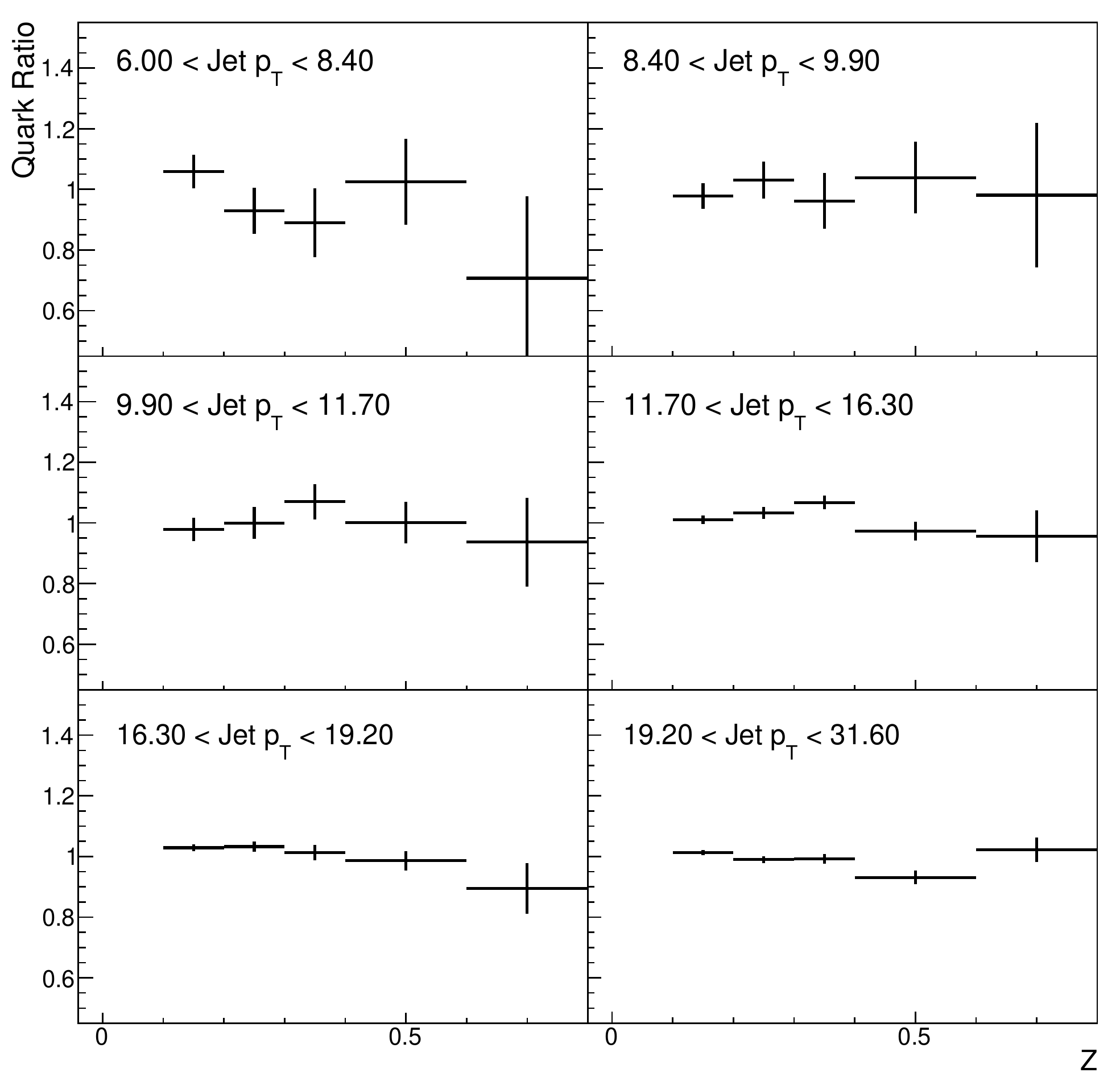}{$R^{quark}$ vs. $z$ in Jet $p_T$ Ranges}{The ratio of the detector to particle level quark fractions as a function of $z$, given for the final kinematic ranges of jet $p_T$.}{0.8}

\figuremacroW{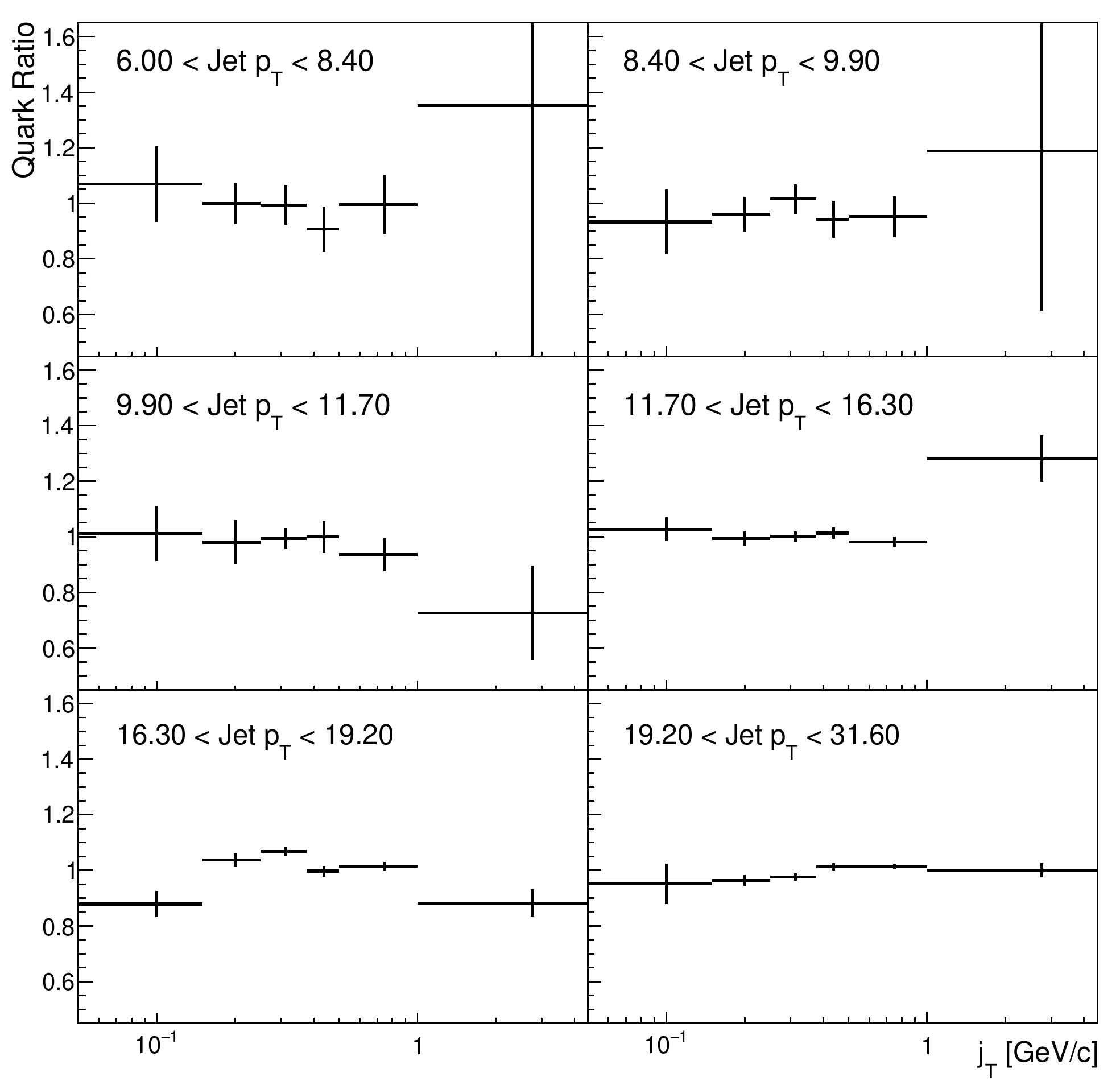}{$R^{quark}$ vs. $j_T$ in Jet $p_T$ Ranges}{The ratio of the detector to particle level quark fractions as a function of $j_T$, given for the final kinematic ranges of jet $p_T$.}{0.8}

\figuremacroW{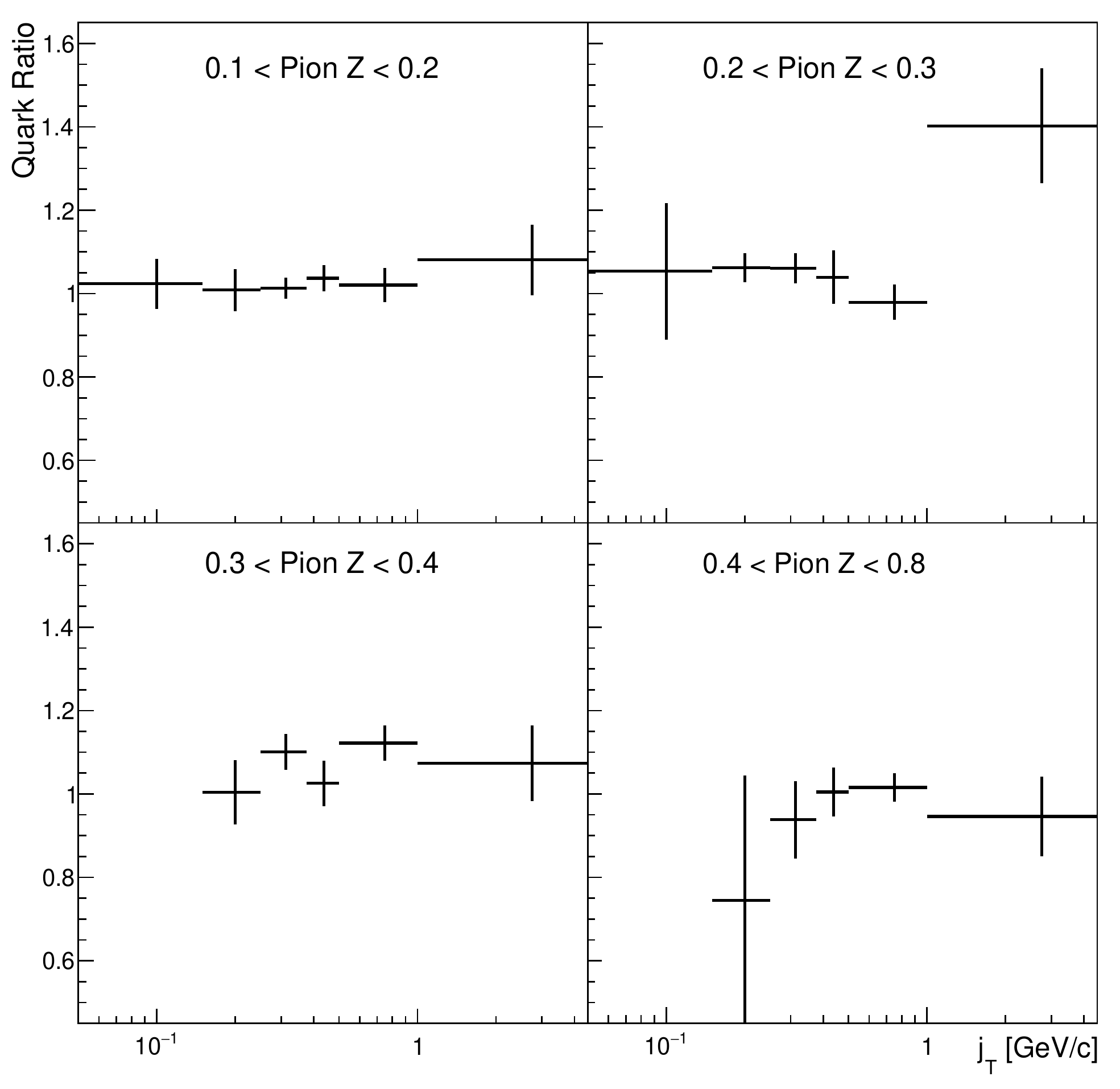}{$R^{quark}$ vs. $j_T$ in Pion $z$ Ranges}{The ratio of the detector to particle level quark fractions as a function of $j_T$, given for the final kinematic ranges of pion $z$.}{0.8}

\begin{table}[htp]
\begin{center}
\begin{tabular}{|c|c|c|}
\hline
Jet $p_T$ Range [GeV/c]& Trigger Bias Error (\%)\\
\hline
6.0 - 8.4 & 1.362 \\
\hline
8.4 - 9.9 & 2.682\\
\hline
9.9 - 11.7 & 1.959\\
\hline
11.7 - 16.3 & 0.955\\
\hline
16.3 - 19.2 & 2.250\\
\hline
19.2 - 31.6 & 0.203\\
\hline
\end{tabular}
\end{center}
\caption[Jet $p_T$ Range Trigger Bias]{\textbf{Jet $p_T$ Range Trigger Bias} - Value of trigger bias error that will be assigned to each of the asymmetry points within the quoted jet $p_T$ regions.}
\label{table:ptRangeTB}
\end{table}%

\begin{table}[htp]
\begin{center}
\begin{tabular}{|c|c|c|}
\hline
Pion $z$ Range& Trigger Bias Error (\%)\\
\hline
0.1 - 0.2 & 2.323 \\
\hline
0.2 - 0.3 & 4.370\\
\hline
0.3 - 0.4 & 9.052\\
\hline
0.4 - 0.8 & 0.179\\
\hline
\end{tabular}
\end{center}
\caption[Pion $z$ Range Trigger Bias]{\textbf{Pion $z$ Range Trigger Bias} - Value of trigger bias error that will be assigned to each of the asymmetry points within the quoted pion $z$ regions.}
\label{table:zRangeTB}
\end{table}%

The errors are calculated as a percentage of the actual asymmetry point for each kinematic bin. The errors are assigned as follows:

\begin{itemize}
\item If $R^{quark} > 1$ then an asymmetric error bar will be applied to each asymmetry point on the side of decreasing magnitude.
\item If $R^{quark} < 1$ then an asymmetric error bar will be applied to each asymmetry point on the side of increasing magnitude.
\end{itemize}

\noindent The reason for the two distinctions is because if $R^{quark} < 1$ then quark events are being suppressed at the detector level, and our final asymmetry is too small because of the dilutions from gluon enhancement. However, if $R^{quark} > 1$ then quarks are enhanced at the detector level and the final asymmetries are too large because we should have more gluon events which would dilute the asymmetry value down. In the example of jet $p_T$ above, the error will be applied towards a decreasing asymmetry value by the amount of $8.34\%$ of each asymmetry point, thus the error bar grows with an increasing asymmetry.

\subsection{Polarization Uncertainty}
The polarization measurements that are taken during each fill have an associated uncertainty that must be propagated and reported. A relative scale uncertainty is reported by the polarimetry group for the polarization during each running year, and for 2012 the value reported was $3.4\%$. However, this value needs to be adjusted if the number of RHIC fills worth of data used for analysis is not the total number used for the polarization determination, and for the error in the beam intensity profile. The calculation of the adjustment factor is straightforward and is outlined in Ref. \cite{ref:PolarizationUncert}. First, the weighted average polarization is calculated by summing over all fills used for analysis and applying

\begin{equation}
\bar{P}=\frac{\sum\limits_f^{N_F} w_f*P_f}{\sum\limits_f^{N_{F}} w_f}
\end{equation}

\noindent where the weight, $w_f$, is the number of events in each fill and $P_f$ is the polarization for each fill. Then, the error in this mean is calculated similarly, only we use the error on the polarization measurement, by

\begin{equation}
\delta\bar{P}=\sqrt{\frac{\sum\limits_f^{N_F} w_f^2*\sigma_f^2}{\sum\limits_f^{N_{F}} w_f^2}}
\end{equation}

\noindent Finally, as noted in Ref. \cite{ref:PolarizationUncert}, we must include a factor $\sqrt{1-M/N}$ where $M$ is the number of fills we use data from and $N$ is the total number of fills to calculate the $3.4\%$ relative scale uncertainty. In the case of the 2012 $\sqrt{s}=200$ GeV dataset, $N$ is $56$ and we use $47$ of these fills for analysis. Thus, due to fill-to-fill uncertainties there is a correction to the $3.4\%$ relative scale uncertainty calculated as

\begin{equation}
\delta P_{scale} = \frac{\delta\bar{P}}{\bar{P}}*\sqrt{1-\frac{47}{56}}
\end{equation}

\noindent This correction is calculated to be $3.00\%$, and will be folded into the total.

The error beam intensity and polarization profile, which is used for calculating the reported $3.4\%$ uncertainty, must also be taken into account. This is a simple correction factor of $2.2\%/\sqrt{M}$, where $M$ is the same number of used fills ($47$) as above. Therefore, this correction comes out to be $0.32\%$.

Using this information, the final scale uncertainty is calculated by adding all sources of error in quadrature:

\begin{equation}
\delta P_{absolute} = \sqrt{\delta P^2_{relative} + \delta P^2_{fills} + \delta P^2_{profile}} = 4.55\%
\end{equation}

\noindent For this analysis, the absolute scale uncertainty comes out to be $4.55\%$ due to uncertainties in the polarization. This number is not folded into the overall systematic uncertainties, rather reported as an overall scale uncertainty which could result in an increase or decrease in the value of each asymmetry point.

\subsection{Spin Angle Offset}
When the asymmetry is measured in the analysis the spin angle is assumed to be perfectly perpendicular to the proton momentum, and is set to $\pm \pi$ for calculation of the azimuthal angle $\phi_S$. However, if the polarization picks up some offset, $\delta$, then $\phi_S$ picks up the offset and propagates it to the Collins angle, $\phi_C$. In this case, it will also appear in the sinusoidal modulation (assuming the asymmetry is A): 

\begin{equation}
A\sin\left(\phi_C + \delta\right) = A\sin\left(\phi_C\right)\cos\left(\delta\right) + A\cos\left(\phi_C\right)\sin\left(\delta\right)
\end{equation}

The form of the fit used to extract the asymmetry from the cross ratio calculation is

\begin{equation}
\epsilon\left(\phi_C\right) = p_0 + p_1 \sin\left(\phi_C\right)
\end{equation}

\noindent Comparing this to what happens above when there is a spin angle offset for the polarization, clearly the result of this offset would be a dilution to the asymmetry value with the size $\cos\left(\delta\right)\approx 1-\delta^2/2$. The STAR ZDC Scaler Polarimetry group \cite{ref:zdcPolarimetry} reports values for $\delta$ for each running period, with $0.05$ radians being a conservative estimate for 2012. This estimate for $\delta$ leads to a $0.063\%$ reduction of the magnitude of the asymmetry value which is negligible when compared to the other systematic errors. Therefore, this error is not folded into the results.


\chapter{Results and Conclusions} 
\label{chapter:Results}

\ifpdf
    \graphicspath{{ch7_Results/figures/PNG/}{ch7_Results/figures/PDF/}{ch7_Results/figures/}}
\else
    \graphicspath{{ch7_Results/figures/EPS/}{ch7_Results/figures/}}
\fi

Now that the asymmetry analysis of Chapter \ref{chapter:AsymmetryAnalysis} has been completed and the corrections and systematic errors discussed in Chapter \ref{ch:Systematics} have all been calculated, we can finally bring it all together. This chapter presents the final corrected asymmetries with their assigned systematic errors.

\section{Inclusive Jet Asymmetry}
\label{sec:InclusiveJetAsym}
The inclusive jet asymmetry, $A_{UT}^{\sin\left(\phi_S\right)}$, is sensitive to the initial state twist-3 collinear quark-gluon correlators, which are related to the TMD Sivers function. Figure \ref{SiversAsymmetryPt} shows $A_{UT}^{\sin\left(\phi_S\right)}$ from the 2012 STAR $\sqrt{s}=200$ GeV $p^\uparrow +p$ dataset for jets that scatter forward ($x_F>0$) and backward ($x_F<0$) with respect to the polarized beam. The presented asymmetries have been corrected for finite bin width dilution (Section \ref{sec:finiteBins}). Also, the jet $p_T$ shifts have been applied so the points are plotted at the true value of jet $p_T$ with the error along the x-axis representing the systematic error on the correction. These plots are analogous to the result presented in Ref. \cite{ref:StarSivers2006} and Figure \ref{StarSiversPRD} with the results integrated into two bins of $\eta$. 

\figuremacroW{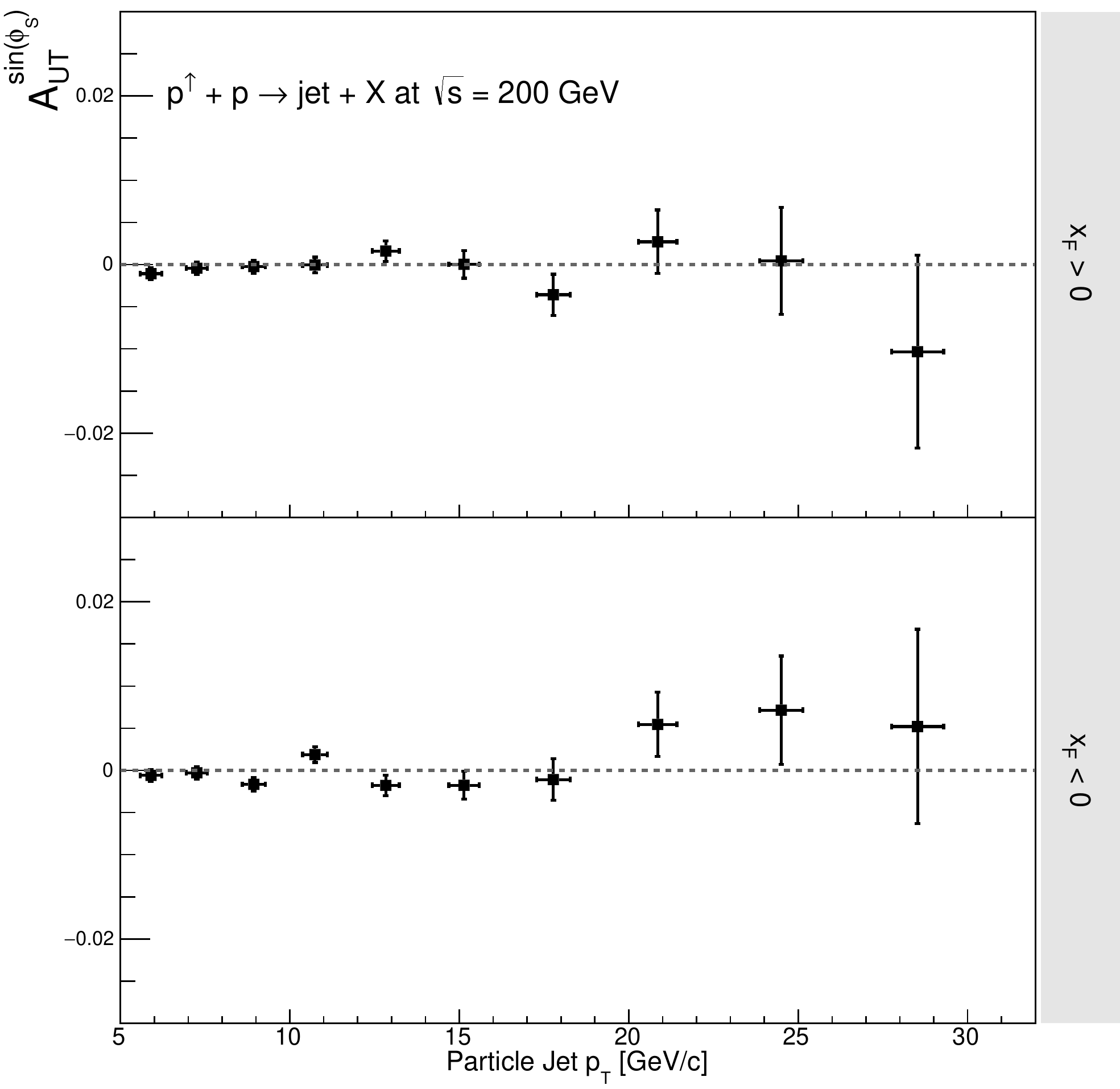}{Inclusive Jet $A_N$ vs. Jet $p_T$}{The inclusive jet asymmetry plotted as a function of the jet $p_T$ for both jet scattering states.}{0.8}

The asymmetries presented in Figure \ref{SiversAsymmetryPt} are consistent with zero. The $A_{UT}^{\sin\left(\phi_S\right)}$ results from the 2012 dataset are consistent with those measured at STAR using the 2006 dataset in Figure \ref{StarSiversPRD} and agree well with the predictions made in Figure \ref{Kanazawa_TwistThreeStar} from Ref. \cite{ref:KanazawaTwist3}.

\section{Collins Asymmetries}
\label{sec:CollinsAsym}
The Collins asymmetry, $A_{UT}^{\sin\left(\phi_S-\phi_H\right)}$, is sensitive to contributions from the transversity parton distribution function and the Collins fragmentation function. Results for the Collins asymmetry for the 2012 STAR $\sqrt{s}=200$ GeV $p^\uparrow +p$ data are shown for all binning arrangements in Figures \ref{CollinsAsymmetryPt} and \ref{CollinsAsymmetryZ_ptBins} - \ref{CollinsAsymmetryJt_zBins} for both $x_F>0$ and $x_F<0$ jet scattering states. These results are binned multidimensionally in the kinematic variables jet $p_T$, $z$, and $j_T$ to map out asymmetry dependencies. The jet $p_T$ accesses the hard scale of the scattering, and $z$ and $j_T$ access the longitudinal and transverse momentum, respectively, in the fragmentation process.

The corrections discussed in Chapter \ref{ch:Systematics} have been applied to all of these asymmetries. The systematic errors on the asymmetries are indicated by the height of the grey shaded bar, while the width represents the systematic error on the correction of the kinematic variable. In some cases the systematic and statistical errors are so small that they are difficult to discern on the plot. Let us look at each plot individually.

\figuremacroW{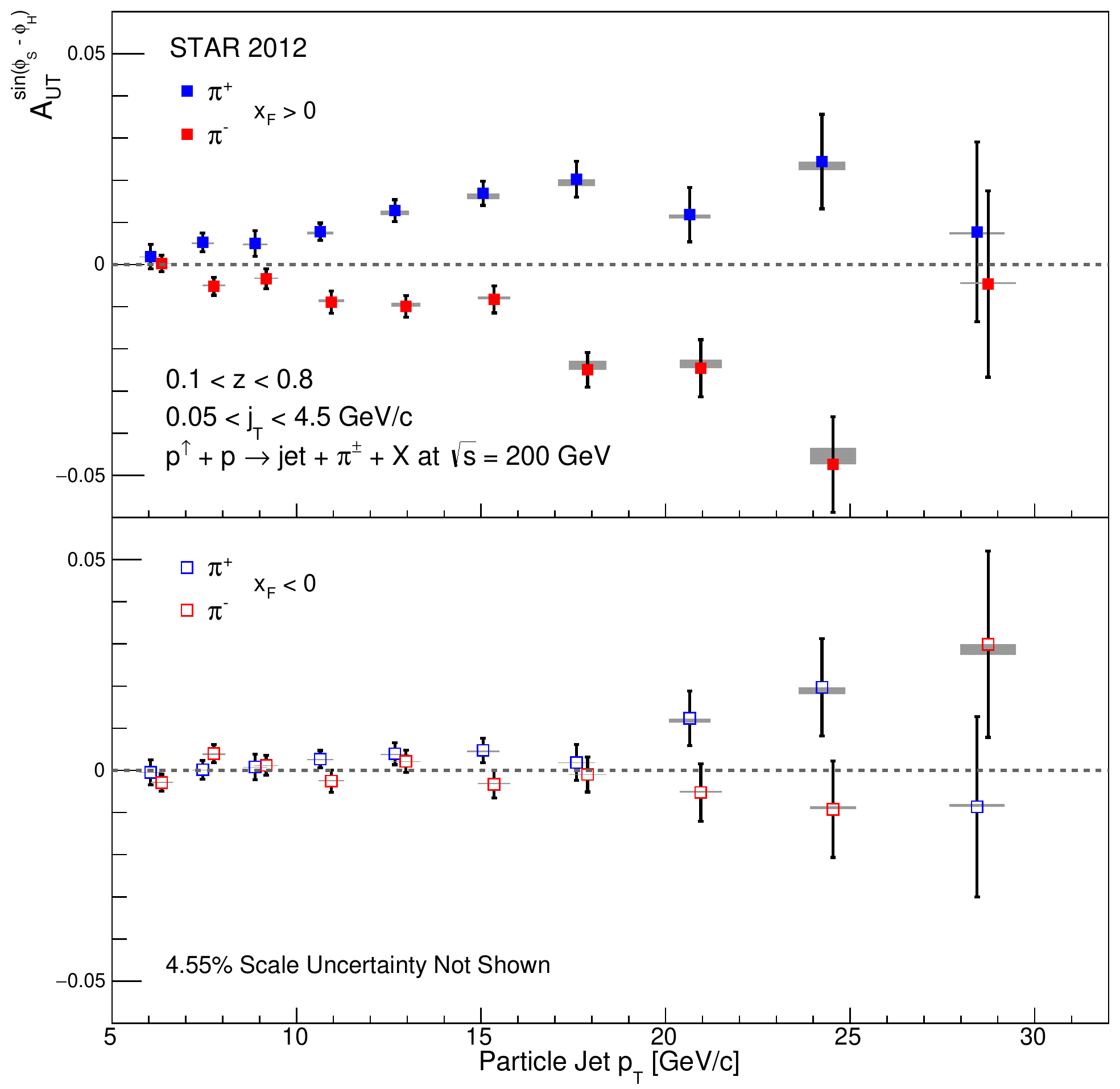}{$A_{UT}^{\sin\left(\phi_S-\phi_H\right)}$ vs. Jet $p_T$}{The Collins asymmetry plotted as a function of the jet $p_T$, for $x_F>0$ and $x_F<0$. This result integrates over the entire range of $z$ and $j_T$.}{0.8}

Figure \ref{CollinsAsymmetryPt} shows $A_{UT}^{\sin\left(\phi_S-\phi_H\right)}$ plotted as a function of jet $p_T$, a surrogate for the momentum transfer $Q$. The asymmetry is very clearly positive for $\pi^+$ and negative for $\pi^-$ which means the two charge states favor asymmetric distributions on opposite sides of the jet. Following the Collins convention that pion production toward the left (right) side of the jet results in positive (negative) asymmetries, the data trends reflect our current understanding of the transversity distributions functions for up and down quarks. Assuming favored fragmentation, the $\pi^+$ is produced from the up quark, which is likely to be polarized vertically and then fragment preferentially to the left, producing positive asymmetries. The down quark transversity distributions are negative and therefore the asymmetries should (and do) reverse for the $\pi^-$. This charge separation is persistent and the same sign on average throughout all of the results presented here.

Plotting for both the forward and backward scattered jets with respect to the polarized beam accesses different average values of $x$, as shown in Figure \ref{PionX}. Accessing lower values of $x$ means the backwards scattered jets are sampling a different region of the transversity distribution than the forward scattered jets. Jets which scatter backwards with respect to the polarized beam are scattering forward with respect to the \textit{unpolarized beam}. Assuming that the parton from each beam scatters forward, then the $x_F<0$ jets are accessing partons from the spin averaged beam. Averaging over the spin of the beam results in averaging out the spin dependent effects in the cross section. It is no surprise, then, that the measured signal goes away for the $x_F<0$ asymmetries. 

\figuremacroW{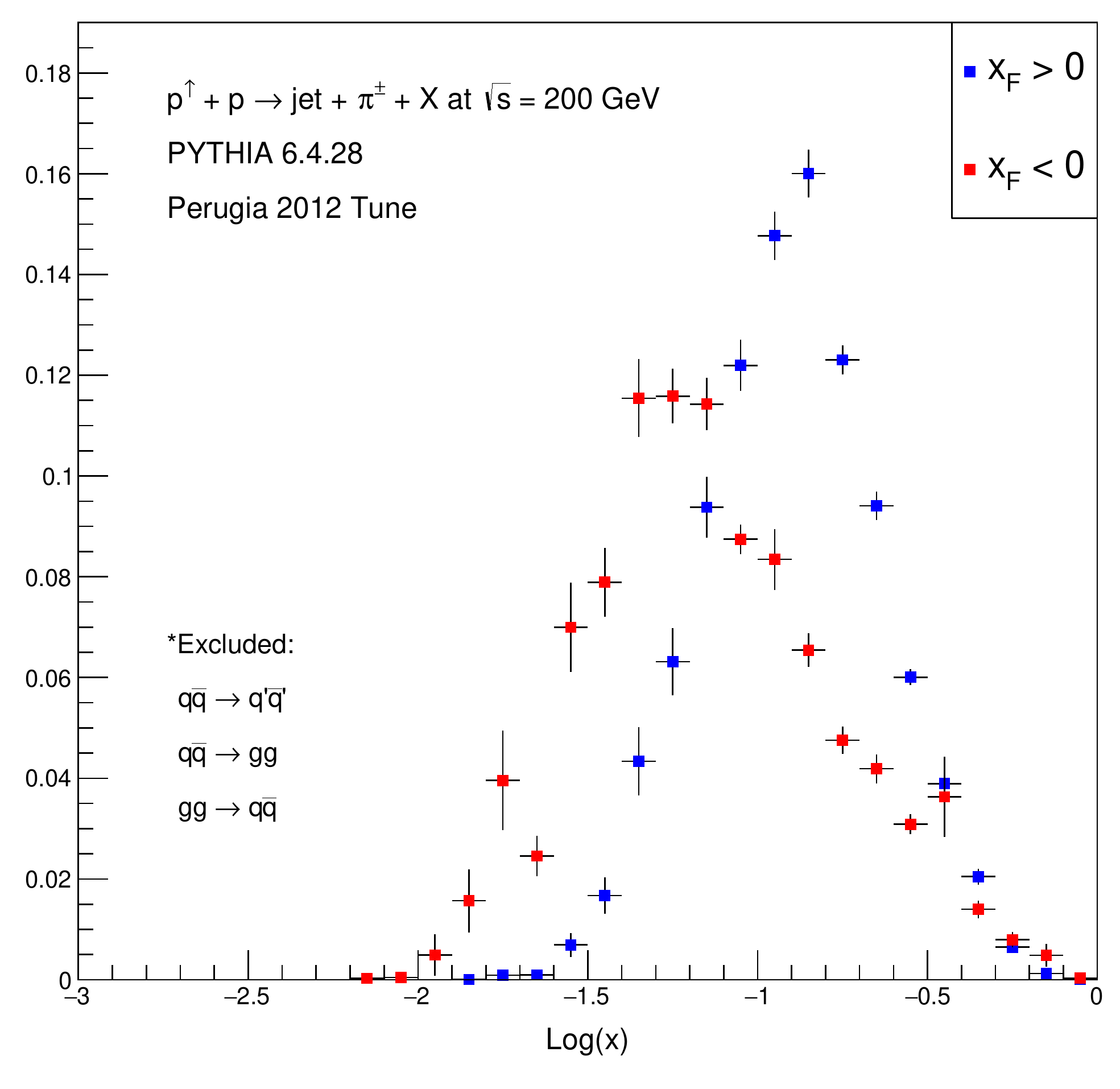}{Accessed $x$ Distributions}{Distributions of the values of $x$ accessed in the analysis. On average, the forward scattered jets access a slightly higher value of $x$ than the backward scattered jets.}{0.8}

$A_{UT}^{\sin\left(\phi_S-\phi_H\right)}$ is plotted as a function of $z$ and $j_T$ for average values of jet $p_T$ in Figures \ref{CollinsAsymmetryZ_ptBins} and \ref{CollinsAsymmetryJt_ptBins} respectively. It is interesting to compare these results to the previous 2006 result presented in Ref \cite{ref:StarCollinsPrelim} and Figure \ref{StarCollinsPrelim_2006}. The immediate conclusion is that the additional statistics have helped pull out the signal that was previously found to be consistent with zero. It is clear from Figure \ref{CollinsAsymmetryPt} the signal will start to really pop out above $p_T\approx 10$ GeV/c, and that is exactly what these results show. In fact, the magnitude of the asymmetry is continually growing in each $\langle p_T\rangle$ bin, except for the highest $\langle p_T\rangle$ bin where the statistics are becoming sparse. An interesting note for these plots is the shapes of both the $z$ and $j_T$ dependencies, they share similar features. The asymmetry as a function of $j_T$ turns over for every significant signal region, a dependence that can be found to an extent as a function of $z$ as well. These two kinematic variables are positively correlated, so it would seem that they are driving each other's shape. We will explore this concept more in the following section.

The Collins fragmentation function depends upon the kinematic variables $z$ and $j_T$. So plotting the asymmetry results for several average values of the hard scale $p_T$ allows for mapping out how the fragmentation function is varying as a function of the hard scale. Then, plotting the asymmetry as a function of $j_T$ for different average values of $z$ allows for disentangling the dependencies of the fragmentation function upon the kinematic variables. Figure \ref{CollinsAsymmetryJt_zBins} shows the $j_T$ dependence of $A_{UT}^{\sin\left(\phi_S-\phi_H\right)}$ for four bins of $\langle z \rangle$ in the region where the asymmetries are large, $9.9 < p_T < 31.6$ GeV/c. Apart from the first, and lowest, $\langle z\rangle$ bin the average asymmetries are similar.

\figuremacroW{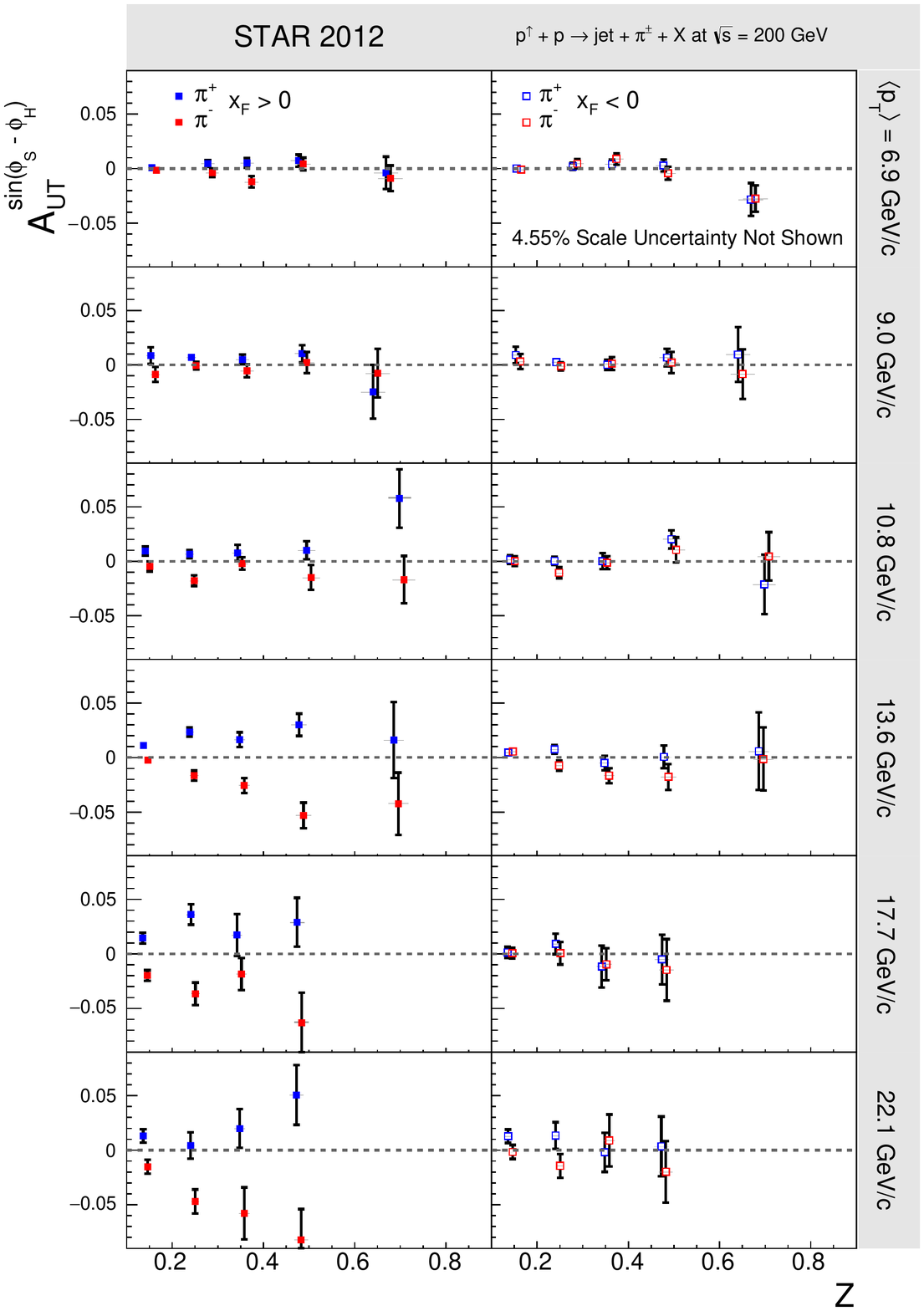}{$A_{UT}^{\sin\left(\phi_S-\phi_H\right)}$ vs. $z$}{The Collins asymmetry plotted as a function of $z$, for $x_F>0$ and $x_F<0$ and six average values of jet $p_T$ to map out the dependence of the fragmentation function on z and the hard scale.}{0.8}

Once the fragmentation is mapped out well as a function of the hard scale, then the last piece of the puzzle is to gain an understanding of the transversity distribution. Going back to the hard scale dependence in Figure \ref{CollinsAsymmetryPt} and introducing the now understood fragmentation functions should allow for transversity to be understood, and hopefully extracted, as a function of the hard scale.

\figuremacroW{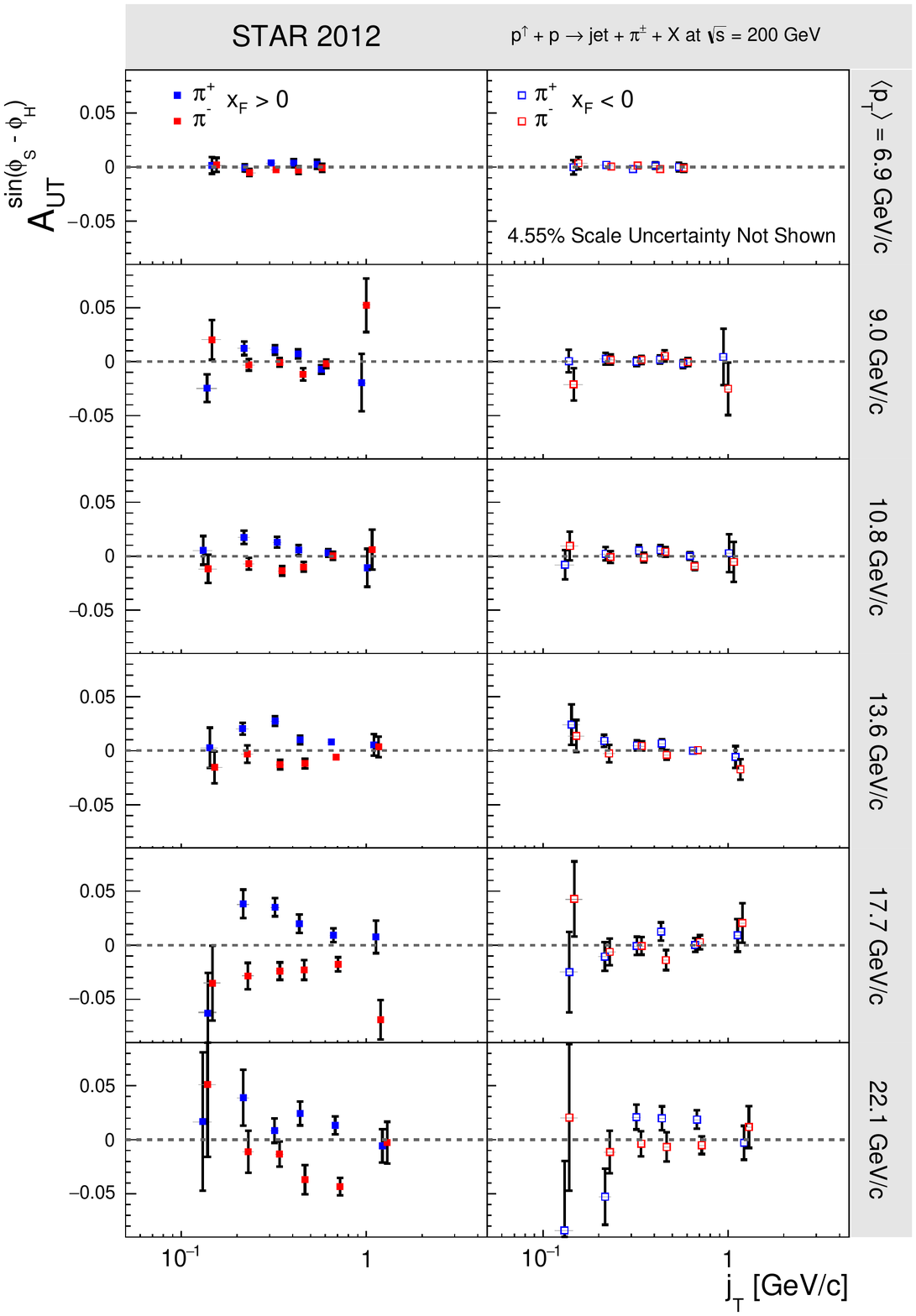}{$A_{UT}^{\sin\left(\phi_S-\phi_H\right)}$ vs. $j_T$}{The Collins asymmetry plotted as a function of $j_T$, for $x_F>0$ and $x_F<0$ and six average values of jet $p_T$ to map out the dependence of the fragmentation function on $j_T$ and the hard scale.}{0.8}

\figuremacroW{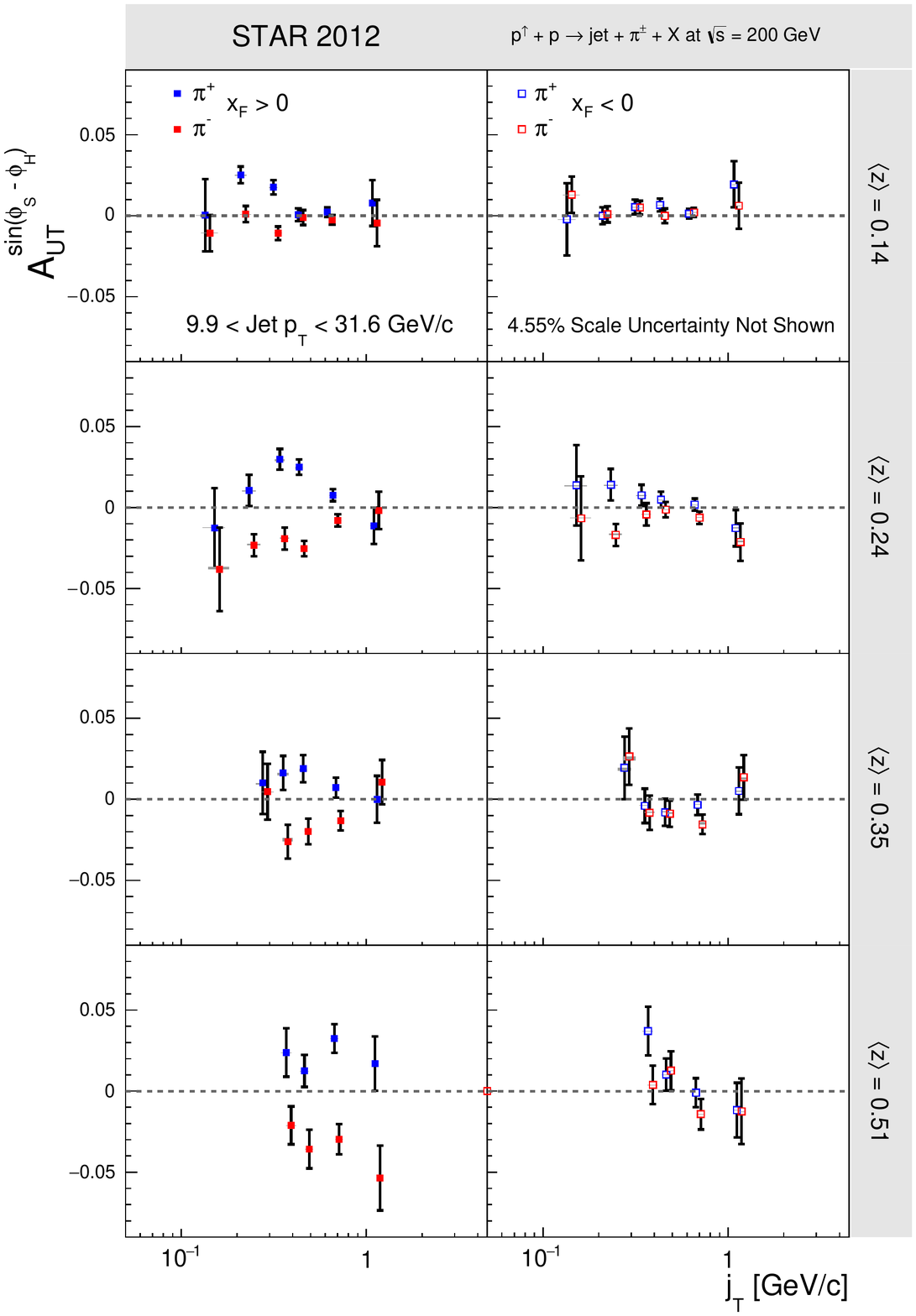}{$A_{UT}^{\sin\left(\phi_S-\phi_H\right)}$ vs. $j_T$}{The Collins asymmetry plotted as a function of $j_T$, for $x_F>0$ and $x_F<0$ and four average values of $z$ to help disentangle the fragmentation function kinematic dependencies.}{0.8}

\section{Tests of Universality and Evolution}
In 2011 RHIC collided protons at $\sqrt{s}=500$ GeV, and STAR collected transversely polarized collision data. From this data 25 $pb^{-1}$ were analyzed to extract the Collins asymmetry. Having the result at a much higher center of mass energy samples different subprocesses and a much wider range of the hard scale, $p_T$. It also provides the opportunity for a cross check when the same kinematics are sampled between the analyses. If the values of $x_T=2\langle p_T\rangle /\sqrt{s}$ are the same for both $\sqrt{s}=200$ GeV and $\sqrt{s}=500$ GeV, then the subprocess fractions are also the same and the $\sqrt{s}$ dependence is largely removed.

The results from the $\sqrt{s}  = 200$ GeV \cite{ref:adkinsSpin2014} and the $\sqrt{s} = 500$ GeV \cite{ref:drachenbergDIS2015} analyses were compared using the preliminary release data for both. For this comparison, $p_T > 10$ GeV/c and  $0.125 < j_T < 4.5$ GeV/c cuts were placed on the $200$ GeV data, resulting in $\langle p_T\rangle = 12.9$ GeV/c and therefore $x_T = 0.129$. To hit a similar $x_T$, the $500$ GeV results were plotted with $\langle p_T\rangle=31.0$ GeV/c leading to $x_T=0.124$. Matching up the $x_T$ values does not guarantee that the sampled kinematics are identical, however. The $\pi^\pm$ kinematics are dictated by the $\Delta R_{min}$ cut that is applied to each pion within the jet. The $\langle j_T\rangle_{min}$ is related to the $\Delta R_{min}$ cut through the following relationship:

\begin{equation}
\langle j_T\rangle_{min} \approx z\times\Delta R_{min}\times\langle p_{T,jet}\rangle
\end{equation}

\noindent This relationship has the important consequence that if the $\Delta R_{min}$ grows too large then $\langle j_T\rangle_{min}$ also grows large, and can become so large that the $j_T$ asymmetry has gone to zero, thus no asymmetry would be observed as a function of $z$. To exhibit this behavior, asymmetries are plotted as a function of $z$ for two different values of $\Delta R_{min}$ in Figure \ref{CollinsZ_200v500}. In an effort to match the sampled kinematics between the 200 GeV and 500 GeV data the top panel uses $\Delta R > 0.1$ for $\sqrt{s}=200$ GeV and $\Delta R>0.04$ for $\sqrt{s}=500$ GeV, and the bottom panel uses $\Delta R > 0.25$ for $\sqrt{s}=200$ GeV and $\Delta R>0.1$ for $\sqrt{s}=500$ GeV. 

\figuremacroW{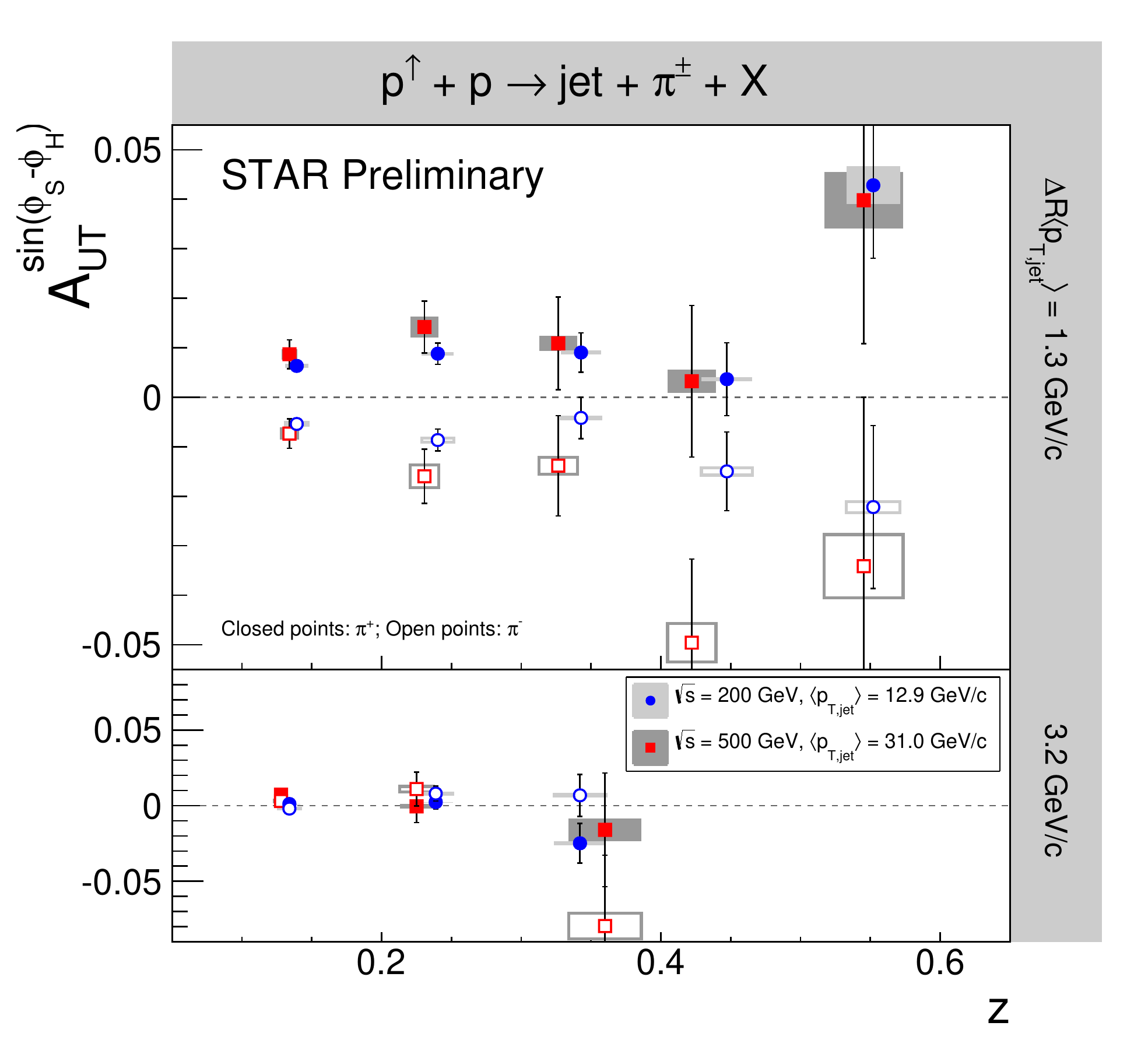}{$A_{UT}^{\sin\left(\phi_S-\phi_H\right)}$ Comparison at $\sqrt{s}=200$ GeV and $500$ GeV}{The Collins asymmetry from the 2011 $\sqrt{s}=500$ GeV and 2012 $\sqrt{s}=200$ GeV analyses. The results are plotted for identical kinematics by applying appropriate $\Delta R_{min}$ cuts and for the same value of $x_T$.}{0.9}

The results for $\sqrt{s}=200$ GeV and $500$ GeV match each other very well in Figure \ref{CollinsZ_200v500} for both $\Delta R_{min}$ cuts, meaning the results satisfy $x_T$ scaling. In the bottom panel, where both $\Delta R_{min}$ cuts are large, the asymmetry completely goes away as a function of $z$. However, when this cut is lowered back down and we are sampling a lower $\langle j_T\rangle_{min}$, the asymmetry pops back out with the same clear charge distinction we saw in the previous asymmetry plots. Because of the strong dependence on the pion $j_T$ which we saw in Figure \ref{CollinsAsymmetryJt_ptBins}, the asymmetries depend strongly on the $\Delta R_{min}$ cut. This result also demonstrates why the shapes of the $z$ and $j_T$ dependent asymmetries are very similar, it is because their correlation is driving the shape of the $z$ asymmetry.

A recent theoretical investigation used transversity and Collins fragmentation function extractions from global analyses of SIDIS and $e^+e^-$ annihilation results to calculate the Collins asymmetry \cite{ref:kangStarPrelim}. In Figures \ref{KangCalc_WithoutEvo} and \ref{KangCalc_WithEvo} the results of these calculations are compared to the $\sqrt{s}=200$ GeV and $\sqrt{s}=500$ GeV preliminary Collins asymmetries, and they are consistent. The distinction between the two sets of theory curves is those in Figure \ref{KangCalc_WithoutEvo} include no TMD evolution and Figure \ref{KangCalc_WithoutEvo} does include the effects of TMD evolution. In both cases, the calculated curves match the STAR results very well for both $\langle p_{T,jet}\rangle$ values, pointing to a slow evolution with $Q^2$. TMD evolution contains a non-perturbative piece that cannot be calculated from first principles. This data will provide some of the first direct input on this non-perturbative piece of TMD evolution.

The agreement between the theory curves and the STAR results is important because the transversity and Collins fragmentation functions are a product of SIDIS and $e^+e^-$ experimental data. Since the calculations of the Collins asymmetry using these extractions agree well with the results from the STAR jet analyses, the conclusion is that within the precision of our error budget we see no effects of factorization breaking for Collins asymmetries of hadrons in jets. Another new paper confirms this conclusion, and explicitly states that factorization holds for this Collins channel \cite{ref:kangUniversality}. This is a critical conclusion because it means moving forward the STAR Collins asymmetry measurements may be used to extract TMD functions such as the transversity distribution and Collins fragmentation function. 

\figuremacroW{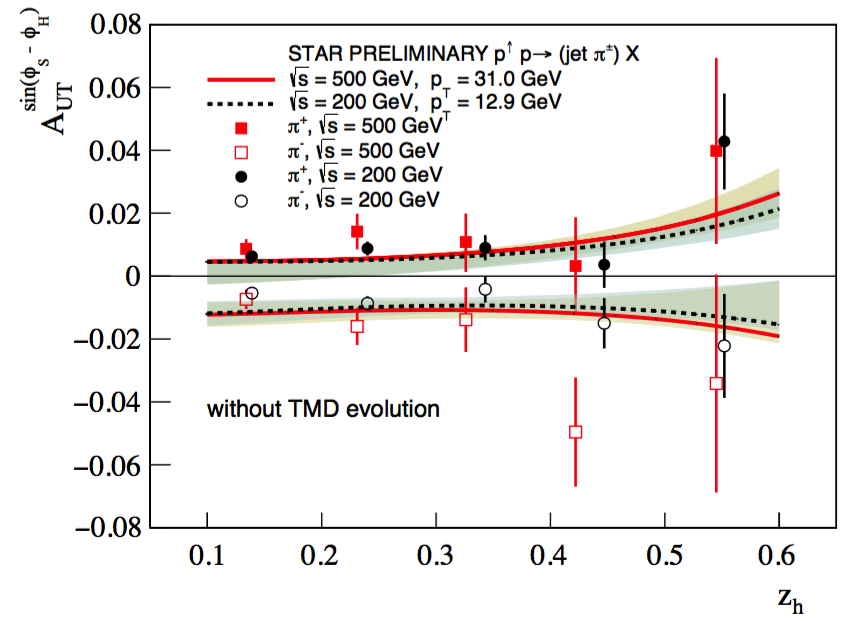}{Theory Comparison Without TMD Evolution}{Results from the theoretical Collins asymmetry calculations compared to the $\sqrt{s}=200$ GeV and 500 GeV STAR preliminary result \cite{ref:kangStarPrelim}. The TMD functions in this calculation include no evolution.}{0.9}

\figuremacroW{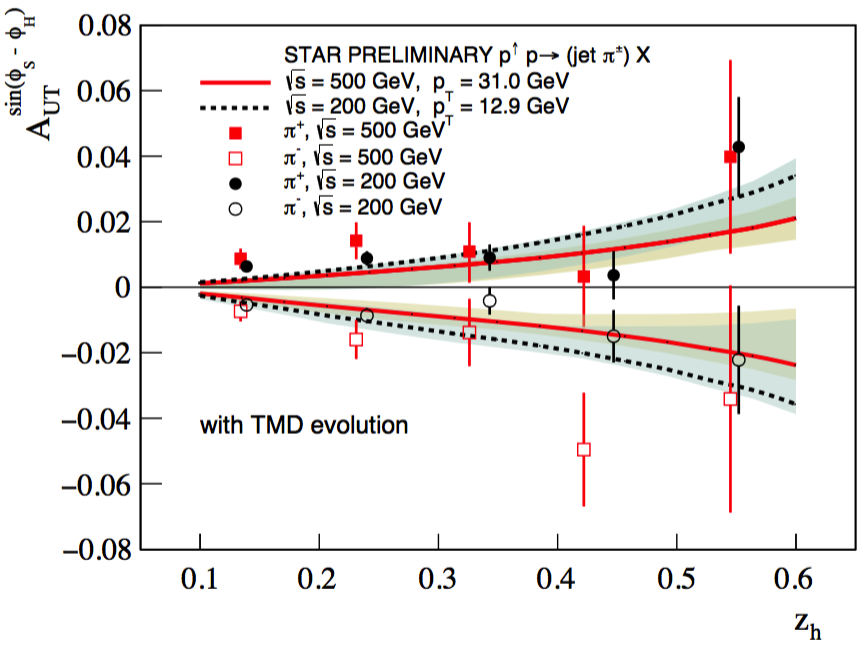}{Theory Comparison With TMD Evolution}{Results from the theoretical Collins asymmetry calculations compared to the $\sqrt{s}=200$ GeV and 500 GeV STAR preliminary result \cite{ref:kangStarPrelim}. The TMD functions in this calculation include effects from TMD evolution.}{0.9}

\section{Conclusions}
The asymmetries presented in this thesis represent the first statistically significant non-zero Collins asymmetries measured in $\sqrt{s}=200$ GeV hadronic collisions. Recent theoretical developments prove that factorization holds for the Collins asymmetry of hadrons in jets. As a result, these data may be included in global analyses aimed at extracting TMD functions. The implementation of multidimensional binning for the asymmetry results should prove valuable to map out the kinematic dependencies of TMD PDFs and FFs.

The Collins asymmetry is a unique channel to access the transversity distribution. Using jets to measure the asymmetry in $p^\uparrow p$ collisions provides the special ability to map out the $j_T$ dependence and further investigate the universality of TMD observables. This analysis, and the complementary one at $\sqrt{s}=500$ GeV provide the opportunity to gain deeper insight into the $Q^2$ evolution of the TMD functions. The $Q^2$ values sampled in the existing SIDIS data are an order of magnitude smaller than those sampled in the $e^+e^-$ annihilation data, requiring assumptions about TMD evolution in order to extract TMD functions. In contrast, the STAR data presented in this work sample a $Q^2$ region comparable to the $e^+e^-$ data, providing the first glimpse into the size of TMD evolution effects. The fact that the asymmetries measured at STAR agree quite well with predictions from global analyses using only SIDIS and $e^+e^-$ data indicate that these evolution effects are not large. This is supported by the excellent agreement between the $\sqrt{s} =$ 200 GeV and 500 GeV STAR Collins asymmetries. Most importantly, these results will provide some of the first direct input on the non-perturbative part of TMD evolution which cannot be calculated from first principles.

During the 2015 running period RHIC again collided protons at $\sqrt{s}=200$ GeV, and STAR collected more than twice the transversely polarized proton collision statistics presented in the results of Section \ref{sec:CollinsAsym}. The extra statistics should allow us to report asymmetries using a more finely binned multidimensional analysis. These extra bins will lead to more accurate extractions of the transversity distribution and Collins fragmentation function.



\chapter{Appendix} 


\ifpdf
    \graphicspath{{ch8_Appendix/figures/PNG/}{ch8_Appendix/figures/PDF/}{ch8_Appendix/figures/}}
\else
    \graphicspath{{ch8_Appendix/figures/EPS/}{ch8_Appendix/figures/}}
\fi

\section{Additional Embedding QA Plots}
The plots shown here are embedding QA histograms that supplement those presented in Section \ref{sec:embedQA} that were produced to test that the detector level embedding output agrees well with the data.

\figuremacroW{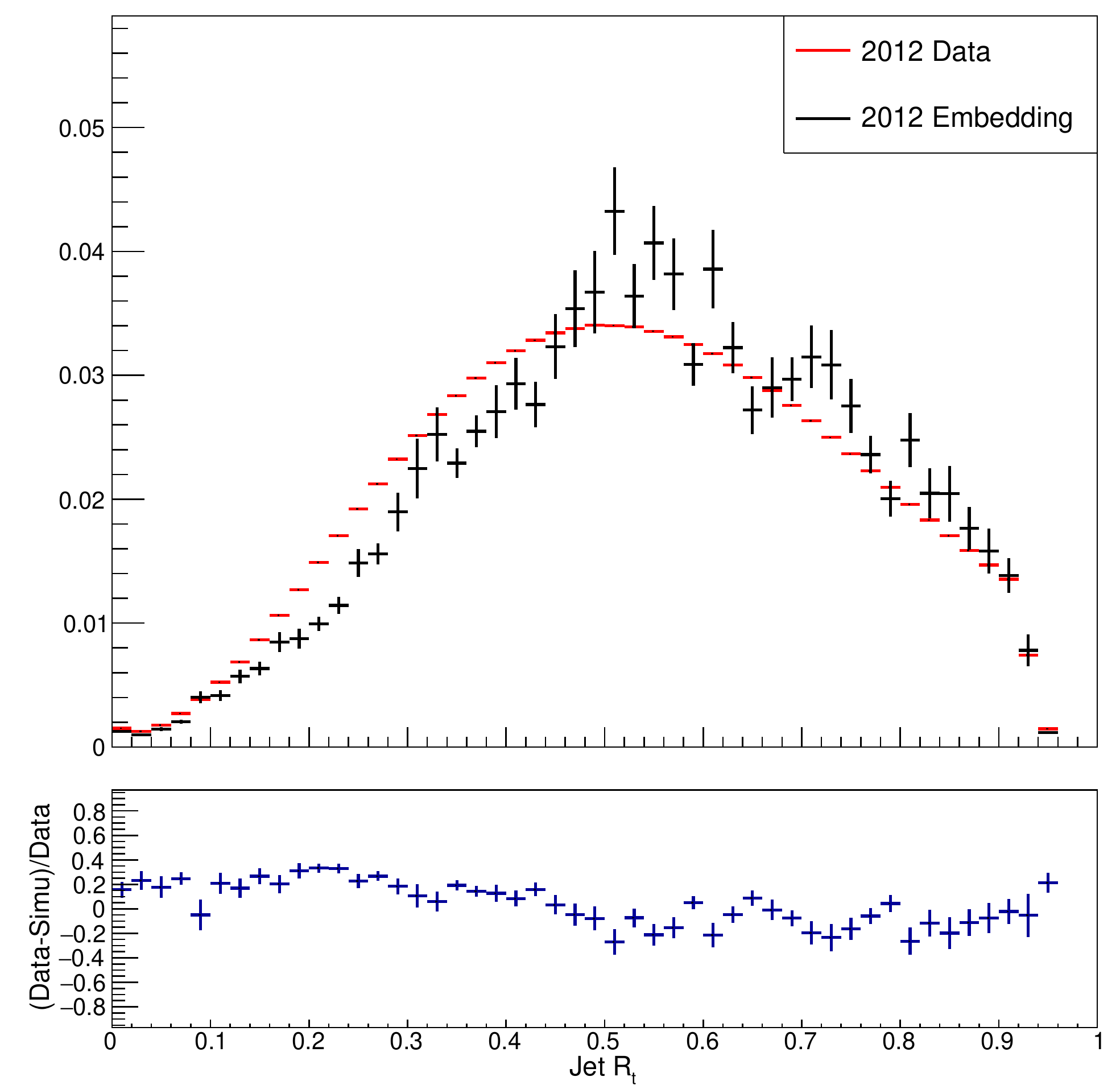}{Combined Triggers Jet $R_t$}{Comparing the jet $R_t$ distributions for the combined triggers in data and embedding.}{0.6}

\figuremacroW{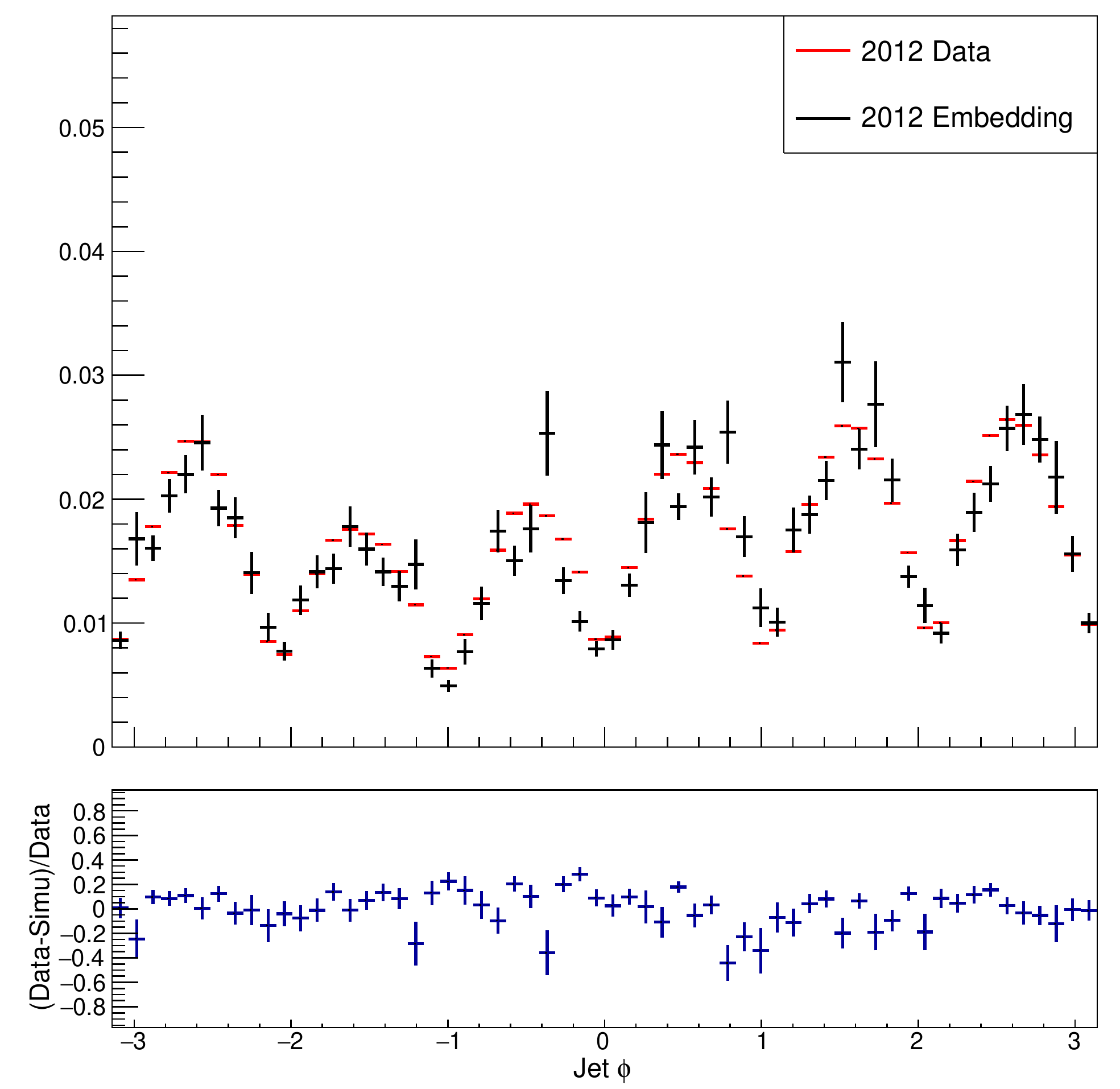}{Combined Triggers Jet $\phi$}{Comparing the jet $\phi$ distributions for the combined triggers in data and embedding.}{0.6}

\figuremacroW{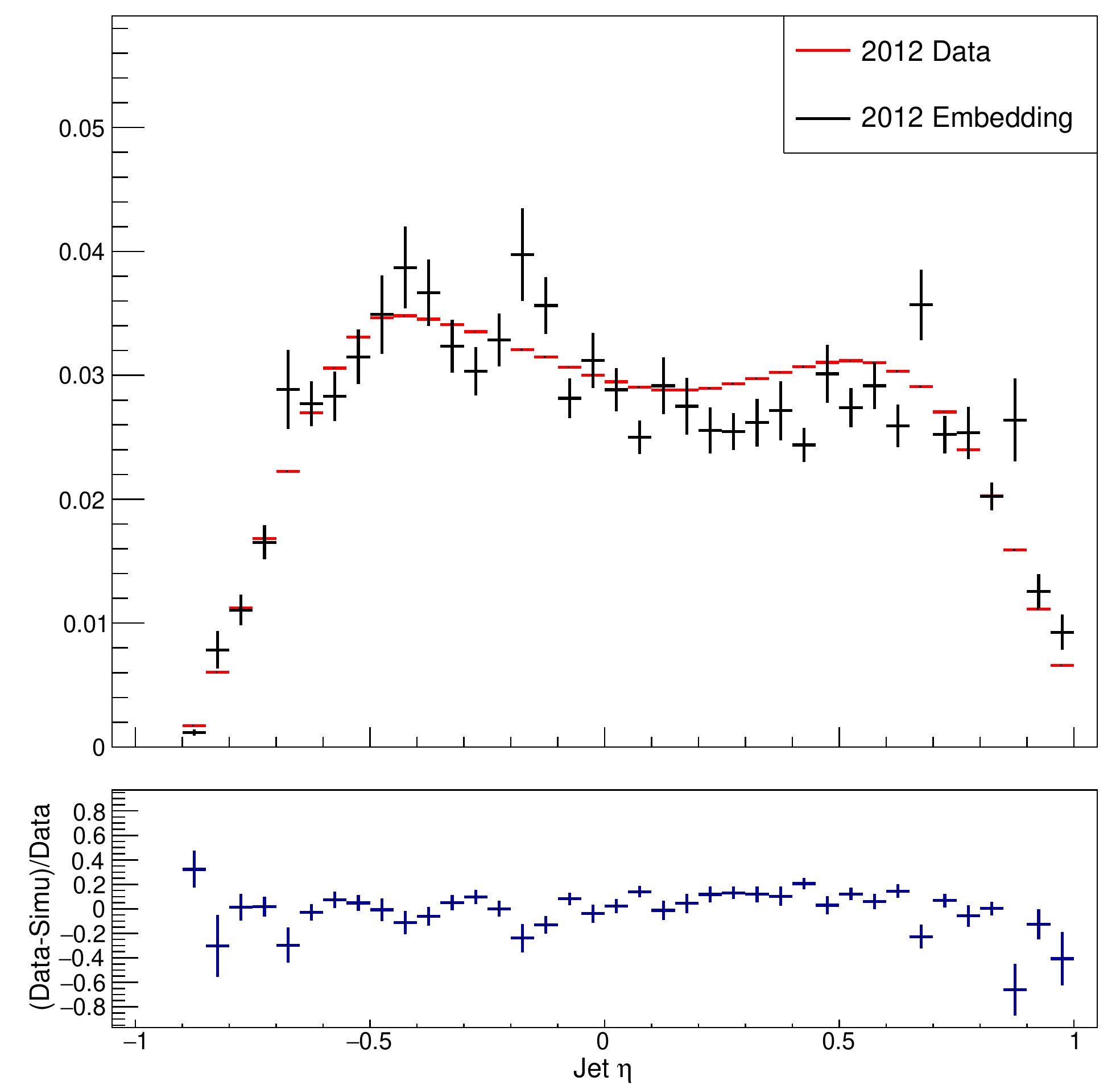}{Combined Triggers Jet $\eta$}{Comparing the jet $\eta$ distributions for the combined triggers in data and embedding.}{0.6}

\figuremacroW{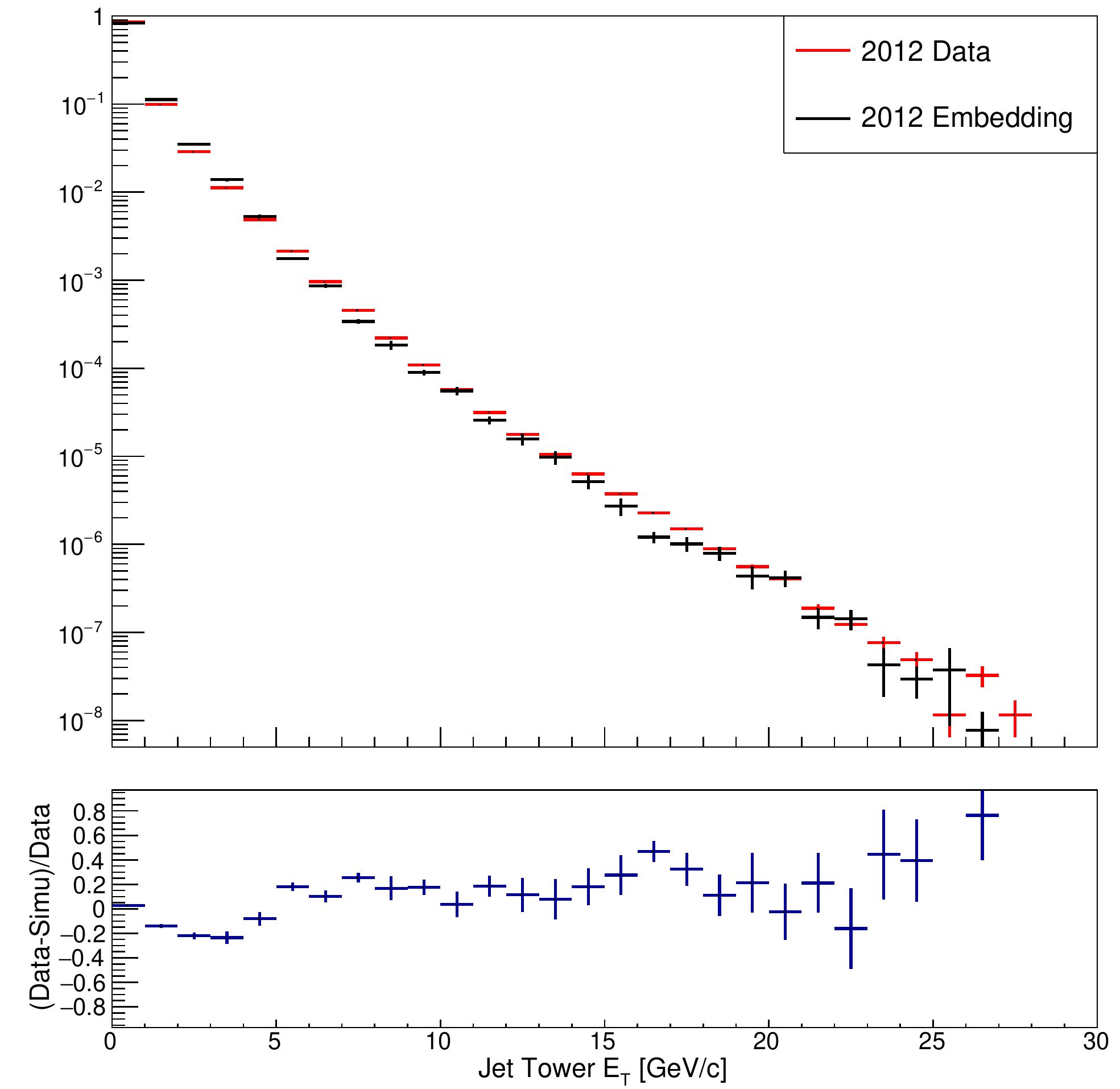}{Combined Triggers Tower $E_T$}{Comparing the tower $E_T$ distributions for the combined triggers in data and embedding.}{0.6}

\figuremacroW{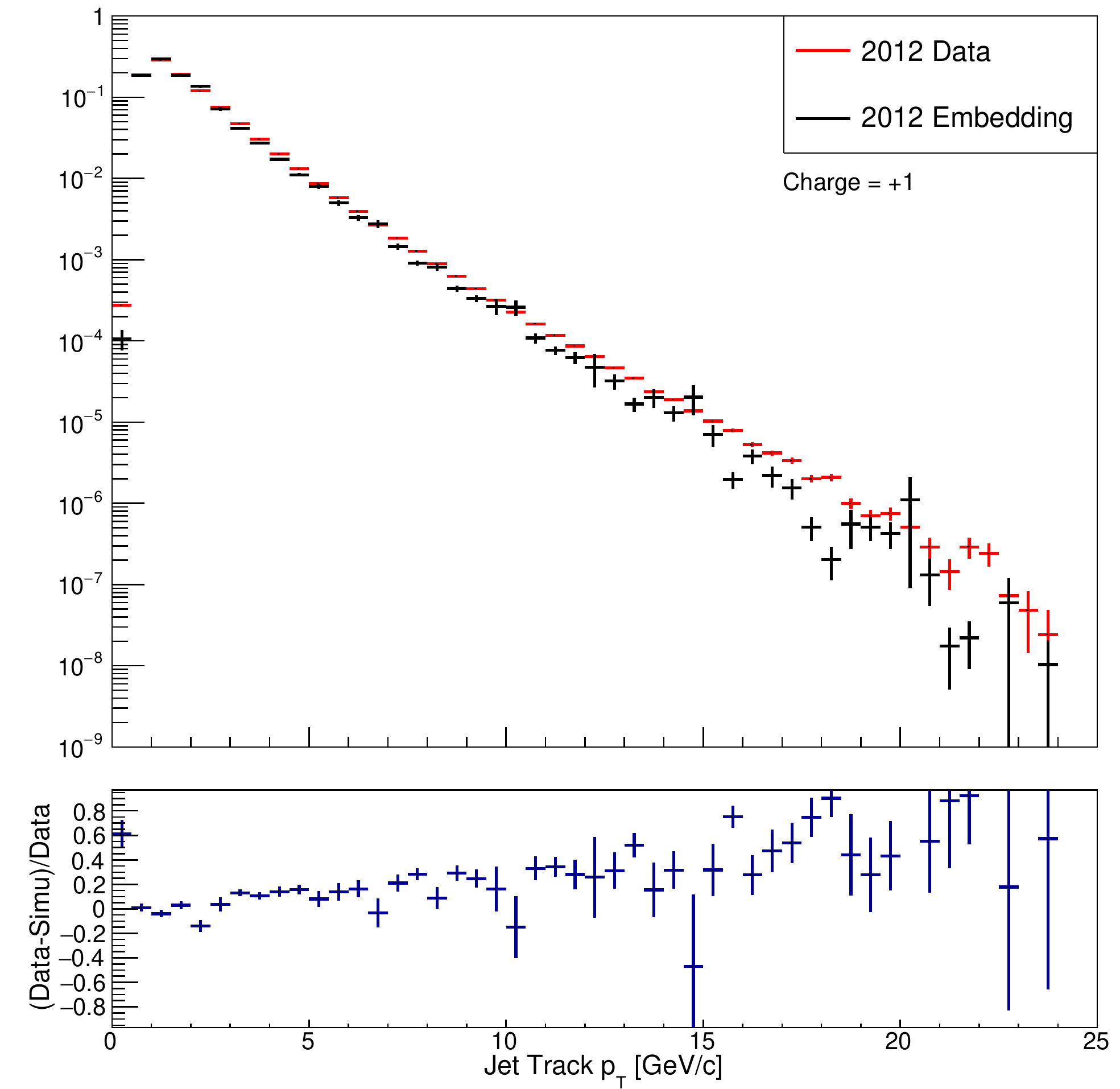}{Combined Triggers Track $p_T$}{Comparing the track $p_T$ distributions for the combined triggers in data and embedding.}{0.6}

\section{Tables of Results and Errors}
\subsection{Inclusive Jet Asymmetry}
The tables presented here are the final results that are presented in Section \ref{sec:InclusiveJetAsym}. There are several columns in these tables which give the data and statistical errors in addition to corrections applied to the data and their systematic errors:

\begin{itemize}
\item $p_{T,true}$ and $\delta p_{T,true}$ are the corrected value of the jet $p_T$ and its associated systematic error, respectively, that are used for plotting on the x-axis. The extraction of these corrected values is discussed in Section \ref{sec:kinShifts}. The calculation of the associated errors is discussed in detail in Section \ref{sec:kinShiftError}. Note that the bin edges for each bin are given in Table \ref{table:jetPtBins}.
\item $A_{UT,raw}^{\sin\left(\phi_S\right)}$ and $\delta A_{UT,raw}^{\sin\left(\phi_S\right)}$ denotes the raw value of the inclusive jet asymmetry and its statistical error, respectively. These values include no corrections.
\item The finite bin dilution correction value, which is 0.989 for all bins, is not shown since it is a constant value, but is corrected for. This correction is described in Section \ref{sec:finiteBins}. 
\item $A_{UT,final}^{\sin\left(\phi_S\right)}$ and $A_{UT,final}^{\sin\left(\phi_S\right)}$ are the values of the asymmetry and statistical error which have been corrected for finite bin width dilution. These are the values that appear on the plots.
\end{itemize}

\begin{table}[htbp]
\centering

\begin{tabular}{|c|c|c|c|c|c|}
\hline
$p_{T,true}$ & $\delta p_{T,true}$ & $A_{UT,raw}^{\sin\left(\phi_S-\phi_H\right)}$ & $\delta A_{UT,raw}^{\sin\left(\phi_S-\phi_H\right)}$ & $A_{UT,final}^{\sin\left(\phi_S-\phi_H\right)}$ & $\delta A_{UT,final}^{\sin\left(\phi_S-\phi_H\right)}$ \\
\hline
    5.90  & 0.32  & -1.07E-03 & 6.70E-04 & -1.08E-03 & 6.78E-04 \\
    \hline
    7.25  & 0.32  & -4.23E-04 & 7.23E-04 & -4.28E-04 & 7.31E-04 \\
    \hline
    8.94  & 0.35  & -2.27E-04 & 7.63E-04 & -2.29E-04 & 7.71E-04 \\
    \hline
    10.74 & 0.37  & -4.22E-05 & 9.14E-04 & -4.27E-05 & 9.24E-04 \\
    \hline
    12.83 & 0.40  & 1.57E-03 & 1.19E-03 & 1.59E-03 & 1.20E-03 \\
    \hline
    15.14 & 0.45  & 1.77E-05 & 1.63E-03 & 1.79E-05 & 1.65E-03 \\
    \hline
    17.78 & 0.50  & -3.54E-03 & 2.42E-03 & -3.58E-03 & 2.44E-03 \\
    \hline
    20.85 & 0.57  & 2.70E-03 & 3.74E-03 & 2.73E-03 & 3.78E-03 \\
    \hline
    24.50 & 0.64  & 4.25E-04 & 6.31E-03 & 4.30E-04 & 6.38E-03 \\
    \hline
    28.52 & 0.77  & -1.03E-02 & 1.13E-02 & -1.04E-02 & 1.15E-02 \\
    \hline
\end{tabular}%
\caption{Inclusive jet asymmetry results for $x_F > 0$. }

\vspace{20pt}

\begin{tabular}{|c|c|c|c|c|c|}
\hline
$p_{T,true}$ & $\delta p_{T,true}$ & $A_{UT,raw}^{\sin\left(\phi_S-\phi_H\right)}$ & $\delta A_{UT,raw}^{\sin\left(\phi_S-\phi_H\right)}$ & $A_{UT,final}^{\sin\left(\phi_S-\phi_H\right)}$ & $\delta A_{UT,final}^{\sin\left(\phi_S-\phi_H\right)}$ \\
\hline
    5.90  & 0.32  & -5.85E-04 & 6.79E-04 & -5.91E-04 & 6.87E-04 \\
    \hline
    7.25  & 0.32  & -2.84E-04 & 7.31E-04 & -2.87E-04 & 7.39E-04 \\
    \hline
    8.94  & 0.35  & -1.63E-03 & 7.71E-04 & -1.65E-03 & 7.79E-04 \\
    \hline
    10.74 & 0.37  & 1.86E-03 & 9.23E-04 & 1.88E-03 & 9.33E-04 \\
    \hline
    12.83 & 0.40  & -1.76E-03 & 1.20E-03 & -1.78E-03 & 1.21E-03 \\
    \hline
    15.14 & 0.45  & -1.75E-03 & 1.65E-03 & -1.77E-03 & 1.67E-03 \\
    \hline
    17.78 & 0.50  & -1.08E-03 & 2.44E-03 & -1.09E-03 & 2.47E-03 \\
    \hline
    20.85 & 0.57  & 5.42E-03 & 3.78E-03 & 5.48E-03 & 3.82E-03 \\
    \hline
    24.50 & 0.64  & 7.06E-03 & 6.37E-03 & 7.14E-03 & 6.44E-03 \\
    \hline
    28.52 & 0.77  & 5.06E-03 & 1.14E-02 & 5.12E-03 & 1.16E-02 \\
    \hline
\end{tabular}%
\caption{Inclusive jet asymmetry results for $x_F < 0$. }
  \end{table}

\newpage
\subsection{Collins Asymmetries}
The tables presented here are the final results that are presented in Section \ref{sec:CollinsAsym}. There are several columns in these tables which give the data and statistical errors in addition to corrections applied to the data and systematic errors:

\begin{itemize}
\item $p_{T,true}$, $z_{true}$ and $j_{T,true}$ are the corrected value of the kinematic variable that are used for plotting on the x-axis. The extraction of these corrected values is discussed in Section \ref{sec:kinShifts}. Note that the bin edges for each bin are given in Tables \ref{table:jetPtBins} - \ref{table:pionJtBins}.
\item $\delta p_{T,true}$, $\delta z_{true}$ and $\delta j_{T,true}$ are the errors in the corrected value of the kinematic variables and show up as errors along the x-axis. The calculation of these errors is discussed in detail in Section \ref{sec:kinShiftError}.
\item $A_{UT,raw}^{\sin\left(\phi_S-\phi_H\right)}$ and $\delta A_{UT,raw}^{\sin\left(\phi_S-\phi_H\right)}$ denote the raw value of the Collins asymmetry and its statistical error, respectively. These values include no corrections.
\item $A_{UT,\pi corr}^{\sin\left(\phi_S-\phi_H\right)}$ and $\delta A_{UT,\pi corr}^{\sin\left(\phi_S-\phi_H\right)}$ represent the values of the asymmetry and statistical error that have been corrected for contaminations to the pion sample. This correction is discussed in Section \ref{sec:pionCorrection}.
\item $D_{\phi_C}$ is the value of the dilution to the asymmetry due to the $\phi_C$ resolution. If this value is $1$, the statistics were not sufficient to calculate a dilution value, and thus the asymmetry for this bin is not plotted.
\item The finite bin dilution correction value, which is 0.989 for all bins, is not shown since it is a constant value, but is corrected for. This correction is described in Section \ref{sec:finiteBins}. 
\item $A_{UT,final}^{\sin\left(\phi_S-\phi_H\right)}$ and $\delta A_{UT,final}^{\sin\left(\phi_S-\phi_H\right)}$ are the asymmetry and statistical error values which are plotted. These values have been corrected for pion sample contamination, $\phi_C$ resolution, and finite bin width dilution.
\item ``TB syst.'' is the value of the trigger bias systematic error, applied as described in Section \ref{sec:triggerBias}. The sign of this error is significant since it is applied asymmetrically. If the sign of the error is negative, then the error is applied towards decreasing the magnitude of the asymmetry. If the sign is positive, then the error is applied towards increasing the magnitude of the asymmetry.
\end{itemize}

\begin{sidewaystable}[htbp]
\centering
\begin{tabular}{|c|c|c|c|c|c|c|c|c|c|}
\hline
$p_{T,true}$ & $\delta p_{T,true}$ & $A_{UT,raw}^{\sin\left(\phi_S-\phi_H\right)}$ & $\delta A_{UT,raw}^{\sin\left(\phi_S-\phi_H\right)}$& $A_{UT,\pi corr}^{\sin\left(\phi_S-\phi_H\right)}$ & $\delta A_{UT,\pi corr}^{\sin\left(\phi_S-\phi_H\right)}$ & $D_{\phi_C}$ & $A_{UT,final}^{\sin\left(\phi_S-\phi_H\right)}$ & $\delta A_{UT,final}^{\sin\left(\phi_S-\phi_H\right)}$ & TB syst. \\
\hline
    6.20  & 0.30  & 1.38E-03 & 9.23E-04 & 1.00E-03 & 1.56E-03 & 0.535 & 1.89E-03 & 2.94E-03 & -1.57E-04 \\
    \hline
    7.61  & 0.30  & 2.58E-03 & 9.63E-04 & 3.83E-03 & 1.62E-03 & 0.731 & 5.30E-03 & 2.24E-03 & -4.42E-04 \\
    \hline
    9.03  & 0.33  & 2.27E-03 & 1.02E-03 & 2.61E-03 & 1.59E-03 & 0.526 & 5.02E-03 & 3.05E-03 & -4.18E-04 \\
    \hline
    10.79 & 0.35  & 4.89E-03 & 1.19E-03 & 6.49E-03 & 1.69E-03 & 0.840 & 7.81E-03 & 2.04E-03 & -6.52E-04 \\
    \hline
    12.81 & 0.39  & 8.50E-03 & 1.54E-03 & 9.80E-03 & 1.97E-03 & 0.773 & 1.28E-02 & 2.58E-03 & -1.07E-03 \\
    \hline
    15.20 & 0.44  & 1.32E-02 & 2.11E-03 & 1.42E-02 & 2.42E-03 & 0.851 & 1.69E-02 & 2.88E-03 & -1.41E-03 \\
    \hline
    17.74 & 0.50  & 1.67E-02 & 3.14E-03 & 1.65E-02 & 3.46E-03 & 0.825 & 2.03E-02 & 4.25E-03 & -1.69E-03 \\
    \hline
    20.80 & 0.57  & 9.08E-03 & 4.94E-03 & 9.46E-03 & 5.14E-03 & 0.806 & 1.19E-02 & 6.45E-03 & -9.89E-04 \\
    \hline
    24.39 & 0.63  & 2.05E-02 & 8.57E-03 & 2.07E-02 & 9.50E-03 & 0.857 & 2.44E-02 & 1.12E-02 & -2.04E-03 \\
    \hline
    28.59 & 0.75  & 8.79E-03 & 1.59E-02 & 6.63E-03 & 1.83E-02 & 0.866 & 7.74E-03 & 2.13E-02 & -6.46E-04 \\
    \hline
    \end{tabular}%
\caption{$\pi^+$ Collins asymmetry for $x_F>0$ as a function of jet $p_T$ integrated over all $z$ and $j_T$.}

\vspace{20pt}

\begin{tabular}{|c|c|c|c|c|c|c|c|c|c|}
\hline
$p_{T,true}$ & $\delta p_{T,true}$ & $A_{UT,raw}^{\sin\left(\phi_S-\phi_H\right)}$ & $\delta A_{UT,raw}^{\sin\left(\phi_S-\phi_H\right)}$& $A_{UT,\pi corr}^{\sin\left(\phi_S-\phi_H\right)}$ & $\delta A_{UT,\pi corr}^{\sin\left(\phi_S-\phi_H\right)}$ & $D_{\phi_C}$ & $A_{UT,final}^{\sin\left(\phi_S-\phi_H\right)}$ & $\delta A_{UT,final}^{\sin\left(\phi_S-\phi_H\right)}$ & TB syst. \\
\hline
    6.20  & 0.30  & 2.62E-04 & 9.36E-04 & 1.72E-04 & 1.34E-03 & 0.695 & 2.50E-04 & 1.96E-03 & -2.09E-05 \\
    \hline
    7.61  & 0.30  & -2.57E-03 & 9.83E-04 & -3.92E-03 & 1.63E-03 & 0.769 & -5.15E-03 & 2.14E-03 & -4.30E-04 \\
    \hline
    9.03  & 0.33  & -2.13E-03 & 1.04E-03 & -2.26E-03 & 1.60E-03 & 0.678 & -3.38E-03 & 2.38E-03 & -2.82E-04 \\
    \hline
    10.79 & 0.35  & -5.26E-03 & 1.23E-03 & -5.93E-03 & 1.74E-03 & 0.670 & -8.94E-03 & 2.63E-03 & -7.46E-04 \\
    \hline
    12.81 & 0.39  & -7.30E-03 & 1.59E-03 & -7.70E-03 & 2.02E-03 & 0.784 & -9.93E-03 & 2.60E-03 & -8.28E-04 \\
    \hline
    15.20 & 0.44  & -7.52E-03 & 2.18E-03 & -6.41E-03 & 2.49E-03 & 0.785 & -8.26E-03 & 3.20E-03 & -6.89E-04 \\
    \hline
    17.74 & 0.50  & -2.03E-02 & 3.26E-03 & -2.12E-02 & 3.52E-03 & 0.861 & -2.49E-02 & 4.13E-03 & -2.08E-03 \\
    \hline
    20.80 & 0.57  & -1.93E-02 & 5.15E-03 & -2.01E-02 & 5.57E-03 & 0.828 & -2.46E-02 & 6.80E-03 & -2.05E-03 \\
    \hline
    24.39 & 0.63  & -3.79E-02 & 8.91E-03 & -4.07E-02 & 9.72E-03 & 0.867 & -4.74E-02 & 1.13E-02 & -3.95E-03 \\
    \hline
    28.59 & 0.75  & -4.12E-03 & 1.69E-02 & -3.91E-03 & 1.88E-02 & 0.857 & -4.61E-03 & 2.21E-02 & -3.84E-04 \\
    \hline
    \end{tabular}%
\caption{$\pi^-$ Collins asymmetry for $x_F>0$ as a function of jet $p_T$ integrated over all $z$ and $j_T$.}
  \end{sidewaystable}

\begin{sidewaystable}[htbp]
\centering
\begin{tabular}{|c|c|c|c|c|c|c|c|c|c|}
\hline
$p_{T,true}$ & $\delta p_{T,true}$ & $A_{UT,raw}^{\sin\left(\phi_S-\phi_H\right)}$ & $\delta A_{UT,raw}^{\sin\left(\phi_S-\phi_H\right)}$& $A_{UT,\pi corr}^{\sin\left(\phi_S-\phi_H\right)}$ & $\delta A_{UT,\pi corr}^{\sin\left(\phi_S-\phi_H\right)}$ & $D_{\phi_C}$ & $A_{UT,final}^{\sin\left(\phi_S-\phi_H\right)}$ & $\delta A_{UT,final}^{\sin\left(\phi_S-\phi_H\right)}$ & TB syst. \\
\hline
    6.20  & 0.30  & -1.33E-04 & 9.28E-04 & -2.56E-04 & 1.57E-03 & 0.535 & -4.84E-04 & 2.96E-03 & -4.04E-05 \\
    \hline
    7.61  & 0.30  & -2.70E-05 & 9.66E-04 & 1.07E-04 & 1.63E-03 & 0.731 & 1.49E-04 & 2.25E-03 & -1.24E-05 \\
    \hline
    9.03  & 0.33  & 3.76E-04 & 1.02E-03 & 4.23E-04 & 1.59E-03 & 0.526 & 8.14E-04 & 3.07E-03 & -6.79E-05 \\
    \hline
    10.79 & 0.35  & 1.48E-03 & 1.20E-03 & 2.23E-03 & 1.70E-03 & 0.840 & 2.68E-03 & 2.04E-03 & -2.24E-04 \\
    \hline
    12.81 & 0.39  & 1.93E-03 & 1.54E-03 & 3.01E-03 & 1.98E-03 & 0.773 & 3.93E-03 & 2.59E-03 & -3.28E-04 \\
    \hline
    15.20 & 0.44  & 3.25E-03 & 2.12E-03 & 4.01E-03 & 2.43E-03 & 0.851 & 4.77E-03 & 2.89E-03 & -3.97E-04 \\
    \hline
    17.74 & 0.50  & 1.52E-03 & 3.16E-03 & 1.52E-03 & 3.48E-03 & 0.825 & 1.86E-03 & 4.27E-03 & -1.55E-04 \\
    \hline
    20.80 & 0.57  & 9.86E-03 & 4.97E-03 & 9.84E-03 & 5.17E-03 & 0.806 & 1.23E-02 & 6.49E-03 & -1.03E-03 \\
    \hline
    24.39 & 0.63  & 1.43E-02 & 8.81E-03 & 1.67E-02 & 9.77E-03 & 0.857 & 1.97E-02 & 1.15E-02 & -1.64E-03 \\
    \hline
    28.59 & 0.75  & -6.97E-03 & 1.60E-02 & -7.41E-03 & 1.83E-02 & 0.866 & -8.65E-03 & 2.14E-02 & -7.21E-04 \\
\hline
\end{tabular}%
\caption{$\pi^+$ Collins asymmetry for $x_F<0$ as a function of jet $p_T$ integrated over all $z$ and $j_T$.}

\vspace{20pt}

\begin{tabular}{|c|c|c|c|c|c|c|c|c|c|}
\hline
$p_{T,true}$ & $\delta p_{T,true}$ & $A_{UT,raw}^{\sin\left(\phi_S-\phi_H\right)}$ & $\delta A_{UT,raw}^{\sin\left(\phi_S-\phi_H\right)}$& $A_{UT,\pi corr}^{\sin\left(\phi_S-\phi_H\right)}$ & $\delta A_{UT,\pi corr}^{\sin\left(\phi_S-\phi_H\right)}$ & $D_{\phi_C}$ & $A_{UT,final}^{\sin\left(\phi_S-\phi_H\right)}$ & $\delta A_{UT,final}^{\sin\left(\phi_S-\phi_H\right)}$ & TB syst. \\
\hline
    6.20  & 0.30  & -1.58E-03 & 9.40E-04 & -2.01E-03 & 1.35E-03 & 0.695 & -2.92E-03 & 1.97E-03 & -2.44E-04 \\
    \hline
    7.61  & 0.30  & 1.51E-03 & 9.85E-04 & 3.04E-03 & 1.64E-03 & 0.769 & 3.99E-03 & 2.15E-03 & -3.33E-04 \\
    \hline
    9.03  & 0.33  & 3.12E-04 & 1.05E-03 & 8.03E-04 & 1.60E-03 & 0.678 & 1.20E-03 & 2.39E-03 & -1.00E-04 \\
    \hline
    10.79 & 0.35  & -1.23E-03 & 1.23E-03 & -1.69E-03 & 1.75E-03 & 0.670 & -2.55E-03 & 2.64E-03 & -2.12E-04 \\
    \hline
    12.81 & 0.39  & 7.86E-04 & 1.59E-03 & 1.66E-03 & 2.03E-03 & 0.784 & 2.14E-03 & 2.61E-03 & -1.78E-04 \\
    \hline
    15.20 & 0.44  & -2.23E-03 & 2.19E-03 & -2.56E-03 & 2.49E-03 & 0.785 & -3.30E-03 & 3.21E-03 & -2.75E-04 \\
    \hline
    17.74 & 0.50  & -1.27E-03 & 3.27E-03 & -8.28E-04 & 3.54E-03 & 0.861 & -9.72E-04 & 4.15E-03 & -8.10E-05 \\
    \hline
    20.80 & 0.57  & -5.18E-03 & 5.18E-03 & -4.30E-03 & 5.60E-03 & 0.828 & -5.26E-03 & 6.84E-03 & -4.38E-04 \\
    \hline
    24.39 & 0.63  & -8.47E-03 & 8.98E-03 & -7.91E-03 & 9.79E-03 & 0.867 & -9.22E-03 & 1.14E-02 & -7.69E-04 \\
    \hline
    28.59 & 0.75  & 2.15E-02 & 1.69E-02 & 2.54E-02 & 1.87E-02 & 0.857 & 2.99E-02 & 2.21E-02 & -2.49E-03 \\
\hline
\end{tabular}%
\caption{$\pi^-$ Collins asymmetry for $x_F<0$ as a function of jet $p_T$ integrated over all $z$ and $j_T$.}
  \end{sidewaystable}
  \clearpage
\begin{sidewaystable}[htbp]
\centering
\begin{tabular}{|c|c|c|c|c|c|c|c|c|c|}
\hline
$z_{true}$ & $\delta z_{true}$ & $A_{UT,raw}^{\sin\left(\phi_S-\phi_H\right)}$ & $\delta A_{UT,raw}^{\sin\left(\phi_S-\phi_H\right)}$& $A_{UT,\pi corr}^{\sin\left(\phi_S-\phi_H\right)}$ & $\delta A_{UT,\pi corr}^{\sin\left(\phi_S-\phi_H\right)}$ & $D_{\phi_C}$ & $A_{UT,final}^{\sin\left(\phi_S-\phi_H\right)}$ & $\delta A_{UT,final}^{\sin\left(\phi_S-\phi_H\right)}$ & TB syst. \\
\hline
    0.16  & 0.01  & 1.49E-03 & 8.27E-04 & 5.52E-04 & 1.87E-03 & 0.646 & 8.65E-04 & 2.92E-03 & 1.18E-05 \\
    \hline
    0.28  & 0.01  & 2.74E-03 & 1.41E-03 & 3.67E-03 & 2.47E-03 & 0.794 & 4.67E-03 & 3.15E-03 & 6.36E-05 \\
    \hline
    0.37  & 0.02  & 2.94E-03 & 2.39E-03 & 3.13E-03 & 2.62E-03 & 0.606 & 5.21E-03 & 4.37E-03 & 7.09E-05 \\
    \hline
    0.48  & 0.02  & 4.21E-03 & 3.24E-03 & 4.77E-03 & 3.55E-03 & 0.656 & 7.35E-03 & 5.47E-03 & 9.99E-05 \\
    \hline
    0.67  & 0.03  & -3.54E-03 & 8.66E-03 & -2.37E-03 & 9.04E-03 & 0.616 & -3.89E-03 & 1.48E-02 & 5.29E-05 \\
    \hline
    \end{tabular}%
\caption{$\pi^+$ Collins asymmetry for $x_F>0$ as a function of $z$ for $\langle p_T\rangle=6.9$ GeV/c and integrated over all $j_T$.}

\vspace{20pt}

\begin{tabular}{|c|c|c|c|c|c|c|c|c|c|}
\hline
$z_{true}$ & $\delta z_{true}$ & $A_{UT,raw}^{\sin\left(\phi_S-\phi_H\right)}$ & $\delta A_{UT,raw}^{\sin\left(\phi_S-\phi_H\right)}$& $A_{UT,\pi corr}^{\sin\left(\phi_S-\phi_H\right)}$ & $\delta A_{UT,\pi corr}^{\sin\left(\phi_S-\phi_H\right)}$ & $D_{\phi_C}$ & $A_{UT,final}^{\sin\left(\phi_S-\phi_H\right)}$ & $\delta A_{UT,final}^{\sin\left(\phi_S-\phi_H\right)}$ & TB syst. \\
\hline
    0.16  & 0.01  & 7.16E-04 & 1.25E-03 & 6.18E-03 & 5.46E-03 & 0.729 & 8.57E-03 & 7.56E-03 & 2.30E-04 \\
    \hline
    0.25  & 0.01  & 5.89E-03 & 2.21E-03 & 5.75E-03 & 2.46E-03 & 0.834 & 6.97E-03 & 2.99E-03 & 1.87E-04 \\
    \hline
    0.36  & 0.01  & 3.90E-03 & 3.64E-03 & 4.25E-03 & 3.90E-03 & 0.864 & 4.98E-03 & 4.56E-03 & 1.33E-04 \\
    \hline
    0.49  & 0.02  & 7.59E-03 & 4.96E-03 & 7.49E-03 & 5.67E-03 & 0.731 & 1.04E-02 & 7.84E-03 & 2.78E-04 \\
    \hline
    0.65  & 0.03  & -1.37E-02 & 1.41E-02 & -1.53E-02 & 1.53E-02 & 0.628 & -2.47E-02 & 2.46E-02 & 6.62E-04 \\
    \hline
    \end{tabular}%
\caption{$\pi^+$ Collins asymmetry for $x_F>0$ as a function of $z$ for $\langle p_T\rangle=9.0$ GeV/c and integrated over all $j_T$.}

\vspace{20pt}
\begin{tabular}{|c|c|c|c|c|c|c|c|c|c|}
\hline
$z_{true}$ & $\delta z_{true}$ & $A_{UT,raw}^{\sin\left(\phi_S-\phi_H\right)}$ & $\delta A_{UT,raw}^{\sin\left(\phi_S-\phi_H\right)}$& $A_{UT,\pi corr}^{\sin\left(\phi_S-\phi_H\right)}$ & $\delta A_{UT,\pi corr}^{\sin\left(\phi_S-\phi_H\right)}$ & $D_{\phi_C}$ & $A_{UT,final}^{\sin\left(\phi_S-\phi_H\right)}$ & $\delta A_{UT,final}^{\sin\left(\phi_S-\phi_H\right)}$ & TB syst. \\
\hline
    0.15  & 0.01  & 4.77E-03 & 1.47E-03 & 8.07E-03 & 3.48E-03 & 0.867 & 9.41E-03 & 4.06E-03 & 1.85E-04 \\
    \hline
    0.24  & 0.01  & 4.59E-03 & 2.59E-03 & 4.86E-03 & 2.79E-03 & 0.756 & 6.50E-03 & 3.73E-03 & 1.27E-04 \\
    \hline
    0.35  & 0.01  & 4.07E-03 & 4.22E-03 & 4.85E-03 & 4.69E-03 & 0.648 & 7.56E-03 & 7.32E-03 & 1.48E-04 \\
    \hline
    0.50  & 0.02  & 6.44E-03 & 5.76E-03 & 8.12E-03 & 6.64E-03 & 0.818 & 1.00E-02 & 8.21E-03 & 1.97E-04 \\
    \hline
    0.70  & 0.03  & 3.79E-02 & 1.68E-02 & 4.04E-02 & 1.88E-02 & 0.709 & 5.76E-02 & 2.68E-02 & 1.13E-03 \\
    \hline
    \end{tabular}%
\caption{$\pi^+$ Collins asymmetry for $x_F>0$ as a function of $z$ for $\langle p_T\rangle=10.8$ GeV/c and integrated over all $j_T$.}
  \end{sidewaystable}

\begin{sidewaystable}[htbp]
\centering
\begin{tabular}{|c|c|c|c|c|c|c|c|c|c|}
\hline
$z_{true}$ & $\delta z_{true}$ & $A_{UT,raw}^{\sin\left(\phi_S-\phi_H\right)}$ & $\delta A_{UT,raw}^{\sin\left(\phi_S-\phi_H\right)}$& $A_{UT,\pi corr}^{\sin\left(\phi_S-\phi_H\right)}$ & $\delta A_{UT,\pi corr}^{\sin\left(\phi_S-\phi_H\right)}$ & $D_{\phi_C}$ & $A_{UT,final}^{\sin\left(\phi_S-\phi_H\right)}$ & $\delta A_{UT,final}^{\sin\left(\phi_S-\phi_H\right)}$ & TB syst. \\
\hline
    0.14  & 0.01  & 7.59E-03 & 1.54E-03 & 8.87E-03 & 2.17E-03 & 0.812 & 1.10E-02 & 2.70E-03 & -1.06E-04 \\
    \hline
    0.24  & 0.01  & 1.56E-02 & 2.67E-03 & 1.64E-02 & 2.84E-03 & 0.706 & 2.35E-02 & 4.07E-03 & -2.25E-04 \\
    \hline
    0.35  & 0.01  & 1.15E-02 & 4.33E-03 & 1.21E-02 & 4.90E-03 & 0.743 & 1.65E-02 & 6.67E-03 & -1.58E-04 \\
    \hline
    0.48  & 0.02  & 1.85E-02 & 5.85E-03 & 2.05E-02 & 6.96E-03 & 0.691 & 3.01E-02 & 1.02E-02 & -2.89E-04 \\
    \hline
    0.69  & 0.02  & 8.93E-03 & 1.76E-02 & 9.52E-03 & 2.07E-02 & 0.601 & 1.60E-02 & 3.47E-02 & -1.54E-04 \\
    \hline
    \end{tabular}%
\caption{$\pi^+$ Collins asymmetry for $x_F>0$ as a function of $z$ for $\langle p_T\rangle=13.6$ GeV/c and integrated over all $j_T$.}

\vspace{20pt}

\begin{tabular}{|c|c|c|c|c|c|c|c|c|c|}
\hline
$z_{true}$ & $\delta z_{true}$ & $A_{UT,raw}^{\sin\left(\phi_S-\phi_H\right)}$ & $\delta A_{UT,raw}^{\sin\left(\phi_S-\phi_H\right)}$& $A_{UT,\pi corr}^{\sin\left(\phi_S-\phi_H\right)}$ & $\delta A_{UT,\pi corr}^{\sin\left(\phi_S-\phi_H\right)}$ & $D_{\phi_C}$ & $A_{UT,final}^{\sin\left(\phi_S-\phi_H\right)}$ & $\delta A_{UT,final}^{\sin\left(\phi_S-\phi_H\right)}$ & TB syst. \\
\hline
    0.14  & 0.01  & 1.25E-02 & 3.91E-03 & 1.22E-02 & 4.21E-03 & 0.848 & 1.45E-02 & 5.02E-03 & -3.27E-04 \\
    \hline
    0.25  & 0.01  & 2.88E-02 & 6.72E-03 & 2.86E-02 & 7.43E-03 & 0.800 & 3.61E-02 & 9.39E-03 & -8.13E-04 \\
    \hline
    0.35  & 0.01  & 1.11E-02 & 1.09E-02 & 1.06E-02 & 1.17E-02 & 0.616 & 1.74E-02 & 1.92E-02 & -3.91E-04 \\
    \hline
    0.48  & 0.02  & 1.92E-02 & 1.45E-02 & 2.10E-02 & 1.63E-02 & 0.733 & 2.90E-02 & 2.25E-02 & -6.52E-04 \\
    \hline
    0.69  & 0.03  & 1.02E-01 & 4.39E-02 & 1.15E-01 & 4.83E-02 & 0.819 & 1.42E-01 & 5.96E-02 & -3.19E-03 \\
    \hline
    \end{tabular}%
\caption{$\pi^+$ Collins asymmetry for $x_F>0$ as a function of $z$ for $\langle p_T\rangle=17.7$ GeV/c and integrated over all $j_T$.}

\vspace{20pt}
\begin{tabular}{|c|c|c|c|c|c|c|c|c|c|}
\hline
$z_{true}$ & $\delta z_{true}$ & $A_{UT,raw}^{\sin\left(\phi_S-\phi_H\right)}$ & $\delta A_{UT,raw}^{\sin\left(\phi_S-\phi_H\right)}$& $A_{UT,\pi corr}^{\sin\left(\phi_S-\phi_H\right)}$ & $\delta A_{UT,\pi corr}^{\sin\left(\phi_S-\phi_H\right)}$ & $D_{\phi_C}$ & $A_{UT,final}^{\sin\left(\phi_S-\phi_H\right)}$ & $\delta A_{UT,final}^{\sin\left(\phi_S-\phi_H\right)}$ & TB syst. \\
\hline
    0.14  & 0.01  & 1.11E-02 & 5.13E-03 & 1.15E-02 & 5.40E-03 & 0.888 & 1.31E-02 & 6.15E-03 & -2.62E-05 \\
    \hline
    0.25  & 0.01  & 4.85E-03 & 8.89E-03 & 3.47E-03 & 9.93E-03 & 0.829 & 4.23E-03 & 1.21E-02 & -8.45E-06 \\
    \hline
    0.35  & 0.01  & 1.59E-02 & 1.44E-02 & 1.71E-02 & 1.54E-02 & 0.876 & 1.98E-02 & 1.78E-02 & -3.95E-05 \\
    \hline
    0.48  & 0.02  & 3.37E-02 & 1.91E-02 & 4.36E-02 & 2.35E-02 & 0.871 & 5.06E-02 & 2.73E-02 & -1.01E-04 \\
    \hline
    0.65  & 0.02  & 1.28E-01 & 5.80E-02 & 1.27E-01 & 6.64E-02 & 0.705 & 1.82E-01 & 9.53E-02 & -3.65E-04 \\
    \hline
    \end{tabular}%
\caption{$\pi^+$ Collins asymmetry for $x_F>0$ as a function of $z$ for $\langle p_T\rangle=22.1$ GeV/c and integrated over all $j_T$.}
  \end{sidewaystable}

\begin{sidewaystable}[htbp]
\centering
\begin{tabular}{|c|c|c|c|c|c|c|c|c|c|}
\hline
$z_{true}$ & $\delta z_{true}$ & $A_{UT,raw}^{\sin\left(\phi_S-\phi_H\right)}$ & $\delta A_{UT,raw}^{\sin\left(\phi_S-\phi_H\right)}$& $A_{UT,\pi corr}^{\sin\left(\phi_S-\phi_H\right)}$ & $\delta A_{UT,\pi corr}^{\sin\left(\phi_S-\phi_H\right)}$ & $D_{\phi_C}$ & $A_{UT,final}^{\sin\left(\phi_S-\phi_H\right)}$ & $\delta A_{UT,final}^{\sin\left(\phi_S-\phi_H\right)}$ & TB syst. \\
\hline
    0.16  & 0.01  & -6.15E-04 & 8.45E-04 & -1.15E-03 & 1.12E-03 & 0.796 & -1.47E-03 & 1.42E-03 & 1.99E-05 \\
    \hline
    0.28  & 0.01  & -1.24E-03 & 1.40E-03 & -2.29E-03 & 2.52E-03 & 0.617 & -3.75E-03 & 4.13E-03 & 5.11E-05 \\
    \hline
    0.37  & 0.02  & -5.89E-03 & 2.44E-03 & -6.55E-03 & 2.76E-03 & 0.548 & -1.21E-02 & 5.10E-03 & 1.65E-04 \\
    \hline
    0.48  & 0.02  & 1.93E-03 & 3.40E-03 & 2.89E-03 & 3.85E-03 & 0.682 & 4.29E-03 & 5.71E-03 & 5.83E-05 \\
    \hline
    0.67  & 0.03  & -6.90E-03 & 9.31E-03 & -7.31E-03 & 9.70E-03 & 0.837 & -8.84E-03 & 1.17E-02 & 1.20E-04 \\
    \hline
    \end{tabular}%
\caption{$\pi^-$ Collins asymmetry for $x_F>0$ as a function of $z$ for $\langle p_T\rangle=6.9$ GeV/c and integrated over all $j_T$.}

\vspace{20pt}

\begin{tabular}{|c|c|c|c|c|c|c|c|c|c|}
\hline
$z_{true}$ & $\delta z_{true}$ & $A_{UT,raw}^{\sin\left(\phi_S-\phi_H\right)}$ & $\delta A_{UT,raw}^{\sin\left(\phi_S-\phi_H\right)}$& $A_{UT,\pi corr}^{\sin\left(\phi_S-\phi_H\right)}$ & $\delta A_{UT,\pi corr}^{\sin\left(\phi_S-\phi_H\right)}$ & $D_{\phi_C}$ & $A_{UT,final}^{\sin\left(\phi_S-\phi_H\right)}$ & $\delta A_{UT,final}^{\sin\left(\phi_S-\phi_H\right)}$ & TB syst. \\
\hline
    0.16  & 0.01  & -2.55E-03 & 1.28E-03 & -6.39E-03 & 5.02E-03 & 0.745 & -8.68E-03 & 6.82E-03 & 2.33E-04 \\
    \hline
    0.25  & 0.01  & -8.10E-04 & 2.26E-03 & -3.94E-04 & 2.55E-03 & 0.713 & -5.58E-04 & 3.61E-03 & 1.50E-05 \\
    \hline
    0.36  & 0.01  & -3.84E-03 & 3.81E-03 & -3.72E-03 & 4.12E-03 & 0.703 & -5.36E-03 & 5.93E-03 & 1.44E-04 \\
    \hline
    0.49  & 0.02  & 1.61E-03 & 5.32E-03 & 1.38E-03 & 6.15E-03 & 0.644 & 2.17E-03 & 9.66E-03 & 5.82E-05 \\
    \hline
    0.65  & 0.03  & -4.81E-03 & 1.55E-02 & -5.86E-03 & 1.71E-02 & 0.775 & -7.64E-03 & 2.23E-02 & 2.05E-04 \\
    \hline
    \end{tabular}%
\caption{$\pi^-$ Collins asymmetry for $x_F>0$ as a function of $z$ for $\langle p_T\rangle=9.0$ GeV/c and integrated over all $j_T$.}

\vspace{20pt}
\begin{tabular}{|c|c|c|c|c|c|c|c|c|c|}
\hline
$z_{true}$ & $\delta z_{true}$ & $A_{UT,raw}^{\sin\left(\phi_S-\phi_H\right)}$ & $\delta A_{UT,raw}^{\sin\left(\phi_S-\phi_H\right)}$& $A_{UT,\pi corr}^{\sin\left(\phi_S-\phi_H\right)}$ & $\delta A_{UT,\pi corr}^{\sin\left(\phi_S-\phi_H\right)}$ & $D_{\phi_C}$ & $A_{UT,final}^{\sin\left(\phi_S-\phi_H\right)}$ & $\delta A_{UT,final}^{\sin\left(\phi_S-\phi_H\right)}$ & TB syst. \\
\hline
    0.15  & 0.01  & -3.92E-03 & 1.51E-03 & -3.71E-03 & 3.59E-03 & 0.792 & -4.74E-03 & 4.58E-03 & 9.29E-05 \\
    \hline
    0.24  & 0.01  & -9.80E-03 & 2.68E-03 & -1.04E-02 & 2.89E-03 & 0.591 & -1.77E-02 & 4.94E-03 & 3.48E-04 \\
    \hline
    0.35  & 0.01  & -1.67E-03 & 4.44E-03 & -1.83E-03 & 4.97E-03 & 0.880 & -2.10E-03 & 5.71E-03 & 4.11E-05 \\
    \hline
    0.50  & 0.02  & -9.09E-03 & 6.19E-03 & -9.50E-03 & 7.23E-03 & 0.641 & -1.50E-02 & 1.14E-02 & 2.93E-04 \\
    \hline
    0.70  & 0.03  & -1.38E-02 & 1.88E-02 & -1.64E-02 & 2.11E-02 & 0.978 & -1.69E-02 & 2.18E-02 & 3.31E-04 \\
    \hline
    \end{tabular}%
\caption{$\pi^-$ Collins asymmetry for $x_F>0$ as a function of $z$ for $\langle p_T\rangle=10.8$ GeV/c and integrated over all $j_T$.}
  \end{sidewaystable}

\begin{sidewaystable}[htbp]
\centering
\begin{tabular}{|c|c|c|c|c|c|c|c|c|c|}
\hline
$z_{true}$ & $\delta z_{true}$ & $A_{UT,raw}^{\sin\left(\phi_S-\phi_H\right)}$ & $\delta A_{UT,raw}^{\sin\left(\phi_S-\phi_H\right)}$& $A_{UT,\pi corr}^{\sin\left(\phi_S-\phi_H\right)}$ & $\delta A_{UT,\pi corr}^{\sin\left(\phi_S-\phi_H\right)}$ & $D_{\phi_C}$ & $A_{UT,final}^{\sin\left(\phi_S-\phi_H\right)}$ & $\delta A_{UT,final}^{\sin\left(\phi_S-\phi_H\right)}$ & TB syst. \\
\hline
    0.14  & 0.01  & -3.35E-03 & 1.58E-03 & -1.94E-03 & 2.18E-03 & 0.831 & -2.36E-03 & 2.65E-03 & -2.26E-05 \\
    \hline
    0.24  & 0.01  & -1.04E-02 & 2.76E-03 & -1.06E-02 & 2.95E-03 & 0.649 & -1.66E-02 & 4.59E-03 & -1.59E-04 \\
    \hline
    0.35  & 0.01  & -1.95E-02 & 4.59E-03 & -1.96E-02 & 5.21E-03 & 0.775 & -2.56E-02 & 6.80E-03 & -2.46E-04 \\
    \hline
    0.48  & 0.02  & -3.10E-02 & 6.41E-03 & -3.46E-02 & 7.66E-03 & 0.658 & -5.31E-02 & 1.18E-02 & -5.10E-04 \\
    \hline
    0.69  & 0.02  & -3.52E-02 & 2.04E-02 & -3.52E-02 & 2.37E-02 & 0.840 & -4.24E-02 & 2.85E-02 & -4.07E-04 \\
    \hline
    \end{tabular}%
\caption{$\pi^-$ Collins asymmetry for $x_F>0$ as a function of $z$ for $\langle p_T\rangle=13.6$ GeV/c and integrated over all $j_T$.}

\vspace{20pt}

\begin{tabular}{|c|c|c|c|c|c|c|c|c|c|}
\hline
$z_{true}$ & $\delta z_{true}$ & $A_{UT,raw}^{\sin\left(\phi_S-\phi_H\right)}$ & $\delta A_{UT,raw}^{\sin\left(\phi_S-\phi_H\right)}$& $A_{UT,\pi corr}^{\sin\left(\phi_S-\phi_H\right)}$ & $\delta A_{UT,\pi corr}^{\sin\left(\phi_S-\phi_H\right)}$ & $D_{\phi_C}$ & $A_{UT,final}^{\sin\left(\phi_S-\phi_H\right)}$ & $\delta A_{UT,final}^{\sin\left(\phi_S-\phi_H\right)}$ & TB syst. \\
\hline
    0.14  & 0.01  & -1.75E-02 & 4.01E-03 & -1.78E-02 & 4.30E-03 & 0.912 & -1.98E-02 & 4.77E-03 & -4.44E-04 \\
    \hline
    0.25  & 0.01  & -2.59E-02 & 7.01E-03 & -2.76E-02 & 7.78E-03 & 0.762 & -3.67E-02 & 1.03E-02 & -8.25E-04 \\
    \hline
    0.35  & 0.01  & -1.40E-02 & 1.17E-02 & -1.54E-02 & 1.21E-02 & 0.844 & -1.85E-02 & 1.45E-02 & -4.16E-04 \\
    \hline
    0.48  & 0.02  & -4.06E-02 & 1.63E-02 & -4.46E-02 & 1.95E-02 & 0.714 & -6.32E-02 & 2.76E-02 & -1.42E-03 \\
    \hline
    0.69  & 0.03  & -1.16E-01 & 5.31E-02 & -1.35E-01 & 7.13E-02 & 0.918 & -1.49E-01 & 7.86E-02 & -3.35E-03 \\
    \hline
    \end{tabular}%
\caption{$\pi^-$ Collins asymmetry for $x_F>0$ as a function of $z$ for $\langle p_T\rangle=17.7$ GeV/c and integrated over all $j_T$.}

\vspace{20pt}
\begin{tabular}{|c|c|c|c|c|c|c|c|c|c|}
\hline
$z_{true}$ & $\delta z_{true}$ & $A_{UT,raw}^{\sin\left(\phi_S-\phi_H\right)}$ & $\delta A_{UT,raw}^{\sin\left(\phi_S-\phi_H\right)}$& $A_{UT,\pi corr}^{\sin\left(\phi_S-\phi_H\right)}$ & $\delta A_{UT,\pi corr}^{\sin\left(\phi_S-\phi_H\right)}$ & $D_{\phi_C}$ & $A_{UT,final}^{\sin\left(\phi_S-\phi_H\right)}$ & $\delta A_{UT,final}^{\sin\left(\phi_S-\phi_H\right)}$ & TB syst. \\
\hline
    0.14  & 0.01  & -1.31E-02 & 5.26E-03 & -1.32E-02 & 5.49E-03 & 0.871 & -1.53E-02 & 6.37E-03 & -3.06E-05 \\
    \hline
    0.25  & 0.01  & -4.09E-02 & 9.39E-03 & -4.23E-02 & 9.83E-03 & 0.910 & -4.70E-02 & 1.09E-02 & -9.40E-05 \\
    \hline
    0.35  & 0.01  & -3.43E-02 & 1.57E-02 & -4.52E-02 & 1.86E-02 & 0.789 & -5.79E-02 & 2.38E-02 & -1.16E-04 \\
    \hline
    0.48  & 0.02  & -6.62E-02 & 2.22E-02 & -7.01E-02 & 2.40E-02 & 0.862 & -8.22E-02 & 2.82E-02 & -1.64E-04 \\
    \hline
    0.65  & 0.02  & -5.82E-02 & 7.94E-02 & -1.11E-01 & 1.09E-01 & 0.626 & -1.80E-01 & 1.77E-01 & -3.60E-04 \\
    \hline
    \end{tabular}%
\caption{$\pi^-$ Collins asymmetry for $x_F>0$ as a function of $z$ for $\langle p_T\rangle=22.1$ GeV/c and integrated over all $j_T$.}
  \end{sidewaystable}
  
  \begin{sidewaystable}[htbp]
\centering
\begin{tabular}{|c|c|c|c|c|c|c|c|c|c|}
\hline
$z_{true}$ & $\delta z_{true}$ & $A_{UT,raw}^{\sin\left(\phi_S-\phi_H\right)}$ & $\delta A_{UT,raw}^{\sin\left(\phi_S-\phi_H\right)}$& $A_{UT,\pi corr}^{\sin\left(\phi_S-\phi_H\right)}$ & $\delta A_{UT,\pi corr}^{\sin\left(\phi_S-\phi_H\right)}$ & $D_{\phi_C}$ & $A_{UT,final}^{\sin\left(\phi_S-\phi_H\right)}$ & $\delta A_{UT,final}^{\sin\left(\phi_S-\phi_H\right)}$ & TB syst. \\
\hline
    0.16  & 0.01  & -9.66E-04 & 8.29E-04 & 1.40E-04 & 1.88E-03 & 0.646 & 2.20E-04 & 2.94E-03 & 2.99E-06 \\
    \hline
    0.28  & 0.01  & 1.68E-03 & 1.41E-03 & 1.64E-03 & 2.49E-03 & 0.794 & 2.09E-03 & 3.18E-03 & 2.84E-05 \\
    \hline
    0.37  & 0.02  & 2.58E-03 & 2.41E-03 & 2.42E-03 & 2.65E-03 & 0.606 & 4.03E-03 & 4.41E-03 & 5.49E-05 \\
    \hline
    0.48  & 0.02  & 1.50E-03 & 3.29E-03 & 1.92E-03 & 3.60E-03 & 0.656 & 2.96E-03 & 5.55E-03 & 4.03E-05 \\
    \hline
    0.67  & 0.03  & -1.58E-02 & 8.81E-03 & -1.72E-02 & 9.19E-03 & 0.616 & -2.83E-02 & 1.51E-02 & 3.85E-04 \\
    \hline
    \end{tabular}%
\caption{$\pi^+$ Collins asymmetry for $x_F<0$ as a function of $z$ for $\langle p_T\rangle=6.9$ GeV/c and integrated over all $j_T$.}

\vspace{20pt}

\begin{tabular}{|c|c|c|c|c|c|c|c|c|c|}
\hline
$z_{true}$ & $\delta z_{true}$ & $A_{UT,raw}^{\sin\left(\phi_S-\phi_H\right)}$ & $\delta A_{UT,raw}^{\sin\left(\phi_S-\phi_H\right)}$& $A_{UT,\pi corr}^{\sin\left(\phi_S-\phi_H\right)}$ & $\delta A_{UT,\pi corr}^{\sin\left(\phi_S-\phi_H\right)}$ & $D_{\phi_C}$ & $A_{UT,final}^{\sin\left(\phi_S-\phi_H\right)}$ & $\delta A_{UT,final}^{\sin\left(\phi_S-\phi_H\right)}$ & TB syst. \\
\hline
    0.16  & 0.01  & -6.18E-04 & 1.25E-03 & 6.54E-03 & 5.50E-03 & 0.729 & 9.07E-03 & 7.63E-03 & 2.43E-04 \\
    \hline
    0.25  & 0.01  & 2.62E-03 & 2.21E-03 & 2.08E-03 & 2.47E-03 & 0.834 & 2.52E-03 & 3.00E-03 & 6.75E-05 \\
    \hline
    0.36  & 0.01  & 3.38E-04 & 3.67E-03 & 1.76E-04 & 3.93E-03 & 0.864 & 2.06E-04 & 4.60E-03 & 5.52E-06 \\
    \hline
    0.49  & 0.02  & 3.81E-03 & 5.02E-03 & 4.83E-03 & 5.73E-03 & 0.731 & 6.68E-03 & 7.93E-03 & 1.79E-04 \\
    \hline
    0.65  & 0.03  & 7.47E-03 & 1.43E-02 & 5.94E-03 & 1.56E-02 & 0.628 & 9.55E-03 & 2.51E-02 & 2.56E-04 \\
    \hline
    \end{tabular}%
\caption{$\pi^+$ Collins asymmetry for $x_F<0$ as a function of $z$ for $\langle p_T\rangle=9.0$ GeV/c and integrated over all $j_T$.}

\vspace{20pt}
\begin{tabular}{|c|c|c|c|c|c|c|c|c|c|}
\hline
$z_{true}$ & $\delta z_{true}$ & $A_{UT,raw}^{\sin\left(\phi_S-\phi_H\right)}$ & $\delta A_{UT,raw}^{\sin\left(\phi_S-\phi_H\right)}$& $A_{UT,\pi corr}^{\sin\left(\phi_S-\phi_H\right)}$ & $\delta A_{UT,\pi corr}^{\sin\left(\phi_S-\phi_H\right)}$ & $D_{\phi_C}$ & $A_{UT,final}^{\sin\left(\phi_S-\phi_H\right)}$ & $\delta A_{UT,final}^{\sin\left(\phi_S-\phi_H\right)}$ & TB syst. \\
\hline
    0.15  & 0.01  & 1.54E-03 & 1.47E-03 & 1.25E-03 & 3.50E-03 & 0.867 & 1.45E-03 & 4.08E-03 & 2.85E-05 \\
    \hline
    0.24  & 0.01  & -9.05E-05 & 2.60E-03 & 1.82E-04 & 2.80E-03 & 0.756 & 2.44E-04 & 3.74E-03 & 4.78E-06 \\
    \hline
    0.35  & 0.01  & -2.87E-04 & 4.24E-03 & 3.23E-05 & 4.72E-03 & 0.648 & 5.05E-05 & 7.36E-03 & 9.89E-07 \\
    \hline
    0.50  & 0.02  & 1.38E-02 & 5.80E-03 & 1.63E-02 & 6.70E-03 & 0.818 & 2.02E-02 & 8.28E-03 & 3.96E-04 \\
    \hline
    0.70  & 0.03  & -1.21E-02 & 1.70E-02 & -1.49E-02 & 1.91E-02 & 0.709 & -2.13E-02 & 2.72E-02 & 4.17E-04 \\
    \hline
    \end{tabular}%
\caption{$\pi^+$ Collins asymmetry for $x_F<0$ as a function of $z$ for $\langle p_T\rangle=10.8$ GeV/c and integrated over all $j_T$.}
  \end{sidewaystable}

\begin{sidewaystable}[htbp]
\centering
\begin{tabular}{|c|c|c|c|c|c|c|c|c|c|}
\hline
$z_{true}$ & $\delta z_{true}$ & $A_{UT,raw}^{\sin\left(\phi_S-\phi_H\right)}$ & $\delta A_{UT,raw}^{\sin\left(\phi_S-\phi_H\right)}$& $A_{UT,\pi corr}^{\sin\left(\phi_S-\phi_H\right)}$ & $\delta A_{UT,\pi corr}^{\sin\left(\phi_S-\phi_H\right)}$ & $D_{\phi_C}$ & $A_{UT,final}^{\sin\left(\phi_S-\phi_H\right)}$ & $\delta A_{UT,final}^{\sin\left(\phi_S-\phi_H\right)}$ & TB syst. \\
\hline
    0.14  & 0.01  & 2.52E-03 & 1.54E-03 & 3.83E-03 & 2.18E-03 & 0.812 & 4.77E-03 & 2.71E-03 & -4.58E-05 \\
    \hline
    0.24  & 0.01  & 4.80E-03 & 2.68E-03 & 5.29E-03 & 2.85E-03 & 0.706 & 7.57E-03 & 4.09E-03 & -7.27E-05 \\
    \hline
    0.35  & 0.01  & -3.15E-03 & 4.36E-03 & -3.66E-03 & 4.94E-03 & 0.743 & -4.98E-03 & 6.72E-03 & -4.78E-05 \\
    \hline
    0.48  & 0.02  & -1.21E-03 & 5.90E-03 & 4.36E-04 & 7.02E-03 & 0.691 & 6.38E-04 & 1.03E-02 & -6.12E-06 \\
    \hline
    0.69  & 0.02  & 1.03E-03 & 1.79E-02 & 3.46E-03 & 2.11E-02 & 0.601 & 5.81E-03 & 3.55E-02 & -5.58E-05 \\
    \hline
    \end{tabular}%
\caption{$\pi^+$ Collins asymmetry for $x_F<0$ as a function of $z$ for $\langle p_T\rangle=13.6$ GeV/c and integrated over all $j_T$.}

\vspace{20pt}

\begin{tabular}{|c|c|c|c|c|c|c|c|c|c|}
\hline
$z_{true}$ & $\delta z_{true}$ & $A_{UT,raw}^{\sin\left(\phi_S-\phi_H\right)}$ & $\delta A_{UT,raw}^{\sin\left(\phi_S-\phi_H\right)}$& $A_{UT,\pi corr}^{\sin\left(\phi_S-\phi_H\right)}$ & $\delta A_{UT,\pi corr}^{\sin\left(\phi_S-\phi_H\right)}$ & $D_{\phi_C}$ & $A_{UT,final}^{\sin\left(\phi_S-\phi_H\right)}$ & $\delta A_{UT,final}^{\sin\left(\phi_S-\phi_H\right)}$ & TB syst. \\
\hline
    0.14  & 0.01  & 1.33E-03 & 3.93E-03 & 1.08E-03 & 4.23E-03 & 0.848 & 1.29E-03 & 5.05E-03 & -2.90E-05 \\
    \hline
    0.25  & 0.01  & 5.47E-03 & 6.77E-03 & 7.25E-03 & 7.48E-03 & 0.800 & 9.16E-03 & 9.45E-03 & -2.06E-04 \\
    \hline
    0.35  & 0.01  & -5.91E-03 & 1.10E-02 & -7.10E-03 & 1.18E-02 & 0.616 & -1.17E-02 & 1.93E-02 & -2.62E-04 \\
    \hline
    0.48  & 0.02  & -2.92E-03 & 1.46E-02 & -3.67E-03 & 1.65E-02 & 0.733 & -5.07E-03 & 2.27E-02 & -1.14E-04 \\
    \hline
    0.69  & 0.03  & 9.49E-03 & 4.40E-02 & 7.24E-03 & 4.82E-02 & 0.819 & 8.93E-03 & 5.94E-02 & -2.01E-04 \\
    \hline
    \end{tabular}%
\caption{$\pi^+$ Collins asymmetry for $x_F<0$ as a function of $z$ for $\langle p_T\rangle=17.7$ GeV/c and integrated over all $j_T$.}

\vspace{20pt}
\begin{tabular}{|c|c|c|c|c|c|c|c|c|c|}
\hline
$z_{true}$ & $\delta z_{true}$ & $A_{UT,raw}^{\sin\left(\phi_S-\phi_H\right)}$ & $\delta A_{UT,raw}^{\sin\left(\phi_S-\phi_H\right)}$& $A_{UT,\pi corr}^{\sin\left(\phi_S-\phi_H\right)}$ & $\delta A_{UT,\pi corr}^{\sin\left(\phi_S-\phi_H\right)}$ & $D_{\phi_C}$ & $A_{UT,final}^{\sin\left(\phi_S-\phi_H\right)}$ & $\delta A_{UT,final}^{\sin\left(\phi_S-\phi_H\right)}$ & TB syst. \\
\hline
    0.14  & 0.01  & 1.11E-02 & 5.16E-03 & 1.13E-02 & 5.43E-03 & 0.888 & 1.28E-02 & 6.19E-03 & -2.57E-05 \\
    \hline
    0.25  & 0.01  & 1.05E-02 & 8.96E-03 & 1.10E-02 & 1.00E-02 & 0.829 & 1.34E-02 & 1.22E-02 & -2.68E-05 \\
    \hline
    0.35  & 0.01  & -2.85E-03 & 1.45E-02 & -1.68E-03 & 1.56E-02 & 0.876 & -1.94E-03 & 1.80E-02 & -3.87E-06 \\
    \hline
    0.48  & 0.02  & 4.96E-03 & 1.92E-02 & 2.96E-03 & 2.36E-02 & 0.871 & 3.44E-03 & 2.74E-02 & -6.87E-06 \\
    \hline
    0.65  & 0.02  & 4.91E-02 & 5.82E-02 & 3.97E-02 & 6.69E-02 & 0.705 & 5.69E-02 & 9.59E-02 & -1.14E-04 \\
    \hline
    \end{tabular}%
\caption{$\pi^+$ Collins asymmetry for $x_F<0$ as a function of $z$ for $\langle p_T\rangle=22.1$ GeV/c and integrated over all $j_T$.}
  \end{sidewaystable}

\begin{sidewaystable}[htbp]
\centering
\begin{tabular}{|c|c|c|c|c|c|c|c|c|c|}
\hline
$z_{true}$ & $\delta z_{true}$ & $A_{UT,raw}^{\sin\left(\phi_S-\phi_H\right)}$ & $\delta A_{UT,raw}^{\sin\left(\phi_S-\phi_H\right)}$& $A_{UT,\pi corr}^{\sin\left(\phi_S-\phi_H\right)}$ & $\delta A_{UT,\pi corr}^{\sin\left(\phi_S-\phi_H\right)}$ & $D_{\phi_C}$ & $A_{UT,final}^{\sin\left(\phi_S-\phi_H\right)}$ & $\delta A_{UT,final}^{\sin\left(\phi_S-\phi_H\right)}$ & TB syst. \\
\hline
    0.16  & 0.01  & -6.23E-04 & 8.47E-04 & -7.42E-04 & 1.12E-03 & 0.796 & -9.41E-04 & 1.43E-03 & 1.28E-05 \\
    \hline
    0.28  & 0.01  & 7.58E-04 & 1.40E-03 & 2.78E-03 & 2.54E-03 & 0.617 & 4.55E-03 & 4.16E-03 & 6.19E-05 \\
    \hline
    0.37  & 0.02  & 4.31E-03 & 2.46E-03 & 4.82E-03 & 2.78E-03 & 0.548 & 8.89E-03 & 5.14E-03 & 1.21E-04 \\
    \hline
    0.48  & 0.02  & -2.60E-03 & 3.45E-03 & -2.87E-03 & 3.91E-03 & 0.682 & -4.26E-03 & 5.80E-03 & 5.80E-05 \\
    \hline
    0.67  & 0.03  & -2.21E-02 & 9.49E-03 & -2.27E-02 & 9.90E-03 & 0.837 & -2.74E-02 & 1.20E-02 & 3.73E-04 \\
    \hline
    \end{tabular}%
\caption{$\pi^-$ Collins asymmetry for $x_F<0$ as a function of $z$ for $\langle p_T\rangle=6.9$ GeV/c and integrated over all $j_T$.}

\vspace{20pt}

\begin{tabular}{|c|c|c|c|c|c|c|c|c|c|}
\hline
$z_{true}$ & $\delta z_{true}$ & $A_{UT,raw}^{\sin\left(\phi_S-\phi_H\right)}$ & $\delta A_{UT,raw}^{\sin\left(\phi_S-\phi_H\right)}$& $A_{UT,\pi corr}^{\sin\left(\phi_S-\phi_H\right)}$ & $\delta A_{UT,\pi corr}^{\sin\left(\phi_S-\phi_H\right)}$ & $D_{\phi_C}$ & $A_{UT,final}^{\sin\left(\phi_S-\phi_H\right)}$ & $\delta A_{UT,final}^{\sin\left(\phi_S-\phi_H\right)}$ & TB syst. \\
\hline
    0.16  & 0.01  & 6.32E-04 & 1.28E-03 & 2.38E-03 & 5.05E-03 & 0.745 & 3.23E-03 & 6.86E-03 & 8.65E-05 \\
    \hline
    0.25  & 0.01  & -8.40E-04 & 2.27E-03 & -8.83E-04 & 2.55E-03 & 0.713 & -1.25E-03 & 3.62E-03 & 3.35E-05 \\
    \hline
    0.36  & 0.01  & 8.12E-04 & 3.83E-03 & 9.44E-04 & 4.15E-03 & 0.703 & 1.36E-03 & 5.97E-03 & 3.64E-05 \\
    \hline
    0.49  & 0.02  & 5.04E-04 & 5.38E-03 & 1.41E-03 & 6.21E-03 & 0.644 & 2.21E-03 & 9.76E-03 & 5.93E-05 \\
    \hline
    0.65  & 0.03  & -5.48E-03 & 1.58E-02 & -6.43E-03 & 1.74E-02 & 0.775 & -8.38E-03 & 2.27E-02 & 2.25E-04 \\
    \hline
    \end{tabular}%
\caption{$\pi^-$ Collins asymmetry for $x_F<0$ as a function of $z$ for $\langle p_T\rangle=9.0$ GeV/c and integrated over all $j_T$.}

\vspace{20pt}
\begin{tabular}{|c|c|c|c|c|c|c|c|c|c|}
\hline
$z_{true}$ & $\delta z_{true}$ & $A_{UT,raw}^{\sin\left(\phi_S-\phi_H\right)}$ & $\delta A_{UT,raw}^{\sin\left(\phi_S-\phi_H\right)}$& $A_{UT,\pi corr}^{\sin\left(\phi_S-\phi_H\right)}$ & $\delta A_{UT,\pi corr}^{\sin\left(\phi_S-\phi_H\right)}$ & $D_{\phi_C}$ & $A_{UT,final}^{\sin\left(\phi_S-\phi_H\right)}$ & $\delta A_{UT,final}^{\sin\left(\phi_S-\phi_H\right)}$ & TB syst. \\
\hline
    0.15  & 0.01  & -3.03E-04 & 1.51E-03 & 1.37E-04 & 3.61E-03 & 0.792 & 1.75E-04 & 4.61E-03 & 3.42E-06 \\
    \hline
    0.24  & 0.01  & -5.79E-03 & 2.68E-03 & -6.17E-03 & 2.90E-03 & 0.591 & -1.05E-02 & 4.96E-03 & 2.07E-04 \\
    \hline
    0.35  & 0.01  & -8.13E-04 & 4.47E-03 & -1.12E-03 & 5.01E-03 & 0.880 & -1.29E-03 & 5.76E-03 & 2.53E-05 \\
    \hline
    0.50  & 0.02  & 6.57E-03 & 6.27E-03 & 6.60E-03 & 7.33E-03 & 0.641 & 1.04E-02 & 1.16E-02 & 2.04E-04 \\
    \hline
    0.70  & 0.03  & 4.65E-03 & 1.91E-02 & 4.32E-03 & 2.15E-02 & 0.978 & 4.47E-03 & 2.22E-02 & 8.76E-05 \\
    \hline
    \end{tabular}%
\caption{$\pi^-$ Collins asymmetry for $x_F<0$ as a function of $z$ for $\langle p_T\rangle=10.8$ GeV/c and integrated over all $j_T$.}
  \end{sidewaystable}

\begin{sidewaystable}[htbp]
\centering
\begin{tabular}{|c|c|c|c|c|c|c|c|c|c|}
\hline
$z_{true}$ & $\delta z_{true}$ & $A_{UT,raw}^{\sin\left(\phi_S-\phi_H\right)}$ & $\delta A_{UT,raw}^{\sin\left(\phi_S-\phi_H\right)}$& $A_{UT,\pi corr}^{\sin\left(\phi_S-\phi_H\right)}$ & $\delta A_{UT,\pi corr}^{\sin\left(\phi_S-\phi_H\right)}$ & $D_{\phi_C}$ & $A_{UT,final}^{\sin\left(\phi_S-\phi_H\right)}$ & $\delta A_{UT,final}^{\sin\left(\phi_S-\phi_H\right)}$ & TB syst. \\
\hline
    0.14  & 0.01  & 2.98E-03 & 1.58E-03 & 4.54E-03 & 2.19E-03 & 0.831 & 5.52E-03 & 2.66E-03 & -5.30E-05 \\
    \hline
    0.24  & 0.01  & -4.54E-03 & 2.78E-03 & -4.68E-03 & 2.96E-03 & 0.649 & -7.29E-03 & 4.62E-03 & -7.00E-05 \\
    \hline
    0.35  & 0.01  & -1.12E-02 & 4.63E-03 & -1.27E-02 & 5.26E-03 & 0.775 & -1.65E-02 & 6.86E-03 & -1.59E-04 \\
    \hline
    0.48  & 0.02  & -9.51E-03 & 6.47E-03 & -1.15E-02 & 7.73E-03 & 0.658 & -1.77E-02 & 1.19E-02 & -1.70E-04 \\
    \hline
    0.69  & 0.02  & -2.74E-04 & 2.08E-02 & -1.18E-03 & 2.40E-02 & 0.840 & -1.42E-03 & 2.89E-02 & -1.36E-05 \\
    \hline
    \end{tabular}%
\caption{$\pi^-$ Collins asymmetry for $x_F<0$ as a function of $z$ for $\langle p_T\rangle=13.6$ GeV/c and integrated over all $j_T$.}

\vspace{20pt}

\begin{tabular}{|c|c|c|c|c|c|c|c|c|c|}
\hline
$z_{true}$ & $\delta z_{true}$ & $A_{UT,raw}^{\sin\left(\phi_S-\phi_H\right)}$ & $\delta A_{UT,raw}^{\sin\left(\phi_S-\phi_H\right)}$& $A_{UT,\pi corr}^{\sin\left(\phi_S-\phi_H\right)}$ & $\delta A_{UT,\pi corr}^{\sin\left(\phi_S-\phi_H\right)}$ & $D_{\phi_C}$ & $A_{UT,final}^{\sin\left(\phi_S-\phi_H\right)}$ & $\delta A_{UT,final}^{\sin\left(\phi_S-\phi_H\right)}$ & TB syst. \\
\hline
    0.14  & 0.01  & 4.62E-04 & 4.02E-03 & 6.10E-04 & 4.31E-03 & 0.912 & 6.77E-04 & 4.78E-03 & -1.52E-05 \\
    \hline
    0.25  & 0.01  & -9.83E-04 & 7.05E-03 & 4.81E-04 & 7.82E-03 & 0.762 & 6.38E-04 & 1.04E-02 & -1.44E-05 \\
    \hline
    0.35  & 0.01  & -7.70E-03 & 1.18E-02 & -8.05E-03 & 1.23E-02 & 0.844 & -9.64E-03 & 1.47E-02 & -2.17E-04 \\
    \hline
    0.48  & 0.02  & -1.62E-02 & 1.66E-02 & -1.04E-02 & 2.00E-02 & 0.714 & -1.47E-02 & 2.83E-02 & -3.31E-04 \\
    \hline
    0.69  & 0.03  & -3.68E-02 & 5.43E-02 & -6.25E-02 & 7.28E-02 & 0.918 & -6.88E-02 & 8.02E-02 & -1.55E-03 \\
    \hline
    \end{tabular}%
\caption{$\pi^-$ Collins asymmetry for $x_F<0$ as a function of $z$ for $\langle p_T\rangle=17.7$ GeV/c and integrated over all $j_T$.}

\vspace{20pt}
\begin{tabular}{|c|c|c|c|c|c|c|c|c|c|}
\hline
$z_{true}$ & $\delta z_{true}$ & $A_{UT,raw}^{\sin\left(\phi_S-\phi_H\right)}$ & $\delta A_{UT,raw}^{\sin\left(\phi_S-\phi_H\right)}$& $A_{UT,\pi corr}^{\sin\left(\phi_S-\phi_H\right)}$ & $\delta A_{UT,\pi corr}^{\sin\left(\phi_S-\phi_H\right)}$ & $D_{\phi_C}$ & $A_{UT,final}^{\sin\left(\phi_S-\phi_H\right)}$ & $\delta A_{UT,final}^{\sin\left(\phi_S-\phi_H\right)}$ & TB syst. \\
\hline
    0.14  & 0.01  & -1.91E-03 & 5.28E-03 & -1.43E-03 & 5.52E-03 & 0.871 & -1.66E-03 & 6.41E-03 & -3.32E-06 \\
    \hline
    0.25  & 0.01  & -1.32E-02 & 9.46E-03 & -1.30E-02 & 9.91E-03 & 0.910 & -1.44E-02 & 1.10E-02 & -2.88E-05 \\
    \hline
    0.35  & 0.01  & 1.05E-03 & 1.58E-02 & 6.95E-03 & 1.86E-02 & 0.789 & 8.91E-03 & 2.38E-02 & -1.78E-05 \\
    \hline
    0.48  & 0.02  & -1.62E-02 & 2.23E-02 & -1.69E-02 & 2.40E-02 & 0.862 & -1.98E-02 & 2.82E-02 & -3.96E-05 \\
    \hline
    0.65  & 0.02  & 8.22E-02 & 7.95E-02 & 6.85E-02 & 1.10E-01 & 0.626 & 1.11E-01 & 1.77E-01 & -2.21E-04 \\
    \hline
    \end{tabular}%
\caption{$\pi^-$ Collins asymmetry for $x_F<0$ as a function of $z$ for $\langle p_T\rangle=22.1$ GeV/c and integrated over all $j_T$.}
  \end{sidewaystable}
\clearpage
\begin{sidewaystable}[htbp]
\centering
\begin{tabular}{|c|c|c|c|c|c|c|c|c|c|}
\hline
$j_{T,true}$ & $\delta j_{T,true}$ & $A_{UT,raw}^{\sin\left(\phi_S-\phi_H\right)}$ & $\delta A_{UT,raw}^{\sin\left(\phi_S-\phi_H\right)}$& $A_{UT,\pi corr}^{\sin\left(\phi_S-\phi_H\right)}$ & $\delta A_{UT,\pi corr}^{\sin\left(\phi_S-\phi_H\right)}$ & $D_{\phi_C}$ & $A_{UT,final}^{\sin\left(\phi_S-\phi_H\right)}$ & $\delta A_{UT,final}^{\sin\left(\phi_S-\phi_H\right)}$ & TB syst. \\
\hline
    0.15  & 0.02  & 2.23E-03 & 1.90E-03 & 9.93E-04 & 4.79E-03 & 0.624 & 1.61E-03 & 7.77E-03 & 2.19E-05 \\
    \hline
    0.23  & 0.02  & 1.36E-03 & 1.35E-03 & -5.03E-04 & 2.71E-03 & 0.806 & -6.31E-04 & 3.40E-03 & 8.58E-06 \\
    \hline
    0.32  & 0.02  & 3.06E-03 & 1.16E-03 & 3.11E-03 & 2.31E-03 & 0.824 & 3.82E-03 & 2.83E-03 & 5.19E-05 \\
    \hline
    0.42  & 0.02  & 1.26E-03 & 1.52E-03 & 2.97E-03 & 2.92E-03 & 0.817 & 3.68E-03 & 3.62E-03 & 5.00E-05 \\
    \hline
    0.56  & 0.03  & 9.81E-04 & 2.04E-03 & 1.83E-03 & 3.09E-03 & 0.742 & 2.49E-03 & 4.21E-03 & 3.39E-05 \\
    \hline
    1.01  & 0.09  & 0.00E+00 & 0.00E+00 & 0.00E+00 & 0.00E+00 & 0.965 & 0.00E+00 & 0.00E+00 & 0.00E+00 \\
    \hline
    \end{tabular}%
\caption{$\pi^+$ Collins asymmetry for $x_F>0$ as a function of $j_T$ for $\langle p_T\rangle=6.9$ GeV/c and integrated over all $z$.}

\vspace{20pt}

\begin{tabular}{|c|c|c|c|c|c|c|c|c|c|}
\hline
$j_{T,true}$ & $\delta j_{T,true}$ & $A_{UT,raw}^{\sin\left(\phi_S-\phi_H\right)}$ & $\delta A_{UT,raw}^{\sin\left(\phi_S-\phi_H\right)}$& $A_{UT,\pi corr}^{\sin\left(\phi_S-\phi_H\right)}$ & $\delta A_{UT,\pi corr}^{\sin\left(\phi_S-\phi_H\right)}$ & $D_{\phi_C}$ & $A_{UT,final}^{\sin\left(\phi_S-\phi_H\right)}$ & $\delta A_{UT,final}^{\sin\left(\phi_S-\phi_H\right)}$ & TB syst. \\
\hline
    0.14  & 0.02  & -5.04E-03 & 3.52E-03 & -1.81E-02 & 9.40E-03 & 0.745 & -2.45E-02 & 1.28E-02 & 6.57E-04 \\
    \hline
    0.23  & 0.02  & 7.77E-03 & 2.30E-03 & 8.96E-03 & 4.55E-03 & 0.737 & 1.23E-02 & 6.24E-03 & 3.29E-04 \\
    \hline
    0.33  & 0.02  & 4.86E-03 & 1.89E-03 & 8.49E-03 & 3.42E-03 & 0.796 & 1.08E-02 & 4.34E-03 & 2.89E-04 \\
    \hline
    0.44  & 0.02  & 2.20E-03 & 2.12E-03 & 6.28E-03 & 3.79E-03 & 0.890 & 7.13E-03 & 4.31E-03 & 1.91E-04 \\
    \hline
    0.59  & 0.03  & -3.59E-03 & 2.29E-03 & -5.82E-03 & 3.20E-03 & 0.827 & -7.11E-03 & 3.91E-03 & 1.90E-04 \\
    \hline
    0.97  & 0.05  & -1.25E-02 & 1.93E-02 & -1.65E-02 & 2.25E-02 & 0.856 & -1.95E-02 & 2.66E-02 & 5.22E-04 \\
    \hline
    \end{tabular}%
\caption{$\pi^+$ Collins asymmetry for $x_F>0$ as a function of $j_T$ for $\langle p_T\rangle=9.0$ GeV/c and integrated over all $z$.}

\vspace{20pt}
\begin{tabular}{|c|c|c|c|c|c|c|c|c|c|}
\hline
$j_{T,true}$ & $\delta j_{T,true}$ & $A_{UT,raw}^{\sin\left(\phi_S-\phi_H\right)}$ & $\delta A_{UT,raw}^{\sin\left(\phi_S-\phi_H\right)}$& $A_{UT,\pi corr}^{\sin\left(\phi_S-\phi_H\right)}$ & $\delta A_{UT,\pi corr}^{\sin\left(\phi_S-\phi_H\right)}$ & $D_{\phi_C}$ & $A_{UT,final}^{\sin\left(\phi_S-\phi_H\right)}$ & $\delta A_{UT,final}^{\sin\left(\phi_S-\phi_H\right)}$ & TB syst. \\
\hline
    0.14  & 0.02  & 6.85E-03 & 4.70E-03 & 4.38E-03 & 1.09E-02 & 0.825 & 5.37E-03 & 1.33E-02 & 1.05E-04 \\
    \hline
    0.22  & 0.02  & 8.50E-03 & 2.87E-03 & 1.25E-02 & 4.51E-03 & 0.732 & 1.73E-02 & 6.23E-03 & 3.40E-04 \\
    \hline
    0.34  & 0.02  & 6.75E-03 & 2.31E-03 & 8.98E-03 & 3.41E-03 & 0.707 & 1.28E-02 & 4.88E-03 & 2.52E-04 \\
    \hline
    0.44  & 0.02  & 3.30E-03 & 2.52E-03 & 4.51E-03 & 3.75E-03 & 0.805 & 5.67E-03 & 4.71E-03 & 1.11E-04 \\
    \hline
    0.64  & 0.03  & 2.06E-03 & 2.34E-03 & 2.57E-03 & 2.95E-03 & 0.879 & 2.95E-03 & 3.39E-03 & 5.79E-05 \\
    \hline
    1.04  & 0.04  & -1.09E-02 & 1.30E-02 & -9.53E-03 & 1.56E-02 & 0.895 & -1.08E-02 & 1.76E-02 & 2.11E-04 \\
    \hline
    \end{tabular}%
\caption{$\pi^+$ Collins asymmetry for $x_F>0$ as a function of $j_T$ for $\langle p_T\rangle=10.8$ GeV/c and integrated over all $z$.}
  \end{sidewaystable}

\begin{sidewaystable}[htbp]
\centering
\begin{tabular}{|c|c|c|c|c|c|c|c|c|c|}
\hline
$j_{T,true}$ & $\delta j_{T,true}$ & $A_{UT,raw}^{\sin\left(\phi_S-\phi_H\right)}$ & $\delta A_{UT,raw}^{\sin\left(\phi_S-\phi_H\right)}$& $A_{UT,\pi corr}^{\sin\left(\phi_S-\phi_H\right)}$ & $\delta A_{UT,\pi corr}^{\sin\left(\phi_S-\phi_H\right)}$ & $D_{\phi_C}$ & $A_{UT,final}^{\sin\left(\phi_S-\phi_H\right)}$ & $\delta A_{UT,final}^{\sin\left(\phi_S-\phi_H\right)}$ & TB syst. \\
\hline
    0.15  & 0.02  & 6.97E-03 & 5.94E-03 & 1.63E-03 & 1.14E-02 & 0.618 & 2.67E-03 & 1.86E-02 & -2.56E-05 \\
    \hline
    0.22  & 0.02  & 1.32E-02 & 3.22E-03 & 1.65E-02 & 4.40E-03 & 0.815 & 2.04E-02 & 5.46E-03 & -1.96E-04 \\
    \hline
    0.33  & 0.02  & 1.68E-02 & 2.50E-03 & 1.84E-02 & 3.13E-03 & 0.677 & 2.74E-02 & 4.68E-03 & -2.63E-04 \\
    \hline
    0.45  & 0.02  & 7.35E-03 & 2.67E-03 & 8.39E-03 & 3.33E-03 & 0.842 & 1.01E-02 & 4.00E-03 & -9.67E-05 \\
    \hline
    0.67  & 0.03  & 6.56E-03 & 2.20E-03 & 7.05E-03 & 2.60E-03 & 0.877 & 8.12E-03 & 3.00E-03 & -7.80E-05 \\
    \hline
    1.13  & 0.05  & 2.58E-03 & 8.06E-03 & 4.98E-03 & 9.38E-03 & 0.947 & 5.32E-03 & 1.00E-02 & -5.11E-05 \\
    \hline
    \end{tabular}%
\caption{$\pi^+$ Collins asymmetry for $x_F>0$ as a function of $j_T$ for $\langle p_T\rangle=13.6$ GeV/c and integrated over all $z$.}

\vspace{20pt}

\begin{tabular}{|c|c|c|c|c|c|c|c|c|c|}
\hline
$j_{T,true}$ & $\delta j_{T,true}$ & $A_{UT,raw}^{\sin\left(\phi_S-\phi_H\right)}$ & $\delta A_{UT,raw}^{\sin\left(\phi_S-\phi_H\right)}$& $A_{UT,\pi corr}^{\sin\left(\phi_S-\phi_H\right)}$ & $\delta A_{UT,\pi corr}^{\sin\left(\phi_S-\phi_H\right)}$ & $D_{\phi_C}$ & $A_{UT,final}^{\sin\left(\phi_S-\phi_H\right)}$ & $\delta A_{UT,final}^{\sin\left(\phi_S-\phi_H\right)}$ & TB syst. \\
\hline
    0.14  & 0.02  & -3.01E-02 & 2.08E-02 & -3.93E-02 & 2.33E-02 & 0.632 & -6.29E-02 & 3.73E-02 & -1.42E-03 \\
    \hline
    0.22  & 0.02  & 2.89E-02 & 8.98E-03 & 2.79E-02 & 9.56E-03 & 0.736 & 3.83E-02 & 1.31E-02 & -8.61E-04 \\
    \hline
    0.33  & 0.02  & 2.88E-02 & 6.60E-03 & 2.93E-02 & 7.00E-03 & 0.842 & 3.51E-02 & 8.41E-03 & -7.91E-04 \\
    \hline
    0.45  & 0.02  & 1.81E-02 & 6.84E-03 & 1.77E-02 & 7.47E-03 & 0.899 & 1.99E-02 & 8.40E-03 & -4.49E-04 \\
    \hline
    0.68  & 0.03  & 8.67E-03 & 5.22E-03 & 8.53E-03 & 5.85E-03 & 0.933 & 9.25E-03 & 6.34E-03 & -2.08E-04 \\
    \hline
    1.16  & 0.04  & 7.45E-03 & 1.36E-02 & 7.28E-03 & 1.42E-02 & 0.951 & 7.74E-03 & 1.51E-02 & -1.74E-04 \\
    \hline
    \end{tabular}%
\caption{$\pi^+$ Collins asymmetry for $x_F>0$ as a function of $j_T$ for $\langle p_T\rangle=17.7$ GeV/c and integrated over all $z$.}

\vspace{20pt}
\begin{tabular}{|c|c|c|c|c|c|c|c|c|c|}
\hline
$j_{T,true}$ & $\delta j_{T,true}$ & $A_{UT,raw}^{\sin\left(\phi_S-\phi_H\right)}$ & $\delta A_{UT,raw}^{\sin\left(\phi_S-\phi_H\right)}$& $A_{UT,\pi corr}^{\sin\left(\phi_S-\phi_H\right)}$ & $\delta A_{UT,\pi corr}^{\sin\left(\phi_S-\phi_H\right)}$ & $D_{\phi_C}$ & $A_{UT,final}^{\sin\left(\phi_S-\phi_H\right)}$ & $\delta A_{UT,final}^{\sin\left(\phi_S-\phi_H\right)}$ & TB syst. \\
\hline
    0.13  & 0.01  & 1.37E-02 & 4.20E-02 & 1.32E-02 & 5.01E-02 & 0.791 & 1.68E-02 & 6.41E-02 & -3.36E-05 \\
    \hline
    0.22  & 0.02  & 2.06E-02 & 1.33E-02 & 2.13E-02 & 1.41E-02 & 0.553 & 3.89E-02 & 2.57E-02 & -7.79E-05 \\
    \hline
    0.33  & 0.02  & 7.56E-03 & 9.03E-03 & 7.08E-03 & 9.42E-03 & 0.842 & 8.51E-03 & 1.13E-02 & -1.70E-05 \\
    \hline
    0.45  & 0.02  & 1.95E-02 & 9.09E-03 & 2.16E-02 & 9.77E-03 & 0.898 & 2.43E-02 & 1.10E-02 & -4.86E-05 \\
    \hline
    0.70  & 0.03  & 1.15E-02 & 6.62E-03 & 1.16E-02 & 7.34E-03 & 0.883 & 1.33E-02 & 8.41E-03 & -2.66E-05 \\
    \hline
    1.26  & 0.04  & -5.68E-03 & 1.40E-02 & -5.44E-03 & 1.49E-02 & 0.971 & -5.67E-03 & 1.56E-02 & -1.13E-05 \\
    \hline
    \end{tabular}%
\caption{$\pi^+$ Collins asymmetry for $x_F>0$ as a function of $j_T$ for $\langle p_T\rangle=22.1$ GeV/c and integrated over all $z$.}
  \end{sidewaystable}

\begin{sidewaystable}[htbp]
\centering
\begin{tabular}{|c|c|c|c|c|c|c|c|c|c|}
\hline
$j_{T,true}$ & $\delta j_{T,true}$ & $A_{UT,raw}^{\sin\left(\phi_S-\phi_H\right)}$ & $\delta A_{UT,raw}^{\sin\left(\phi_S-\phi_H\right)}$& $A_{UT,\pi corr}^{\sin\left(\phi_S-\phi_H\right)}$ & $\delta A_{UT,\pi corr}^{\sin\left(\phi_S-\phi_H\right)}$ & $D_{\phi_C}$ & $A_{UT,final}^{\sin\left(\phi_S-\phi_H\right)}$ & $\delta A_{UT,final}^{\sin\left(\phi_S-\phi_H\right)}$ & TB syst. \\
\hline
    0.15  & 0.02  & -8.22E-04 & 1.93E-03 & 1.16E-03 & 3.81E-03 & 0.586 & 1.99E-03 & 6.58E-03 & 2.71E-05 \\
    \hline
    0.23  & 0.02  & -2.81E-03 & 1.36E-03 & -3.22E-03 & 2.11E-03 & 0.643 & -5.06E-03 & 3.32E-03 & 6.88E-05 \\
    \hline
    0.32  & 0.02  & -9.23E-04 & 1.18E-03 & -1.79E-03 & 2.22E-03 & 0.735 & -2.46E-03 & 3.05E-03 & 3.35E-05 \\
    \hline
    0.42  & 0.02  & 5.44E-04 & 1.54E-03 & -2.34E-03 & 2.77E-03 & 0.832 & -2.85E-03 & 3.37E-03 & 3.87E-05 \\
    \hline
    0.56  & 0.03  & -1.13E-03 & 2.11E-03 & -5.47E-04 & 3.09E-03 & 0.790 & -7.00E-04 & 3.95E-03 & 9.52E-06 \\
    \hline
    1.01  & 0.09  & 0.00E+00 & 0.00E+00 & 0.00E+00 & 0.00E+00 & 0.873 & 0.00E+00 & 0.00E+00 & 0.00E+00 \\
    \hline
    \end{tabular}%
\caption{$\pi^-$ Collins asymmetry for $x_F>0$ as a function of $j_T$ for $\langle p_T\rangle=6.9$ GeV/c and integrated over all $z$.}

\vspace{20pt}

\begin{tabular}{|c|c|c|c|c|c|c|c|c|c|}
\hline
$j_{T,true}$ & $\delta j_{T,true}$ & $A_{UT,raw}^{\sin\left(\phi_S-\phi_H\right)}$ & $\delta A_{UT,raw}^{\sin\left(\phi_S-\phi_H\right)}$& $A_{UT,\pi corr}^{\sin\left(\phi_S-\phi_H\right)}$ & $\delta A_{UT,\pi corr}^{\sin\left(\phi_S-\phi_H\right)}$ & $D_{\phi_C}$ & $A_{UT,final}^{\sin\left(\phi_S-\phi_H\right)}$ & $\delta A_{UT,final}^{\sin\left(\phi_S-\phi_H\right)}$ & TB syst. \\
\hline
    0.14  & 0.02  & 1.31E-03 & 3.59E-03 & 1.08E-02 & 9.84E-03 & 0.544 & 2.01E-02 & 1.83E-02 & 5.38E-04 \\
    \hline
    0.23  & 0.02  & -3.97E-03 & 2.36E-03 & -2.42E-03 & 4.26E-03 & 0.818 & -2.99E-03 & 5.27E-03 & 8.02E-05 \\
    \hline
    0.33  & 0.02  & -1.30E-03 & 1.94E-03 & -6.19E-04 & 3.40E-03 & 0.850 & -7.36E-04 & 4.05E-03 & 1.97E-05 \\
    \hline
    0.44  & 0.02  & -3.73E-03 & 2.17E-03 & -7.78E-03 & 3.84E-03 & 0.673 & -1.17E-02 & 5.78E-03 & 3.14E-04 \\
    \hline
    0.59  & 0.03  & -1.76E-03 & 2.36E-03 & -1.54E-03 & 3.21E-03 & 0.803 & -1.94E-03 & 4.04E-03 & 5.21E-05 \\
    \hline
    0.97  & 0.05  & 3.79E-02 & 2.00E-02 & 4.78E-02 & 2.27E-02 & 0.928 & 5.21E-02 & 2.47E-02 & 1.40E-03 \\
    \hline
    \end{tabular}%
\caption{$\pi^-$ Collins asymmetry for $x_F>0$ as a function of $j_T$ for $\langle p_T\rangle=9.0$ GeV/c and integrated over all $z$.}

\vspace{20pt}
\begin{tabular}{|c|c|c|c|c|c|c|c|c|c|}
\hline
$j_{T,true}$ & $\delta j_{T,true}$ & $A_{UT,raw}^{\sin\left(\phi_S-\phi_H\right)}$ & $\delta A_{UT,raw}^{\sin\left(\phi_S-\phi_H\right)}$& $A_{UT,\pi corr}^{\sin\left(\phi_S-\phi_H\right)}$ & $\delta A_{UT,\pi corr}^{\sin\left(\phi_S-\phi_H\right)}$ & $D_{\phi_C}$ & $A_{UT,final}^{\sin\left(\phi_S-\phi_H\right)}$ & $\delta A_{UT,final}^{\sin\left(\phi_S-\phi_H\right)}$ & TB syst. \\
\hline
    0.14  & 0.02  & -2.23E-03 & 4.81E-03 & -9.46E-03 & 1.06E-02 & 0.813 & -1.18E-02 & 1.32E-02 & 2.31E-04 \\
    \hline
    0.22  & 0.02  & -5.92E-03 & 2.96E-03 & -5.91E-03 & 4.62E-03 & 0.857 & -6.97E-03 & 5.45E-03 & 1.37E-04 \\
    \hline
    0.34  & 0.02  & -9.41E-03 & 2.38E-03 & -1.07E-02 & 3.48E-03 & 0.803 & -1.35E-02 & 4.38E-03 & 2.65E-04 \\
    \hline
    0.44  & 0.02  & -7.34E-03 & 2.59E-03 & -8.28E-03 & 3.73E-03 & 0.837 & -1.00E-02 & 4.50E-03 & 1.96E-04 \\
    \hline
    0.64  & 0.03  & 2.18E-04 & 2.42E-03 & 3.60E-04 & 3.01E-03 & 0.846 & 4.30E-04 & 3.60E-03 & 8.43E-06 \\
    \hline
    1.04  & 0.04  & 3.87E-03 & 1.36E-02 & 5.17E-03 & 1.59E-02 & 0.872 & 5.99E-03 & 1.85E-02 & 1.17E-04 \\
    \hline
    \end{tabular}%
\caption{$\pi^-$ Collins asymmetry for $x_F>0$ as a function of $j_T$ for $\langle p_T\rangle=10.8$ GeV/c and integrated over all $z$.}
  \end{sidewaystable}

\begin{sidewaystable}[htbp]
\centering
\begin{tabular}{|c|c|c|c|c|c|c|c|c|c|}
\hline
$j_{T,true}$ & $\delta j_{T,true}$ & $A_{UT,raw}^{\sin\left(\phi_S-\phi_H\right)}$ & $\delta A_{UT,raw}^{\sin\left(\phi_S-\phi_H\right)}$& $A_{UT,\pi corr}^{\sin\left(\phi_S-\phi_H\right)}$ & $\delta A_{UT,\pi corr}^{\sin\left(\phi_S-\phi_H\right)}$ & $D_{\phi_C}$ & $A_{UT,final}^{\sin\left(\phi_S-\phi_H\right)}$ & $\delta A_{UT,final}^{\sin\left(\phi_S-\phi_H\right)}$ & TB syst. \\
\hline
    0.15  & 0.02  & -1.09E-02 & 6.09E-03 & -1.21E-02 & 1.15E-02 & 0.794 & -1.54E-02 & 1.46E-02 & -1.48E-04 \\
    \hline
    0.22  & 0.02  & -5.88E-03 & 3.32E-03 & -1.72E-03 & 4.56E-03 & 0.567 & -3.07E-03 & 8.12E-03 & -2.94E-05 \\
    \hline
    0.33  & 0.02  & -9.21E-03 & 2.58E-03 & -9.81E-03 & 3.26E-03 & 0.771 & -1.29E-02 & 4.27E-03 & -1.23E-04 \\
    \hline
    0.45  & 0.02  & -8.88E-03 & 2.76E-03 & -8.86E-03 & 3.43E-03 & 0.760 & -1.18E-02 & 4.56E-03 & -1.13E-04 \\
    \hline
    0.67  & 0.03  & -5.96E-03 & 2.27E-03 & -5.48E-03 & 2.66E-03 & 0.926 & -5.99E-03 & 2.90E-03 & -5.75E-05 \\
    \hline
    1.13  & 0.05  & 3.19E-03 & 8.39E-03 & 3.37E-03 & 8.79E-03 & 0.943 & 3.61E-03 & 9.42E-03 & -3.47E-05 \\
    \hline
    \end{tabular}%
\caption{$\pi^-$ Collins asymmetry for $x_F>0$ as a function of $j_T$ for $\langle p_T\rangle=13.6$ GeV/c and integrated over all $z$.}

\vspace{20pt}

\begin{tabular}{|c|c|c|c|c|c|c|c|c|c|}
\hline
$j_{T,true}$ & $\delta j_{T,true}$ & $A_{UT,raw}^{\sin\left(\phi_S-\phi_H\right)}$ & $\delta A_{UT,raw}^{\sin\left(\phi_S-\phi_H\right)}$& $A_{UT,\pi corr}^{\sin\left(\phi_S-\phi_H\right)}$ & $\delta A_{UT,\pi corr}^{\sin\left(\phi_S-\phi_H\right)}$ & $D_{\phi_C}$ & $A_{UT,final}^{\sin\left(\phi_S-\phi_H\right)}$ & $\delta A_{UT,final}^{\sin\left(\phi_S-\phi_H\right)}$ & TB syst. \\
\hline
    0.14  & 0.02  & -2.08E-02 & 2.13E-02 & -2.38E-02 & 2.34E-02 & 0.683 & -3.53E-02 & 3.46E-02 & -7.93E-04 \\
    \hline
    0.22  & 0.02  & -2.30E-02 & 9.35E-03 & -2.33E-02 & 9.94E-03 & 0.825 & -2.85E-02 & 1.22E-02 & -6.42E-04 \\
    \hline
    0.33  & 0.02  & -2.07E-02 & 6.83E-03 & -2.13E-02 & 7.14E-03 & 0.897 & -2.40E-02 & 8.04E-03 & -5.39E-04 \\
    \hline
    0.45  & 0.02  & -1.84E-02 & 7.10E-03 & -1.89E-02 & 7.60E-03 & 0.837 & -2.28E-02 & 9.18E-03 & -5.14E-04 \\
    \hline
    0.68  & 0.03  & -1.59E-02 & 5.40E-03 & -1.62E-02 & 5.96E-03 & 0.927 & -1.77E-02 & 6.51E-03 & -3.98E-04 \\
    \hline
    1.16  & 0.04  & -5.16E-02 & 1.43E-02 & -6.47E-02 & 1.71E-02 & 0.949 & -6.90E-02 & 1.82E-02 & -1.55E-03 \\
    \hline
    \end{tabular}%
\caption{$\pi^-$ Collins asymmetry for $x_F>0$ as a function of $j_T$ for $\langle p_T\rangle=17.7$ GeV/c and integrated over all $z$.}

\vspace{20pt}
\begin{tabular}{|c|c|c|c|c|c|c|c|c|c|}
\hline
$j_{T,true}$ & $\delta j_{T,true}$ & $A_{UT,raw}^{\sin\left(\phi_S-\phi_H\right)}$ & $\delta A_{UT,raw}^{\sin\left(\phi_S-\phi_H\right)}$& $A_{UT,\pi corr}^{\sin\left(\phi_S-\phi_H\right)}$ & $\delta A_{UT,\pi corr}^{\sin\left(\phi_S-\phi_H\right)}$ & $D_{\phi_C}$ & $A_{UT,final}^{\sin\left(\phi_S-\phi_H\right)}$ & $\delta A_{UT,final}^{\sin\left(\phi_S-\phi_H\right)}$ & TB syst. \\
\hline
    0.13  & 0.01  & 3.14E-02 & 4.26E-02 & 3.56E-02 & 4.68E-02 & 0.706 & 5.11E-02 & 6.70E-02 & -1.02E-04 \\
    \hline
    0.22  & 0.02  & -8.43E-03 & 1.38E-02 & -8.19E-03 & 1.44E-02 & 0.745 & -1.11E-02 & 1.95E-02 & -2.22E-05 \\
    \hline
    0.33  & 0.02  & -1.17E-02 & 9.41E-03 & -1.10E-02 & 9.68E-03 & 0.843 & -1.33E-02 & 1.16E-02 & -2.65E-05 \\
    \hline
    0.45  & 0.02  & -2.72E-02 & 9.43E-03 & -2.81E-02 & 1.02E-02 & 0.770 & -3.69E-02 & 1.35E-02 & -7.38E-05 \\
    \hline
    0.70  & 0.03  & -3.71E-02 & 6.92E-03 & -4.04E-02 & 7.63E-03 & 0.943 & -4.34E-02 & 8.18E-03 & -8.67E-05 \\
    \hline
    1.26  & 0.04  & 3.38E-03 & 1.48E-02 & -2.36E-03 & 1.76E-02 & 0.922 & -2.59E-03 & 1.93E-02 & -5.18E-06 \\
    \hline
    \end{tabular}%
\caption{$\pi^-$ Collins asymmetry for $x_F>0$ as a function of $j_T$ for $\langle p_T\rangle=22.1$ GeV/c and integrated over all $z$.}
  \end{sidewaystable}

  \begin{sidewaystable}[htbp]
\centering
\begin{tabular}{|c|c|c|c|c|c|c|c|c|c|}
\hline
$j_{T,true}$ & $\delta j_{T,true}$ & $A_{UT,raw}^{\sin\left(\phi_S-\phi_H\right)}$ & $\delta A_{UT,raw}^{\sin\left(\phi_S-\phi_H\right)}$& $A_{UT,\pi corr}^{\sin\left(\phi_S-\phi_H\right)}$ & $\delta A_{UT,\pi corr}^{\sin\left(\phi_S-\phi_H\right)}$ & $D_{\phi_C}$ & $A_{UT,final}^{\sin\left(\phi_S-\phi_H\right)}$ & $\delta A_{UT,final}^{\sin\left(\phi_S-\phi_H\right)}$ & TB syst. \\
\hline
    0.15  & 0.02  & -6.26E-04 & 1.91E-03 & -6.90E-05 & 4.03E-03 & 0.624 & -1.12E-04 & 6.54E-03 & 1.52E-06 \\
    \hline
    0.23  & 0.02  & -1.20E-04 & 1.35E-03 & 1.64E-03 & 2.37E-03 & 0.806 & 2.06E-03 & 2.98E-03 & 2.80E-05 \\
    \hline
    0.32  & 0.02  & -3.55E-04 & 1.16E-03 & -1.48E-03 & 2.02E-03 & 0.824 & -1.82E-03 & 2.47E-03 & 2.47E-05 \\
    \hline
    0.42  & 0.02  & 6.44E-04 & 1.52E-03 & 1.14E-03 & 2.57E-03 & 0.817 & 1.41E-03 & 3.18E-03 & 1.92E-05 \\
    \hline
    0.56  & 0.03  & 6.07E-04 & 2.04E-03 & 1.57E-04 & 2.89E-03 & 0.742 & 2.13E-04 & 3.93E-03 & 2.90E-06 \\
    \hline
    1.01  & 0.09  & 0.00E+00 & 0.00E+00 & 0.00E+00 & 0.00E+00 & 0.965 & 0.00E+00 & 0.00E+00 & 0.00E+00 \\
    \hline
    \end{tabular}%
\caption{$\pi^+$ Collins asymmetry for $x_F<0$ as a function of $j_T$ for $\langle p_T\rangle=6.9$ GeV/c and integrated over all $z$.}

\vspace{20pt}

\begin{tabular}{|c|c|c|c|c|c|c|c|c|c|}
\hline
$j_{T,true}$ & $\delta j_{T,true}$ & $A_{UT,raw}^{\sin\left(\phi_S-\phi_H\right)}$ & $\delta A_{UT,raw}^{\sin\left(\phi_S-\phi_H\right)}$& $A_{UT,\pi corr}^{\sin\left(\phi_S-\phi_H\right)}$ & $\delta A_{UT,\pi corr}^{\sin\left(\phi_S-\phi_H\right)}$ & $D_{\phi_C}$ & $A_{UT,final}^{\sin\left(\phi_S-\phi_H\right)}$ & $\delta A_{UT,final}^{\sin\left(\phi_S-\phi_H\right)}$ & TB syst. \\
\hline
    0.14  & 0.02  & 1.88E-03 & 3.54E-03 & 3.12E-04 & 7.69E-03 & 0.745 & 4.24E-04 & 1.04E-02 & 1.14E-05 \\
    \hline
    0.23  & 0.02  & 1.74E-03 & 2.31E-03 & 1.95E-03 & 4.00E-03 & 0.737 & 2.68E-03 & 5.48E-03 & 7.17E-05 \\
    \hline
    0.33  & 0.02  & 1.13E-03 & 1.90E-03 & -1.34E-05 & 3.05E-03 & 0.796 & -1.70E-05 & 3.88E-03 & 4.56E-07 \\
    \hline
    0.44  & 0.02  & -1.19E-03 & 2.12E-03 & 1.64E-03 & 3.38E-03 & 0.890 & 1.87E-03 & 3.84E-03 & 5.00E-05 \\
    \hline
    0.59  & 0.03  & -9.29E-04 & 2.28E-03 & -1.72E-03 & 3.03E-03 & 0.827 & -2.10E-03 & 3.70E-03 & 5.62E-05 \\
    \hline
    0.97  & 0.05  & 3.93E-03 & 1.91E-02 & 3.70E-03 & 2.20E-02 & 0.856 & 4.37E-03 & 2.60E-02 & 1.17E-04 \\
    \hline
    \end{tabular}%
\caption{$\pi^+$ Collins asymmetry for $x_F<0$ as a function of $j_T$ for $\langle p_T\rangle=9.0$ GeV/c and integrated over all $z$.}

\vspace{20pt}
%
%
\caption{$\pi^+$ Collins asymmetry for $x_F<0$ as a function of $j_T$ for $\langle p_T\rangle=10.8$ GeV/c and integrated over all $z$.}
  \end{sidewaystable}

\begin{sidewaystable}[htbp]
\centering
%
%
\caption{$\pi^+$ Collins asymmetry for $x_F<0$ as a function of $j_T$ for $\langle p_T\rangle=13.6$ GeV/c and integrated over all $z$.}

\vspace{20pt}

%
%
\caption{$\pi^+$ Collins asymmetry for $x_F<0$ as a function of $j_T$ for $\langle p_T\rangle=22.1$ GeV/c and integrated over all $z$.}
  \end{sidewaystable}

\begin{sidewaystable}[htbp]
\centering
%
%
\caption{$\pi^-$ Collins asymmetry for $x_F<0$ as a function of $j_T$ for $\langle p_T\rangle=6.9$ GeV/c and integrated over all $z$.}

\vspace{20pt}

%
%
\caption{$\pi^-$ Collins asymmetry for $x_F<0$ as a function of $j_T$ for $\langle p_T\rangle=9.0$ GeV/c and integrated over all $z$.}

\vspace{20pt}
%
%
\caption{$\pi^-$ Collins asymmetry for $x_F<0$ as a function of $j_T$ for $\langle p_T\rangle=10.8$ GeV/c and integrated over all $z$.}
  \end{sidewaystable}

\begin{sidewaystable}[htbp]
\centering
%
%
\caption{$\pi^-$ Collins asymmetry for $x_F<0$ as a function of $j_T$ for $\langle p_T\rangle=13.6$ GeV/c and integrated over all $z$.}

\vspace{20pt}

%
%
\caption{$\pi^-$ Collins asymmetry for $x_F<0$ as a function of $j_T$ for $\langle p_T\rangle=22.1$ GeV/c and integrated over all $z$.}
  \end{sidewaystable}    
\clearpage
\begin{sidewaystable}[htbp]
\centering
%
%
\caption{$\pi^+$ Collins asymmetry for $x_F>0$ as a function of $j_T$ for $\langle z\rangle=0.14$ and $p_T > 9.9$ GeV/c.}

\vspace{20pt}

%
%
\caption{$\pi^+$ Collins asymmetry for $x_F>0$ as a function of $j_T$ for $\langle z\rangle=0.24$ and $p_T>9.9$ GeV/c.}
  \end{sidewaystable}

\begin{sidewaystable}[htbp]
\centering
%
%
\caption{$\pi^+$ Collins asymmetry for $x_F>0$ as a function of $j_T$ for $\langle z\rangle=0.35$ and $p_T > 9.9$ GeV/c.}

\vspace{20pt}

%
%
\caption{$\pi^+$ Collins asymmetry for $x_F>0$ as a function of $j_T$ for $\langle z\rangle=0.51$ and $p_T>9.9$ GeV/c.}
  \end{sidewaystable}
  
  \begin{sidewaystable}[htbp]
\centering
%
%
\caption{$\pi^-$ Collins asymmetry for $x_F>0$ as a function of $j_T$ for $\langle z\rangle=0.14$ and $p_T > 9.9$ GeV/c.}

\vspace{20pt}

%
%
\caption{$\pi^-$ Collins asymmetry for $x_F>0$ as a function of $j_T$ for $\langle z\rangle=0.24$ and $p_T>9.9$ GeV/c.}
  \end{sidewaystable}

\begin{sidewaystable}[htbp]
\centering
%
%
\caption{$\pi^-$ Collins asymmetry for $x_F>0$ as a function of $j_T$ for $\langle z\rangle=0.35$ and $p_T > 9.9$ GeV/c.}

\vspace{20pt}

%
%
\caption{$\pi^-$ Collins asymmetry for $x_F>0$ as a function of $j_T$ for $\langle z\rangle=0.51$ and $p_T>9.9$ GeV/c.}
  \end{sidewaystable}
  \begin{sidewaystable}[htbp]
\centering
%
%
\caption{$\pi^+$ Collins asymmetry for $x_F<0$ as a function of $j_T$ for $\langle z\rangle=0.14$ and $p_T > 9.9$ GeV/c.}

\vspace{20pt}

%
%
\caption{$\pi^+$ Collins asymmetry for $x_F<0$ as a function of $j_T$ for $\langle z\rangle=0.24$ and $p_T>9.9$ GeV/c.}
  \end{sidewaystable}

\begin{sidewaystable}[htbp]
\centering
%
%
\caption{$\pi^+$ Collins asymmetry for $x_F<0$ as a function of $j_T$ for $\langle z\rangle=0.35$ and $p_T > 9.9$ GeV/c.}

\vspace{20pt}

%
%
\caption{$\pi^+$ Collins asymmetry for $x_F<0$ as a function of $j_T$ for $\langle z\rangle=0.51$ and $p_T>9.9$ GeV/c.}
  \end{sidewaystable}
  
  \begin{sidewaystable}[htbp]
\centering
%
%
\caption{$\pi^-$ Collins asymmetry for $x_F<0$ as a function of $j_T$ for $\langle z\rangle=0.14$ and $p_T > 9.9$ GeV/c.}

\vspace{20pt}

%
%
\caption{$\pi^-$ Collins asymmetry for $x_F<0$ as a function of $j_T$ for $\langle z\rangle=0.24$ and $p_T>9.9$ GeV/c.}
  \end{sidewaystable}

\begin{sidewaystable}[htbp]
\centering
%
%
\caption{$\pi^-$ Collins asymmetry for $x_F<0$ as a function of $j_T$ for $\langle z\rangle=0.35$ and $p_T > 9.9$ GeV/c.}

\vspace{20pt}

%
%
\caption{$\pi^-$ Collins asymmetry for $x_F<0$ as a function of $j_T$ for $\langle z\rangle=0.51$ and $p_T>9.9$ GeV/c.}
  \end{sidewaystable}

\section{Run List}
13047003 13047004 13047022 13047023 13047024 13047026 13047027 13047028\\
13047029 13047030 13047031 13047032 13047033 13047034 13047035 13047036\\
13047037 13047039 13047041 13047042 13047043 13047044 13047122 13047123\\
13047124 13047126 13048009 13048010 13048011 13048012 13048013 13048014\\
13048015 13048016 13048017 13048018 13048019 13048030 13048031 13048032\\
13048040 13048041 13048042 13048043 13048044 13048045 13048046 13048049\\
13048050 13048051 13048052 13048053 13048087 13048088 13048089 13048090\\
13048091 13048092 13048093 13049004 13049005 13049006 13049007 13049031\\
13049032 13049035 13049039 13049041 13049042 13049044 13049045 13049046\\
13049047 13049048 13049049 13049050 13049072 13049073 13049080 13049081\\
13049082 13049086 13049087 13049088 13049089 13049093 13049094 13049096\\
13049098 13049099 13049101 13050001 13050006 13050007 13050009 13050011\\
13050012 13050015 13050016 13050020 13050022 13050023 13050025 13050026\\
13050027 13050028 13050029 13050031 13050032 13050033 13050036 13050037\\
13050038 13050039 13050041 13050042 13050043 13050044 13050046 13050047\\
13050049 13050050 13051006 13051007 13051008 13051009 13051010 13051011\\
13051012 13051015 13051016 13051017 13051019 13051020 13051021 13051022\\
13051023 13051024 13051026 13051028 13051068 13051069 13051070 13051071\\
13051072 13051073 13051074 13051080 13051081 13051083 13051085 13051086\\
13051087 13051088 13051092 13051093 13051095 13052001 13052002 13052003\\
13052004 13052005 13052009 13052010 13052011 13052012 13052013 13052014\\
13052015 13052016 13052017 13052018 13052019 13052020 13052021 13052022\\
13052036 13052037 13052039 13052042 13052043 13052045 13052048 13052050\\
13052051 13052052 13052053 13052054 13052056 13052061 13052088 13053004\\
13053005 13053006 13053007 13053012 13053013 13053015 13053027 13053028\\
13054022 13054023 13054044 13054045 13054060 13054061 13054062 13054063\\
13054064 13054065 13054066 13054068 13054069 13054070 13054084 13054085\\
13055001 13055004 13055006 13055007 13055008 13055009 13055010 13055011\\
13055014 13055015 13055016 13055017 13055018 13055019 13055020 13055021\\
13055022 13055023 13055024 13055035 13055036 13055037 13055038 13055039\\
13055068 13055070 13055072 13055073 13055075 13055076 13055080 13055081\\
13055082 13055086 13055087 13056005 13056007 13056008 13056020 13056021\\
13056022 13056023 13056024 13056025 13056026 13056027 13056028 13056029\\
13056030 13056031 13056033 13056034 13056035 13056037 13056038 13056039\\
13057005 13057006 13057007 13057008 13057009 13057010 13057011 13057014\\
13057015 13057016 13057017 13057018 13057019 13057021 13057022 13057023\\
13057024 13057025 13057026 13057027 13057044 13057045 13057046 13057047\\
13057048 13057049 13057050 13057051 13057052 13057053 13057055 13057056\\
13057057 13057058 13058002 13058008 13058015 13058016 13058017 13058018\\
13058025 13058026 13058028 13058029 13058030 13058031 13058032 13059005\\
13059006 13059007 13059008 13059009 13059010 13059011 13059012 13059013\\
13059014 13059015 13059016 13059017 13059018 13059019 13059020 13059021\\
13059022 13059023 13059025 13059026 13059027 13059076 13059077 13059078\\
13059079 13059080 13059082 13059083 13059084 13059085 13059086 13059087\\
13060001 13060002 13060003 13060008 13060009 13060010 13060011 13060012\\
13061024 13061025 13061026 13061029 13061030 13061031 13061035 13061054\\
13061055 13061059 13061060 13061061 13062001 13062002 13062004 13062005\\
13062006 13062007 13062013 13062025 13062026 13062028 13062029 13062049\\
13062050 13062052 13062059 13062060 13062061 13062062 13062063 13063009\\
13063010 13063011 13063020 13063022 13063023 13063030 13063031 13063032\\
13063033 13063034 13063035 13063036 13063053 13063054 13063062 13063063\\
13063065 13063067 13063068 13063071 13063072 13063073 13063074 13063076\\
13064001 13064002 13064003 13064004 13064005 13064006 13064012 13064014\\
13064020 13064021 13064022 13064023 13064024 13064025 13064026 13064028\\
13064029 13064030 13064031 13064032 13064052 13064055 13064056 13064057\\
13064059 13064061 13064064 13064065 13064066 13064068 13064070 13064074\\
13064075 13065005 13065006 13065007 13065008 13065009 13065013 13065014\\
13065015 13065016 13065017 13065018 13065019 13065020 13065021 13065022\\
13065046 13065047 13065048 13065049 13065050 13065052 13065053 13065055\\
13065056 13065058 13065059 13065060 13066021 13066022 13066023 13066024\\
13066025 13066026 13066027 13066028 13066029 13066030 13066031 13066033\\
13066034 13066035 13066036 13066101 13066102 13066104 13066109 13066110\\
13067001 13067002 13067003 13067010 13067011 13067012 13067013 13067014\\
13067015 13067017 13068017 13068022 13068027 13068029 13068034 13068036\\
13068037 13068084 13068085 13068086 13068087 13068088 13068090 13069001\\
13069002 13069003 13069004 13069005 13069006 13069007 13069008 13069013\\
13069014 13069016 13069017 13069018 13069020 13069021 13069022 13069023\\
13069024 13069026 13069027 13069029 13069030 13069031 13069035 13069036\\
13069066 13069067 13069068 13069069 13069073 13070006 13070008 13070010\\
13070011 13070012 13070014 13070015 13070016 13070017 13070018 13070019\\
13070020 13070021 13070022 13070024 13070025 13070026 13070027 13070056\\
13070057 13070058 13070059 13070060 13070061 13071003 13071004 13071005\\
13071006 13071008 13071009 13071010 13071011 13071012 13071064 13072001\\
13072002 13072003 13072004 13072005 13072006 13072007 13072008 13072009\\
13072010 13072011 13072014 13072015 13072016 13072017 13072018 13072019\\
13072020


\end{DoubleSpace}

\vfill
\backmatter
\bibliographystyle{ieeetr}		
\bibliography{ch9_Backmatter/references}
\cleardoublepage
\phantomsection

\addcontentsline{toc}{chapter}{Vita}
\begin{center}
  \textbf{VITA}\\
   \bigskip
   \bigskip
   \bigskip
   \bigskip
  \textbf{James Kevin Adkins}\\
\end{center}

   \bigskip

\begin{tabbing}
\textbf{Place of Birth:}\\     
\\  
\hspace{0.7 in}   
\= Paintsville, Kentucky\\
\\
\\

\textbf{Education:}\\    
\\
\>  University of Kentucky \\
\>  Lexington, KY\\
\>  M.S., Physics\\
\>  May 2014\\
\\
\>  Morehead State University\\
\>  Morehead, KY\\
\>  B.S., Physics and B.S., Mathematics\\
\>  May 2011\\
\\
\\

\textbf{Professional Positions:}\\
\\
\> University of Kentucky\\
\> Research Assistant\\
\> January 2012 - Present\\
\\
\> University of Kentucky\\
\> Teaching Assistant\\
\> August 2011 - December 2011\\
\\
\> Morehead State University\\
\> Undergraduate Research Fellow\\
\> August 2009 - July 2011\\
\\
\\
\\
\\
\\
\\
\textbf{Publications:}\\
\\
\>  

\begin{minipage}[t]{378 pt}
``Azimuthal Single-Spin Asymmetries of Charged Pions in Jets in $\sqrt{s}=200$ GeV $p^\uparrow p$ Collisions at STAR'',
{}J.K.~Adkins and J.L.~Drachenberg,
\textit{Int. J. Mod. Phys. Conf. Ser.}, \textbf{40} (2016). \\

``A Simple, Portable Apparatus to Measure Night Sky Brightness at Various Zenith Angles",
 {}J.J.~Birriel and J.K.~Adkins,
 \textit{The Journal of the American Association of Variable Star Observers}, \textbf{38}, 221 (2010).\\
\end{minipage}

\end{tabbing}

\end{document}